\documentclass[useAMS,usenatbib]{mn2e}
\usepackage[dvips]{graphicx}
\usepackage{amsmath}
\usepackage{hyperref}
\usepackage{xcolor}
\usepackage{caption}
\newcommand{\bz}{$\langle B_z \rangle$}

\newcommand{\vsini}{$v_{\rm rot} \sin i$}
\newcommand{\kms}{km\,s$^{-1}$}

\newcommand{\mdot}{$\dot{M}$}
\newcommand{\vinf}{$v_\infty$}
\newcommand{\rsun}{R$_\odot$}
\newcommand{\msun}{M$_\odot$}

\newcommand{\teff}{$T_{\rm eff}$}
\newcommand{\ra}{$R_{\rm A}$}
\newcommand{\rk}{$R_{\rm K}$}
\newcommand{\rark}{$\log{R_{\rm A}/R_{\rm K}}$}

\newcommand{\mnras}{$MNRAS$}
\newcommand{\apj}{$ApJ$}
\newcommand{\apjl}{$ApJL$}
\newcommand{\aj}{$AJ$}
\newcommand{\aap}{A\&A}
\newcommand{\aapr}{A\&AR}
\newcommand{\aaps}{A\&AS}

\newcommand{\pasp}{PASP}
\newcommand{\apjs}{ApJS}

\newcommand{\apss}{ApSS}

\newcommand{\nat}{Nature}

\newcommand{\physscr}{Phys.\ Scr.}

\title[Plasma transport in centrifugal magnetospheres]{The Magnetic Early B-type Stars IV: Breakout or Leakage? H$\alpha$ emission as a diagnostic of plasma transport in centrifugal magnetospheres}
\author[M. E. Shultz]
{M.\ E.\ Shultz$^{1}$\thanks{E-mail: mshultz@udel.edu},
S.\ Owocki$^1$
Th.\ Rivinius$^{2}$,
G.\ A. Wade$^{3}$,
C.\ Neiner$^4$,
E.\ Alecian$^5$,
\newauthor{O.\ Kochukhov$^6$,
D.\ Bohlender$^7$,
A.\ ud-Doula$^8$,
J.\ D.\ Landstreet$^{9,10}$,
J.\ Sikora$^{11}$,
}
\newauthor{
A.\ David-Uraz$^1$,
V.\ Petit$^1$,
P.\ Cerraho\u{g}lu$^1$,
R.\ Fine$^1$,
G.\ Henson$^{12}$,
}
\newauthor{
and the MiMeS and BinaMIcS Collaborations
}
\\
$^1$Department of Physics and Astronomy, and SARA, University of Delaware, 217 Sharp Lab, Newark, Delaware, 19716, USA\\
$^2$ESO - European Organisation for Astronomical Research in the Southern Hemisphere, Casilla 19001, Santiago 19, Chile\\
$^3$Department of Physics and Space Science, Royal Military College of Canada, Kingston, Ontario K7K 7B4, Canada\\
$^4$LESIA, Observatoire de Paris, PSL Research University, CNRS, Sorbonne Universit\'es, UPMC Univ. Paris 06, Univ. Paris Diderot,\\
Sorbonne Paris Cit\'e, 5 place Jules Janssen, F-92195 Meudon, France\\
$^5$Universit\'e Grenoble Alpes, IPAG, F-38000 Grenoble, France\\
$^6$Department of Physics and Astronomy, Uppsala University, Box 516, Uppsala 75120, Sweden \\
$^7$National Research Council of Canada, Herzberg Astronomy and Astrophysics Research Centre, 5071 West Saanich Road, Victoria, BC V9E 2E7\\
$^8$Department of Physics, Penn State Scranton, Dunmore, PA 18512, USA\\
$^9$Armagh Observatory and Planetarium, College Hill, Armagh BT61 9DG, UK\\
$^{10}$University of Western Ontario, London, Ontario, N6A 3K7, Canada\\
$^{11}$Department of Physics and Astronomy, Bishop's University, Sherbrooke, Qu{\'e}bec, Canada, J1M 1Z7\\
$^{12}$Physics and Astronomy, and SARA, East Tennessee State University, PO Box 70300, Johnson City, TN 37614, USA\\
}
\begin{document}

\date{}

\pagerange{\pageref{firstpage}--\pageref{lastpage}} \pubyear{2018}

\maketitle

\label{firstpage}

\begin{abstract}
Rapidly rotating early-type stars with strong magnetic fields frequently show H$\alpha$ emission originating in Centrifugal Magnetospheres (CMs), circumstellar structures in which centrifugal support due to magnetically enforced corotation of the magnetically confined plasma enables it to accumulate to high densities. It is not currently known whether the CM plasma escapes via Centrifugal Breakout (CB), or by an unidentified leakage mechanism. We have conducted the first comprehensive examination of the H$\alpha$ emission properties of all stars currently known to display CM-pattern emission. We find that the onset of emission is dependent primarily on the area of the CM, which can be predicted simply by the value $B_{\rm K}$ of the magnetic field at the Kepler corotation radius $R_{\rm K}$. Emission strength is strongly sensitive to both CM area and $B_{\rm K}$. Emission onset and strength are {\em not} dependent on effective temperature, luminosity, or mass-loss rate. These results all favour a CB scenario, however the lack of intrinsic variability in any CM diagnostics indicates that CB must be an essentially continuous process, i.e.\ it effectively acts as a leakage mechanism. We also show that the emission profile shapes are approximately scale-invariant, i.e.\ they are broadly similar across a wide range of emission strengths and stellar parameters. While the radius of maximum emission correlates closely as expected to $R_{\rm K}$, it is always larger, contradicting models that predict that emission should peak at $R_{\rm K}$. 
\end{abstract}

\section{Introduction}

When confined by a strong magnetic field, the ionized wind plasma of hot stars is forced to corotate with the photospheric magnetic field, leading to the formation of a stellar magnetosphere. Magnetically channeled flows from opposite magnetic latitudes collide at the tops of magnetic loops, leading to the concentration of plasma around the magnetic equator \citep[e.g.][]{lb1978, ud2002,2017AN....338..868K,ud2013}. In the simplest case of a star in which rotation is dynamically unimportant, the magnetospheric plasma is pulled back to the star by gravity. As a result, plasma persists within the magnetosphere only over the free-fall timescale. These {\em dynamical magnetospheres} (DMs) are not detectable unless the mass-loss rate is high enough to replenish the magnetosphere on dynamical timescales. For the most part, such magnetospheres are detectable in H$\alpha$ (the primary visible magnetospheric diagnostic) only in O-type stars. Indeed, H$\alpha$ emission consistent with an origin in a DM is essentially ubiquitous amongst the magnetic O-type stars \citep{petit2013}, but has been seen in only one magnetic B-type star, the strongly magnetized and very luminous B0 star $\xi^1$ CMa \citep{2017MNRAS.471.2286S}.

If a star is rapidly rotating, magnetically enforced corotation of the plasma can lead to extremely high rotational velocities within the magnetosphere, generating centrifugal forces strong enough to counteract gravitational infall. In this case the star forms a {\em Centrifugal Magnetosphere} (CM). Below the Kepler corotation radius \rk~(the point of balance between gravitational and centrifugal force), the star retains its DM. Above \rk~and extending out to the Alfv\'en radius \ra, plasma is unable to fall back to the star, while at the same time being compressed into a thin disk by the extreme centrifugal force \citep{town2005c,town2007}. These effects combine to produce very high densities within the CM as compared to a DM. Since magnetic braking tends to quickly remove angular momentum from the system \citep{wd1967,ud2009}, the requirement that the magnetic field be very strong (several kG), but that the rotation be simultaneously very rapid ($P_{\rm rot} \sim 1$~d), in practice means that CM-type emission is seen only in fairly young stars \citep{2019MNRAS.tmp.2196S}. It is furthermore predominantly the magnetic B-type stars that display emission lines consistent with a CM \citep{petit2013,2019MNRAS.tmp.2196S}, first since they spin down much more slowly than the magnetic O-type stars, and second since the weaker winds of B-type stars mean that, without a CM, they do not show any emission at all \citep[with the notable exception of classical Be stars, whose decretion disks are in any case unrelated to stellar winds, see e.g.][]{2013AARv..21...69R}. 

If the magnetic field is extremely strong -- a necessity if corotation is to be maintained out to tens of stellar radii -- the CM can be modelled by means of the Rigidly Rotating Magnetosphere \citep[RRM;][]{town2005c} model\footnote{Visualizations of the RRM model for a variety of magnetic geometries can be found at \url{http://www.astro.wisc.edu/~townsend/static.php?ref=rrm-movies}.}. The RRM model assumes that the magnetic field is unaffected by the motion of the plasma -- i.e. that the magnetic field is perfectly rigid -- and that plasma will accumulate at the local minima of the gravitocentrifugal potential along each field line. Locating these potential minima maps out an accumulation surface. The plasma is assumed to settle within the accumulation surface in hydrostatic equilibrium along each field line, and the equivalent width is found via tuning the central density as a free parameter. For an oblique dipole magnetic field geometry the RRM model predicts that the plasma collects in a warped disk, with the two densest regions near the intersections of the magnetic and rotational equatorial planes. In the limits of a magnetic tilt angle $\beta=0^\circ$ and $90^\circ$, the plasma respectively accumulates in a continuous torus in the magnetic equatorial plane, or two distinct clouds of material in the magnetic equatorial plane. The corotation of the plasma means that its line of sight velocity $v_{\rm r}$ is simply a linear function of its projected distance from the centre of the star. When the CM is seen face-on, the typical RRM geometry produces a double-humped emission profile: since there is no significant material located below \rk, there is no or very little emission at velocities below $(R_{\rm K}/R_*)$\vsini, with any emission that does occur within this boundary corresponding to material projected above or below the equatorial plane (which may or may not be present depending on the magnetic geometry). The emission then sharply peaks at \rk, and falls off thereafter. Rotation of the star modulates the emission profile, with the emission bumps decreasing in strength as they move closer to the line core, corresponding to the decrease in the projected distance of the clouds from the star and the simultaneous decrease in projected area as the clouds change from face-on to edge-on. If the clouds pass in front of the star, they will produce an eclipse leading to a strong increase in absorption in H$\alpha$ \citep[often also detectable in photometric light curves, e.g.][]{1976Natur.262..116H,town2013}.

This basic emission morphology and pattern of variation has been reported for several individual stars, most notably the extensively studied $\sigma$ Ori E, which was the first star in which H$\alpha$ emission was associated with a magnetosphere \citep{lb1978}, which inspired the RRM model \citep{town2005b,oks2012} as well as its immediate precursors \citep{1985ApSS.116..285N,2004AA...417..987P}, and which remains the only star for which custom RRM models have been calculated via extrapolation of Zeeman Doppler Imaging maps \citep[][]{2015MNRAS.451.2015O}. Other stars shown to display CM-type H$\alpha$ emission include HD\,36485 \citep{leone2010}, HD\,182180 \citep{2010MNRAS.405L..51O,2010MNRAS.405L..46R,rivi2013}, HD\,176582 \citep{bohl2011}, HD\,142184 \citep{grun2012}, HD\,23478 \citep{2014ApJ...784L..30E,2015MNRAS.451.1928S,2015A&A...578L...3H,2015ApJ...811L..26W}, HD\,35502 \citep{2016MNRAS.460.1811S}, HD\,345439 \citep{2014ApJ...784L..30E,2015A&A...578L...3H,2015ApJ...811L..26W}, ALS\,3694 \citep{2016ASPC..506..305S}, HD\,164492C \citep{2017MNRAS.465.2517W,2017MNRAS.467..437G}, and CPD$-62^\circ 2124$ \citep{2017A&A...597L...6C,2017MNRAS.472..400H}. CM-type emission has also been found in a tidally locked binary star, HD\,156324 \citep{2018MNRAS.475..839S}, which shows a single, rather than double, humped emission line morphology consistent with distortion of the gravitocentrifugal potential by the presence of a close companion. These studies have been descriptive in character, aimed at characterizing the fundamental parameters of the host stars and exploring the qualitative properties of their variable emission, and have generally found that the RRM model provides a reasonable description of their H$\alpha$ emission.

While it is understood how plasma enters into and is retained within the CM, the question of how material leaves is still a matter of debate. There are two mechanisms under consideration. The first, {\em centrifugal breakout}, may occur when the plasma density exceeds the capacity of the magnetic field to confine it, following which the magnetic field lines are ruptured and the plasma is ejected away from the star due to centrifugal force. This mechanism emerges naturally within 2D magnetohydrodynamic (MHD) simulations \citep{ud2006,ud2008}, and can also be derived from a consideration of first principles \citep{1984AA...138..421H,town2005c}. However, the centrifugal breakout narrative was challenged by the failure by \cite{town2013} to detect any change, over about twenty rotational cycles, in the light curve of $\sigma$ Ori E obtained with the Microvariability and Oscillations in STars \citep[MOST;][]{2003PASP..115.1023W} space telescope. RRM modelling of the MOST light curve furthermore established an upper limit for the CM mass nearly a factor of 50 below the mass required for breakout. Broadband polarimetric observations of $\sigma$ Ori E analyzed by \cite{carc2013} confirmed this mass estimate, and furthermore suggested that the CM plasma is more tightly concentrated than predicted by the RRM model, making breakout even more difficult to reconcile with observations. 

These considerations led \cite{2018MNRAS.474.3090O} to develop a second, alternative mechanism, a {\em diffusion/drift} model. In this scenario, plasma escapes in both directions via diffusion, and escapes away from the star via drift. This then affects both the shape of the emission profile -- leading to a more rounded profile -- and can displace the radius of maximum emission out to higher radii than the Kepler radius (where it naturally occurs in the RRM model). 

So far no comparative study involving a detailed examination of the emission properties of the members of this class of stars has been carried out. It is the aim of this paper to provide the first such study, and to use the comparison to distinguish between the centrifugal breakout and diffusion/drift scenarios. Our sample is drawn from the population of magnetic early B-type stars. High-resolution magnetometry and rotation periods for the stars were presented by \citet[Paper I]{2018MNRAS.475.5144S}. A compilation of atmospheric parameters for the stars collected from the literature or, in several cases, determined for the first time, was presented by \citet[Paper II]{2019MNRAS.485.1508S}. Fundamental parameters, dipolar oblique rotator models, and magnetospheric parameters were determined by \citet[Paper III]{2019MNRAS.tmp.2196S}. In the present work we concern ourselves only with those stars showing detectable H$\alpha$ emission, and with those stars without detectable emisison but having similar rotational and magnetic properties. The properties of these stars were presented in Papers I--III. 

In \S~\ref{sec:obs} a brief overview of the properties of the sample is provided. \S~\ref{sec:onset} looks at the conditions required for the onset of detectable emission. An analysis of the properties of the emission profiles is performed in \S~\ref{sec:emprofs}. The results are discussed in \S~\ref{sec:discussion}, and the conclusions summarized in \S~\ref{sec:conclusion}. Magnetic analyses of 6 new stars included in the sample are described in Appendix \ref{appendix:newstars}. Notes on individual stars with and without H$\alpha$ emission are respectively provided in the Appendices \ref{appendix:indstars} and \ref{appendix:noem}, available online.

\section{Sample and observations}\label{sec:obs}

\begin{table*}
\caption[]{Table of parameters and measurements for H$\alpha$-bright stars. The top row gives parameters determined in Papers I to III. From left to right, these are: the stellar designation; the bolometric luminosity $\log{L}$; the effective temperature \teff; the Kepler corotation radius \rk; the inclination angle $i_{\rm rot}$ of the rotational axis from the line of sight; the obliquity angle $\beta$ of the magnetic axis from the rotation axis; the surface strength of the magnetic field at the magnetic pole $B_{\rm d}$; the Alfv\'en radius \ra; the ratio \rark; and the strength $B_{\rm K}$ of the magnetic field at \rk. Parantheses next to the star name indicate the star is part of a spectroscopic binary, and which component hosts the magnetic field. The bottom row gives the H$\alpha$ measurements presented here, as well as corrections for area deprojection and dilution by light from companion stars. From left to right, the columns of the bottom row are: an alternative stellar designation; the radius of maximum emission $r_{\rm max}$; the outermost radius of emission $r_{\rm out}$; the ratio $r_{\rm max}/R_{\rm K}$; the minimum angle $\alpha_{\rm D}$ between the disk and the line of sight; the ratio $f_*/f_{\rm tot}$ of the magnetic star to its binary companion(s); the maximum measured emission equivalent width $W_{\lambda, \rm obs}$; the maximum emission equivalent width $W_{\lambda, \rm fit}$ obtained from a harmonic fit to the equivalent width curve; and the ratio $V/R$ of the blue to red equivalent width at maximum emission.}
\label{obstab}
\resizebox{18.5 cm}{!}{
\begin{tabular}{l | r r r r r r r r r r}
\hline\hline
Star      & $\log{L/L_\odot}$ & \teff~(kK) & \rk~($R_*$) & $i_{\rm rot}$~($^\circ$) & $\beta$~($^\circ$) & $B_{\rm d}$~(kG) & \ra~($R_*$) & \rark & $\log{(B_{\rm K} / {\rm G})}$\\
Alt. Name & $r_{\rm max}$~($R_*$) & $r_{\rm out}$~($R_*$) & $r_{\rm max}/R_{\rm K}$ & $\cos{\alpha_{\rm D}}$ & $f_*/f_{\rm tot}$ & $W_{\lambda, \rm obs}$~(nm) & $W_{\lambda, \rm fit}$ (nm) & $V/R$ & -- \\
\hline
HD\,23478 &  $3.2 \pm 0.2$ & $20 \pm 2$ &  $2.5 \pm 0.1$ & $56 \pm 5$ &  $4 \pm 2$ & $10 \pm 2$ & $30 \pm 10$ & $1.3 \pm 0.1$ & $2.58 \pm 0.05$\\
ALS\,14589 & $2.8 \pm 0.2$ & $ 6.3 \pm  0.3$ & $1.19 \pm 0.10$ & 0.60 &  -- & $0.257 \pm 0.001$ & $0.256 \pm 0.012$ & $1.0 \pm 0.1$ & \\
\hline
HD\,35502 (A)  &  $3.0 \pm 0.1$ & $18.4 \pm 0.6$ &  $2.22 \pm 0.05$ & $26 \pm 1$ & $70 \pm 1$ &  $7.3 0.5$ & $26 \pm 1$ & $1.07 \pm 0.02$ & $2.55 \pm 0.05$ \\
BD$-02\,1241$ & $4.1 \pm 0.3$ & $ 6.4 \pm  0.4$ & $1.9 \pm 0.2$ & 0.81 & 0.71 & $0.150 \pm 0.001$ & $0.142 \pm 0.006$ & $1.1 \pm 0.1$ & \\
\hline
HD\,36485 (A)  &  $3.1 \pm 0.2$ & $20 \pm 2$ &  $3.35 \pm 0.06$ & $19 \pm 1$ &  $4 \pm 2$ &  $8.9 \pm 0.2$ & $25 \pm 1$ & $0.88 \pm 0.02$ & $2.07 \pm 0.03$ \\
$\delta$\,Ori\,C & $4.7 \pm 0.3$ & $ 6.7 \pm  0.4$ & $1.18 \pm 0.09$ & 0.96 & 0.85 & $0.058 \pm 0.001$ & $0.055 \pm 0.004$ & $0.6 \pm 0.1$ & \\
\hline
HD\,37017 (A)  &  $3.4 \pm 0.2$ & $21 \pm 2$ &  $2.09 \pm 0.06$ & $39 \pm 2$ & $57 \pm 3$ &  $6.3 \pm 0.9$ & $24 \pm 3$ & $1.09 \pm 0.05$ & $2.52 \pm 0.09$ \\
V\,1046\,Ori & $3.8 \pm 0.4$ & $ 6.7 \pm  0.8$ & $1.8 \pm 0.3$ & 0.99 & 0.78 & $0.088 \pm 0.001$ & $0.084 \pm 0.006$ & $0.29 \pm 0.08$ & \\
\hline
HD\,37479 &  $3.5 \pm 0.2$ & $23 \pm 2$ &  $2.69 \pm 0.03$ & $767 \pm 4$ & $38 \pm 9$ & $10.5 \pm 1.5$ & $36 \pm 9$ & $1.18 \pm 0.07$ & $2.46 \pm 0.08$ \\
$\sigma$\,Ori\,E & $3.0 \pm 0.1$ & $ 6.9 \pm  0.3$ & $1.31 \pm 0.07$ & 0.65 &  -- & $0.346 \pm 0.001$ & $0.342 \pm 0.016$ & $1.6 \pm 0.2$ & \\
\hline
HD\,37776 &  $3.3 \pm 0.2$ & $22 \pm 1$ &  $3.13 \pm 0.08$ & $57 \pm 4$ & $47 \pm 9$ &  $6.1 \pm 0.7$ & $24 \pm 4$ & $0.93 \pm 0.05$ & $1.99 \pm 0.06$ \\
V\,901\,Ori & $4.0 \pm 0.2$ & $ 5.9 \pm  0.2$ & $1.26 \pm 0.08$ & 0.93 &  -- & $0.069 \pm 0.001$ & $0.057 \pm 0.005$ & $1.6 \pm 0.5$ & \\
\hline
HD\,64740 &  $3.8 \pm 0.2$ & $24.5 \pm 1.0$ &  $2.1 \pm 0.3$ & $412 \pm 8$ & $72 \pm 5$ &  $3.0 \pm 0.5$ & $12 \pm 1$ & $0.8 \pm 0.1$ & $2.1 \pm 0.2$ \\
HR\,3089 & $2.81 \pm 0.07$ & $4.4 \pm 0.1$ & $1.3 \pm 0.2$ & 0.93 &  -- & $0.041 \pm 0.001$ & $0.039 \pm 0.008$ & $1.7 \pm 0.7$ & \\
\hline
HD\,66765 &  $3.4 \pm 0.2$ & $20 \pm 2$ &  $2.8 \pm 0.4$ & $44 \pm 9$ & $73 \pm 5$ &  $2.8 \pm 0.5$ & $13^{+8}_{-1}$ & $0.9 \pm 0.1$ & $1.8 \pm 0.2$ \\
ALS\,14050 & $3.7 \pm 0.3$ & $4.7 \pm 0.3$ & $1.3 \pm 0.3$ & 0.93 &  -- & $0.023 \pm 0.001$ & $0.023 \pm 0.011$ & $ 0 \pm  1$ & \\
\hline
HD\,142184 &  $2.8 \pm 0.1$ & $18.5 \pm 0.5$ &  $1.5 \pm 0.1$ & $64 \pm 6$ &  $9 \pm 3$ &  $9 \pm 3$ & $31 \pm 3$ & $1.34 \pm 0.03$ & $3.25 \pm 0.04$ \\
HR\,5907 & $1.84 \pm 0.04$ & $4.9 \pm 0.1$ & $1.20 \pm 0.09$ & 0.53 &  -- & $0.229 \pm 0.002$ & $0.228 \pm 0.004$ & $2.0 \pm 0.2$ & \\
\hline
HD\,142990 &  $2.9 \pm 0.1$ & $18.0 \pm 0.5$ &  $2.58 \pm 0.04$ & $55 \pm 2$ & $84 \pm 3$ &  $4.7 \pm 0.4$ & $21 \pm 1$ & $0.93 \pm 0.02$ & $2.12 \pm 0.05$ \\
V\,913\,Sco & $3.03 \pm 0.05$ & $4.51 \pm 0.07$ & $1.18 \pm 0.04$ & 0.90 &  -- & $0.038 \pm 0.001$ & $0.040 \pm 0.028$ & $ 0 \pm  1$ & \\
\hline
HD\,156324 (Aa)  &  $3.8 \pm 0.6$ & $22 \pm 3$ &  $2.9 \pm 0.3$ & $21 \pm 4$ & $75 \pm 4$ & $14 \pm 3$ & $30 \pm 8$ & $1.1 \pm 0.1$ & $2.4 \pm 0.2$ \\
ALS\,4060 & $  5 \pm   1$ & $  7 \pm   1$ & $1.8 \pm 0.5$ & 0.68 & 0.78 & $0.138 \pm 0.001$ & $0.107 \pm 0.011$ & $2.5 \pm 0.5$ & \\
\hline
HD\,164492C  (A)  &  $4.1 \pm 0.3$ & $26 \pm 2$ &  $2.83 \pm 0.06$ & $62 \pm 5$ & $35 \pm 7$ &  $6.6 \pm 0.8$ & $15 \pm 2$ & $0.77 \pm 0.05$ & $2.22 \pm 0.04$ \\
EM*\,LkHA\,123 & $3.99 \pm 0.06$ & $10.9 \pm  0.2$ & $1.41 \pm 0.05$ & 0.80 & 0.55 & $0.234 \pm 0.002$ & $0.180 \pm 0.019$ & $1.2 \pm 0.2$ & \\
\hline
HD\,176582 &  $2.9 \pm 0.1$ & $17.6 \pm 0.4$ &  $3.09 \pm 0.08$ & $84 \pm 2$ & $89 \pm 1$ &  $5.4 \pm 0.2$ & $24 \pm 1$ & $0.89 \pm 0.04$ & $1.95 \pm 0.04$ \\
HR\,7185 & $4.9 \pm 0.3$ & $ 6.0 \pm  0.4$ & $1.6 \pm 0.1$ & 0.99 &  -- & $0.046 \pm 0.001$ & $0.043 \pm 0.004$ & $1.3 \pm 0.2$ & \\
\hline
HD\,182180 &  $3.1 \pm 0.2$ & $19.8 \pm 1.4$ &  $1.45 \pm 0.09$ & $53 \pm 7$ & $82 \pm 4$ &  $9.5 \pm 0.6$ & $24^{+18}_{-1}$ & $1.3 \pm 0.2$ & $3.2 \pm 0.1$ \\
HR\,7355 & $1.96 \pm 0.03$ & $4.25 \pm 0.07$ & $1.4 \pm 0.1$ & 0.91 &  -- & $0.264 \pm 0.001$ & $0.259 \pm 0.014$ & $0.84 \pm 0.06$ & \\
\hline
HD\,189775 & $2.9 \pm 0.1$ & $17.5 \pm 0.6$ & $4.2 \pm 0.3$ & $59 \pm 9$ & $43 \pm 11$ & $4.3 \pm 0.7$ & $21 \pm 2$ & $0.73 \pm 0.04$ & $1.51 \pm 0.08$ & \\
V2100\,Vyg & $4.8 \pm 0.2$ & $6.6 \pm 0.3$ & $1.14 \pm 0.06$ & 0.89 & -- & $0.017 \pm 0.002$ & $0.013 \pm 0.002$ & $0.9 \pm 0.4$ & \\
\hline
HD\,345439 &  $4.0 \pm 0.3$ & $23 \pm 2$ &  $1.6 \pm 0.2$ & $59 \pm 10$ & $46 \pm 13$ &  $9 \pm 1$ & $21 \pm 7$ & $1.2 \pm 0.1$ & $3.0 \pm 0.1$ \\
ALS\,10681 & $2.0 \pm 0.4$ & $4.4 \pm 0.6$ & $1.2 \pm 0.4$ & 0.91 &  -- & $0.478 \pm 0.009$ & $0.503 \pm 0.122$ & $0.6 \pm 0.4$ & \\
\hline
 ALS\,3694  &  $3.8 \pm 0.2$ & $25 \pm 1$ &  $3.0 \pm 0.2$ & $22 \pm 2$ & $26 \pm 11$ & $12 \pm 3$ & $27 \pm 6$ & $1.0 \pm 0.1$ & $2.4 \pm 0.2$ \\
CPD\,$-48^\circ~8684$ & $4.6 \pm 0.3$ & $ 6.2 \pm  0.4$ & $1.5 \pm 0.2$ & 0.99 &  -- & $0.065 \pm 0.002$ & $0.064 \pm 0.005$ & $0.6 \pm 0.1$ & \\
\hline
 ALS\,2394  &  $3.8 \pm 2$ & $23.6 \pm 0.3$ &  $4.2 \pm 0.4$ & $26 \pm 3$ & $26 \pm 11$ & $23 \pm 1$ & $41 \pm 2$ & $1.01 \pm 0.03$ & $2.2 \pm 0.1$ \\
CPD\,$-62^\circ~2124$ & $  8 \pm   1$ & $ 15 \pm   2$ & $1.9 \pm 0.4$ & 0.99 &  -- & $0.320 \pm 0.003$ & -- & $0.4 \pm 0.2$ & \\
\hline
\hline
\end{tabular}
}
\end{table*}

The sample consists of all magnetic early B-type stars exhibiting H$\alpha$ emission consistent with an origin in a CM, i.e.\ with emission profiles peaking at velocities higher than \vsini~and modulated with the rotation period. In total there are 20 stars from Papers I to III that display H$\alpha$ emission. One star, HD\,46328, is consistent with emission originating from a DM \citep{2017MNRAS.471.2286S}, and is not considered here. HD\,37061 C shows some signs of H$\alpha$ emission, however these are somewhat marginal as the spectrum is dominated by the non-magnetic B0 primary and there is not enough data for a clear detection \citep{2019MNRAS.482.3950S}; therefore this star was dropped from the sample. The H$\alpha$ emission of HD\,156424, while in some ways consistent with a CM, is highly anomalous (its emission apparently peaks at about 40$R_*$, close to twice the value of its Alfv\'{e}n radius), and was therefore removed from the sample pending a closer investigation. Upon close examination we have classified HD\,189775, not previously reported to display H$\alpha$ emission, as an H$\alpha$-bright star with extremely weak emission. The final sample therefore includes 18 H$\alpha$-bright stars, which are listed in Table \ref{obstab}. In addition to these stars, we analyze the H$\alpha$ profiles of stars without detected H$\alpha$ emission, but having relatively rapid rotation and strong magnetic fields\footnote{Formally, having a strength of the magnetic field at the Kepler radius $\log{(B_{\rm K}/{\rm G})} > 1.5$.}, in order to place upper limits on their emission strength. Including 6 additional magnetic B-type stars that have been added to the sample (see below), we examine the H$\alpha$ profiles of 15 stars without emission. The remaining 9 stars (HD\,35298, HD\,36526, HD\,43317, HD\,55522, HD\,61556, HD\,105382, HD\,121743, HD\,130807, and HD\,175362) were already included in the sample; their parameters are provided in Papers I to III.

The spectropolarimetric and spectroscopic datasets, magnetic field curves, rotational periods, and \vsini~values of the sample were presented in Paper I. We have also included the high-resolution UVES spectroscopy available for HR\,5907 \citep{grun2012}, HR\,7355 \citep{rivi2013}, and HD\,164492C \citep{2017MNRAS.465.2517W}, and the ARCES spectroscopy for HD\,345439 \citep{2015ApJ...811L..26W}. In addition to this we have collected archival spectra for: HD\,36485 \citep[CFHT, DDO, DAO, published by][]{leone2010}; HD\,142990 \citep[CFHT and ESO, published by ][]{2004A&A...421..203S}; previously unpublished spectra of HD\,64740 obtained with the echelle spectrograph at the Los Cumbres Observatory 2.5 m telescope \citep[for a description of the instrument, see][]{2013A&A...552A...7S}; HD\,35502 \citep[DAO, previously published by][]{2016MNRAS.460.1811S}; and HD\,176582, \citep[DAO, previously published by][]{bohl2011}.

We also obtained 6 new spectroscopic observations of HD\,23478 with the Southeastern Association for Research in Astronomy \citep[SARA;][]{2017PASP..129a5002K} telescope at Kitt Peak. The starlight from the 0.9~m telescope is fiber-fed to a $R=20,000$ echelle spectrograph. The spectra were extracted with custom software based on the optimal extraction method of \cite{2002A&A...385.1095P}\footnote{\url{https://www.astro.uu.se/~piskunov/RESEARCH/REDUCE/reduce.pdf}}. Biases and dark frames were removed from the science data and the orders were traced through flat fields. The science frames were not corrected for pixel-to-pixel variation via the flat, as the cross-dispersion profiles of the flat frames have the same shape as that of the star frames, hence introducing too much noise at the edges of the trace \citep[see discussion in][]{2002A&A...385.1095P}. Scattered light was subtracted by a smoothed, 2D interpolation of regions located between the orders. Scattered light was mostly significant for the flat frames and for very bright stars. The wavelength calibration was done via a Thorium Argon lamp. The extracted spectra for the two orders containing H$\alpha$ were normalized to the continuum via polynomial fitting of individual orders, which were then merged.

   \begin{figure}
   \centering
   \includegraphics[trim=40 20 30 30, width=\hsize]{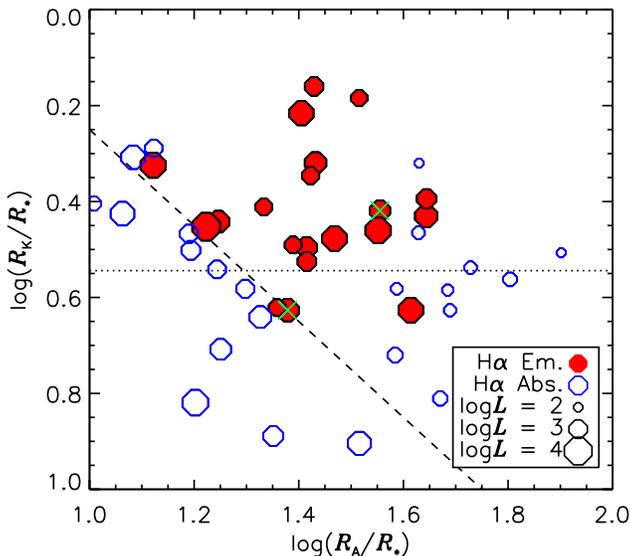} 
      \caption[]{The rotation-magnetic confinement diagram in the vicinity of the H$\alpha$-bright stars. Green crosses indicate stars excluded from the analysis (see text). The horizontal dotted and diagonal dashed lines are empirical divisions between stars with and without H$\alpha$ emission.}
         \label{ipod_zoom}
   \end{figure}

Atmospheric parameters for the sample stars and (for the spectroscopic binaries) for companion stars, were given in Paper II. Paper III derived fundamental stellar parameters, oblique rotator model parameters, and magnetospheric parameters. The datasets and properties of a few stars, added after Paper I, were given in Paper II. The fundamental results of Paper III are summarized in Fig.\ \ref{ipod_zoom}, which shows the rotation-magnetic confinement diagram zoomed in to the region containing the H$\alpha$-bright stars. The stars dropped from the analysis are indicated with green crosses (with the exception of HD\,46328, which does not appear on this diagram).

The original sample described in Paper I had a \teff~cutoff of 15 kK, roughly corresponding to a minimum luminosity of $\log{(L/L_\odot)} = 2.5$. In order to explore the presence or absence of H$\alpha$ emission at low luminosities, we expanded the sample to include rapidly rotating ($P_{\rm rot} < 2$~d), strongly magnetic (longitudinal magnetic field \bz$_{\rm max} > 0.5$ kG) stars with $\log{(L/L_\odot)}$ between about 2.0 and 2.5. This was done by cross-referencing the \bz~curve catalogue of \cite{2005AA...430.1143B} (which includes \bz~measurements and rotation periods) with the catalogues published by \cite{land2007} and \cite{2017MNRAS.468.2745N} (which provide luminosities and, in the former case, r.m.s. \bz~measurements). Candidate stars were then cross-referenced with the online PolarBase archive of ESPaDOnS and Narval observations \citep{d1997,2014PASP..126..469P}\footnote{Available at \url{http://polarbase.irap.omp.eu/citation}}. In the end 6 stars were selected for inclusion: HD\,19832, HD\,22470, HD\,45583, HD\,142301, HD\,144334, and HD\,145501C. The magnetic, rotational, and magnetospheric analysis of these stars, which closely follows the methodology described in Papers I and III, is presented in Appendix \ref{appendix:newstars}. Their parameters are summarized in Table \ref{newstarstab}.

   \begin{figure}
   \centering
   \includegraphics[trim=0 0 0 0, width=\hsize]{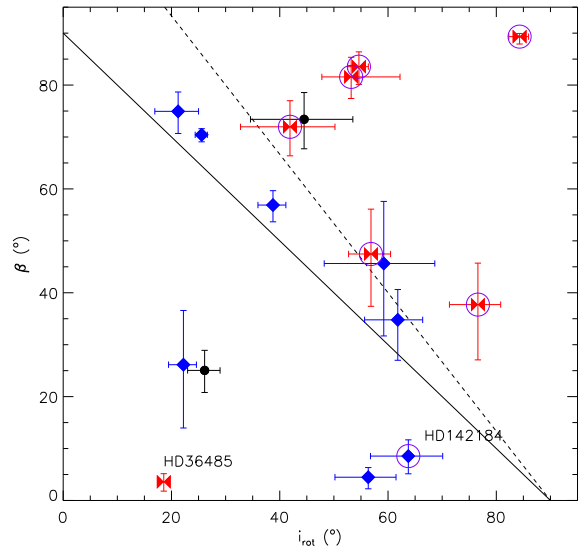}
      \caption[]{Geometrical characterization of H$\alpha$ equivalent width (EW) variability on the $i_{\rm rot}-\beta$ plane. Blue diamonds: single-wave variation; red bowties: double-wave variation; purple circles: eclipses; black circles, insufficient data. Labelled stars are discussed in the text. The dashed line is an empirical division between stars with single- and double-wave EW variations, while the solid line shows $i + \beta = 90^\circ$.}
         \label{cmstars_geom_var}
   \end{figure}

As can be seen from Fig.\ \ref{ipod_zoom}, all of the stars with H$\alpha$-producing CMs exhibit very rapid rotation (Kepler corotation radii $R_{\rm K}$ below about 4 $R_*$) and very strong magnetic confinement (Alfv\'en radii $R_{\rm A}$ extending to at least 10 $R_*$). The dashed diagonal line in the figure indicates $R_{\rm A} = 6 R_{\rm K}$; above this line, most stars show emission. However, along the border are stars both with and without emission. As noted in Paper III, in this liminal region the more luminous stars are more likely to display emission. This suggests that while the extent of the CM is an important threshold, the mass-loading rate from the stellar wind might also play an important role.

Since the CM is to first order a warped disk approximately perpendicular to the magnetic axis, we expect the equivalent width (EW) variability of CMs to occur along two basic patterns, depending on the geometry and projection of the star's surface magnetic field. In cases when only one of the magnetic poles is visible during a rotational cycle (i.e.\ when the sum of the rotational axis inclination angle $i_{\rm rot}$ and the magnetic obliquity angle $\beta$ is less than $90^\circ$), the EW should show a single-wave variation. A single-wave variation shows a single emission peak when the magnetic pole is closest to the line of sight and the projected area of the CM is at a maximum. When both magnetic poles are visible across a rotational cycle we expect a double-wave variation, with each peak again occuring when one of the magnetic poles is closest to the line of sight. Fig.\ \ref{cmstars_geom_var} shows the sample stars on the $i_{\rm rot}-\beta$ plane, with the stars characterized according to whether their H$\alpha$ EW curves are single- or double-wave. With only one exception, HD\,36485, single- and double-wave variations occur when $i_{\rm rot}$ and $\beta$ are large. The diagonal dashed line shows the division between these regimes. In the case of HD\,36485, the double-wave variation is probably a consequence of a magnetic field with significant departures from a dipolar geometry. This possibility was suggested by \cite{leone2010}, and as demonstrated in Paper I there is some indication in \bz~that the magnetic field is indeed complex. 

When $i$ and/or $\beta$ are large, we also expect that the star will be periodically eclipsed by the CM. These eclipses manifest either photometrically \citep[e.g.][]{lb1978,town2008} and/or as enhanced absorption in the rotationally broadened core of the line \citep[e.g.][]{town2005b,bohl2011,grun2012,rivi2013}. Stars exhibiting H$\alpha$ eclipses are also indicated in Fig.\ \ref{cmstars_geom_var}, and eclipses indeed occur almost exclusively amongst stars with double-wave variations. The one exception to this, HD\,142184, is seen at a relatively high inclination but has a very small $\beta$, and its eclipse is fairly shallow \citep{grun2012}. It should be noted that eclipses can be very rapid events, having a duration of much less than 0.1 of a rotational cycle, particularly when the corotating magnetosphere is relatively far from the star. Not every dataset samples the rotational phase curve densely enough to detect eclipses. 

\section{Emission onset}\label{sec:onset}

   \begin{figure*}
   \centering
\begin{tabular}{ccc}
   \includegraphics[trim=50 20 20 0, width=0.3\textwidth]{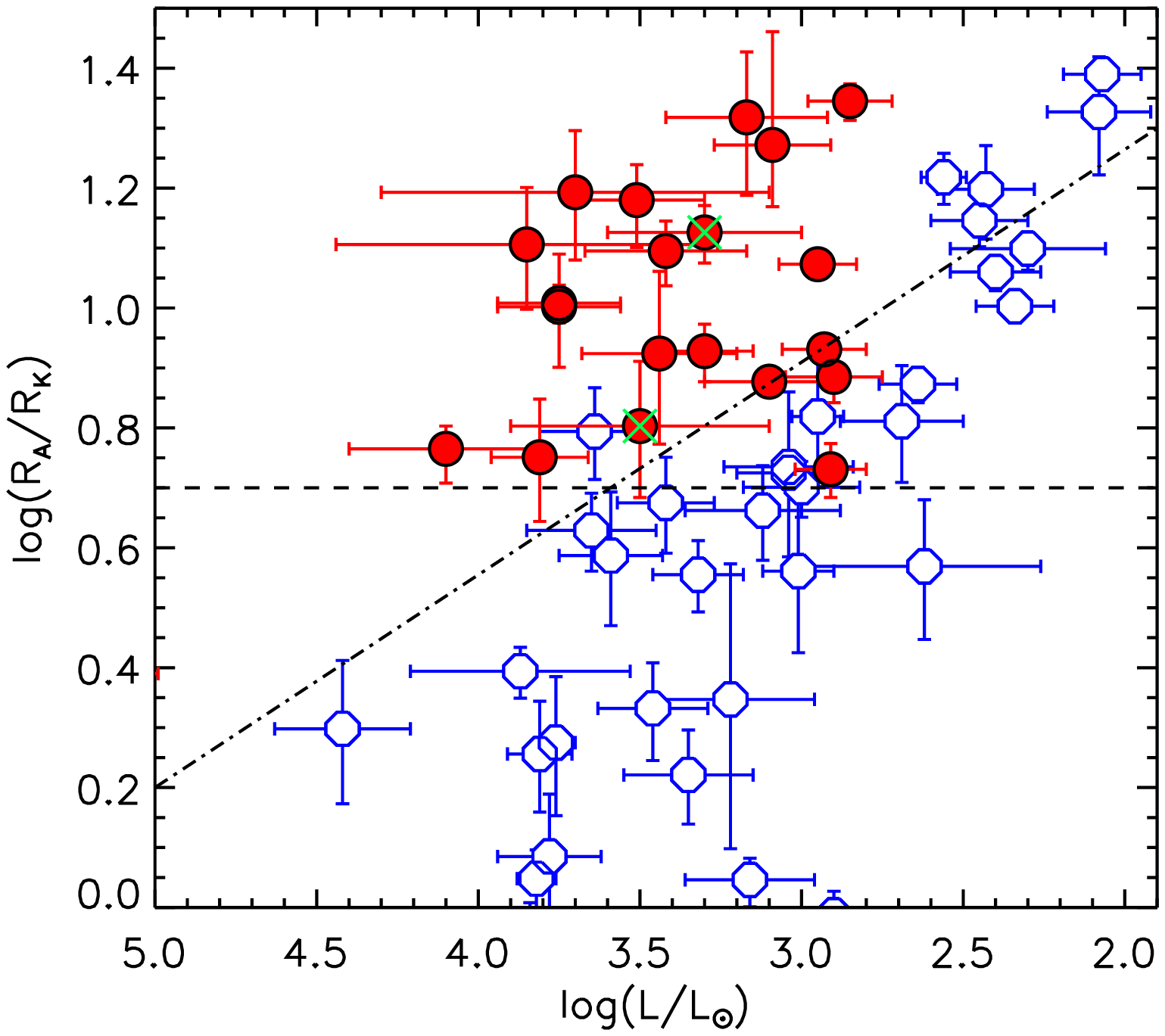} &
   \includegraphics[trim=50 20 20 0, width=0.3\textwidth]{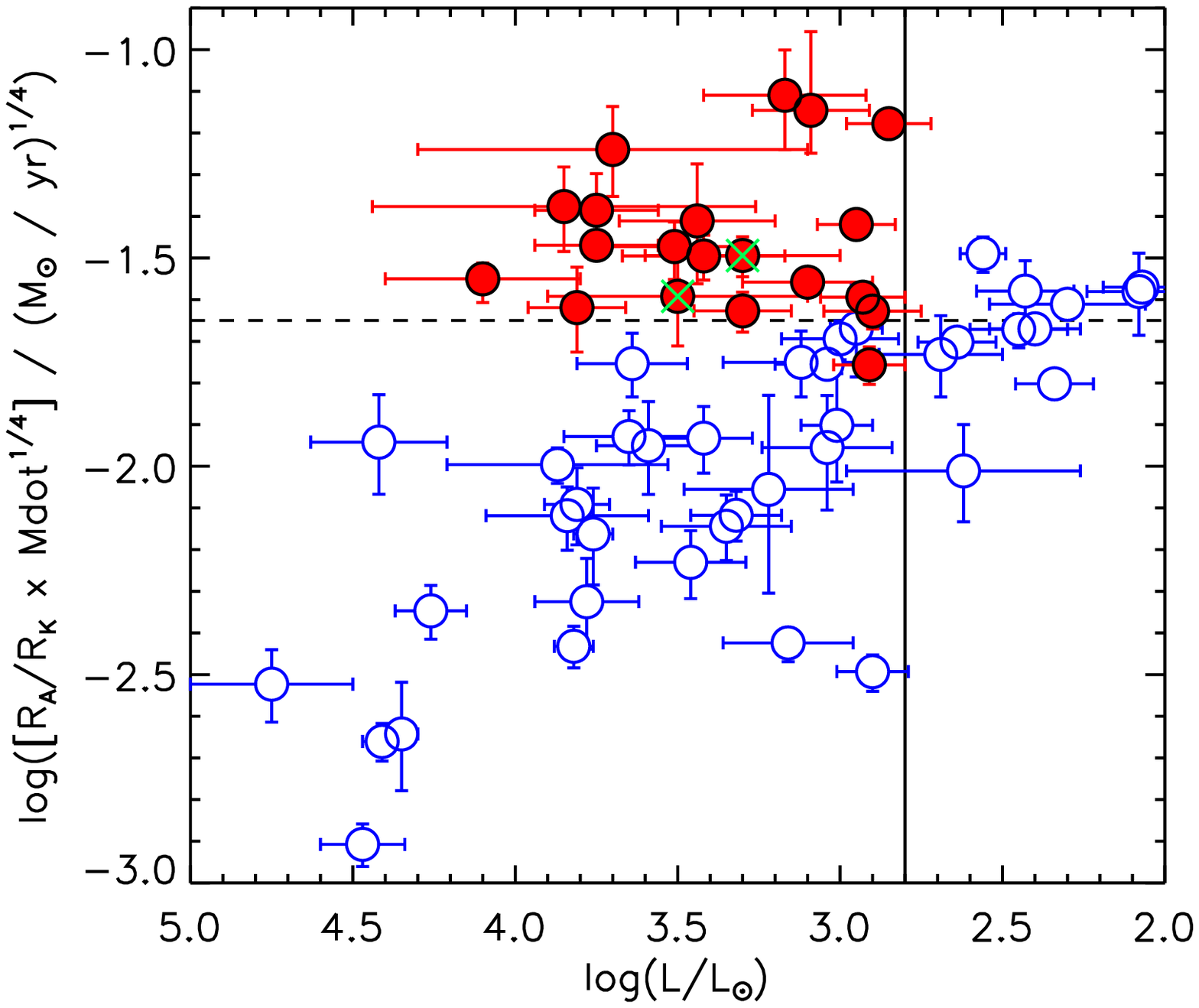} &
   \includegraphics[trim=50 20 20 0, width=0.3\textwidth]{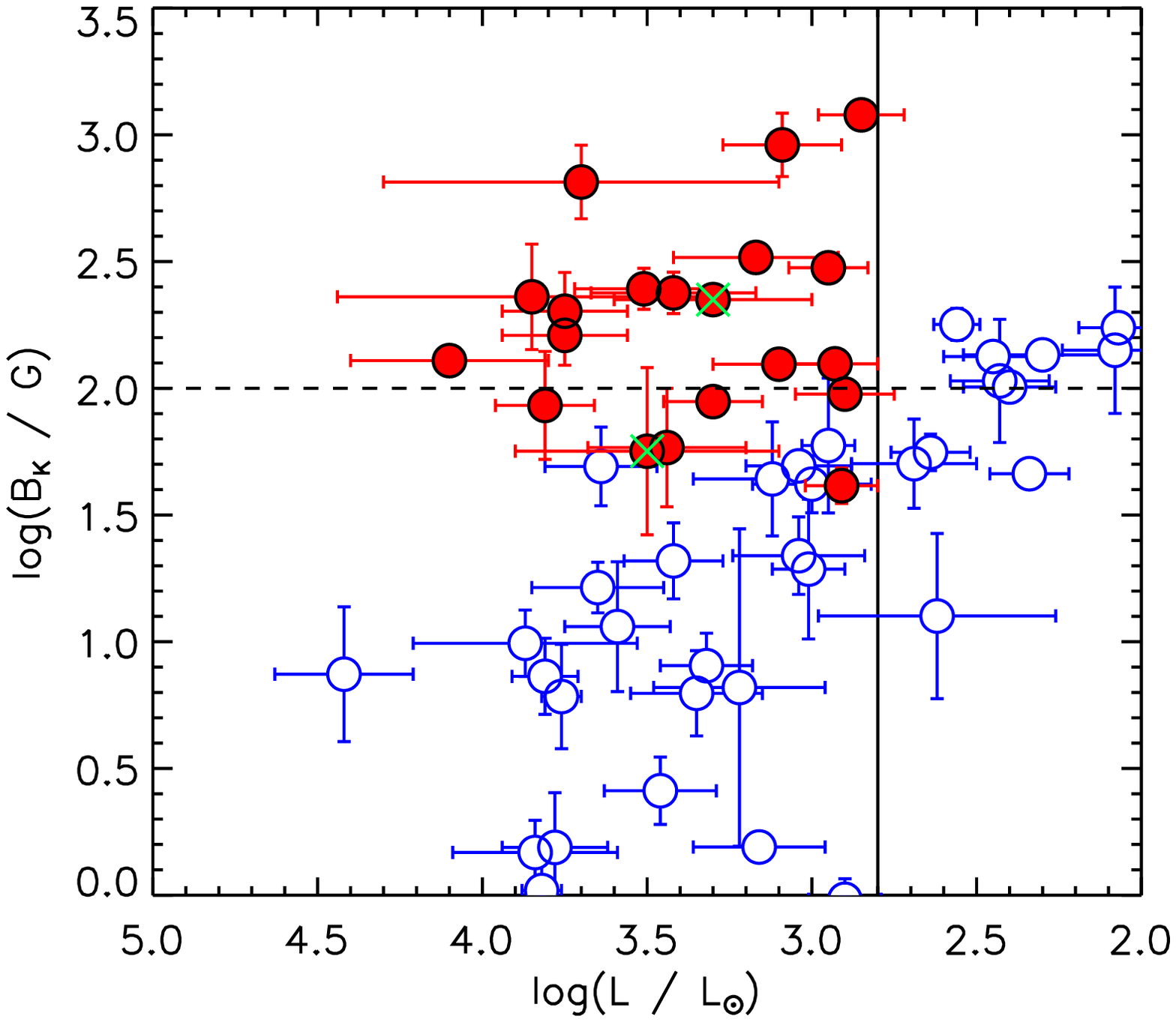} \\
\end{tabular}
      \caption[]{{\em Left}: Centrifugal magnetosphere dimensionless area proxy $\log{(R_{\rm A}/R_{\rm K})}$ as a function of bolometric luminosity $\log{L}$. Filled red circles indicate H$\alpha$-bright stars; open blue circles represent stars with H$\alpha$ in absorption. Green crosses indicate stars dropped from the analysis. The dashed and dot-dashed lines relate to the centrifugal breakout and leakage plasma transport scenarios (see text). {\em Middle}: CM area with \mdot~dependence of \ra~removed, as a function of $\log{L}$. The horizontal dashed line shows the approximate division between stars with and without H$\alpha$ emission {\em Right}: Equatorial magnetic field strength at the Kepler radius $B_{\rm K}$ as a function of $\log{L}$. The horizontal and vertical dashed lines indicate the divisions between stars with and without H$\alpha$ emission.}
         \label{lum_rark2}
   \end{figure*}

The left panel of Fig.\ \ref{lum_rark2} shows the dimensionless CM area proxy $\log{(R_{\rm A}/R_{\rm K})}$ as a function of stellar luminosity. The Alfv\'en radius \ra~is the maximum extent of magnetic confinement, and can be obtained from the wind magnetic confinement parameter $\eta_*$ via \citep{ud2008}

\begin{equation}\label{ra}
\frac{R_{\rm A}}{R_*} \sim 0.3 + (\eta_* + 0.4)^{1/4},
\end{equation}

\noindent where

\begin{equation}\label{etastar}
\eta_* \equiv \frac{B_{\rm eq}^2 R_*^2}{\dot{M}v_\infty},
\end{equation}

\noindent where $B_{\rm eq} = B_{\rm d}/2$ is the equatorial surface strength of the magnetic dipole, $B_{\rm d}$ is the surface magnetic dipole strength, $R_*$ is the stellar radius, $\dot{M}$ is the mass-loss rate, and $v_\infty$ is the terminal velocity of the wind \citep{ud2002}. Throughout this paper, including Fig. \ref{lum_rark2}, we use \cite{vink2001} mass-loss rates (see Paper III), as the \cite{krticka2014} mass-loss rates are not defined below 15 kK.

Eqn.\ \ref{etastar} is calculated under the assumption of a dipolar magnetic field. While stellar magnetic fields are not in general perfect dipoles, the majority of early-type stars whose magnetic fields have been mapped via Zeeman Doppler Imaging \citep[ZDI;][]{pk2002} are well-described by `twisted dipoles' \citep{2019A&A...621A..47K} for which the dipolar approximation gives a good approximation of the surface magnetic field strength and geometry. Some of the stars in this sample, notably HD\,37776 \citep{koch2011}, have very complex surface magnetic fields dominated by higher-order components. Surface dipole magnetic field strengths for these stars were derived from the first two terms of harmonic fits to their longitudinal magnetic field curves (see Papers I and III). Since higher-order components of the magnetic field fall off much faster with distance than the dipolar component, the maximum extent of magnetic confinement should be determined primarily by the dipolar component even for a star with a relatively complex magnetic field \citep[e.g.][]{ud2002}. 

The Kepler corotation radius \rk, defined as the point at which gravitational and centrifugal forces are balanced, can be obtained from the rotation parameter $W$ as $R_{\rm K} / R_* = W^{-2/3}$, with

\begin{equation}\label{wrot}
W \equiv \frac{v_{\rm rot}}{v_{\rm orb}},
\end{equation}

\noindent where $v_{\rm rot}$ is the equatorial rotational velocity, and $v_{\rm orb} = \sqrt{G M_*/R_*}$ is the orbital speed at the stellar surface \citep{ud2008}. As described in Paper III, $W$ was determined taking into account the rotational oblateness of the star. 

Two lines are shown to divide the diagram in the left panel of Fig.\ \ref{lum_rark2}: a horizontal dashed line, and a diagonal dot-dashed line. \cite{petit2013} suggested that these lines could be used to distinguish between the breakout and leakage scenarios. If plasma transport is governed by breakout then emission onset should be sensitive solely to the area of the CM. In this case, all stars above the horizontal line should display emission. On the other hand, in a leakage scenario mass-loading from the wind must compete with whatever mechanism is responsible for leakage; in this case, a diagonal line should divide stars with and without emission, since stars with higher mass-loss rates should be able to fill CMs more rapidly (we make the assumption that the mass-loss rate increases with bolometric luminosity). Looking only at the left panel of Fig.\ \ref{lum_rark2}, the data apparently support the latter scenario, at least for stars with $\log{L/L_\odot} > 2.8$ .

The Alfv\'en radius is a function of $\dot{M}$ (Eqns. \ref{ra} and \ref{etastar}). To check whether the apparent diagonal threshold in emission onset is due to this dependency, the middle panel of Fig.\ \ref{lum_rark2} shows $R_{\rm A}/R_{\rm K}$ multiplied by $\dot{M}^{1/4}$, i.e.\ approximately removing the dependence of $R_{\rm A}$ on $\dot{M}$ ($v_\infty$ is nearly constant over this regime, and so does not strongly affect the results). This follows from Eqn.\ \ref{ra} in the limit of $\eta_* \gg 1$. This yields a nearly flat division between stars with and without emission. The diagonal relationship in the left panel of Fig.\ \ref{lum_rark2} is therefore a consequence of the dependence of \ra~on \mdot. This indicates that the decisive factor governing the appearance of emission around a given star is the area of its CM, i.e. it is unrelated to the star's mass-loss rate.

In a breakout scenario, emission onset should be governed primarily by the ability of the magnetic field to confine the CM plasma. To investigate this, the right panel of Fig.\ \ref{lum_rark2} shows the strength of the equatorial magnetic field at \rk, $B_{\rm K} = B_{\rm d} / 2R_{\rm K}^3$ (with \rk~in units of $R_*$)\footnote{Note that while $B_{\rm K}$ and $R_{\rm A}/R_{\rm K}$ are similar expressions, they are not identical, since $R_{\rm A}/R_{\rm K} \propto \sqrt{B_{\rm d}R_*/2}/R_{\rm K}$.}. All stars with $B_{\rm K} > 100$~G are in emission, while stars below this value are in absorption. This simple parameter does almost as well as $\log{[(R_{\rm A}/R_{\rm K}) \times \dot{M}^{1/4}}]$, and its success in dividing stars with and without emission is evidence for a breakout scenario: as discussed by \cite{town2005c} in their Appendix A, the breakout density of the CM is a function of the magnetic field strength and the Kepler radius, and is independent of the mass-loss rate.

Below $\log{(L/L_\odot)} \sim 2.8$ there are no stars with H$\alpha$ emission regardless of the value of $B_{\rm K}$. This may suggest that at very low mass-loss rates (using Vink mass-loss, about $\log{\dot{M}} \sim -10.5$ at $\log{L/L_\odot} = 2.8$ and the corresponding main sequence \teff~of 15 kK), the wind is no longer able to fill the CM to the degree needed for the plasma to become optically thick, regardless of the strength of magnetic confinement. Possible causes for the low-luminosity emission cutoff are examined in \S~\ref{subsec:lowlumcutoff}. 

\section{Emission profiles}\label{sec:emprofs}

   \begin{figure}
   \centering
\begin{tabular}{cc}
   \includegraphics[trim=40 20 30 30, height=5.5cm]{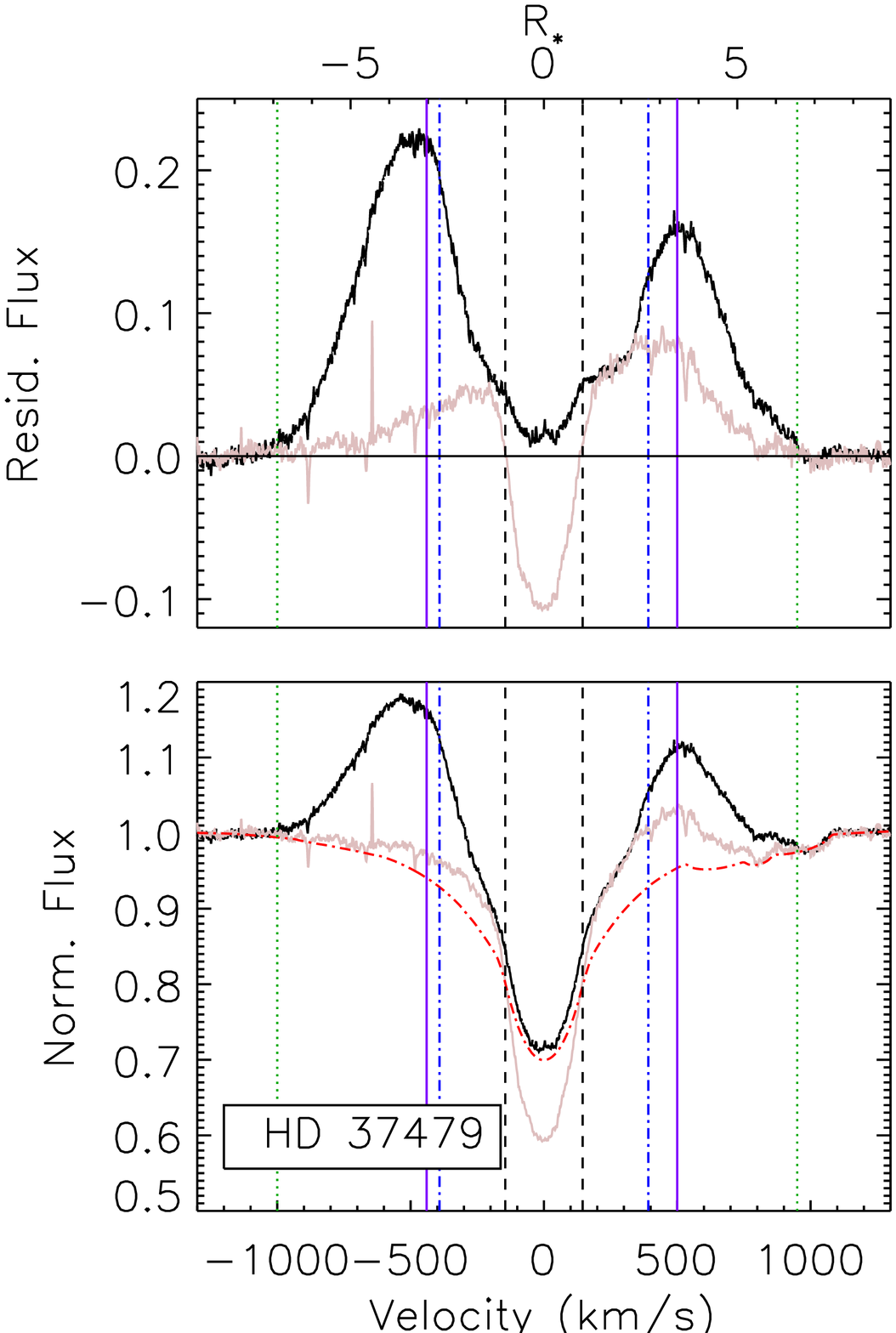} & 
   \includegraphics[trim=20 20 30 30, height=5.5cm]{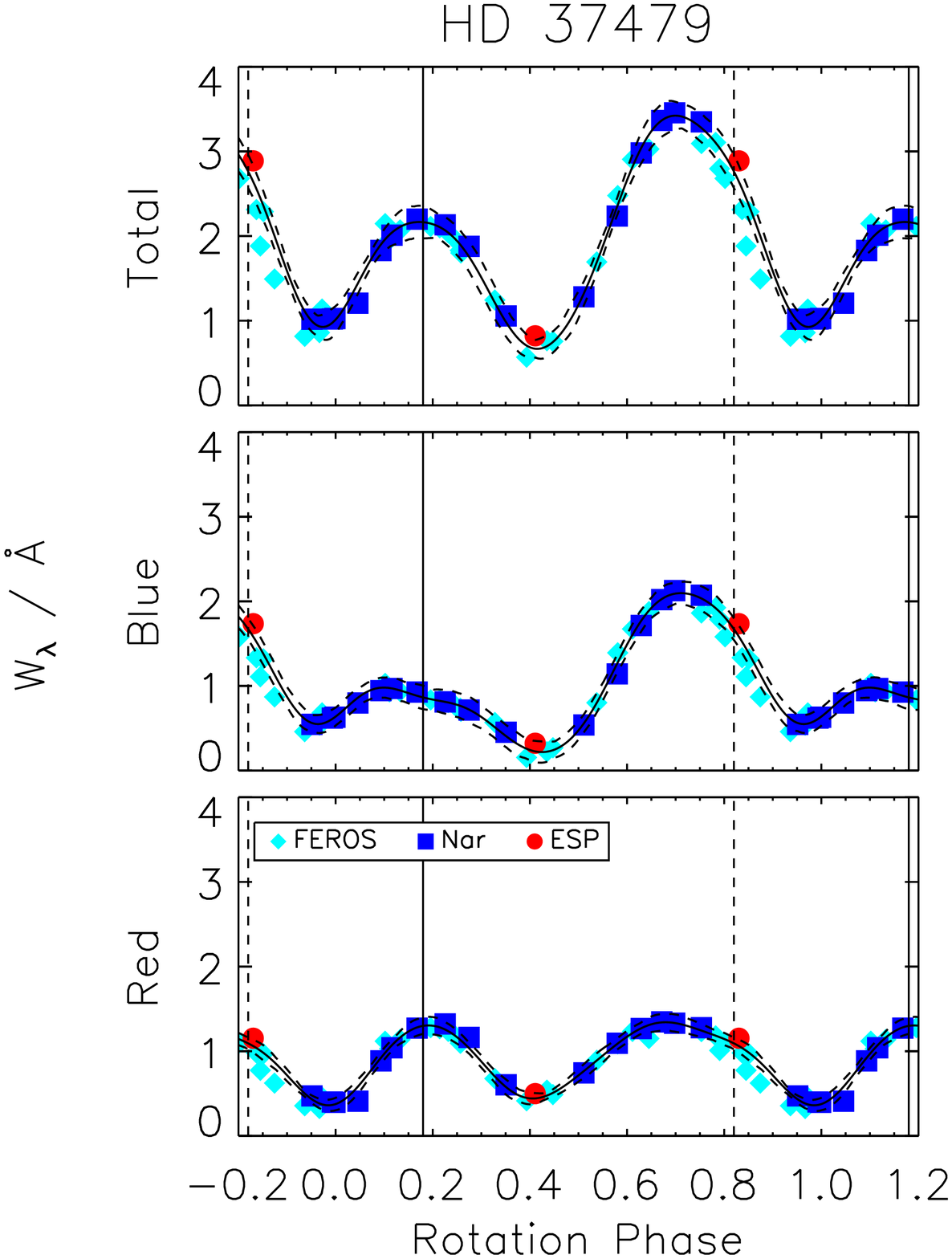} \\ 
\end{tabular}
      \caption[]{{\em Left panels}: (Top) H$\alpha$ residual flux profile of HD\,37479 after subtraction of a synthetic spectrum from the observed flux (Bottom). Emission maximum (black); emission minimum (grey); synthetic photospheric spectrum (dot-dashed red). Vertical black dashed lines show $\pm$\vsini; dot-dashed blue lines show $\pm R_{\rm K}$; solid purple lines shown $\pm r_{\rm max}$; dotted green lines show $\pm r_{\rm rout}$. {\em Right panels}: Emission equivalent width $W_\lambda$ folded with the rotation period for HD\,37479 in (top to bottom) the combined red and blue wings, the blue wing, and the red wing. The legend in the bottom panel indicates the instrument with which each measurement was obtained. Solid and dashed curves show the harmonic fits and fit uncertainties. Solid and dashed vertical lines indicate the negative and positive \bz~extrema, respectively.}
         \label{sigOriE_halpha_minmax}
   \end{figure}

We are interested in three quantities for each star: the radius of maximum emission $r_{\rm max}$, the outermost extent of emission $r_{\rm out}$, and the maximum emission equivalent width $W_\lambda$. To obtain these we began by comparing the observed H$\alpha$ profiles to synthetic H$\alpha$ profiles, calculated using disk-integrated synthetic spectra obtained using either LTE ATLAS9 model atmospheres or from the BSTAR2006 library of synthetic spectra calculated from NLTE TLUSTY models \citep{lanzhubeny2007}. TLUSTY models, which give a better reproducton of the NLTE H$\alpha$ core, were used for single stars with moderate rotation, for which gravity darkening can be neglected; ATLAS9 models were used when rotational distortion becomes important (HD\,142184, HD\,182180, and HD\,345439), or for binary stars with companions below the 15 kK cutoff of the BSTAR2006 library (HD\,35502, HD\,36485, HD\,37017, and HD\,156324). Synthetic spectra obtained from ATLAS9 models were those calculated for use with the {\sc bruce-kylie} spectrum synthesis suite \citep{2014ascl.soft12005T}, and include spectra determined for 20 different limb angles. Limb darkening is thus automatically accounted for in the LTE spectra; for TLUSTY spectra, we used the \cite{2016MNRAS.456.1294R} limb darkening coefficients adopted in Paper III. For stars with critical rotation fractions $\omega > 0.8$, for which rotational distortion becomes salient, oblateness and gravity darkening were accounted for in the disk integration (e.g. Paper II); this was necessary only for HD\,142184, HD\,182180, and HD\,345439. For binary systems, synthetic spectra were calculated individually for each observation, with all stellar components moved to their individual radial velocities. The atmospheric parameters for the binary companions are also given in Paper II. 

The characteristic radii and method of emission equivalent width measurement are illustrated in Fig.\ \ref{sigOriE_halpha_minmax} for the case of $\sigma$ Ori E; the H$\alpha$ profiles of the remaining stars are shown in Figs.\ \ref{halpha_ind1} and \ref{halpha_ind2}. Model parameters were obtained from the atmospheric parameters presented in Paper II. 

Due to the rigid-body rotation of the CM, the line of sight velocity $v_r$ is directly proportional to the projected distance from the star, thus $v_r/v_{\rm rot}\sin{i} = r/R_*$. Using the spectrum identified as having the maximum total emission, the residual flux was used to identify $r_{\rm max}$ and $r_{\rm out}$ by eye (see Fig.\ \ref{sigOriE_halpha_minmax}). These radii were identified by eye since an automated method can yield incorrect results if there are significant differences between the actual surface abundances and the solar metallicity assumed in the models, especially in the vicinity of the C~{\sc ii} lines in the red wing. Uncertainties in these parameters were propagated from uncertainties in \vsini. 

As can be seen in Fig.\ \ref{sigOriE_halpha_minmax}, the red and blue emission bumps do not necessarily yield the same values of $r_{\rm max}$ and $r_{\rm out}$. This can be a consequence of asymmetry in the emission profile due to a magnetic field that is not purely dipolar \citep[e.g.][]{2015MNRAS.451.2015O}. When the two halves of the line yield different characteristic radii, the value corresponding to the strongest emission bump was adopted. 

Equivalent widths $W_\lambda$ were measured in three regions: from the blue edge of emission to $-$\vsini, between $\pm$\vsini, and from $+$\vsini~to the red edge of emission. The blue and red $W_\lambda$ were then combined into a total $W_\lambda$. The region within $\pm$\vsini~was excluded for two reasons. First, several of the stars display enhanced absorption in this region at some phases due to eclipsing of the star by the CM. Second, the shape of the CM accumulation surface is strongly dependent on $\beta$: at small $\beta$ the CM is essentially a ring evenly distributed around the star (since the rotational and magnetic equators are similar), while at large $\beta$ the plasma is strongly concentrated at the intersections of the magnetic and rotational equatorial planes. A non-eclipsing star with small $\beta$ will therefore have emission at all velocities across the line profile, whereas a non-eclipsing star with large $\beta$ will, at maximum emission, display emission only outside of $\pm R_{\rm K}$. Therefore, by considering only the region outside of $\pm$\vsini, the `extra' emission possessed by low-$\beta$ stars is excluded, automatically accounting for any differences in total emission strength introduced by magnetospheric geometry. A third reason is that LTE spectra significantly underestimate the depth of the H$\alpha$ core; while non-LTE TLUSTY \citep{lanzhubeny2007} synthetic spectra were used for the majority of the stars, this was not possible for some of the binary stars with companions cooler than 15 kK \citep[HD\,35502, HD\,36485, and HD\,37017][]{2016MNRAS.460.1811S,leone2010,1998AA...337..183B}, or for the 6 stars analyzed in Appendix \ref{appendix:newstars}, which also have \teff~below this threshold. 

$W_\lambda$ was turned into an emission EW by subtracting the $W_\lambda$ measured from the synthetic spectra in the same line regions; for binaries, this was done for each individual spectrum. Note that $W_\lambda$ is defined as a positive number, such that higher values indicate stronger emission. The peak emission strength was then determined by phasing the blue, red, and total emission $W_\lambda$ with the stellar rotation periods (from Paper I), fitting a harmonic function, and determining the peak value of this function. This is illustrated in Fig.\ \ref{sigOriE_halpha_minmax} for $\sigma$ Ori E; the remaining stars' $W_\lambda$ curves are shown in Appendix B. Fits were used in place of measured values in order to avoid biasing the results due to noise introduced by e.g.\ telluric contamination, which is significant in some of the data. 

The emission of binary stars is diluted by the contribution of the non-magnetic companions to the total flux. The degree of dilution in the vicinity of H$\alpha$ was estimated from the synthetic spectra SEDs, multiplied by the area of each star, in order to determine the fraction $f_*/f_{\rm tot}$ of the magnetic star's flux to the total flux (see Paper II for the atmospheric parameters of companion stars). $W_\lambda$ was then corrected by multiplying it by the inverse of this ratio. Binary corrections range from an 18\% to an 82\% increase in $W_\lambda$.

Since $W_\lambda$ varies as a function of rotational phase, it is clear that the emitting plasma must be opaque for at least some, and possibly all, projection angles. Discounting the variability caused by mutual eclipses of CM plasma and star, the variation of emission strength indicates that it must be the change in projected area that leads to variable emission, since the volume of magnetically confined plasma is constant\footnote{Note, however, that the CM plasma is almost geometrically thin, having a scale height on the order of 0.1 $R_*$ as compared to a radial extent of tens of stellar radii \citep{town2005c}, and it cannot at this stage of the analysis be discounted that CMs may {\em not} be optically thick when viewed face on.}. The phase of maximum emission occurs when the the magnetic equatorial plane is closest to parallel with the plane of the sky (magnetic axis aligned with the line of sight), i.e.\ when the projected area of the CM is at a maximum. If the magnetic pole does not become perfectly aligned with the line of sight, this maximum $W_\lambda$ may not reflect the true maximum emission strength. Since the minimum value of the angle $\alpha$ between the magnetic axis and the line of sight is not the same for all stars, this is a source of scatter in the measurements. The minimum value $\alpha_{\rm min}$ is given by

\begin{equation}\label{cosalpha}
\cos{\alpha_{\rm min}} = \sin{\beta}\sin{i_{\rm rot}} + \cos{\beta}\cos{i_{\rm rot}}.
\end{equation}

In the RRM model, the warped disk of the CM has an axis roughly halfway between the magnetic and rotational equatorial planes. In the appendix published by \cite{town2008} an expression was developed for the approximate angle $\nu$ between the disk normal and the rotational axis:

\begin{equation}\label{nu}
\nu = \beta - \tan^{-1}\left(\frac{\sin{2\beta}}{5 + \cos{2\beta}}\right),
\end{equation}

\noindent resulting in $\nu$ equal to or somewhat less than $\beta$, with the largest difference at intermediate angles. While $\nu$ is not constant across the disk due to the disk's warping, it is accurate at the intersections of the magnetic and rotational equatorial planes i.e.\ precisely where the strongest emission is expected. We therefore calculate the factor $\cos{\alpha_{\rm D}}$ (the angle between the line of sight and the CM) by substituting $\nu$ for $\beta$ in Eqn.\ \ref{cosalpha}.

Normalizing the projected area of the CM to its value when $\cos{\alpha_{\rm D}} = 1$, i.e.\ when the CM normal is perfectly aligned with the line of sight, the correction factor for $\alpha_{\rm D}$ is then simply $1/\cos{\alpha_{\rm D}}$\footnote{While this simple treatment would result in an infinite correction if $\cos{\alpha_{\rm D}} = 0$, in practice none of the stars have such a small value. A more accurate correction should of course include the finite thickness of the CM.}. $\cos{\alpha_{\rm D}}$ is given in Table \ref{obstab}. The maximum value of this correction is an increase of about 90\% for HD\,142184, with a mean correction of 20\%.

In addition to measuring emission strength for stars with detectable H$\alpha$ emission, we also determined upper limits for stars without H$\alpha$ emission but nevertheless having $\log{B_{\rm K}} > 1.5$. The same methodology was employed in determining upper limits on emission strength for stars without H$\alpha$ emission, with the obvious exception that characteristic radii could not be determined. As with the H$\alpha$-bright stars, positive equivalent width is defined as emission (or in this case pseudo-emission) above the synthetic reference spectrum. Integration ranges were either between \vsini~and $10\times$\vsini, or, for stars with very broad spectral lines, \vsini~and the red edge of the C~{\sc ii} 657.8 nm line; the latter was chosen so as to avoid contaminating the results with line profile variability from this line. As is demonstrated in Appendix \ref{appendix:noem}, the H$\alpha$ lines of many of these stars show EW variability that is coherent with the rotation period, with a typical amplitude of around 0.01 nm. This variability is almost certainly unrelated to circumstellar emission, but instead due to modifications to the Stark broadening due to e.g.\ changes in the partial pressure of H due to He abundance patches \citep[e.g.][]{2015MNRAS.447.1418Y,2015MNRAS.449.3945S}, magnetic pressure broadening due to the Lorentz force \citep[e.g.][]{shulyak2007,2010A&A...509A..28S}, or chemical spots \citep[e.g.][]{2009A&A...499..567K}. Other stars show H$\alpha$ variations that are not coherent with the rotation period, but instead are likely due to pulsations. Determining the sources of these variations in these cases is beyond the scope of this work. However, we use the maximum pseudo-emission of the H$\alpha$-absorption stars for the upper limits, which due to these sources of variability are dominated by systematics rather than photon noise. The variability of stars without H$\alpha$ emission is typically an order of magnitude less than the variability of emission-line stars. There is no reason to expect that it is not also present in stars with emission, although the effect of this should be negligible due to the large differences in amplitude.

\subsection{Emission strength}

   \begin{figure*}
   \centering
   \includegraphics[trim=0 0 0 0, width=\hsize]{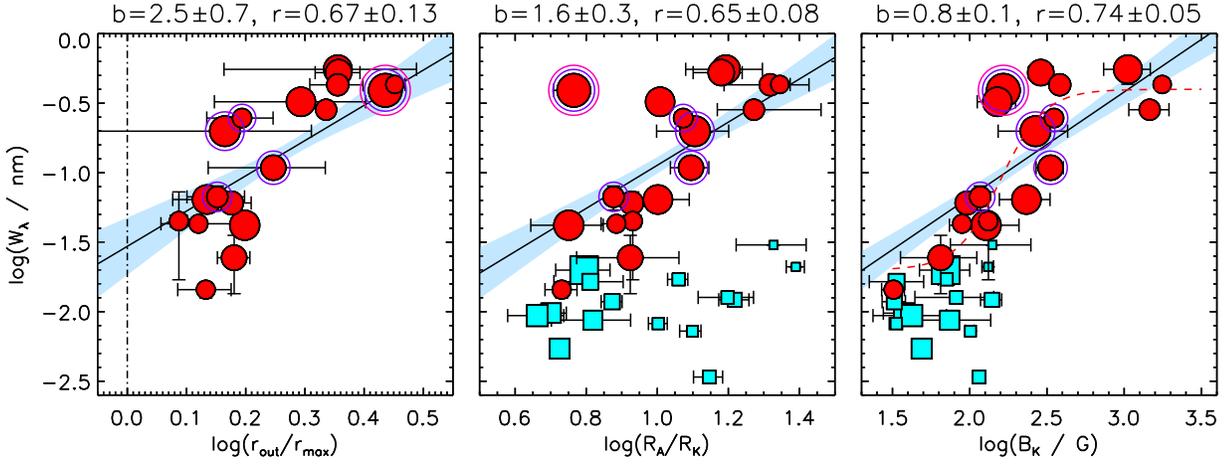} 
      \caption[]{Maximum emission strength $\log{W_{\lambda}}$ as a function of (left to right) $\log{(r_{\rm out}/r_{\rm max})}$, $\log{(R_{\rm A}/R_{\rm K})}$, and $\log{B_{\rm K}}$. Symbol size is proportional to luminosity. HD\,164492C is highlighted with a large pink circle. Spectroscopic binaries are highlighted with purple circles. Solid lines and shaded regions indicate the best-fit lines and the 1$\sigma$ uncertainties. Blue squares indicate emission strength upper limits for stars without detected emision. Slopes $b$ and correlation coefficients $r$ are given in the titles. The dash-dotted line in the first panel indicates $r_{\rm out} = r_{\rm max}$, the threshold below which emission is impossible. In the third panel a sigmoid curve is shown with a red dashed line.}
         \label{wlam_rark_bk}
   \end{figure*}

   \begin{figure*}
   \centering
   \includegraphics[trim=0 0 0 0, width=\hsize]{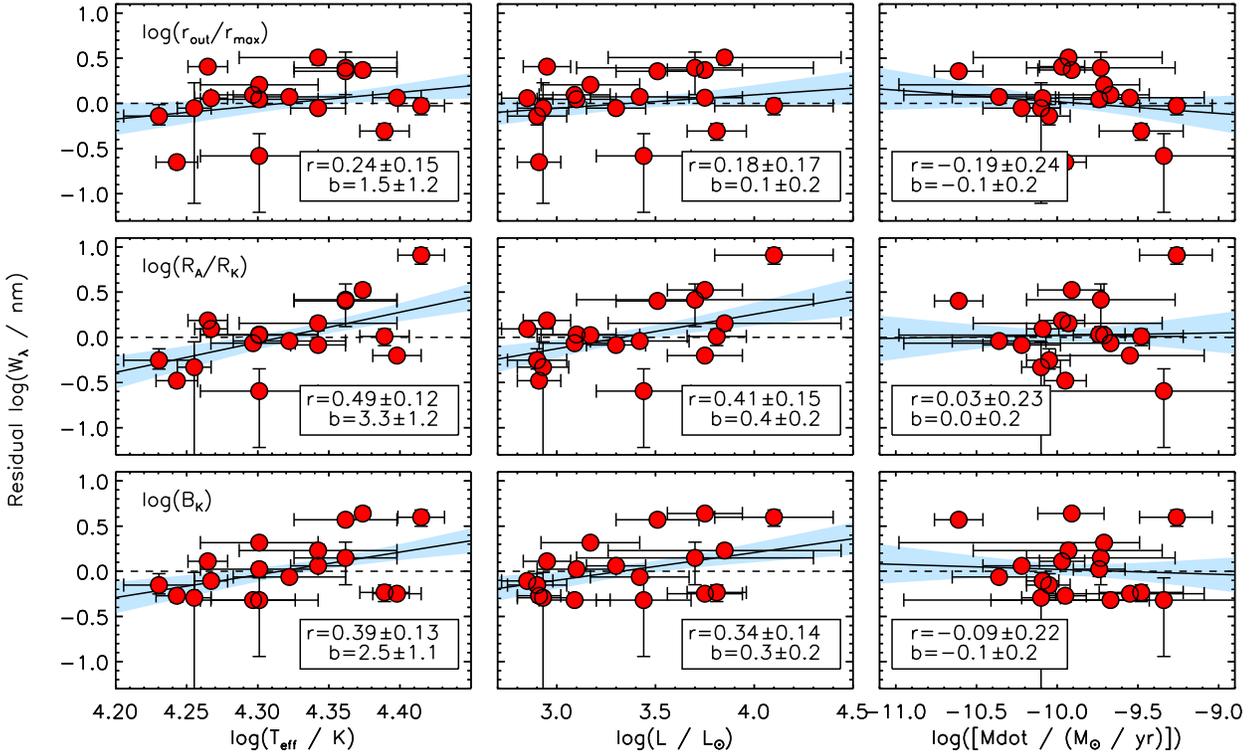} 
      \caption[]{Residual $\log{W_{\lambda}}$ for stars with emission after detrending with the best-fit lines in Fig.\ \ref{wlam_rark_bk} for (top to bottom) $\log{(r_{\rm out}/r_{\rm max})}$, $\log{(R_{\rm A}/R_{\rm K})}$, and $\log{B_{\rm K}}$, as a function of $\log{T_{\rm eff}}$, $\log{L}$, and $\log{\dot{M}}$. Solid lines and shaded regions indicate regressions and uncertainties; dashed lines indicate $\log{W_{\rm lambda}} = 0$. Legends give correlation coefficients $r$ and regression slopes $b$.}
         \label{wlam_rark_bk_resid}
   \end{figure*}

The maximum emission strength $\log{W_\lambda}$ is shown as a function of $\log{(r_{\rm out}/r_{\rm max})}$, which serves as an observational proxy to the area of the CM analagous to the theoretical quantity \rark, in the left panel of Fig.\ \ref{wlam_rark_bk}. The Pearson's correlation coefficient \citep[PCC;][]{1895RSPS...58..240P} is $r \sim 0.7$, indicating a significant correlation. The slope of the relationship is $2.5 \pm 0.7$, suggesting that emission strength goes approximately as the square or cube of $r_{\rm out}/r_{\rm max}$. This indicates that emission strength varies as either the observed area or volume of the CM. The former is consistent with the CM being optically thick even when seen face-on.

In the middle panel of Fig.\ \ref{wlam_rark_bk} $\log{W_\lambda}$ is shown as a function of $\log{R_{\rm A}/R_{\rm K}}$, the theoretical size of the CM relative to the host star. The correlation coefficient is lower than achieved for $\log{r_{\rm out}/r_{\rm max}}$, although the difference is not statistically significant. However, this is largely due to the influence of HD\,164492C (highlighted with a pink circle), which is a clear outlier from the general trend. This may be because HD\,164492C is the hottest star in the sample, with the strongest wind. If this star is removed, $r$ rises to $0.79 \pm 0.07$ and the slope rises to $2.0 \pm 0.4$, i.e.\ slightly better results than achieved using $\log{(r_{\rm out}/r_{\rm max})}$ (since HD\,164492C is not an outlier in the first case, removing it does not change the results). The slope is still consistent with the emission strength being a function of the area of the CM; in this case, however, it is inconsistent with the volume. 

The right panel of Fig.\ \ref{wlam_rark_bk} shows emission strength as a function of $\log{B_{\rm K}}$, which yields a tighter relationsip than is achieved for either $\log{(r_{\rm out}/r_{\rm max})}$ or $\log{(R_{\rm A}/R_{\rm K})}$ despite being wholly ignorant as to the area of the CM. HD\,164492C is less of an outlier in this case as compared to $\log{(R_{\rm A}/R_{\rm K})}$, although it still lies above the best fit; removing it improves the correlation coefficient to $r = 0.78 \pm 0.05$, but the difference is not statistically significant.

In all three panels of Fig.\ \ref{wlam_rark_bk} symbol size is proportional to $\log{L}$. In a leakage scenario in which the mass-loading rate by the wind is relevant to the amount of material in the CM and, hence, the emission strength, we might expect to see that the stars with the strongest emission are also the stars with the highest luminosities. To the contrary, no such difference is discernable. 

The middle and right panels of Fig.\ \ref{wlam_rark_bk} include the upper limits on $\log{W_\lambda}$ for stars without H$\alpha$ emission. These are comparable to or below the weakest emission strengths in the H$\alpha$-bright sample. The more luminous stars without emission have values of $\log{(R_{\rm A}/R_{\rm K})}$ and $\log{B_{\rm K}}$ comparable to the values of the stars with the weakest emission. The least luminous stars have $\log{(R_{\rm A}/R_{\rm K})}$ comparable to H$\alpha$-bright stars with the strongest emission, and $\log{B_{\rm K}}$ values consistent with weak-to-intermediate emission strength.

While a linear regression is the easiest model to test, and does a reasonable job of reproducing the trends in emission strength with CM area and the intensity of magnetic confinement, it is not necessarily the correct model. The rapidity of emission onset with increasing $B_{\rm K}$ -- with emission abruptly appearing at about 100 G, as though a switch were flipped -- is suggestive of an extremely rapid increase in emission strength once the threshold in $B_{\rm K}$ has been reached. However, the clear dependence of $\log{W_\lambda}$ on CM area suggests that once the region near \rk~has become optically thick, increasing the emission strength requires the CM to be optically thick over an increasing area. Due to the $1/r^3$ dependence of the magnetic field it is increasingly difficult to confine plasma further from the star, so we might expect emission strength to be only a weak function of $B_{\rm K}$ once the central region has become optically thick. This concept is illustrated with the sigmoid curve in the right panel of Fig.\ \ref{wlam_rark_bk} (red dashed line), where we used a function of the form $y = A / (1 + e^{B(x + C)}) + D$. In this scenario, once the threshold for optical thickness is reached, the emission strength rapidly increases, after which its growth levels off. While there are insufficient points to provide good constraints to a fit, a sigmoid does appear to provide a better fit to $W_\lambda - B_{\rm K}$ relationship. 

If a leakage mechanism operating according to diffusion or drift is responsible for mass balancing within the CM, then there should also be some dependence of the emission strength on the mass-loss rate \mdot, or on a proxy for \mdot~such as \teff~or $\log{(L/L_\odot)}$. There is no correlation of $W_\lambda$ with any of these quantities (the respective correlation coefficients of \mdot, $\log({T_{\rm eff}/{\rm K})}$, and $\log{(L/L_\odot)}$ are $-0.1 \pm 0.2$, $0.30 \pm 0.12$, and $0.21 \pm 0.14$), but it is possible that there might be some residual dependence once the dependence on $\log{(R_{\rm A}/R_{\rm K})}$ or $\log{B_{\rm K}}$ is removed. Fig.\ \ref{wlam_rark_bk_resid} shows the residual $\log{W_\lambda}$, after de-trending with the best-fit lines in the top panels, as a function of these three quantities. There is no consistent trend according to the best-fit lines: while the residual equivalent width increases slightly with increasing \teff~and $\log{L}$, it decreases slightly with increasing \mdot, which is the precise opposite of what is expected if emission strength depends on a competition between mass-loading via the wind and a diffusion/drift leakage mechanism. Correlation coefficients are always low: the highest is $0.49 \pm 0.12$ (for \teff~and the \rark~residuals). None of the regression slopes are significant at the 3$\sigma$ level, and most are around 1$\sigma$. The potentially significant result using \teff~and the \rark~residuals is not robust against removal of HD\,164492C, which reduces the correlation coefficient to $r = 0.41 \pm 0.14$, below $3\sigma$ significance. We conclude that there is no evidence for any dependence of the emission strength on the properties of the stellar wind. These results are unchanged if the same tests are performed using \cite{krticka2014} mass-loss rates instead of \cite{vink2001} mass-loss rates.

\subsection{Characteristic radii}

   \begin{figure}
   \centering
   \includegraphics[trim=0 0 25 0, width=\hsize]{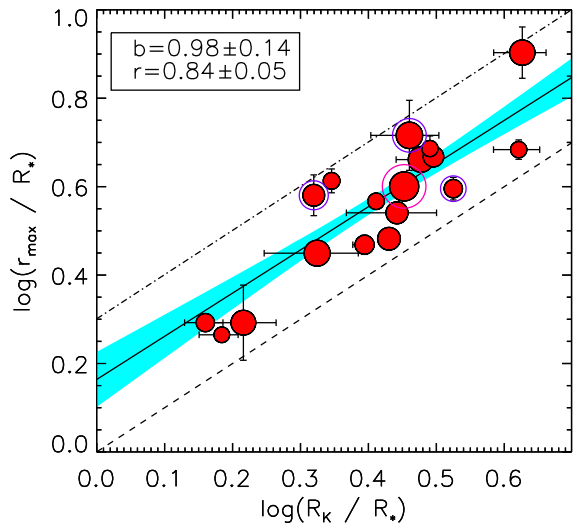} 
      \caption[]{As Fig.\ \ref{wlam_rark_bk} for the radius of maximum emission $r_{\rm max}$ as a function of the Kepler corotation radius \rk. The dashed line shows $r_{\rm max} = R_{\rm K}$,and the dot-dashed line shows $r_{\rm max} = 2R_{\rm K}$.}
         \label{rmax_rk}
   \end{figure}

The radius of maximum emission $r_{\rm max}$ should be a proxy to the Kepler radius. Fig.\ \ref{rmax_rk} compares these two quantities, and demonstrates that the correlation is indeed highly significant ($r=0.84$), with a slope consistent with unity. However, the stars are distributed between $r_{\rm max} = R_{\rm K}$ and $r_{\rm max} = 2R_{\rm K}$, with a mean value of about 25\% higher than \rk. 

   \begin{figure}
   \centering
   \includegraphics[trim=0 0 100 0, width=\hsize]{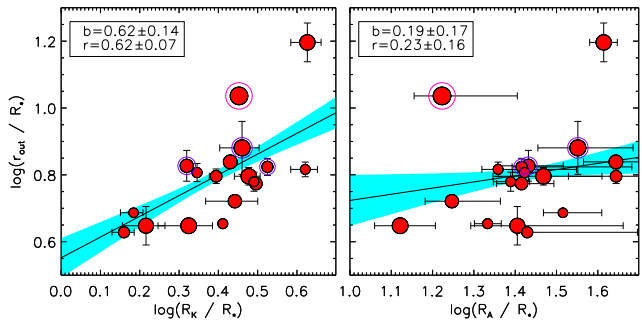}
      \caption[]{As Fig.\ \ref{wlam_rark_bk} for the radius of outermost extent of emission $r_{\rm out}$ as a function of the Kepler corotation radius \rk~and the Alfv\'en radius \ra.}
         \label{r0_rk_ra}
   \end{figure}

Just as $r_{\rm max}$ is expected to be the observational counterpart to \rk, we might expect the outermost extent of emission $r_{\rm out}$ to be a proxy for \ra. Fig.\ \ref{r0_rk_ra} demonstrates that this is only approximately the case: the correlation between $r_{\rm out}$ and \ra~is very weak, and is not significant above the 2$\sigma$ level. By contrast, $r_{\rm out}$ correlates very strongly with \rk. This is probably explained by the necessity that $r_{\rm out}$ be greater than $r_{\rm max}$ (and hence \rk). 

HD\,164492C (highlighted) is also an outlier in Fig.\ \ref{r0_rk_ra}. Removing this star from the regression does not affect the relationship of $r_{\rm out}$ to \rk, but does lead to a more significant relationship between $r_{\rm out}$ and \ra, with a steeper slope (0.5) and a higher correlation coefficient (0.6). 

   \begin{figure}
   \centering
   \includegraphics[trim=0 0 100 0, width=\hsize]{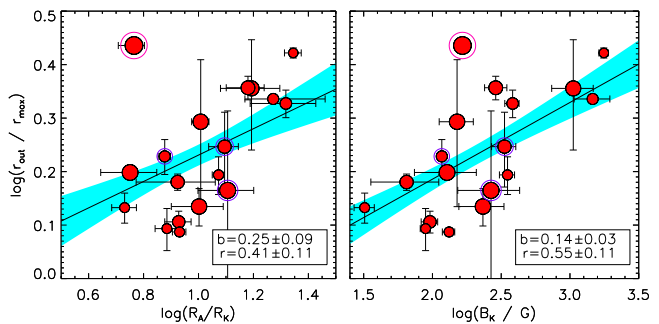}
      \caption[]{As Fig.\ \ref{wlam_rark_bk} for $r_{\rm out}/r_{\rm max}$ as a function of $R_{\rm A}/R_{\rm K}$ and $B_{\rm K}$.}
         \label{r0rmax_rark_bk}
   \end{figure}

Since $r_{\rm out}/r_{\rm max}$ and $R_{\rm A}/R_{\rm K}$ are both measures of the area of the CM, we expect them to correlate. The left panel of Fig.\ \ref{r0rmax_rark_bk} demonstrates that they do show a weak correlation ($r=0.4$), although this is once again affected by HD\,164492C. Removing the outlier from the regression improves the correlation to $r=0.6$ and increases the slope by about a factor of 2. As with emission strength, we again get a better relationship between $B_{\rm K}$ and $r_{\rm out}/r_{\rm max}$, which has a stronger correlation ($r=0.6$); removing HD\,164492C improves the correlation to $r=0.7$ (although the change is within the uncertainty). The slope of the relationship is however quite small, indicating that while an increase in $B_{\rm K}$ drives an increase in $r_{\rm out}/r_{\rm max}$ it does so very slowly. 

\subsection{Emission profile morphology}

   \begin{figure}
   \centering
   \includegraphics[trim=40 20 30 30, width=\hsize]{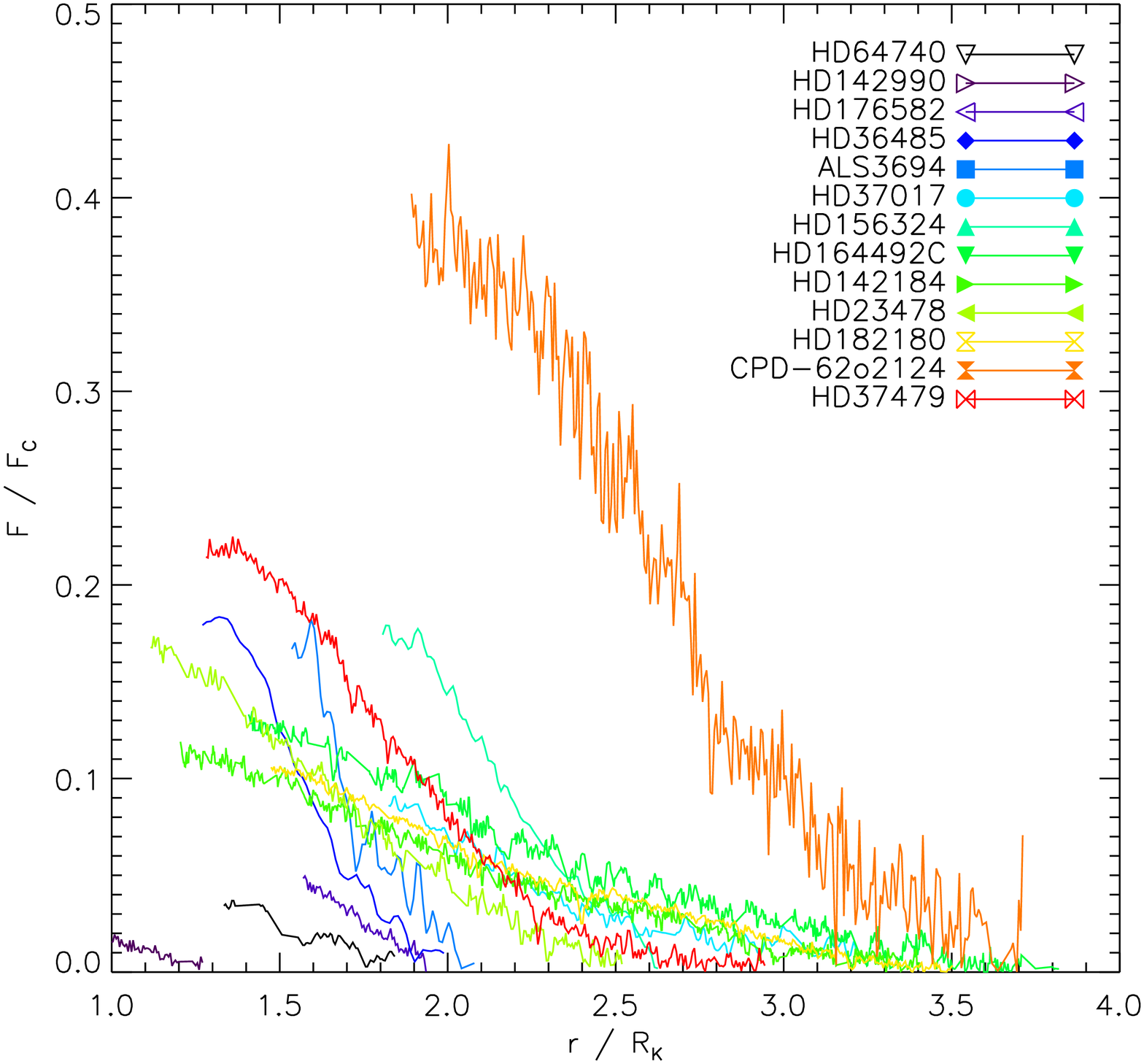} 
   \includegraphics[trim=40 20 30 30, width=\hsize]{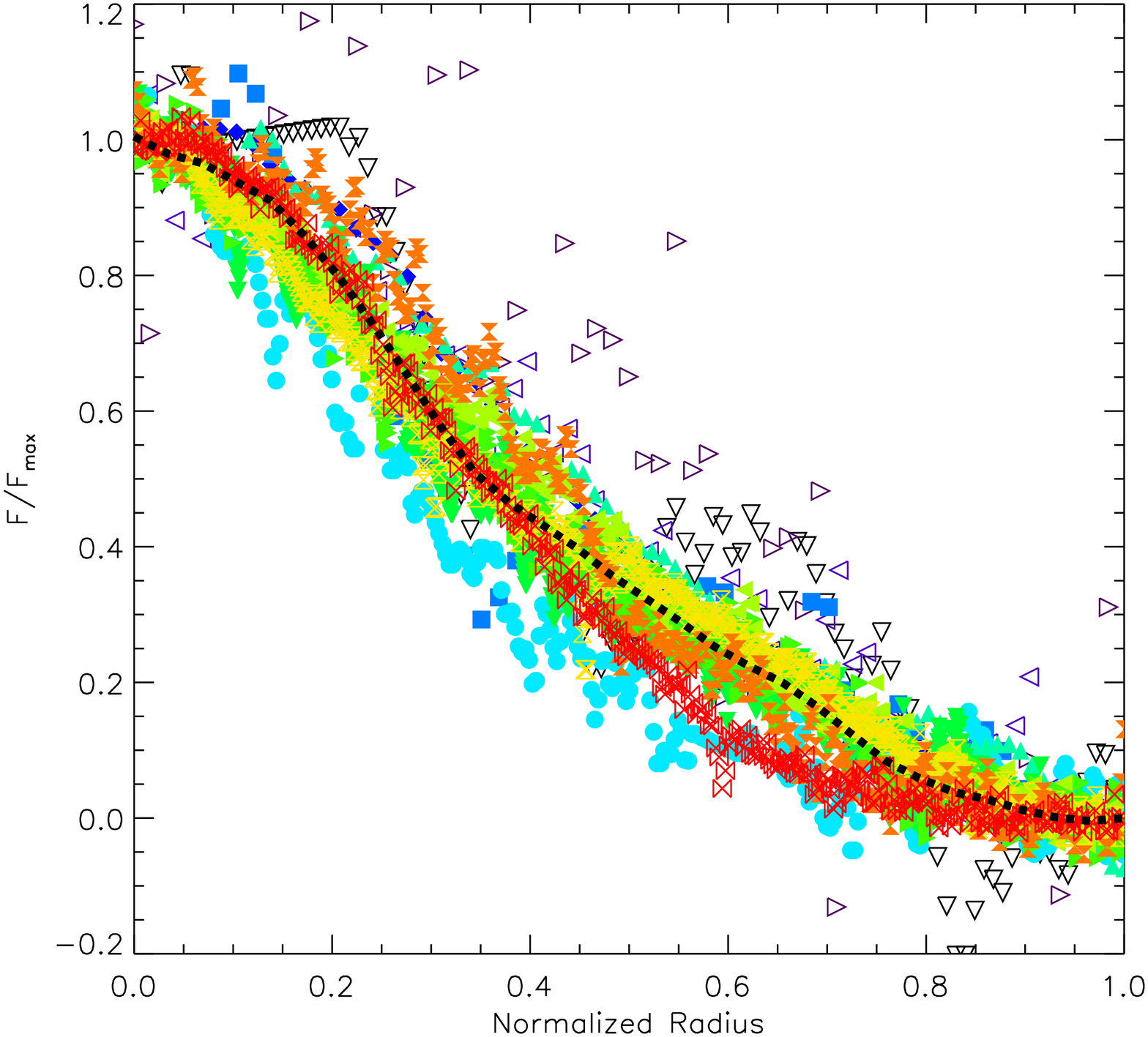}
      \caption[]{{\em Top}: Un-normalized emission wings for the stars with the cleanest emission profiles, as a function of stellar radius, from the observations obtained at maximum emission. Emission wings are shown from $r_{\rm max}$ to $r_{\rm out}$. {\em Bottom}: Emission wings with flux normalized to the peak flux, and radius normalized such that $r_{\rm max} = 0$ and $r_{\rm out}=1$. The thick dotted black line shows the average across all emission wings. The symbols in the bottom panel correspond to those in the legend of the top panel.}
         \label{emission_wings}
   \end{figure}

If mass flow within CMs is governed by similar physics, we should expect that their emission profiles will be similar to one another. This is best examined in the region outside of $r_{\rm max}$, since inside of this radius the different magnetic geometries will combine with projection effects to give rise to a variety of different shapes that might obscure the underlying physical similarity. The top panel of Fig.\ \ref{emission_wings} shows the emission wings of several of the stars with the cleanest emission profiles, measured between $r_{\rm max}$ and $r_{\rm out}$ by subtracting the same synthetic photospheric spectrum used to measure $W_\lambda$. When the red and blue emission lobes at maximum emission were of different strengths, the stronger was chosen. The horizontal axis is scaled to \rk~(i.e.\ $(v_r/v_{\rm rot}\sin{i})/R_{\rm K}$ for the purposes of display. It is immediately apparent that stars with emission extending further out also have stronger peak emission. 

Exact comparison of the emission wings in the top panel of Fig.\ \ref{emission_wings} is precluded by two factors. First, the stars have a variety of different rotational axis inclinations $i_{\rm rot}$. As $i_{\rm rot}$ increases, rotational broadening spreads the emission out over a larger number of velocity bins, depressing the peak emission strength. Second, $r_{\rm max}$ is not usually identical to \rk, and the emission peaks are therefore found at a variety of different radii. To address these difficulties, we renormalized the horizontal scale by setting $r_{\rm max} = 0$ and $r_{\rm out}=1$, and renormalized the vertical scale to be unity at $r_{\rm max}$. The results are shown in the bottom panel of Fig.\ \ref{emission_wings}. The rescaled emission wings of the various stars exhibit a remarkable degree of similarity, with the majority of them corresponding very closely to the mean emission wing profile (black dotted line). In most cases the emission profile is convex close to $r_{\rm max}$, becoming concave at about the mid-point between $r_{\rm max}$ and $r_{\rm out}$. 

If the innermost CM is optically thick and we approximate this region as a circular disk, the emission strength at a given velocity bin is directly proportional to the area of a velocity isocontour of that circle. This directly gives rise to a convex emission wing in the inner region. The transition to concavity about halfway through the emission profile likely reflects the local density falling below the optically thick limit. This model is worked out in detail by \cite{owocki2020}.

\section{Discussion}\label{sec:discussion}

\subsection{Mass Balancing via Centrifugal Leakage}

   \begin{figure}
   \centering
   \includegraphics[trim=40 20 30 30, width=\hsize]{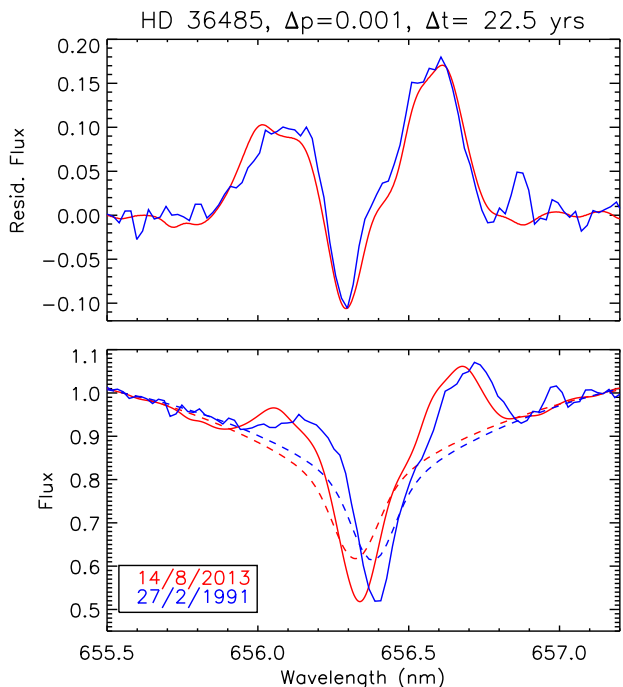} 
      \caption[]{Comparison between H$\alpha$ observations of HD\,36485 obtained at similar phases ($\Delta p = 0.001$) but separated in time by over 20 years. The bottom panel shows observations (solid lines) and synthetic spectra (dashed lines). Note that HD\,36485 is an SB2. The top panel shows residual flux, shifted to the rest frame of the primary. The 1991 DDO spectrum has a lower spectral resolution than the 2013 ESPaDOnS spectrum; the spectral resolution of the latter has therefore been reduced to that of the former for purposes of comparison.}
         \label{HD36485_halpha_deltat}
   \end{figure}

The onset of emission appears to depend almost exclusively on the area of the CM, and indeed can be predicted with remarkable precision by a simple threshold in $B_{\rm K}$ (Fig. \ref{lum_rark2}). Emission strength futhermore depends only on the area of the CM, and is strongly sensitive to $B_{\rm K}$ (Fig.\ \ref{wlam_rark_bk}). By contrast, there is essentially no dependence of either emission onset or strength on the mass-loss rate (Fig.\ \ref{wlam_rark_bk_resid}). This is impossible to explain in terms of a diffusion/drift mechanism, since with such a process the leakage would be competing with mass-feeding by the wind, which would presumably lead to some dependence on the mass-loss rate. On the other hand, all emission properties are very easily explained by centrifugal breakout, since in this process what determines the onset of emission, and the extent over which the CM is optically thick, is purely a function of the capacity of the magnetic field to confine the plasma. 

This leaves us with something of a conundrum, since breakout has never been detected. \cite{town2013} saw no change in the light curve of $\sigma$ Ori E over about 20 rotational cycles of high-precision space photometry, nor has any intrinsic change in  H$\alpha$ emission morphology ever been reported. A striking example is provided by the H$\alpha$ spectra of HD\,36485, which span over 20 years. Two spectra of this star, obtained at similar rotation phases but over 20 years apart, are shown in Fig.\ \ref{HD36485_halpha_deltat}. The residual flux of the two observations is almost indistinguishable within the limits of S/N. Significant changes in H$\alpha$ are therefore not seen at least over a timescale of decades in HD\,36485. Notably, $\log{B_{\rm K}} = 2.07 \pm 0.03$ is quite low; being near the threshold for the onset of detectable H$\alpha$ emission, if any star in the sample were to show intrinsic changes in H$\alpha$ due to large-scale breakout events, HD\,36485 would be one of the best candidates. It is furthermore worth noting that no CM host star has ever been confirmed as an X-ray flare source\footnote{While \cite{2004AA...421..715S} reported X-ray flares from $\sigma$ Ori E, these almost certainly originate from the low-mass companion star discovered by \cite{2009AA...493..931B}. Similarly, \cite{2016A&A...592A..88P} reported X-ray flaring around the magnetic B-type star $\rho$ Oph C, however this star has a K-type companion which is probably the source of the flares.}, which might be an expected consequence of large-scale magnetic reconnection accompanying a breakout event. 

There is therefore no evidence for {\em large-scale} reorganization of the magnetosphere during breakout events. However, the close dependence of H$\alpha$ emission properties on $B_{\rm K}$ can only be consistent with breakout. In their Appendix A \cite{town2005c} demonstrated that the limiting mass of a CM can be predicted on the basis of $B_{\rm K}$, which is the primary factor governing the breakout density. The dependence of emission strength on $B_{\rm K}$ suggests that the magnetospheric masses of the H$\alpha$-bright stars are indeed at their breakout limits. Further, the absence of evidence for intrinsic changes in their magnetospheric diagnostics suggests that their CMs are {\em always} at the breakout limit. If this is the case, breakout must be occurring on a continuous basis, but must also occur on a small enough spatial scale that magnetospheric diagnostics are left undisturbed.

It should be noted that the MHD simulations exploring breakout conducted by \cite{ud2006,ud2008} were performed in 2D, and that it is very possible that, when averaged over the full 3D CM, the breakout events reported by \citeauthor{ud2008} in these studies might become an essentially continuous process. Along these lines, it is also worth noting that the spatially and temporally homogeneous ion source feeding a stellar magnetosphere may mean that the magnetosphere settles into a steady state. In fact, examining the simulations with the strongest magnetic confinement and most rapid rotation in their Fig.\ 9, there seems to already be some support for this in the simulations. This is in contrast to planetary magnetospheres, which have external (stellar wind) and internal (volcanic moons) ion sources, which are a) not isotropic with respect to the planetary magnetic field, b) intrinsically temporally variable, and c) in motion relative to the planetary magnetic field. As a result of this, magnetic reconnection within planetary magnetospheres -- e.g.\ magnetotail reconnection events following solar flares -- are impulsive events following the sudden insertion of a large quantity of plasma into the magnetosphere. In the case of stellar magnetospheres, breakout events in the outermost magnetosphere may be essentially continuous, and in consequence may not lead to intrinsic variation in magnetospheric diagnostics. It is interesting to note that \cite{2014ApJS..215...10N} found that CM stars are systematically about 1 dex more luminous in X-rays than predicted by the X-ray Analytical Dynamical Magnetosphere (XADM) model developed by \cite{ud2014}, and clearly stand out from other magnetic early-type stars which are in general about 1 dex {\em less} luminous than predicted by the XADM model. It could be that this additional X-ray luminosity is the signature of continuous breakout at the edge of the CM.

The analytical treatment of breakout timescales in \cite{town2005c} yielded an infinite breakout time at \rk, becoming finite above this and rapidly declining with increasing distance. They evaluated the breakout time for $\sigma$ Ori E as being about 2 centuries at $r = 2R_{\rm K}$. However, solving their equation A6 using the same parameters as they adopted for $\sigma$ Ori E yields the result that, close to \ra~(about 30 $R_*$, well outside $r_{\rm out} = 7.5 R_*$), the breakout time falls to a few days. This can easily fall to a few hours for slightly different stellar, rotational, or magnetic parameters. However, this assumes that the magnetic field lines close to \ra~remain perfectly rigid, which is unlikely the case given the extreme centrifugal stress they are being subjected to at this distance (the rotational velocity at $\sigma$ Ori E's Alfv\'en radius is about 4500 \kms). Since the magnetic field lines in this regime are probably stretching before they snap, it is not unreasonable to suppose that breakout near the edge of the CM may become essentially continuous.

The resolution of the breakout/leakage debate therefore appears to be that mass balancing is in fact governed by breakout, but that breakout essentially acts as the leakage mechanism. Indeed, since this process must be happening continuously, it is probably more appropriate to refer to it as `centrifugal leakage'. The net breakout rate must therefore be identical to the rate of mass-feeding by the wind, with the CM itself having a constant mass. 

\subsection{What can explain the discrepancy between $r_{\max}$ and \rk?}

   \begin{figure}
   \centering
   \includegraphics[trim=0 0 0 0, width=\hsize]{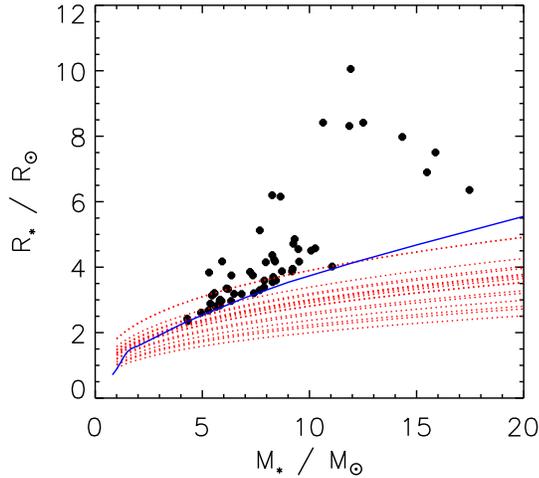}
      \caption[]{Red dotted lines show $R_*$ as a function of $M_*$ if $r_{\rm max} = R_{\rm K}$ for the various stars in this sample. The black dots show the full sample of magnetic B-type stars from the Papers I--III. The solid blue line shows the zero-age main sequence from the \protect\cite{ekstrom2012} models.}
         \label{rk_ms_rs}
   \end{figure}

   \begin{figure}
   \centering
   \includegraphics[trim=0 0 0 0, width=\hsize]{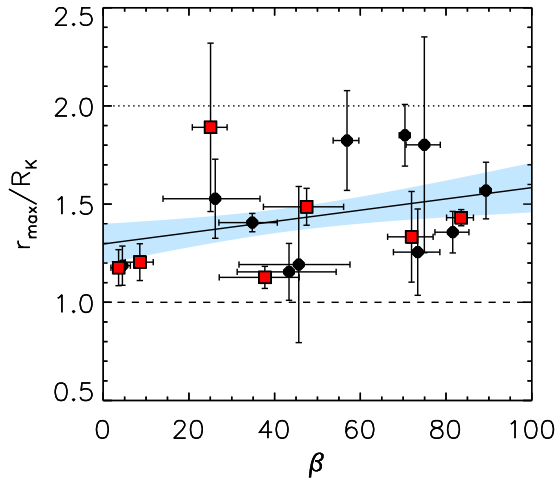}
      \caption[]{Ratio of $r_{\rm max}$ to the Kepler radius \rk~as a function of the magnetic obliquity angle $\beta$. Red squares indicate stars with known or suspected multipolar surface magnetic fields.}
         \label{rmaxrk_beta}
   \end{figure}

It is extremely puzzling that $r_{\rm max}$ is systematically at least 25\% higher than \rk, since both the RRM model and MHD simulations predict that the highest density should be at the Kepler radius, and that therefore the emission should peak exactly at \rk~\citep{town2005c,ud2008}. 

One obvious possibility is that \rk~has simply been incorrectly determined. This could happen if the radius, the mass, or the rotation period of the star is wrong. It could also be a consequence of \vsini~being wrong, since it is \vsini~that determines the scaling of velocity and projected stellar radius. In the majority of cases, $P_{\rm rot}$ and \vsini~are known to very high precision (for instance, in the case of HD\,142184 and HD\,182180, the former is known to within one part in a million and the latter to within a few percent). Errors in these quantities can therefore probably be discarded. To explore the possibility that systematic errors in mass and/or radius are behind the discrepancy, we made the assumption that $r_{\rm max}$ is in fact the true value of \rk~in all cases, and then solved for $R_*$ using Eqn. \ref{wrot}. The results of this test are shown in Fig.\ \ref{rk_ms_rs}. Even under the very conservative assumption that $M_*$ might be anywhere between 1 \msun~and 20 \msun, in the majority of cases fixing $r_{\rm max} = R_{\rm K}$ would require radii smaller than are seen at the zero-age main sequence. Unless the evolutionary tracks of magnetic stars close to the ZAMS are very different from those of non-magnetic stars \citep[which 1D evolutionary models suggest is not the case, e.g.][]{2019MNRAS.485.5843K,2020MNRAS.tmp..227K}, this test suggests that the discrepancy between $r_{\rm max}$ and \rk~cannot be a consequence of inaccurate fundamental parameters, i.e.\ it seems that the discrepancy must be real. 

Since material is apparently failing to accumulate at \rk, we must look for mechanisms that might be able to prevent the predicted accumulation. Revisiting \cite{town2005c}, we see that they assumed that material would accumulate wherever there is a potential minimum along a field line. However, at \rk~the potential is in fact flat, and the potential minima at which the local accumulation surface is located is quite shallow close to \rk. When wind flows from opposite hemispheres collide in a shallow potential, if they are slightly unbalanced their momenta might not exactly cancel, leading to the stronger flow overpowering the weaker flow, pushing it out of the potential minimum, and establishing a siphon flow to the opposite hemisphere. Such imbalances might be caused by either an oblique dipole (which is the case for the majority of stars in the sample), by a surface magnetic field with significant departures from a purely dipolar geometry (which is also not uncommon in this sample), or by surface variations in $\dot{M}$ caused by chemical spots \citep{krticka2014}. Siphon flows were reported in the rigid-field hydrodynamic simulations conducted by \cite{town2007}; while these occurred only below \rk~for the aligned dipole model, they did not comment on whether such flows appeared at or above \rk~for oblique dipoles. It is also worth noting that the MHD simulations of CMs conducted by \cite{ud2008,ud2009} were for aligned dipoles in two dimensions and therefore would not in principle show this phenomenon.

If the discrepancy is indeed driven by departures from an aligned dipole, we might expect there to be a correlation between the magnitude of the discrepancy between \rk~and $r_{\rm max}$ and the magnetic obliquity angle $\beta$. Fig.\ \ref{rmaxrk_beta} shows, although there is a trend of increasing $r_{\rm max}/R_{\rm K}$ with increasing $\beta$, the correlation is not statistically significant (correlation coefficient $r=0.20 \pm 0.15$). Stars with multipolar fields (see Paper I and references therein for how multipolar fields were identified) generally have $r_{\rm max}$ somewhat closer to \rk~than other stars. It is worth noting that multipolar contributions are more easily detected at higher S/N, and that stellar parameters are therefore more precisely determined for these stars; notably, the uncertainties are generally smaller for this subsample. In any case there is no obvious tendency for stars with multipolar magnetic fields to exhibit larger discrepancies in $r_{\rm max}/R_{\rm K}$ than seen in stars with predominantly dipolar magnetic fields.

\subsection{Emission profile asymmetry}

   \begin{figure}
   \centering
   \includegraphics[trim=0 0 0 0, width=\hsize]{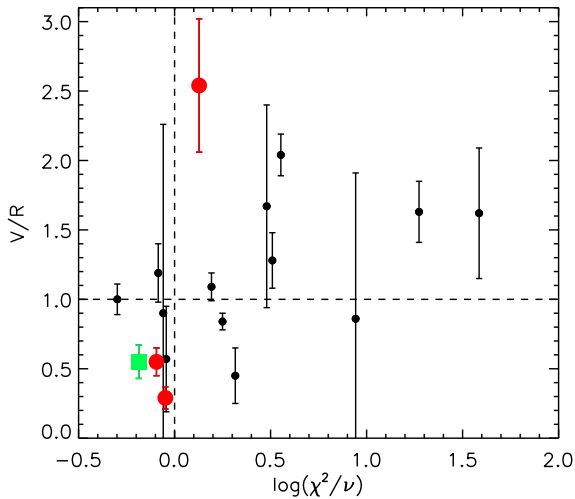}
      \caption[]{H$\alpha$ V/R measurements at maximum emission as a function of the reduced $\chi^2$ of first-order sinusoidal fits to \bz. Red circles indicate close binaries. The green square indicates ALS\,3694 (discussed further in the text).}
         \label{halphavr_bzchi}
   \end{figure}

For a tilted dipole, the RRM model predicts that the circumstellar material will be symmetric across the rotational axis. This should result in emission profiles that share this symmetry at the phase of magnetic maximum when the CM is closest to face-on. On the other hand, the emission profile of $\sigma$ Ori E (the only star for which detailed RRM modelling has been performed) is asymmetric. Using an RRM model extrapolated from a Zeeman Doppler Imaging map \cite{2015MNRAS.451.2015O} demonstrated this asymmetry to be a consequence of a surface magnetic field with significant departures from a pure dipole. Many of the stars in the sample show H$\alpha$ profiles that are asymmetric (see Figs.\ \ref{halpha_ind1}-\ref{halpha_ind2}). Aside from $\sigma$ Ori E, the only star with an available ZDI map is HD\,37776 \citep[which has the most complex magnetic field of any known early-type star][]{koch2011}. However, Paper I provided a simple, albeit less sensitive, diagnostic for the presence of important higher-order components of the surface magnetic field. This is simply the reduced $\chi^2$ of a first-order sinusoidal fit to the \bz~measurements obtained from H lines (a tilted dipole should produce a simple sinusoidal \bz~variation as the star rotates, and H should be relatively free of the chemical spots that can introduce anharmonicity into the \bz~curve that is unrelated to the surface magnetic field). 

As a simple proxy to line profile asymmetry, we calculated $V/R$ at emission maximum, i.e.\ the ratio of the EW in the blue to the red half of the line (see Table \ref{obstab}). While this quantity is generally variable due to rotational modulation, when the CM is face-on it should be close to 1 in the case of a symmetric CM. Fig.\ \ref{halphavr_bzchi} shows $V/R$ as a function of the reduced $\chi^2$, $\log{\chi^2/\nu}$ (where $\nu$ is the number of degrees of freedom). The majority of stars with $\log{\chi^2/\nu}$ close to 0 (i.e.\ for which a dipolar model is a good fit to \bz) also have $V/R$ close to 1. The stars with the largest $V/R$ values also tend to have large $\chi^2/\nu$. 

There are some exceptions to this tendency. The first are the close binaries, indicated in Fig.\ \ref{halphavr_bzchi} with large red circles. All have $\log{\chi^2/\nu}$ consistent with 0, yet two of these stars (HD\,156324 and HD\,37017) have the $V/R$ values with the largest departures from unity. In the case of HD\,156324 this is almost certainly due to modification of the RRM accumulation surface by the presence of its tidally locked companion \citep{2018MNRAS.475..839S}. While HD\,37017 is not tidally locked, the striking similarity of its H$\alpha$ profile to that of HD\,156324 is suggestive that binarity is playing a role here as well. 

The third close binary, HD\,36485, has a relatively long orbit \citep[$\sim 30$~d;][]{leone2010}, and its emission profile is probably not affected by its orbital companion. However, while its \bz~curve is formally consistent with a dipole, there are indications that it may be more complex \citep[see Paper I and][]{leone2010}.  

There is one other obvious outlier in Fig.\ \ref{halphavr_bzchi}. ALS\,3694 (indicated with a green square) has a very small $V/R$ value but no indication of a magnetic field curve more complex than a dipole. In this case, this might plausibly be due to the very large uncertainties in \bz~(see Paper I). 

We conclude that, with the exception of a few special cases, the basic prediction of the RRM model that simple magnetic geometries should correlate to simple magnetospheres is consistent with the observations. 

\subsection{Evolution}

   \begin{figure}
   \centering
   \includegraphics[trim=0 0 0 0, width=\hsize]{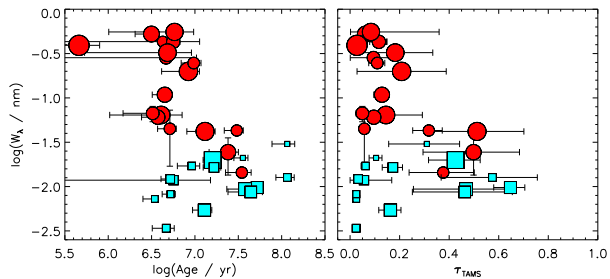}
      \caption[]{Emission strength as a function of log age (left) and fractional main sequence age $\tau_{\rm TAMS}$ (right). Red circles indicate stars with H$\alpha$ emission; blue squares, stars without H$\alpha$ emission.}
         \label{wlam_time}
   \end{figure}

In Paper III it was shown that due to the simultaneous decline in the magnetic field strength with time and rapid magnetic braking, the presence of H$\alpha$ emission is an indicator of youth. Indeed, while H$\alpha$-bright CM host stars are only about 25\% of the magnetic early B-type star population, they form a majority of the stars in the first third of the main sequence. We therefore expect that the emission strength should also decrease over time. Fig.\ \ref{wlam_time} shows emission strength as a function of the absolute and fractional main sequence age. The stars with the strongest emission are amongst the youngest in the population, with ages of a few Myr and fractional ages below 0.2, while the oldest stars are also the stars with the weakest emission. It therefore appears that emission strength indeed declines very rapidly with age, validating the suggestion in Paper III that efforts to expand the sample of H$\alpha$-bright CM host stars be focused on very young stellar clusters. 

\subsection{Predictions of the model for other stars}

   \begin{figure}
   \centering
\begin{tabular}{cc}
   \includegraphics[trim=50 0 25 0, width=0.225\textwidth]{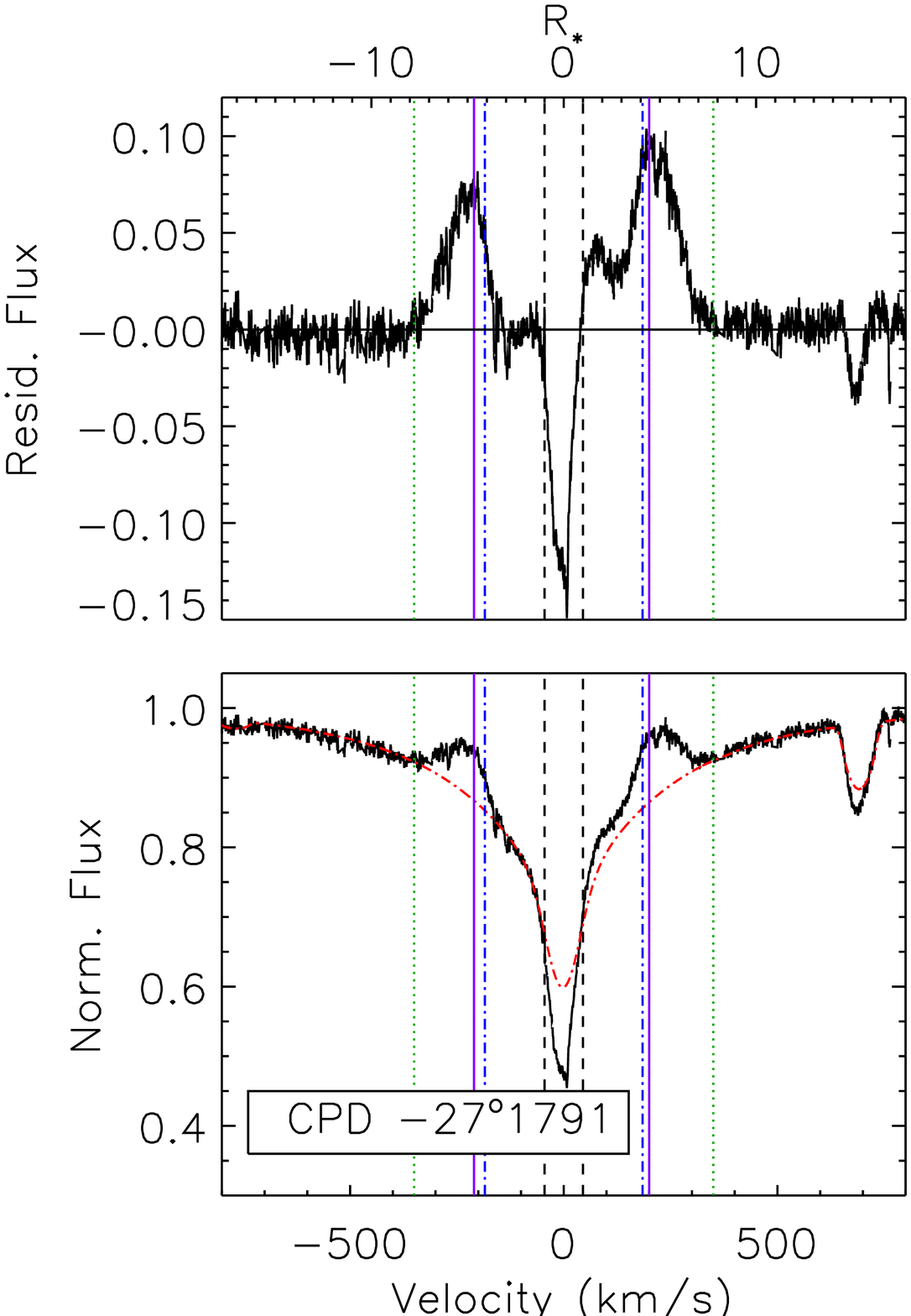} &
   \includegraphics[trim=50 0 25 0, width=0.225\textwidth]{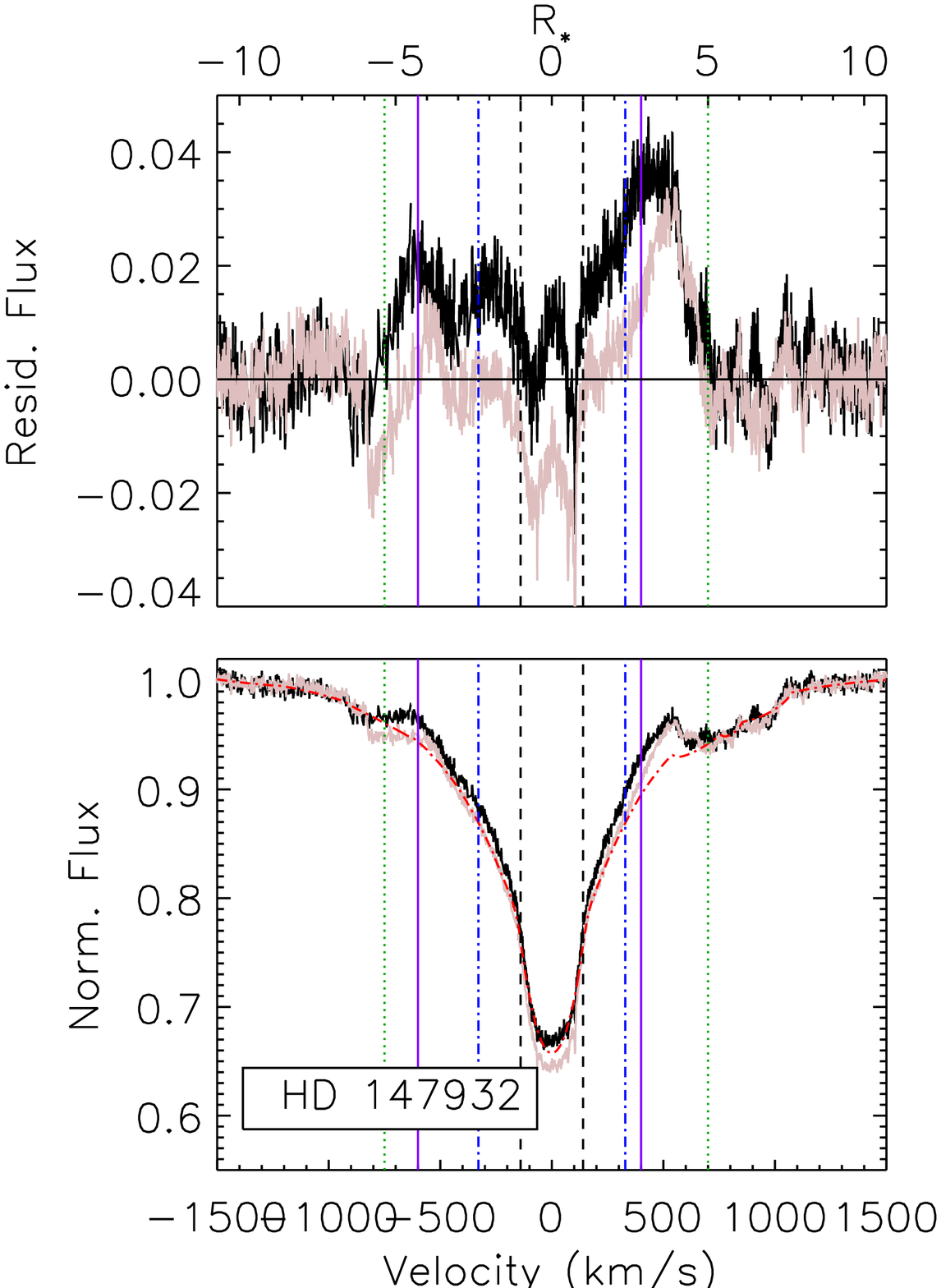} \\
\end{tabular}
      \caption[]{As Fig.\ \protect\ref{sigOriE_halpha_minmax} for CPD\,$-27^\circ 1791$ and HD\,147932. The radial velocity was determined by aligning the synthetic spectrum with the C~{\sc ii} lines in the red wing of $H\alpha$.}
         \label{CPD-271791_halpha_minmax}
   \end{figure}

One application of the results of this paper is that if the rotational properties of a magnetic star with CM-type H$\alpha$ emission are known, these can be used to predict its magnetic properties. Such a case is presented by the He-strong star CPD$-27^\circ 1791$. \cite{2018A&A...618L...2J} detected a strong magnetic field in this star (\bz$\sim$1 kG) using HARPSpol, and reported a relatively short rotational period of about 2.6~d on the basis of ASAS data. While they said nothing about the presence of emission in this star, the rapid rotation and strong \bz~motivated us to examine the star's H$\alpha$ line in the one HARPSpol spectrum, which as can be seen in Fig.\ \ref{CPD-271791_halpha_minmax} plainly displays H$\alpha$ emission. From the star's SIMBAD photometry and Gaia parallax, we inferred a luminosity $\log{(L/L_\odot)}=3.6 \pm 0.15$, which with the published \teff~of $23 \pm 1$~kK implies a radius of $3.8 \pm 0.3$~\rsun~and a mass of $8.8 \pm 0.5$~\msun. This then yields \rk~$=4.1 \pm 0.2~R_*$. In order for the star to be above the $B_{\rm K} = 100$ G threshold for emission, the surface magnetic dipole would need to be a minimum of 13.8 kG, considerably higher than the 4 kG lower limit inferred from the star's single \bz~measurement. 

This prediction can be refined somewhat by looking at the emission EW obtained from the star's residual flux profile (upper panel of Fig.\ \ref{CPD-271791_halpha_minmax}). The emission is fairly weak, with $W_\lambda = 0.043 \pm 0.001$ nm. Comparing to the other stars in the right panel of Fig.\ \ref{wlam_rark_bk}, this would be consistent with the star having $B_{\rm K} \sim 100$~G. CPD\,$-27^\circ 1791$ could therefore be an interesting target as, like CPD\,$-62^\circ 2124$, it has a relatively large Kepler radius, but on the other hand it has much weaker emission. Further observations of this star would therefore be useful to fill out the parameter space of CM properties. 

In order for $B_{\rm d}$ to be so much higher than the lower limit implied by \bz, either the star must have been observed at close to magnetic minimum, and/or it must have very extreme angular parameters. The star's \vsini~($45 $~\kms), rotation period, and radius imply that its rotational axis inclination should be about $i_{\rm rot} \sim 35 \pm 10^\circ$. A large $B_{\rm d}$ therefore requires a large $\beta$. Examining CPD\,$-27^\circ 1791$'s H$\alpha$ emission, we see that it has two well-defined emission bumps. This pattern is similar to that of HD\,176582, which indeed has $\beta$ close to 90$^\circ$. 

In other respects the star's emission properties are consistent with the remainder of the sample. Emission is mostly above \rk, but $r_{\rm max}$ is about 20--30\% higher than \rk. $\log{r_{\rm out}/r_{\rm max}}$ is about 0.2; comparing to Fig.\ \ref{r0rmax_rark_bk}, this ratio also predicts that $\log{B_{\rm K}} \sim 2$. It should of course be pointed out that, since only a snapshot of H$\alpha$ is available, the values for emission strength and characteristic radii are all lower limits. 

Another star for which weak H$\alpha$ emission consistent with an origin in a CM has been reported is HD\,147932 \citep[$\rho$\,Oph\,C][]{alecian2014}. These observations are shown in Fig.\ \ref{CPD-271791_halpha_minmax}. Only two \bz~measurements have been published, both around $-1$~kG and indicating a lower limit to $B_{\rm d}$ of about 3.5~kG. The star has broad spectral lines (\vsini~$\sim 140$~\kms), indicating that it must be a rapid rotator, and indeed its K2 period of about 0.86~d makes it amongst the most rapidly rotating magnetic B-type stars \citep{2018AJ....155..196R}. Using the Gaia DR2 parallax, the \teff~$=17 \pm 1$~kK~determined by \cite{alecian2014}, the bolometric correction from \cite{2008AA...491..545N}, and adopting $E(B-V) \sim 0.5$~mag from comparing its $B$ and $V$ magnitudes to the empirical intrinsic colours given by \cite{2013ApJS..208....9P}, the luminosity is $\log{(L/L_\odot)} = 2.5 \pm 0.2$ and the radius is $2.6 \pm 0.2$~\rsun. Rotating evolutionary tracks \citep{ekstrom2012} give $M_* = 4.8 \pm 0.3$~\msun. Given the K2 period, then $R_{\rm K} \sim 2.4~R_*$ and $\log{B_{\rm K}} > 2.25$. The maximum emission strength measured from the model fit in Fig.\ \ref{CPD-271791_halpha_minmax} is $W_\lambda = 0.047 \pm 0.001$~nm, comparable to that of CPD\,$-27^\circ 1791$. Referring to Fig.\ \ref{wlam_rark_bk}, the emission strength of this star is consistent with the lower limit on $\log{B_{\rm K}}$, so if the period is accurate then it is unlikely that $B_{\rm d}$ differs significantly from 4 kG. Using the red line wing of H$\alpha$ (which has the strongest emission), $\log{r_{\rm out}/r_{\rm max}} \sim 0.25$; from Fig.\ \ref{r0rmax_rark_bk}, this is also consistent with the inferred value of $\log{B_{\rm K}}$. If $B_{\rm d}$ is close to the lower limit, then $i$ and $\beta$ should both be fairly small. The inclination inferred from \vsini, $P_{\rm rot}$, and the fundamental parameters is $i_{\rm rot} = 52^{+8~\circ}_{-6}$, indicating that $\beta$ must be very small. In this case, we should expect that the emission bumps should not be strongly concentrated (unless the surface magnetic field is not a simple dipole). Comparing the H$\alpha$ profiles of HD\,147932 to that of CPD\,$-62^\circ 2124$, the former do seem to be more spread out.

\subsection{The low-luminosity emission cutoff}\label{subsec:lowlumcutoff}

As demonstrated in Figs.\ \ref{lum_rark2} and \ref{wlam_rark_bk}, below about $\log{(L/L_\odot)} = 2.8$ H$\alpha$ emission apparently shuts off, regardless of the magnitude of $B_{\rm K}$. An important caveat to this is that there are currently no known stars with $\log{(L/L_\odot)}$ between and 2 and 2.8, and $\log{B_{\rm K}} \sim 3$; indeed all the stars with luminosities in this range have $\log{B_{\rm K}} \sim 2.2$ or below. It is therefore not entirely certain that H$\alpha$ emission is entirely absent in this luminosity regime, since the most extreme rotational and magnetic parameter space is apparently unexplored. 

There are at least two extremely rapidly rotating stars just below our sample's cutoff of $\log{(L/L_\odot)} = 2$. The least luminous ($\log{(L/L_\odot)} = 1.8 \pm 0.1$, determined via the Gaia DR2 parallax and \teff~$=11.9 \pm 0.2$ kK from \citealt{2012MNRAS.423..328B}) is HD\,133880, an extremely rapid rotator with a strong magnetic field \citep[$P_{\rm rot} = 0.877483$ d, $B_{\rm d} \sim 12$~kG][]{2017A&A...605A..13K}. For this star $\log{B_{\rm K}} \sim 2.5$. The other is CU\,Vir (HD\,124224), an even more rapid rotator \citep[$P_{\rm rot} \sim 0.52069415$ d][]{miku2011} with a somewhat weaker magnetic field \citep[$B_{\rm d} \sim 4$~kG][]{2014A&A...565A..83K}. Using the atmospheric parameters determined by \cite{2019MNRAS.483.3127S} (\teff~$=12.3 \pm 0.2$ kK, $\log{(L/L_\odot)} = 1.93 \pm 0.01$) yields $\log{B_{\rm K}} \sim 2.6$. Notably, neither of these stars display H$\alpha$ emission, despite having relatively high $B_{\rm K}$ values. While neither star probes the maximal $B_{\rm K}$ values seen in the H$\alpha$-bright stars, they are well above the median $B_{\rm K}$ seen in this population, strongly suggesting that H$\alpha$ emission disappears at low luminosities regardless of the rotational or magnetic properties of the star.

One possibility is that these stars are simply too cool to show emission: rather than a lack of excitable atoms, the problem is a lack of exciting photons. A similar phenomenon is seen amongst the A-type shell stars, which possess decretion disks similar to classical Be stars, but are detectable only when the disk is seen edge-on and is eclipsing the star \citep{2013AARv..21...69R}. A suggestive case is presented by the example of HD\,79158 (36\,Lyn), a cool \citep[13 kK;][]{2008AA...491..545N} Bp star that shows no sign of H$\alpha$ emission, yet displays clear eclipse signatures in the core of its H$\alpha$ line \citep{2006AA...458..581S}. However, it should be noted that the classical Be phenomenon is found across the full range of spectral types from B0 to B9 \citep{2013AARv..21...69R}, whereas CM emission seems to disappear around B5. Furthermore, using 36 Lyn's published stellar, magnetic, and rotational parameters \citep{2008AA...491..545N,2018MNRAS.473.3367O}, the star has $\log{B_{\rm K}} = 1.3 \pm 0.1$, well below the range at which H$\alpha$ emission is seen. It is therefore possible that its CM is seen only in occultation because it is only at this phase that the optical depth is greater than unity (although this raises the question of why a similar phenomenon has not been seen in other cool Bp stars with higher values of $B_{\rm K}$). The possible infrared excess around this star \citep{2012MNRAS.427..343M} may however indicate that its circumstellar environment is rather peculiar as compared to similar stars.

Since the principal difference between these stars and stars with H$\alpha$ emission is that they possess much weaker winds, it is possible that the rapid drop-off in mass-loss rates is responsible for the absence of emission. This is curious as we have just shown that H$\alpha$ emission strength is seemingly unrelated to mass-loss rates. The obvious conclusion is that centrifugal breakout ceases to be the dominant plasma transport process at low $\dot{M}$. One possibility is that leakage via diffusion and/or drift, as explored by \cite{2018MNRAS.474.3090O}, empties the magnetospheres of these stars more rapidly than they can be filled by their winds. This implies that leakage rates can be inferred from the mass-loss rates of these stars.

Another possibility is that the winds of these stars might be runaway metallic winds, in which the metal ions decouple from the H and He ions \citep{1992A&A...262..515S,2002ApJ...568..965O}. \cite{1995A&A...301..823B} showed that metallic winds can exist below the classical wind limit. In this case, the absence of H$\alpha$ emission would be a simple consequence of the absence of H in the stellar wind. In this scenario it is still possible for centrifugal breakout to be the dominant plasma transport process, and it may be possible that CM signatures might be detectable in wind-sensitive metallic lines. It is worth noting that UV C~{\sc iv} emission has been reported in several rapidly rotating He-weak magnetic stars \citep[e.g.][]{1990ApJ...365..665S,2004A&A...421..203S}, although the presence of similar UV emission in slowly rotating magnetic B-type stars possessing only dynamical magnetospheres \citep[e.g.][]{smithgroote2001,neiner2003b,henrichs2013,petit2013,2015MNRAS.447.1418Y} suggests that UV emission is not necessarily related to rapid rotation. 

It has been suggested that late B-type stars simply do not have winds at all \citep[e.g.][]{1996A&A...309..867B,krticka2014}. While this would neatly  explain the absence of H$\alpha$ emission at low luminosities, it is inconsistent with the magnetospheric UV emission of some of these stars, as well as the gyrosynchrotron emission observed in many cases \citep[e.g.][]{1987ApJ...322..902D,1992ApJ...393..341L,1996A&A...310..271L}, neither of which can be explained without a source of ions.

\section{Conclusions}\label{sec:conclusion}

We have performed the first comparative study of the emission properties of H$\alpha$-bright magnetic early B-type stars with centrifugal magnetospheres (CMs). We find that the onset of emission is effectively independent of the stellar mass-loss rate, and instead can be predicted with a single parameter, the strength $B_{\rm K}$ of the equatorial magnetic field at the Kepler radius. In particular, stars with $B_{\rm K} \ge 100$~G all display H$\alpha$ emission lines, while such emission is absent beneath this threshold. An emission onset that depends only on this parameter is difficult to explain with a leakage scenario for mass balancing within the CM, since in this case the onset of emission should depend on the competition between feeding via the wind and draining via the leakage mechanism. It is, however, very easy to explain in a centrifugal breakout scenario. 

Emission strength seems to depend only on the area of the CM; this is true whether the area is determined using the observed size of the CM, or its theoretical extent. This strongly suggests that the CM is optically thick in H$\alpha$ even when seen face-on. Since CMs are almost geometrically thin, if they are optically thick when seen face-on their central densities must be very high. This implies that CM densities may well be within the range necessary for centrifugal breakout to take place. Emission strength also rises rapidly with increasing $B_{\rm K}$. While a regression of $\log{W_\lambda}$ vs.\ $\log{B_{\rm K}}$ produces an acceptable fit, there is some indication that a non-linear relationship may be more appropriate. In particular, it may be that emission strength increases extremely rapidly once the 100 G threshold is reached, following which it increases more slowly with increasing $B_{\rm K}$. This would be a straightforward consequence of the rapid decrease in $B$ with distance from the star, and, hence, the difficulty of expanding the optically thick area of the CM. 

In contrast to the strong dependence of emission strength on CM area and $B_{\rm K}$, we find no statistically significant dependence of $W_\lambda$ on the mass-loss rate. This is further evidence that a diffusion/drift leakage mechanism does not play a significant role in magnetospheric mass balancing. To reconcile these results with the failure to detect any large scale change in the CM plasma distribution, we propose that on large scales centrifugal breakout is not a stochastic, eruptive mechanism. Instead, breakout events occur at small spatial scales and are essentially continuous when averaged across the CM. In this limit centrifugal breakout becomes in effect a steady state leakage mechanism -- centrifugal leakage -- with the rate of plasma flow out of the CM exactly balancing the feeding rate from the wind. 

We quantified the extent of observable H$\alpha$ emission using two characteristic radii: the radius of maximum emission $r_{\rm max}$ and the outermost extent of emission $r_{\rm out}$. As expected $r_{\rm max}$ correlates very strongly with \rk. However, $r_{\rm max}$ is systematically larger than \rk, ranging from about 20\% to 80\% greater. This difference cannot be explained as systematic error in stellar masses and radii. One possible explanation of this discrepancy is that plasma does not accumulate in the shallow potential minima at or slightly above \rk, as a consequence of asymmetry in the strength of flows from opposite co-latitudes. If this hypothesis is correct we expect to see an increase in the discrepancy between $r_{\rm max}$ and \rk~with increasing magnetic obliquity, and indeed we find some evidence for this. Further theoretical work, utilizing 3D MHD simulations with tilted dipoles, or rigid-field hydrodynamics simulations \citep{town2007} with tilted dipoles or non-dipolar magnetic fields, is required. 

In contrast to $r_{\rm max}$, $r_{\rm out}$ correlates only weakly to \ra, and in fact correlates more closely to \rk. It is not clear why $r_{\rm out}$ does not correlate well with \ra. The ratio of $r_{\rm out}/r_{\rm max}$ correlates well with both $R_{\rm A}/R_{\rm K}$ and with $B_{\rm K}$, with the latter providing the superior correlation. It should be noted that the relatively poor correlations of $W_\lambda$ and $r_{\rm out}/r_{\rm max}$ with $R_{\rm A}/R_{\rm K}$ are improved if HD\,164492C is dropped from the regression. This suggests that one or more of this star's parameters may be in error. 

Perhaps our most striking finding is that the emission wings -- the regions between $r_{\rm max}$ and $r_{\rm out}$ -- are self-similar across essentially the entire range of parameters spanned by the sample. All follow a pattern in which the emission wing is initially convex, switching over to concave at the approximate half-way mark. It is likely that this reflects a change in the optical depth of the profiles. This self-similarity indicates that CM emission profiles can be reproduced via a simple scaling relationship. 

The central result of this work -- that the emission properties of H$\alpha$-bright CM host stars can only be explained by centrifugal leakage -- is explored analytically by \cite{owocki2020}, who demonstrate that the emission onset, emission strength, and emission profile shapes of CMs can be reproduced not just qualitatively but also quantitatively within this framework.

\section*{Acknowledgements}

This work is based on observations obtained at the Canada-France-Hawaii Telescope (CFHT) which is operated by the National Research Council of Canada, the Institut National des Sciences de l'Univers (INSU) of the Centre National de la Recherche Scientifique (CNRS) of France, and the University of Hawaii; at the La Silla Observatory, ESO Chile with the MPA 2.2 m telescope; at the Observatoire du Pic du Midi (France), operated by the INSU; at the Dominion Astrophysical Observatory's (DAO) 1.8 m Plaskett Telescope and 1.2 m telescope; and on observations obtained with the SARA Observatory 0.9 m telescope at Kitt Peak, which is owned and operated by the Southeastern Association for Research in Astronomy. The authors are honored to be permitted to conduct astronomical research on Iolkam Du’ag (Kitt Peak), a mountain with particular significance to the Tohono O’odham Nation. Based on observations made with ESO Telescopes at the La Silla and Paranal Observatories under programme IDs 092.A-9018(A), 093.D-0267(B), 095.D-0269(A), 095.A-9007(A), 187.D-0917(C), 191.D-0255(G/H/I), 284.D-5058(B), and 383.D-0095(A). MES acknowledges the financial support provided by the European Southern Observatory studentship program in Santiago, Chile; the Natural Sciences and Engineering Research Council (NSERC) Postdoctoral Fellowship program; and the Annie Jump Cannon Fellowship, supported by the University of Delaware and endowed by the Mount Cuba Astronomical Observatory. The MiMeS and BinaMIcS collaborations acknowledge financial support from the Programme National de Physique Stellaire (PNPS) of INSU/CNRS. We acknowledge the Canadian Astronomy Data Centre (CADC). ADU acknowledges support from the NSERC Postdoctoral Fellowship Program. VP acknowledges support from the National Science Foundation under Grant No.\ 1747658. GAW acknowledges support from the Natural Sciences and Engineering Research Council (NSERC) of Canada in the form of a Discovery Grant. The authors thank the referee for taking the time to provide a thorough and very helpful review.

\section*{Data Availability Statement}
Reduced ESPaDOnS spectra are available at the CFHT archive maintained by the CADC at \url{https://www.cadc-ccda.hia-iha.nrc-cnrc.gc.ca/en/}, where they can be found via standard stellar designations. ESPaDOnS and Narval data can also be obtained at the PolarBase archive at \url{http://polarbase.irap.omp.eu/}. ESO data are available in raw form at the ESo archive at \url{http://archive.eso.org/eso/eso_archive_main.html}. Reduced data are available from the authors at request.

\bibliography{bib_dat.bib}{}

\begin{thebibliography}{}
\makeatletter
\relax
\def\mn@urlcharsother{\let\do\@makeother \do\$\do\&\do\#\do\^\do\_\do\%\do\~}
\def\mn@doi{\begingroup\mn@urlcharsother \@ifnextchar [ {\mn@doi@}
  {\mn@doi@[]}}
\def\mn@doi@[#1]#2{\def\@tempa{#1}\ifx\@tempa\@empty \href
  {http://dx.doi.org/#2} {doi:#2}\else \href {http://dx.doi.org/#2} {#1}\fi
  \endgroup}
\def\mn@eprint#1#2{\mn@eprint@#1:#2::\@nil}
\def\mn@eprint@arXiv#1{\href {http://arxiv.org/abs/#1} {{\tt arXiv:#1}}}
\def\mn@eprint@dblp#1{\href {http://dblp.uni-trier.de/rec/bibtex/#1.xml}
  {dblp:#1}}
\def\mn@eprint@#1:#2:#3:#4\@nil{\def\@tempa {#1}\def\@tempb {#2}\def\@tempc
  {#3}\ifx \@tempc \@empty \let \@tempc \@tempb \let \@tempb \@tempa \fi \ifx
  \@tempb \@empty \def\@tempb {arXiv}\fi \@ifundefined
  {mn@eprint@\@tempb}{\@tempb:\@tempc}{\expandafter \expandafter \csname
  mn@eprint@\@tempb\endcsname \expandafter{\@tempc}}}

\bibitem[\protect\citeauthoryear{{Adelman} \& {Boyce}}{{Adelman} \&
  {Boyce}}{1995}]{1995A&AS..114..253A}
{Adelman} S.~J.,  {Boyce} P.~W.,  1995, \aaps, \href
  {https://ui.adsabs.harvard.edu/abs/1995A&AS..114..253A} {114, 253}

\bibitem[\protect\citeauthoryear{{Adelman} \& {Fried}}{{Adelman} \&
  {Fried}}{1993}]{1993AJ....105.1103A}
{Adelman} S.~J.,  {Fried} R.,  1993, \mn@doi [\aj] {10.1086/116497}, \href
  {https://ui.adsabs.harvard.edu/abs/1993AJ....105.1103A} {105, 1103}

\bibitem[\protect\citeauthoryear{{Alecian} et~al.,}{{Alecian}
  et~al.}{2014}]{alecian2014}
{Alecian} E.,  et~al., 2014, \mn@doi [\aap] {10.1051/0004-6361/201323286},
  \href {http://adsabs.harvard.edu/abs/2014A26A...567A..28A} {567, A28}

\bibitem[\protect\citeauthoryear{{Babel}}{{Babel}}{1995}]{1995A&A...301..823B}
{Babel} J.,  1995, \aap, \href
  {https://ui.adsabs.harvard.edu/abs/1995A&A...301..823B} {301, 823}

\bibitem[\protect\citeauthoryear{{Babel}}{{Babel}}{1996}]{1996A&A...309..867B}
{Babel} J.,  1996, \aap, \href
  {https://ui.adsabs.harvard.edu/abs/1996A&A...309..867B} {309, 867}

\bibitem[\protect\citeauthoryear{{Bagnulo}, {Landstreet}, {Mason}, {Andretta},
  {Silaj}  \& {Wade}}{{Bagnulo} et~al.}{2006}]{bagn2006}
{Bagnulo} S.,  {Landstreet} J.~D.,  {Mason} E.,  {Andretta} V.,  {Silaj} J.,
  {Wade} G.~A.,  2006, \mn@doi [\aap] {10.1051/0004-6361:20054223}, \href
  {http://adsabs.harvard.edu/abs/2006A26A...450..777B} {450, 777}

\bibitem[\protect\citeauthoryear{{Bagnulo}, {Fossati}, {Landstreet}  \&
  {Izzo}}{{Bagnulo} et~al.}{2015}]{2015AA...583A.115B}
{Bagnulo} S.,  {Fossati} L.,  {Landstreet} J.~D.,   {Izzo} C.,  2015, \mn@doi
  [\aap] {10.1051/0004-6361/201526497}, \href
  {http://adsabs.harvard.edu/abs/2015A26A...583A.115B} {583, A115}

\bibitem[\protect\citeauthoryear{{Bailey} et~al.,}{{Bailey}
  et~al.}{2012}]{2012MNRAS.423..328B}
{Bailey} J.~D.,  et~al., 2012, \mn@doi [\mnras]
  {10.1111/j.1365-2966.2012.20881.x}, \href
  {http://adsabs.harvard.edu/abs/2012MNRAS.423..328B} {423, 328}

\bibitem[\protect\citeauthoryear{{Baumgardt}, {Dettbarn}  \&
  {Wielen}}{{Baumgardt} et~al.}{2000}]{2000A&AS..146..251B}
{Baumgardt} H.,  {Dettbarn} C.,   {Wielen} R.,  2000, \mn@doi [\aaps]
  {10.1051/aas:2000362}, \href
  {http://adsabs.harvard.edu/abs/2000A%26AS..146..251B} {146, 251}

\bibitem[\protect\citeauthoryear{{Bohlender} \& {Monin}}{{Bohlender} \&
  {Monin}}{2011}]{bohl2011}
{Bohlender} D.~A.,  {Monin} D.,  2011, \mn@doi [\aj]
  {10.1088/0004-6256/141/5/169}, \href
  {http://cdsads.u-strasbg.fr/abs/2011AJ....141..169B} {141, 169}

\bibitem[\protect\citeauthoryear{{Bolton}, {Harmanec}, {Lyons}, {Odell}  \&
  {Pyper}}{{Bolton} et~al.}{1998}]{1998AA...337..183B}
{Bolton} C.~T.,  {Harmanec} P.,  {Lyons} R.~W.,  {Odell} A.~P.,   {Pyper}
  D.~M.,  1998, \aap, \href
  {http://adsabs.harvard.edu/abs/1998A26A...337..183B} {337, 183}

\bibitem[\protect\citeauthoryear{{Borra}, {Landstreet}  \& {Thompson}}{{Borra}
  et~al.}{1983}]{1983ApJS...53..151B}
{Borra} E.~F.,  {Landstreet} J.~D.,   {Thompson} I.,  1983, \mn@doi [\apjs]
  {10.1086/190889}, \href {http://cdsads.u-strasbg.fr/abs/1983ApJS...53..151B}
  {53, 151}

\bibitem[\protect\citeauthoryear{{Bouy} et~al.,}{{Bouy}
  et~al.}{2009}]{2009AA...493..931B}
{Bouy} H.,  et~al., 2009, \mn@doi [\aap] {10.1051/0004-6361:200810267}, \href
  {http://adsabs.harvard.edu/abs/2009A26A...493..931B} {493, 931}

\bibitem[\protect\citeauthoryear{{Buysschaert}, {Neiner}, {Martin}, {Oksala},
  {Aerts}, {Tkachenko}, {Alecian}  \& {MiMeS Collaboration}}{{Buysschaert}
  et~al.}{2019}]{2019A&A...622A..67B}
{Buysschaert} B.,  {Neiner} C.,  {Martin} A.~J.,  {Oksala} M.~E.,  {Aerts} C.,
  {Tkachenko} A.,  {Alecian} E.,   {MiMeS Collaboration} 2019, \mn@doi [\aap]
  {10.1051/0004-6361/201731913}, \href
  {https://ui.adsabs.harvard.edu/abs/2019A&A...622A..67B} {622, A67}

\bibitem[\protect\citeauthoryear{{Bychkov}, {Bychkova}  \& {Madej}}{{Bychkov}
  et~al.}{2005}]{2005AA...430.1143B}
{Bychkov} V.~D.,  {Bychkova} L.~V.,   {Madej} J.,  2005, \mn@doi [\aap]
  {10.1051/0004-6361:20034563}, \href
  {http://adsabs.harvard.edu/abs/2005A26A...430.1143B} {430, 1143}

\bibitem[\protect\citeauthoryear{{Carciofi}, {Faes}, {Townsend}  \&
  {Bjorkman}}{{Carciofi} et~al.}{2013}]{carc2013}
{Carciofi} A.~C.,  {Faes} D.~M.,  {Townsend} R.~H.~D.,   {Bjorkman} J.~E.,
  2013, \mn@doi [\apjl] {10.1088/2041-8205/766/1/L9}, \href
  {http://adsabs.harvard.edu/abs/2013ApJ...766L...9C} {766, L9}

\bibitem[\protect\citeauthoryear{{Castro} et~al.,}{{Castro}
  et~al.}{2017}]{2017A&A...597L...6C}
{Castro} N.,  et~al., 2017, \mn@doi [\aap] {10.1051/0004-6361/201629751}, \href
  {http://adsabs.harvard.edu/abs/2017A%26A...597L...6C} {597, L6}

\bibitem[\protect\citeauthoryear{{Das}, {Chandra}, {Shultz}  \& {Wade}}{{Das}
  et~al.}{2019}]{2019ApJ...877..123D}
{Das} B.,  {Chandra} P.,  {Shultz} M.~E.,   {Wade} G.~A.,  2019, \mn@doi [\apj]
  {10.3847/1538-4357/ab1b12}, \href
  {https://ui.adsabs.harvard.edu/abs/2019ApJ...877..123D} {877, 123}

\bibitem[\protect\citeauthoryear{{Donati}, {Semel}  \& {Rees}}{{Donati}
  et~al.}{1992}]{1992AA...265..669D}
{Donati} J.-F.,  {Semel} M.,   {Rees} D.~E.,  1992, \aap, \href
  {http://adsabs.harvard.edu/abs/1992A26A...265..669D} {265, 669}

\bibitem[\protect\citeauthoryear{{Donati}, {Semel}, {Carter}, {Rees}  \&
  {Collier Cameron}}{{Donati} et~al.}{1997}]{d1997}
{Donati} J.-F.,  {Semel} M.,  {Carter} B.~D.,  {Rees} D.~E.,   {Collier
  Cameron} A.,  1997, MNRAS, \href
  {http://adsabs.harvard.edu/abs/1997MNRAS.291..658D} {291, 658}

\bibitem[\protect\citeauthoryear{{Drake}, {Abbott}, {Bastian}, {Bieging},
  {Churchwell}, {Dulk}  \& {Linsky}}{{Drake}
  et~al.}{1987}]{1987ApJ...322..902D}
{Drake} S.~A.,  {Abbott} D.~C.,  {Bastian} T.~S.,  {Bieging} J.~H.,
  {Churchwell} E.,  {Dulk} G.,   {Linsky} J.~L.,  1987, \mn@doi [\apj]
  {10.1086/165784}, \href {http://adsabs.harvard.edu/abs/1987ApJ...322..902D}
  {322, 902}

\bibitem[\protect\citeauthoryear{{Dubath} et~al.,}{{Dubath}
  et~al.}{2011}]{2011MNRAS.414.2602D}
{Dubath} P.,  et~al., 2011, \mn@doi [\mnras]
  {10.1111/j.1365-2966.2011.18575.x}, \href
  {http://adsabs.harvard.edu/abs/2011MNRAS.414.2602D} {414, 2602}

\bibitem[\protect\citeauthoryear{{Eikenberry} et~al.,}{{Eikenberry}
  et~al.}{2014}]{2014ApJ...784L..30E}
{Eikenberry} S.~S.,  et~al., 2014, \mn@doi [\apjl]
  {10.1088/2041-8205/784/2/L30}, \href
  {http://adsabs.harvard.edu/abs/2014ApJ...784L..30E} {784, L30}

\bibitem[\protect\citeauthoryear{{Ekstr{\"o}m} et~al.,}{{Ekstr{\"o}m}
  et~al.}{2012}]{ekstrom2012}
{Ekstr{\"o}m} S.,  et~al., 2012, \mn@doi [\aap] {10.1051/0004-6361/201117751},
  \href {http://cdsads.u-strasbg.fr/abs/2012A26A...537A.146E} {537, A146}

\bibitem[\protect\citeauthoryear{{Gaia Collaboration} et~al.,}{{Gaia
  Collaboration} et~al.}{2018}]{2018A&A...616A...1G}
{Gaia Collaboration} et~al., 2018, \mn@doi [\aap]
  {10.1051/0004-6361/201833051}, \href
  {http://adsabs.harvard.edu/abs/2018A%26A...616A...1G} {616, A1}

\bibitem[\protect\citeauthoryear{{Glebocki} \& {Gnacinski}}{{Glebocki} \&
  {Gnacinski}}{2005}]{2005yCat.3244....0G}
{Glebocki} R.,  {Gnacinski} P.,  2005, VizieR Online Data Catalog, \href
  {https://ui.adsabs.harvard.edu/abs/2005yCat.3244....0G} {p. III/244}

\bibitem[\protect\citeauthoryear{{Gonz{\'a}lez} et~al.,}{{Gonz{\'a}lez}
  et~al.}{2017}]{2017MNRAS.467..437G}
{Gonz{\'a}lez} J.~F.,  et~al., 2017, \mn@doi [\mnras] {10.1093/mnras/stx105},
  \href {https://ui.adsabs.harvard.edu/abs/2017MNRAS.467..437G} {467, 437}

\bibitem[\protect\citeauthoryear{{Grunhut} et~al.,}{{Grunhut}
  et~al.}{2012}]{grun2012}
{Grunhut} J.~H.,  et~al., 2012, \mn@doi [\mnras]
  {10.1111/j.1365-2966.2011.19824.x}, \href
  {http://cdsads.u-strasbg.fr/abs/2012MNRAS.419.1610G} {419, 1610}

\bibitem[\protect\citeauthoryear{{Havnes} \& {Goertz}}{{Havnes} \&
  {Goertz}}{1984}]{1984AA...138..421H}
{Havnes} O.,  {Goertz} C.~K.,  1984, \aap, \href
  {http://adsabs.harvard.edu/abs/1984A26A...138..421H} {138, 421}

\bibitem[\protect\citeauthoryear{{Henrichs} et~al.,}{{Henrichs}
  et~al.}{2013}]{henrichs2013}
{Henrichs} H.~F.,  et~al., 2013, \mn@doi [\aap] {10.1051/0004-6361/201321584},
  \href {http://adsabs.harvard.edu/abs/2013A26A...555A..46H} {555, A46}

\bibitem[\protect\citeauthoryear{{Hesser}, {Walborn}  \& {Ugarte}}{{Hesser}
  et~al.}{1976}]{1976Natur.262..116H}
{Hesser} J.~E.,  {Walborn} N.~R.,   {Ugarte} P.~P.,  1976, \mn@doi [\nat]
  {10.1038/262116a0}, \href {http://adsabs.harvard.edu/abs/1976Natur.262..116H}
  {262, 116}

\bibitem[\protect\citeauthoryear{{Hubrig} et~al.,}{{Hubrig}
  et~al.}{2014}]{2014AA...564L..10H}
{Hubrig} S.,  et~al., 2014, \mn@doi [\aap] {10.1051/0004-6361/201423490}, \href
  {http://adsabs.harvard.edu/abs/2014A26A...564L..10H} {564, L10}

\bibitem[\protect\citeauthoryear{{Hubrig} et~al.,}{{Hubrig}
  et~al.}{2015}]{2015A&A...578L...3H}
{Hubrig} S.,  et~al., 2015, \mn@doi [\aap] {10.1051/0004-6361/201526262}, \href
  {http://adsabs.harvard.edu/abs/2015A%26A...578L...3H} {578, L3}

\bibitem[\protect\citeauthoryear{{Hubrig}, {Kholtygin}, {Sch{\"o}ller}  \&
  {Ilyin}}{{Hubrig} et~al.}{2017a}]{2017MNRAS.467L..81H}
{Hubrig} S.,  {Kholtygin} A.~F.,  {Sch{\"o}ller} M.,   {Ilyin} I.,  2017a,
  \mn@doi [\mnras] {10.1093/mnrasl/slx005}, \href
  {https://ui.adsabs.harvard.edu/\#abs/2017MNRAS.467L..81H} {467, L81}

\bibitem[\protect\citeauthoryear{{Hubrig}, {Mikul{\'a}{\v s}ek}, {Kholtygin},
  {Ilyin}, {Sch{\"o}ller}, {J{\"a}rvinen}, {Scholz}  \& {Zejda}}{{Hubrig}
  et~al.}{2017b}]{2017MNRAS.472..400H}
{Hubrig} S.,  {Mikul{\'a}{\v s}ek} Z.,  {Kholtygin} A.~F.,  {Ilyin} I.,
  {Sch{\"o}ller} M.,  {J{\"a}rvinen} S.~P.,  {Scholz} R.-D.,   {Zejda} M.,
  2017b, \mn@doi [\mnras] {10.1093/mnras/stx1994}, \href
  {http://adsabs.harvard.edu/abs/2017MNRAS.472..400H} {472, 400}

\bibitem[\protect\citeauthoryear{{J{\"a}rvinen}, {Hubrig}, {Ilyin},
  {Sch{\"o}ller}, {Nieva}, {Przybilla}  \& {Castro}}{{J{\"a}rvinen}
  et~al.}{2018}]{2018A&A...618L...2J}
{J{\"a}rvinen} S.~P.,  {Hubrig} S.,  {Ilyin} I.,  {Sch{\"o}ller} M.,  {Nieva}
  M.~F.,  {Przybilla} N.,   {Castro} N.,  2018, \mn@doi [\aap]
  {10.1051/0004-6361/201833171}, \href
  {http://adsabs.harvard.edu/abs/2018A%26A...618L...2J} {618, L2}

\bibitem[\protect\citeauthoryear{{Keel} et~al.,}{{Keel}
  et~al.}{2017}]{2017PASP..129a5002K}
{Keel} W.~C.,  et~al., 2017, \mn@doi [\pasp]
  {10.1088/1538-3873/129/971/015002}, \href
  {https://ui.adsabs.harvard.edu/abs/2017PASP..129a5002K} {129, 015002}

\bibitem[\protect\citeauthoryear{{Keszthelyi}, {Meynet}, {Georgy}, {Wade},
  {Petit}  \& {David-Uraz}}{{Keszthelyi} et~al.}{2019}]{2019MNRAS.485.5843K}
{Keszthelyi} Z.,  {Meynet} G.,  {Georgy} C.,  {Wade} G.~A.,  {Petit} V.,
  {David-Uraz} A.,  2019, \mn@doi [\mnras] {10.1093/mnras/stz772}, \href
  {http://adsabs.harvard.edu/abs/2019MNRAS.485.5843K} {485, 5843}

\bibitem[\protect\citeauthoryear{{Keszthelyi} et~al.,}{{Keszthelyi}
  et~al.}{2020}]{2020MNRAS.tmp..227K}
{Keszthelyi} Z.,  et~al., 2020, \mn@doi [\mnras] {10.1093/mnras/staa237}, \href
  {https://ui.adsabs.harvard.edu/abs/2020MNRAS.tmp..227K} {p.~227}

\bibitem[\protect\citeauthoryear{{Kharchenko}, {Piskunov}, {R{\"o}ser},
  {Schilbach}  \& {Scholz}}{{Kharchenko} et~al.}{2005}]{2005AA...438.1163K}
{Kharchenko} N.~V.,  {Piskunov} A.~E.,  {R{\"o}ser} S.,  {Schilbach} E.,
  {Scholz} R.-D.,  2005, \mn@doi [\aap] {10.1051/0004-6361:20042523}, \href
  {http://adsabs.harvard.edu/abs/2005A26A...438.1163K} {438, 1163}

\bibitem[\protect\citeauthoryear{{Kochukhov}, {Makaganiuk}  \&
  {Piskunov}}{{Kochukhov} et~al.}{2010}]{koch2010}
{Kochukhov} O.,  {Makaganiuk} V.,   {Piskunov} N.,  2010, \mn@doi [\aap]
  {10.1051/0004-6361/201015429}, \href
  {http://cdsads.u-strasbg.fr/abs/2010A26A...524A...5K} {524, A5}

\bibitem[\protect\citeauthoryear{{Kochukhov}, {Lundin}, {Romanyuk}  \&
  {Kudryavtsev}}{{Kochukhov} et~al.}{2011}]{koch2011}
{Kochukhov} O.,  {Lundin} A.,  {Romanyuk} I.,   {Kudryavtsev} D.,  2011,
  \mn@doi [ApJ] {10.1088/0004-637X/726/1/24}, \href
  {http://adsabs.harvard.edu/abs/2011ApJ...726...24K} {726, 24}

\bibitem[\protect\citeauthoryear{{Kochukhov}, {L{\"u}ftinger}, {Neiner},
  {Alecian}  \& {MiMeS Collaboration}}{{Kochukhov}
  et~al.}{2014}]{2014A&A...565A..83K}
{Kochukhov} O.,  {L{\"u}ftinger} T.,  {Neiner} C.,  {Alecian} E.,   {MiMeS
  Collaboration} 2014, \mn@doi [\aap] {10.1051/0004-6361/201423472}, \href
  {http://adsabs.harvard.edu/abs/2014A%26A...565A..83K} {565, A83}

\bibitem[\protect\citeauthoryear{{Kochukhov}, {Silvester}, {Bailey},
  {Landstreet}  \& {Wade}}{{Kochukhov} et~al.}{2017}]{2017A&A...605A..13K}
{Kochukhov} O.,  {Silvester} J.,  {Bailey} J.~D.,  {Landstreet} J.~D.,   {Wade}
  G.~A.,  2017, \mn@doi [\aap] {10.1051/0004-6361/201730919}, \href
  {http://adsabs.harvard.edu/abs/2017A%26A...605A..13K} {605, A13}

\bibitem[\protect\citeauthoryear{{Kochukhov}, {Shultz}  \&
  {Neiner}}{{Kochukhov} et~al.}{2019}]{2019A&A...621A..47K}
{Kochukhov} O.,  {Shultz} M.,   {Neiner} C.,  2019, \mn@doi [\aap]
  {10.1051/0004-6361/201834279}, \href
  {https://ui.adsabs.harvard.edu/abs/2019A&A...621A..47K} {621, A47}

\bibitem[\protect\citeauthoryear{{Krti{\v c}ka}}{{Krti{\v
  c}ka}}{2014}]{krticka2014}
{Krti{\v c}ka} J.,  2014, \mn@doi [\aap] {10.1051/0004-6361/201321980}, \href
  {http://cdsads.u-strasbg.fr/abs/2014A26A...564A..70K} {564, A70}

\bibitem[\protect\citeauthoryear{{Krti{\v c}ka}, {Mikul{\'a}{\v s}ek}, {Henry},
  {Zverko}, {{\v Z}i{\v z}ovsk{\'y}}, {Skalick{\'y}}  \& {Zv{\v e}{\v
  r}ina}}{{Krti{\v c}ka} et~al.}{2009}]{2009A&A...499..567K}
{Krti{\v c}ka} J.,  {Mikul{\'a}{\v s}ek} Z.,  {Henry} G.~W.,  {Zverko} J.,
  {{\v Z}i{\v z}ovsk{\'y}} J.,  {Skalick{\'y}} J.,   {Zv{\v e}{\v r}ina} P.,
  2009, \mn@doi [\aap] {10.1051/0004-6361/200811123}, \href
  {http://adsabs.harvard.edu/abs/2009A%26A...499..567K} {499, 567}

\bibitem[\protect\citeauthoryear{{Kudryavtsev}, {Romanyuk}, {Elkin}  \&
  {Paunzen}}{{Kudryavtsev} et~al.}{2006}]{2006MNRAS.372.1804K}
{Kudryavtsev} D.~O.,  {Romanyuk} I.~I.,  {Elkin} V.~G.,   {Paunzen} E.,  2006,
  \mn@doi [\mnras] {10.1111/j.1365-2966.2006.10994.x}, \href
  {https://ui.adsabs.harvard.edu/abs/2006MNRAS.372.1804K} {372, 1804}

\bibitem[\protect\citeauthoryear{{K{\"u}ker}}{{K{\"u}ker}}{2017}]{2017AN....338..868K}
{K{\"u}ker} M.,  2017, \mn@doi [Astronomische Nachrichten]
  {10.1002/asna.201713412}, \href
  {https://ui.adsabs.harvard.edu/abs/2017AN....338..868K} {338, 868}

\bibitem[\protect\citeauthoryear{{Kupka}, {Piskunov}, {Ryabchikova}, {Stempels}
   \& {Weiss}}{{Kupka} et~al.}{1999}]{kupka1999}
{Kupka} F.~G.,  {Piskunov} N.,  {Ryabchikova} T.~A.,  {Stempels} H.~C.,
  {Weiss} W.~W.,  1999, \mn@doi [\aaps] {10.1051/aas:1999267}, \href
  {http://adsabs.harvard.edu/abs/1999A26AS..138..119K} {138, 119}

\bibitem[\protect\citeauthoryear{{Kupka}, {Ryabchikova}, {Piskunov}, {Stempels}
   \& {Weiss}}{{Kupka} et~al.}{2000}]{kupka2000}
{Kupka} F.~G.,  {Ryabchikova} T.~A.,  {Piskunov} N.~E.,  {Stempels} H.~C.,
  {Weiss} W.~W.,  2000, Balt.\ Astron., \href
  {http://adsabs.harvard.edu/abs/2000BaltA...9..590K} {9, 590}

\bibitem[\protect\citeauthoryear{{Lamers}, {Snow}  \& {Lindholm}}{{Lamers}
  et~al.}{1995}]{lamers1995}
{Lamers} H.~J.~G.~L.~M.,  {Snow} T.~P.,   {Lindholm} D.~M.,  1995, \mn@doi
  [\apj] {10.1086/176575}, \href
  {http://adsabs.harvard.edu/abs/1995ApJ...455..269L} {455, 269}

\bibitem[\protect\citeauthoryear{{Landstreet} \& {Borra}}{{Landstreet} \&
  {Borra}}{1978}]{lb1978}
{Landstreet} J.~D.,  {Borra} E.~F.,  1978, \mn@doi [\apjl] {10.1086/182746},
  \href {http://cdsads.u-strasbg.fr/abs/1978ApJ...224L...5L} {224, L5}

\bibitem[\protect\citeauthoryear{{Landstreet}, {Borra}  \&
  {Fontaine}}{{Landstreet} et~al.}{1979}]{1979MNRAS.188..609L}
{Landstreet} J.~D.,  {Borra} E.~F.,   {Fontaine} G.,  1979, \mn@doi [\mnras]
  {10.1093/mnras/188.3.609}, \href
  {https://ui.adsabs.harvard.edu/abs/1979MNRAS.188..609L} {188, 609}

\bibitem[\protect\citeauthoryear{{Landstreet}, {Bagnulo}, {Andretta},
  {Fossati}, {Mason}, {Silaj}  \& {Wade}}{{Landstreet} et~al.}{2007}]{land2007}
{Landstreet} J.~D.,  {Bagnulo} S.,  {Andretta} V.,  {Fossati} L.,  {Mason} E.,
  {Silaj} J.,   {Wade} G.~A.,  2007, \mn@doi [\aap]
  {10.1051/0004-6361:20077343}, \href
  {http://adsabs.harvard.edu/abs/2007A26A...470..685L} {470, 685}

\bibitem[\protect\citeauthoryear{{Lanz} \& {Hubeny}}{{Lanz} \&
  {Hubeny}}{2007}]{lanzhubeny2007}
{Lanz} T.,  {Hubeny} I.,  2007, \mn@doi [\apjs] {10.1086/511270}, \href
  {http://adsabs.harvard.edu/abs/2007ApJS..169...83L} {169, 83}

\bibitem[\protect\citeauthoryear{{Leone}}{{Leone}}{1993}]{1993A&A...273..509L}
{Leone} F.,  1993, \aap, \href
  {http://adsabs.harvard.edu/abs/1993A%26A...273..509L} {273, 509}

\bibitem[\protect\citeauthoryear{{Leone}, {Umana}  \& {Trigilio}}{{Leone}
  et~al.}{1996}]{1996A&A...310..271L}
{Leone} F.,  {Umana} G.,   {Trigilio} C.,  1996, \aap, \href
  {https://ui.adsabs.harvard.edu/#abs/1996A&A...310..271L} {310, 271}

\bibitem[\protect\citeauthoryear{{Leone}, {Bohlender}, {Bolton}, {Buemi},
  {Catanzaro}, {Hill}  \& {Stift}}{{Leone} et~al.}{2010}]{leone2010}
{Leone} F.,  {Bohlender} D.~A.,  {Bolton} C.~T.,  {Buemi} C.,  {Catanzaro} G.,
  {Hill} G.~M.,   {Stift} M.~J.,  2010, \mn@doi [\mnras]
  {10.1111/j.1365-2966.2009.15858.x}, \href
  {http://cdsads.u-strasbg.fr/abs/2010MNRAS.401.2739L} {401, 2739}

\bibitem[\protect\citeauthoryear{{Linsky}, {Drake}  \& {Bastian}}{{Linsky}
  et~al.}{1992}]{1992ApJ...393..341L}
{Linsky} J.~L.,  {Drake} S.~A.,   {Bastian} T.~S.,  1992, \mn@doi [\apj]
  {10.1086/171509}, \href {http://adsabs.harvard.edu/abs/1992ApJ...393..341L}
  {393, 341}

\bibitem[\protect\citeauthoryear{{McDonald}, {Zijlstra}  \& {Boyer}}{{McDonald}
  et~al.}{2012}]{2012MNRAS.427..343M}
{McDonald} I.,  {Zijlstra} A.~A.,   {Boyer} M.~L.,  2012, \mn@doi [\mnras]
  {10.1111/j.1365-2966.2012.21873.x}, \href
  {https://ui.adsabs.harvard.edu/abs/2012MNRAS.427..343M} {427, 343}

\bibitem[\protect\citeauthoryear{{Mikul{\'a}{\v s}ek} et~al.,}{{Mikul{\'a}{\v
  s}ek} et~al.}{2011}]{miku2011}
{Mikul{\'a}{\v s}ek} Z.,  et~al., 2011, \mn@doi [\aap]
  {10.1051/0004-6361/201117784}, \href
  {http://adsabs.harvard.edu/abs/2011A26A...534L...5M} {534, L5}

\bibitem[\protect\citeauthoryear{{Nakajima}}{{Nakajima}}{1985}]{1985ApSS.116..285N}
{Nakajima} R.,  1985, \mn@doi [\apss] {10.1007/BF00653783}, \href
  {http://adsabs.harvard.edu/abs/1985Ap26SS.116..285N} {116, 285}

\bibitem[\protect\citeauthoryear{{Naz{\'e}}, {Petit}, {Rinbrand}, {Cohen},
  {Owocki}, {ud-Doula}  \& {Wade}}{{Naz{\'e}}
  et~al.}{2014}]{2014ApJS..215...10N}
{Naz{\'e}} Y.,  {Petit} V.,  {Rinbrand} M.,  {Cohen} D.,  {Owocki} S.,
  {ud-Doula} A.,   {Wade} G.~A.,  2014, \mn@doi [\apjs]
  {10.1088/0067-0049/215/1/10}, \href
  {http://adsabs.harvard.edu/abs/2014ApJS..215...10N} {215, 10}

\bibitem[\protect\citeauthoryear{{Neiner} et~al.,}{{Neiner}
  et~al.}{2003}]{neiner2003b}
{Neiner} C.,  et~al., 2003, \mn@doi [\aap] {10.1051/0004-6361:20031342}, \href
  {http://adsabs.harvard.edu/abs/2003A26A...411..565N} {411, 565}

\bibitem[\protect\citeauthoryear{{Netopil}, {Paunzen}, {Maitzen}, {North}  \&
  {Hubrig}}{{Netopil} et~al.}{2008}]{2008AA...491..545N}
{Netopil} M.,  {Paunzen} E.,  {Maitzen} H.~M.,  {North} P.,   {Hubrig} S.,
  2008, \mn@doi [\aap] {10.1051/0004-6361:200810325}, \href
  {http://cdsads.u-strasbg.fr/abs/2008A26A...491..545N} {491, 545}

\bibitem[\protect\citeauthoryear{{Netopil}, {Paunzen}, {H{\"u}mmerich}  \&
  {Bernhard}}{{Netopil} et~al.}{2017}]{2017MNRAS.468.2745N}
{Netopil} M.,  {Paunzen} E.,  {H{\"u}mmerich} S.,   {Bernhard} K.,  2017,
  \mn@doi [\mnras] {10.1093/mnras/stx674}, \href
  {https://ui.adsabs.harvard.edu/abs/2017MNRAS.468.2745N} {468, 2745}

\bibitem[\protect\citeauthoryear{{North}}{{North}}{1987}]{1987A&AS...69..371N}
{North} P.,  1987, \aaps, \href
  {https://ui.adsabs.harvard.edu/abs/1987A&AS...69..371N} {69, 371}

\bibitem[\protect\citeauthoryear{{Oksala}, {Wade}, {Marcolino}, {Grunhut},
  {Bohlender}, {Manset}, {Townsend}  \& {Mimes Collaboration}}{{Oksala}
  et~al.}{2010}]{2010MNRAS.405L..51O}
{Oksala} M.~E.,  {Wade} G.~A.,  {Marcolino} W.~L.~F.,  {Grunhut} J.,
  {Bohlender} D.,  {Manset} N.,  {Townsend} R.~H.~D.,   {Mimes Collaboration}
  2010, \mn@doi [\mnras] {10.1111/j.1745-3933.2010.00857.x}, \href
  {http://adsabs.harvard.edu/abs/2010MNRAS.405L..51O} {405, L51}

\bibitem[\protect\citeauthoryear{{Oksala}, {Wade}, {Townsend}, {Owocki},
  {Kochukhov}, {Neiner}, {Alecian}  \& {Grunhut}}{{Oksala}
  et~al.}{2012}]{oks2012}
{Oksala} M.~E.,  {Wade} G.~A.,  {Townsend} R.~H.~D.,  {Owocki} S.~P.,
  {Kochukhov} O.,  {Neiner} C.,  {Alecian} E.,   {Grunhut} J.,  2012, \mn@doi
  [MNRAS] {10.1111/j.1365-2966.2011.19753.x}, \href
  {http://adsabs.harvard.edu/abs/2012MNRAS.419..959O} {419, 959}

\bibitem[\protect\citeauthoryear{{Oksala} et~al.,}{{Oksala}
  et~al.}{2015}]{2015MNRAS.451.2015O}
{Oksala} M.~E.,  et~al., 2015, \mn@doi [\mnras] {10.1093/mnras/stv1086}, \href
  {http://adsabs.harvard.edu/abs/2015MNRAS.451.2015O} {451, 2015}

\bibitem[\protect\citeauthoryear{{Oksala}, {Silvester}, {Kochukhov}, {Neiner},
  {Wade}  \& {MiMeS Collaboration}}{{Oksala}
  et~al.}{2018}]{2018MNRAS.473.3367O}
{Oksala} M.~E.,  {Silvester} J.,  {Kochukhov} O.,  {Neiner} C.,  {Wade} G.~A.,
   {MiMeS Collaboration} 2018, \mn@doi [\mnras] {10.1093/mnras/stx2487}, \href
  {https://ui.adsabs.harvard.edu/abs/2018MNRAS.473.3367O} {473, 3367}

\bibitem[\protect\citeauthoryear{{Owocki} \& {Cranmer}}{{Owocki} \&
  {Cranmer}}{2018}]{2018MNRAS.474.3090O}
{Owocki} S.~P.,  {Cranmer} S.~R.,  2018, \mn@doi [\mnras]
  {10.1093/mnras/stx2989}, \href
  {http://adsabs.harvard.edu/abs/2018MNRAS.474.3090O} {474, 3090}

\bibitem[\protect\citeauthoryear{{Owocki} \& {Puls}}{{Owocki} \&
  {Puls}}{2002}]{2002ApJ...568..965O}
{Owocki} S.~P.,  {Puls} J.,  2002, \mn@doi [\apj] {10.1086/339037}, \href
  {https://ui.adsabs.harvard.edu/abs/2002ApJ...568..965O} {568, 965}

\bibitem[\protect\citeauthoryear{{Owocki}, {Shultz}, {ud-Doula}, {Sundqvist},
  {Townsend}  \& {Cranmer}}{{Owocki} et~al.}{2020}]{owocki2020}
{Owocki} S.,  {Shultz} M.~E.,  {ud-Doula} A.,  {Sundqvist} J.~O.,  {Townsend}
  R.~H.~D.,   {Cranmer} S.~R.,  2020, \mnras

\bibitem[\protect\citeauthoryear{{P{\'a}pics} et~al.,}{{P{\'a}pics}
  et~al.}{2012}]{2012A&A...542A..55P}
{P{\'a}pics} P.~I.,  et~al., 2012, \mn@doi [\aap]
  {10.1051/0004-6361/201218809}, \href
  {http://adsabs.harvard.edu/abs/2012A%26A...542A..55P} {542, A55}

\bibitem[\protect\citeauthoryear{{Pearson}}{{Pearson}}{1895}]{1895RSPS...58..240P}
{Pearson} K.,  1895, Proceedings of the Royal Society of London Series I, \href
  {http://adsabs.harvard.edu/abs/1895RSPS...58..240P} {58, 240}

\bibitem[\protect\citeauthoryear{{Pecaut} \& {Mamajek}}{{Pecaut} \&
  {Mamajek}}{2013}]{2013ApJS..208....9P}
{Pecaut} M.~J.,  {Mamajek} E.~E.,  2013, \mn@doi [\apjs]
  {10.1088/0067-0049/208/1/9}, \href
  {http://adsabs.harvard.edu/abs/2013ApJS..208....9P} {208, 9}

\bibitem[\protect\citeauthoryear{{Petit} et~al.,}{{Petit}
  et~al.}{2013}]{petit2013}
{Petit} V.,  et~al., 2013, \mn@doi [\mnras] {10.1093/mnras/sts344}, \href
  {http://adsabs.harvard.edu/abs/2013MNRAS.429..398P} {429, 398}

\bibitem[\protect\citeauthoryear{{Petit}, {Louge}, {Th{\'e}ado}, {Paletou},
  {Manset}, {Morin}, {Marsden}  \& {Jeffers}}{{Petit}
  et~al.}{2014}]{2014PASP..126..469P}
{Petit} P.,  {Louge} T.,  {Th{\'e}ado} S.,  {Paletou} F.,  {Manset} N.,
  {Morin} J.,  {Marsden} S.~C.,   {Jeffers} S.~V.,  2014, \mn@doi [\pasp]
  {10.1086/676976}, \href
  {https://ui.adsabs.harvard.edu/abs/2014PASP..126..469P} {126, 469}

\bibitem[\protect\citeauthoryear{{Pillitteri}, {Wolk}, {Chen}  \&
  {Goodman}}{{Pillitteri} et~al.}{2016}]{2016A&A...592A..88P}
{Pillitteri} I.,  {Wolk} S.~J.,  {Chen} H.~H.,   {Goodman} A.,  2016, \mn@doi
  [\aap] {10.1051/0004-6361/201628284}, \href
  {https://ui.adsabs.harvard.edu/abs/2016A&A...592A..88P} {592, A88}

\bibitem[\protect\citeauthoryear{{Piskunov} \& {Kochukhov}}{{Piskunov} \&
  {Kochukhov}}{2002}]{pk2002}
{Piskunov} N.~E.,  {Kochukhov} O.,  2002, \mn@doi [\aap]
  {10.1051/0004-6361:20011517}, \href
  {http://adsabs.harvard.edu/abs/2002A26A...381..736P} {381, 736}

\bibitem[\protect\citeauthoryear{{Piskunov} \& {Valenti}}{{Piskunov} \&
  {Valenti}}{2002}]{2002A&A...385.1095P}
{Piskunov} N.~E.,  {Valenti} J.~A.,  2002, \mn@doi [\aap]
  {10.1051/0004-6361:20020175}, \href
  {http://adsabs.harvard.edu/abs/2002A%26A...385.1095P} {385, 1095}

\bibitem[\protect\citeauthoryear{{Piskunov}, {Kupka}, {Ryabchikova}, {Weiss}
  \& {Jeffery}}{{Piskunov} et~al.}{1995}]{piskunov1995}
{Piskunov} N.~E.,  {Kupka} F.,  {Ryabchikova} T.~A.,  {Weiss} W.~W.,
  {Jeffery} C.~S.,  1995, \aaps, \href
  {http://adsabs.harvard.edu/abs/1995A26AS..112..525P} {112, 525}

\bibitem[\protect\citeauthoryear{{Prat}, {Mathis}, {Buysschaert}, {Van Beeck},
  {Bowman}, {Aerts}  \& {Neiner}}{{Prat} et~al.}{2019}]{2019A&A...627A..64P}
{Prat} V.,  {Mathis} S.,  {Buysschaert} B.,  {Van Beeck} J.,  {Bowman} D.~M.,
  {Aerts} C.,   {Neiner} C.,  2019, \mn@doi [\aap]
  {10.1051/0004-6361/201935462}, \href
  {https://ui.adsabs.harvard.edu/abs/2019A&A...627A..64P} {627, A64}

\bibitem[\protect\citeauthoryear{{Preuss}, {Sch{\"u}ssler}, {Holzwarth}  \&
  {Solanki}}{{Preuss} et~al.}{2004}]{2004AA...417..987P}
{Preuss} O.,  {Sch{\"u}ssler} M.,  {Holzwarth} V.,   {Solanki} S.~K.,  2004,
  \mn@doi [\aap] {10.1051/0004-6361:20034525}, \href
  {http://adsabs.harvard.edu/abs/2004A26A...417..987P} {417, 987}

\bibitem[\protect\citeauthoryear{{Rebull}, {Stauffer}, {Cody}, {Hillenbrand },
  {David}  \& {Pinsonneault}}{{Rebull} et~al.}{2018}]{2018AJ....155..196R}
{Rebull} L.~M.,  {Stauffer} J.~R.,  {Cody} A.~M.,  {Hillenbrand } L.~A.,
  {David} T.~J.,   {Pinsonneault} M.,  2018, \mn@doi [\aj]
  {10.3847/1538-3881/aab605}, \href
  {https://ui.adsabs.harvard.edu/abs/2018AJ....155..196R} {155, 196}

\bibitem[\protect\citeauthoryear{{Reeve} \& {Howarth}}{{Reeve} \&
  {Howarth}}{2016}]{2016MNRAS.456.1294R}
{Reeve} D.~C.,  {Howarth} I.~D.,  2016, \mn@doi [\mnras]
  {10.1093/mnras/stv2631}, \href
  {http://adsabs.harvard.edu/abs/2016MNRAS.456.1294R} {456, 1294}

\bibitem[\protect\citeauthoryear{{Rivinius}, {Szeifert}, {Barrera}, {Townsend},
  {{\v S}tefl}  \& {Baade}}{{Rivinius} et~al.}{2010}]{2010MNRAS.405L..46R}
{Rivinius} T.,  {Szeifert} T.,  {Barrera} L.,  {Townsend} R.~H.~D.,  {{\v
  S}tefl} S.,   {Baade} D.,  2010, \mn@doi [\mnras]
  {10.1111/j.1745-3933.2010.00856.x}, \href
  {http://adsabs.harvard.edu/abs/2010MNRAS.405L..46R} {405, L46}

\bibitem[\protect\citeauthoryear{{Rivinius}, {Carciofi}  \&
  {Martayan}}{{Rivinius} et~al.}{2013a}]{2013AARv..21...69R}
{Rivinius} T.,  {Carciofi} A.~C.,   {Martayan} C.,  2013a, \mn@doi [\aapr]
  {10.1007/s00159-013-0069-0}, \href
  {http://adsabs.harvard.edu/abs/2013A26ARv..21...69R} {21, 69}

\bibitem[\protect\citeauthoryear{{Rivinius}, {Townsend}, {Kochukhov}, {{\v
  S}tefl}, {Baade}, {Barrera}  \& {Szeifert}}{{Rivinius}
  et~al.}{2013b}]{rivi2013}
{Rivinius} T.,  {Townsend} R.~H.~D.,  {Kochukhov} O.,  {{\v S}tefl} S.,
  {Baade} D.,  {Barrera} L.,   {Szeifert} T.,  2013b, \mn@doi [\mnras]
  {10.1093/mnras/sts323}, \href
  {http://cdsads.u-strasbg.fr/abs/2013MNRAS.429..177R} {429, 177}

\bibitem[\protect\citeauthoryear{{Robichon}, {Arenou}, {Mermilliod}  \&
  {Turon}}{{Robichon} et~al.}{1999}]{1999A&A...345..471R}
{Robichon} N.,  {Arenou} F.,  {Mermilliod} J.~C.,   {Turon} C.,  1999, \aap,
  \href {https://ui.adsabs.harvard.edu/abs/1999A&A...345..471R} {345, 471}

\bibitem[\protect\citeauthoryear{{Romanyuk}, {Semenko}, {Kudryavtsev},
  {Moiseeva}  \& {Yakunin}}{{Romanyuk} et~al.}{2017}]{2017AstBu..72..391R}
{Romanyuk} I.~I.,  {Semenko} E.~A.,  {Kudryavtsev} D.~O.,  {Moiseeva} A.~V.,
  {Yakunin} I.~A.,  2017, \mn@doi [Astrophysical Bulletin]
  {10.1134/S1990341317040046}, \href
  {https://ui.adsabs.harvard.edu/#abs/2017AstBu..72..391R} {72, 391}

\bibitem[\protect\citeauthoryear{{Ryabchikova}, {Piskunov}, {Kupka}  \&
  {Weiss}}{{Ryabchikova} et~al.}{1997}]{ryabchikova1997}
{Ryabchikova} T.~A.,  {Piskunov} N.~E.,  {Kupka} F.,   {Weiss} W.~W.,  1997,
  Balt.\ Astron., \href {http://cdsads.u-strasbg.fr/abs/1997BaltA...6..244R}
  {6, 244}

\bibitem[\protect\citeauthoryear{{Ryabchikova}, {Piskunov}, {Kurucz},
  {Stempels}, {Heiter}, {Pakhomov}  \& {Barklem}}{{Ryabchikova}
  et~al.}{2015}]{2015PhyS...90e4005R}
{Ryabchikova} T.,  {Piskunov} N.,  {Kurucz} R.~L.,  {Stempels} H.~C.,  {Heiter}
  U.,  {Pakhomov} Y.,   {Barklem} P.~S.,  2015, \mn@doi [\physscr]
  {10.1088/0031-8949/90/5/054005}, \href
  {http://adsabs.harvard.edu/abs/2015PhyS...90e4005R} {90, 054005}

\bibitem[\protect\citeauthoryear{{Sanz-Forcada}, {Franciosini}  \&
  {Pallavicini}}{{Sanz-Forcada} et~al.}{2004}]{2004AA...421..715S}
{Sanz-Forcada} J.,  {Franciosini} E.,   {Pallavicini} R.,  2004, \mn@doi [\aap]
  {10.1051/0004-6361:20047159}, \href
  {http://adsabs.harvard.edu/abs/2004A26A...421..715S} {421, 715}

\bibitem[\protect\citeauthoryear{{Scandariato} et~al.,}{{Scandariato}
  et~al.}{2013}]{2013A&A...552A...7S}
{Scandariato} G.,  et~al., 2013, \mn@doi [\aap] {10.1051/0004-6361/201219875},
  \href {https://ui.adsabs.harvard.edu/abs/2013A&A...552A...7S} {552, A7}

\bibitem[\protect\citeauthoryear{{Semenko}, {Kudryavtsev}, {Ryabchikova}  \&
  {Romanyuk}}{{Semenko} et~al.}{2008}]{2008AstBu..63..128S}
{Semenko} E.~A.,  {Kudryavtsev} D.~O.,  {Ryabchikova} T.~A.,   {Romanyuk}
  I.~I.,  2008, \mn@doi [Astrophysical Bulletin] {10.1134/S1990341308020041},
  \href {https://ui.adsabs.harvard.edu/abs/2008AstBu..63..128S} {63, 128}

\bibitem[\protect\citeauthoryear{{Shore} \& {Brown}}{{Shore} \&
  {Brown}}{1990}]{1990ApJ...365..665S}
{Shore} S.~N.,  {Brown} D.~N.,  1990, \mn@doi [\apj] {10.1086/169520}, \href
  {http://adsabs.harvard.edu/abs/1990ApJ...365..665S} {365, 665}

\bibitem[\protect\citeauthoryear{{Shore}, {Bohlender}, {Bolton}, {North}  \&
  {Hill}}{{Shore} et~al.}{2004}]{2004A&A...421..203S}
{Shore} S.~N.,  {Bohlender} D.~A.,  {Bolton} C.~T.,  {North} P.,   {Hill}
  G.~M.,  2004, \mn@doi [\aap] {10.1051/0004-6361:20035612}, \href
  {https://ui.adsabs.harvard.edu/\#abs/2004A&A...421..203S} {421, 203}

\bibitem[\protect\citeauthoryear{{Shultz} et~al.,}{{Shultz}
  et~al.}{2015}]{2015MNRAS.449.3945S}
{Shultz} M.,  et~al., 2015, \mn@doi [\mnras] {10.1093/mnras/stv564}, \href
  {http://adsabs.harvard.edu/abs/2015MNRAS.449.3945S} {449, 3945}

\bibitem[\protect\citeauthoryear{{Shultz}, {Wade}, {Rivinius}, {Sikora}  \&
  {MiMeS Collaboration}}{{Shultz} et~al.}{2016}]{2016ASPC..506..305S}
{Shultz} M.,  {Wade} G.,  {Rivinius} T.,  {Sikora} J.,   {MiMeS Collaboration}
  2016, in {Sigut} T.~A.~A.,  {Jones} C.~E.,  eds,  Astronomical Society of the
  Pacific Conference Series Vol. 506, Bright Emissaries: Be Stars as Messengers
  of Star-Disk Physics. p.~305

\bibitem[\protect\citeauthoryear{{Shultz}, {Wade}, {Rivinius}, {Neiner},
  {Henrichs}, {Marcolino}  \& {MiMeS Collaboration}}{{Shultz}
  et~al.}{2017}]{2017MNRAS.471.2286S}
{Shultz} M.,  {Wade} G.~A.,  {Rivinius} T.,  {Neiner} C.,  {Henrichs} H.,
  {Marcolino} W.,   {MiMeS Collaboration} 2017, \mn@doi [\mnras]
  {10.1093/mnras/stx1632}, \href
  {http://adsabs.harvard.edu/abs/2017MNRAS.471.2286S} {471, 2286}

\bibitem[\protect\citeauthoryear{{Shultz}, {Rivinius}, {Wade}, {Alecian}  \&
  {Petit}}{{Shultz} et~al.}{2018a}]{2018MNRAS.475..839S}
{Shultz} M.,  {Rivinius} T.,  {Wade} G.~A.,  {Alecian} E.,   {Petit} V.,
  2018a, \mn@doi [\mnras] {10.1093/mnras/stx3238}, \href
  {http://adsabs.harvard.edu/abs/2018MNRAS.475..839S} {475, 839}

\bibitem[\protect\citeauthoryear{{Shultz} et~al.,}{{Shultz}
  et~al.}{2018b}]{2018MNRAS.475.5144S}
{Shultz} M.~E.,  et~al., 2018b, \mn@doi [\mnras] {10.1093/mnras/sty103}, \href
  {http://adsabs.harvard.edu/abs/2018MNRAS.475.5144S} {475, 5144}

\bibitem[\protect\citeauthoryear{{Shultz} et~al.,}{{Shultz}
  et~al.}{2019a}]{2019MNRAS.tmp.2196S}
{Shultz} M.~E.,  et~al., 2019a, \mn@doi [\mnras] {10.1093/mnras/stz2551}, \href
  {https://ui.adsabs.harvard.edu/abs/2019MNRAS.tmp.2196S} {}

\bibitem[\protect\citeauthoryear{{Shultz} et~al.,}{{Shultz}
  et~al.}{2019b}]{2019MNRAS.482.3950S}
{Shultz} M.,  et~al., 2019b, \mn@doi [\mnras] {10.1093/mnras/sty2985}, \href
  {http://adsabs.harvard.edu/abs/2019MNRAS.482.3950S} {482, 3950}

\bibitem[\protect\citeauthoryear{{Shultz} et~al.,}{{Shultz}
  et~al.}{2019c}]{2019MNRAS.485.1508S}
{Shultz} M.~E.,  et~al., 2019c, \mn@doi [\mnras] {10.1093/mnras/stz416}, \href
  {https://ui.adsabs.harvard.edu/abs/2019MNRAS.485.1508S} {485, 1508}

\bibitem[\protect\citeauthoryear{{Shulyak} et~al.,}{{Shulyak}
  et~al.}{2007}]{shulyak2007}
{Shulyak} D.,  et~al., 2007, \mn@doi [\aap] {10.1051/0004-6361:20064998}, \href
  {http://cdsads.u-strasbg.fr/abs/2007A26A...464.1089S} {464, 1089}

\bibitem[\protect\citeauthoryear{{Shulyak}, {Kochukhov}, {Valyavin}, {Lee},
  {Galazutdinov}, {Kim}, {Han}  \& {Burlakova}}{{Shulyak}
  et~al.}{2010}]{2010A&A...509A..28S}
{Shulyak} D.,  {Kochukhov} O.,  {Valyavin} G.,  {Lee} B.~C.,  {Galazutdinov}
  G.,  {Kim} K.~M.,  {Han} I.,   {Burlakova} T.,  2010, \mn@doi [\aap]
  {10.1051/0004-6361/200912615}, \href
  {https://ui.adsabs.harvard.edu/abs/2010A&A...509A..28S} {509, A28}

\bibitem[\protect\citeauthoryear{{Sikora} et~al.,}{{Sikora}
  et~al.}{2015}]{2015MNRAS.451.1928S}
{Sikora} J.,  et~al., 2015, \mn@doi [\mnras] {10.1093/mnras/stv1051}, \href
  {http://adsabs.harvard.edu/abs/2015MNRAS.451.1928S} {451, 1928}

\bibitem[\protect\citeauthoryear{{Sikora} et~al.,}{{Sikora}
  et~al.}{2016}]{2016MNRAS.460.1811S}
{Sikora} J.,  et~al., 2016, \mn@doi [\mnras] {10.1093/mnras/stw1077}, \href
  {http://adsabs.harvard.edu/abs/2016MNRAS.460.1811S} {460, 1811}

\bibitem[\protect\citeauthoryear{{Sikora}, {Wade}, {Power}  \&
  {Neiner}}{{Sikora} et~al.}{2019a}]{2019MNRAS.483.2300S}
{Sikora} J.,  {Wade} G.~A.,  {Power} J.,   {Neiner} C.,  2019a, \mn@doi
  [\mnras] {10.1093/mnras/sty3105}, \href
  {http://adsabs.harvard.edu/abs/2019MNRAS.483.2300S} {483, 2300}

\bibitem[\protect\citeauthoryear{{Sikora}, {Wade}, {Power}  \&
  {Neiner}}{{Sikora} et~al.}{2019b}]{2019MNRAS.483.3127S}
{Sikora} J.,  {Wade} G.~A.,  {Power} J.,   {Neiner} C.,  2019b, \mn@doi
  [\mnras] {10.1093/mnras/sty2895}, \href
  {http://adsabs.harvard.edu/abs/2019MNRAS.483.3127S} {483, 3127}

\bibitem[\protect\citeauthoryear{{Smith} \& {Groote}}{{Smith} \&
  {Groote}}{2001}]{smithgroote2001}
{Smith} M.~A.,  {Groote} D.,  2001, \mn@doi [\aap]
  {10.1051/0004-6361:20010472}, \href
  {http://cdsads.u-strasbg.fr/abs/2001A26A...372..208S} {372, 208}

\bibitem[\protect\citeauthoryear{{Smith}, {Wade}, {Bohlender}  \&
  {Bolton}}{{Smith} et~al.}{2006}]{2006AA...458..581S}
{Smith} M.~A.,  {Wade} G.~A.,  {Bohlender} D.~A.,   {Bolton} C.~T.,  2006,
  \mn@doi [\aap] {10.1051/0004-6361:20054760}, \href
  {http://cdsads.u-strasbg.fr/abs/2006A26A...458..581S} {458, 581}

\bibitem[\protect\citeauthoryear{{Springmann} \& {Pauldrach}}{{Springmann} \&
  {Pauldrach}}{1992}]{1992A&A...262..515S}
{Springmann} U.~W.~E.,  {Pauldrach} A.~W.~A.,  1992, \aap, \href
  {https://ui.adsabs.harvard.edu/abs/1992A&A...262..515S} {262, 515}

\bibitem[\protect\citeauthoryear{{Stepien} \& {Czechowski}}{{Stepien} \&
  {Czechowski}}{1993}]{1993A&A...268..187S}
{Stepien} K.,  {Czechowski} W.,  1993, \aap, \href
  {https://ui.adsabs.harvard.edu/abs/1993A&A...268..187S} {268, 187}

\bibitem[\protect\citeauthoryear{{Telting}, {Schrijvers}, {Ilyin},
  {Uytterhoeven}, {De Ridder}, {Aerts}  \& {Henrichs}}{{Telting}
  et~al.}{2006}]{2006AA...452..945T}
{Telting} J.~H.,  {Schrijvers} C.,  {Ilyin} I.~V.,  {Uytterhoeven} K.,  {De
  Ridder} J.,  {Aerts} C.,   {Henrichs} H.~F.,  2006, \mn@doi [\aap]
  {10.1051/0004-6361:20054730}, \href
  {http://cdsads.u-strasbg.fr/abs/2006A26A...452..945T} {452, 945}

\bibitem[\protect\citeauthoryear{{Thompson} \& {Landstreet}}{{Thompson} \&
  {Landstreet}}{1985}]{thom1985}
{Thompson} I.~B.,  {Landstreet} J.~D.,  1985, \mn@doi [ApJL] {10.1086/184424},
  \href {http://cdsads.u-strasbg.fr/abs/1985ApJ...289L...9T} {289, L9}

\bibitem[\protect\citeauthoryear{{Townsend}}{{Townsend}}{2008}]{town2008}
{Townsend} R.~H.~D.,  2008, \mn@doi [\mnras]
  {10.1111/j.1365-2966.2008.13462.x}, \href
  {http://adsabs.harvard.edu/abs/2008MNRAS.389..559T} {389, 559}

\bibitem[\protect\citeauthoryear{{Townsend}}{{Townsend}}{2014}]{2014ascl.soft12005T}
{Townsend} R.,  2014, {BRUCE/KYLIE: Pulsating star spectra synthesizer},
  Astrophysics Source Code Library (\mn@eprint {ascl} {1412.005})

\bibitem[\protect\citeauthoryear{{Townsend} \& {Owocki}}{{Townsend} \&
  {Owocki}}{2005}]{town2005c}
{Townsend} R.~H.~D.,  {Owocki} S.~P.,  2005, \mn@doi [\mnras]
  {10.1111/j.1365-2966.2005.08642.x}, \href
  {http://adsabs.harvard.edu/abs/2005MNRAS.357..251T} {357, 251}

\bibitem[\protect\citeauthoryear{{Townsend}, {Owocki}  \& {Groote}}{{Townsend}
  et~al.}{2005}]{town2005b}
{Townsend} R.~H.~D.,  {Owocki} S.~P.,   {Groote} D.,  2005, \mn@doi [\apjl]
  {10.1086/462413}, \href {http://cdsads.u-strasbg.fr/abs/2005ApJ...630L..81T}
  {630, L81}

\bibitem[\protect\citeauthoryear{{Townsend}, {Owocki}  \&
  {Ud-Doula}}{{Townsend} et~al.}{2007}]{town2007}
{Townsend} R.~H.~D.,  {Owocki} S.~P.,   {Ud-Doula} A.,  2007, \mn@doi [\mnras]
  {10.1111/j.1365-2966.2007.12427.x}, \href
  {http://cdsads.u-strasbg.fr/abs/2007MNRAS.382..139T} {382, 139}

\bibitem[\protect\citeauthoryear{{Townsend}, {Oksala}, {Cohen}, {Owocki}  \&
  {ud-Doula}}{{Townsend} et~al.}{2010}]{town2010}
{Townsend} R.~H.~D.,  {Oksala} M.~E.,  {Cohen} D.~H.,  {Owocki} S.~P.,
  {ud-Doula} A.,  2010, \mn@doi [\apjl] {10.1088/2041-8205/714/2/L318}, \href
  {http://adsabs.harvard.edu/abs/2010ApJ...714L.318T} {714, L318}

\bibitem[\protect\citeauthoryear{{Townsend} et~al.,}{{Townsend}
  et~al.}{2013}]{town2013}
{Townsend} R.~H.~D.,  et~al., 2013, \mn@doi [\apj]
  {10.1088/0004-637X/769/1/33}, \href
  {http://adsabs.harvard.edu/abs/2013ApJ...769...33T} {769, 33}

\bibitem[\protect\citeauthoryear{{Vink}, {de Koter}  \& {Lamers}}{{Vink}
  et~al.}{2001}]{vink2001}
{Vink} J.~S.,  {de Koter} A.,   {Lamers} H.~J.~G.~L.~M.,  2001, \mn@doi [\aap]
  {10.1051/0004-6361:20010127}, \href
  {http://adsabs.harvard.edu/abs/2001A26A...369..574V} {369, 574}

\bibitem[\protect\citeauthoryear{{Wade} et~al.,}{{Wade}
  et~al.}{2017}]{2017MNRAS.465.2517W}
{Wade} G.~A.,  et~al., 2017, \mn@doi [\mnras] {10.1093/mnras/stw2799}, \href
  {http://adsabs.harvard.edu/abs/2017MNRAS.465.2517W} {465, 2517}

\bibitem[\protect\citeauthoryear{{Walborn}}{{Walborn}}{1974}]{walborn1974}
{Walborn} N.~R.,  1974, \mn@doi [\apjl] {10.1086/181558}, \href
  {http://cdsads.u-strasbg.fr/abs/1974ApJ...191L..95W} {191, L95}

\bibitem[\protect\citeauthoryear{{Walker} et~al.,}{{Walker}
  et~al.}{2003}]{2003PASP..115.1023W}
{Walker} G.,  et~al., 2003, \mn@doi [\pasp] {10.1086/377358}, \href
  {https://ui.adsabs.harvard.edu/abs/2003PASP..115.1023W} {115, 1023}

\bibitem[\protect\citeauthoryear{{Weber} \& {Davis}}{{Weber} \&
  {Davis}}{1967}]{wd1967}
{Weber} E.~J.,  {Davis} Jr. L.,  1967, \mn@doi [\apj] {10.1086/149138}, \href
  {http://adsabs.harvard.edu/abs/1967ApJ...148..217W} {148, 217}

\bibitem[\protect\citeauthoryear{{Wisniewski} et~al.,}{{Wisniewski}
  et~al.}{2015}]{2015ApJ...811L..26W}
{Wisniewski} J.~P.,  et~al., 2015, \mn@doi [\apjl]
  {10.1088/2041-8205/811/2/L26}, \href
  {http://adsabs.harvard.edu/abs/2015ApJ...811L..26W} {811, L26}

\bibitem[\protect\citeauthoryear{{Wraight}, {Fossati}, {Netopil}, {Paunzen},
  {Rode-Paunzen}, {Bewsher}, {Norton}  \& {White}}{{Wraight}
  et~al.}{2012}]{2012MNRAS.420..757W}
{Wraight} K.~T.,  {Fossati} L.,  {Netopil} M.,  {Paunzen} E.,  {Rode-Paunzen}
  M.,  {Bewsher} D.,  {Norton} A.~J.,   {White} G.~J.,  2012, \mn@doi [\mnras]
  {10.1111/j.1365-2966.2011.20090.x}, \href
  {http://adsabs.harvard.edu/abs/2012MNRAS.420..757W} {420, 757}

\bibitem[\protect\citeauthoryear{{Yakunin} et~al.,}{{Yakunin}
  et~al.}{2015}]{2015MNRAS.447.1418Y}
{Yakunin} I.,  et~al., 2015, \mn@doi [\mnras] {10.1093/mnras/stu2401}, \href
  {http://adsabs.harvard.edu/abs/2015MNRAS.447.1418Y} {447, 1418}

\bibitem[\protect\citeauthoryear{{de Zeeuw}, {Hoogerwerf}, {de Bruijne},
  {Brown}  \& {Blaauw}}{{de Zeeuw} et~al.}{1999}]{1999AJ....117..354D}
{de Zeeuw} P.~T.,  {Hoogerwerf} R.,  {de Bruijne} J.~H.~J.,  {Brown} A.~G.~A.,
   {Blaauw} A.,  1999, \mn@doi [\aj] {10.1086/300682}, \href
  {http://adsabs.harvard.edu/abs/1999AJ....117..354D} {117, 354}

\bibitem[\protect\citeauthoryear{{ud-Doula} \& {Owocki}}{{ud-Doula} \&
  {Owocki}}{2002}]{ud2002}
{ud-Doula} A.,  {Owocki} S.~P.,  2002, \mn@doi [ApJ] {10.1086/341543}, \href
  {http://adsabs.harvard.edu/abs/2002ApJ...576..413U} {576, 413}

\bibitem[\protect\citeauthoryear{{ud-Doula}, {Townsend}  \&
  {Owocki}}{{ud-Doula} et~al.}{2006}]{ud2006}
{ud-Doula} A.,  {Townsend} R.~H.~D.,   {Owocki} S.~P.,  2006, \mn@doi [ApJl]
  {10.1086/503382}, \href {http://adsabs.harvard.edu/abs/2006ApJ...640L.191U}
  {640, L191}

\bibitem[\protect\citeauthoryear{{ud-Doula}, {Owocki}  \&
  {Townsend}}{{ud-Doula} et~al.}{2008}]{ud2008}
{ud-Doula} A.,  {Owocki} S.~P.,   {Townsend} R.~H.~D.,  2008, \mn@doi [MNRAS]
  {10.1111/j.1365-2966.2008.12840.x}, \href
  {http://adsabs.harvard.edu/abs/2008MNRAS.385...97U} {385, 97}

\bibitem[\protect\citeauthoryear{{ud-Doula}, {Owocki}  \&
  {Townsend}}{{ud-Doula} et~al.}{2009}]{ud2009}
{ud-Doula} A.,  {Owocki} S.~P.,   {Townsend} R.~H.~D.,  2009, \mn@doi [MNRAS]
  {10.1111/j.1365-2966.2008.14134.x}, \href
  {http://adsabs.harvard.edu/abs/2009MNRAS.392.1022U} {392, 1022}

\bibitem[\protect\citeauthoryear{{ud-Doula}, {Sundqvist}, {Owocki}, {Petit}  \&
  {Townsend}}{{ud-Doula} et~al.}{2013}]{ud2013}
{ud-Doula} A.,  {Sundqvist} J.~O.,  {Owocki} S.~P.,  {Petit} V.,   {Townsend}
  R.~H.~D.,  2013, \mn@doi [\mnras] {10.1093/mnras/sts246}, \href
  {http://adsabs.harvard.edu/abs/2013MNRAS.428.2723U} {428, 2723}

\bibitem[\protect\citeauthoryear{{ud-Doula}, {Owocki}, {Townsend}, {Petit}  \&
  {Cohen}}{{ud-Doula} et~al.}{2014}]{ud2014}
{ud-Doula} A.,  {Owocki} S.,  {Townsend} R.,  {Petit} V.,   {Cohen} D.,  2014,
  \mn@doi [\mnras] {10.1093/mnras/stu769}, \href
  {http://cdsads.u-strasbg.fr/abs/2014MNRAS.441.3600U} {441, 3600}

\makeatother
\end{thebibliography}

\appendix

\section{New magnetic analyses}\label{appendix:newstars}

Line lists for Least Squares Deconvolution \citep[LSD;][]{d1997,koch2010} were obtained from the Vienna Atomic Line Database \citep[VALD3;][]{piskunov1995,ryabchikova1997,kupka1999,kupka2000,2015PhyS...90e4005R} using `extract stellar' requests for the \teff~for each star found in the literature (see Table \ref{newstarstab}), $\log{g} = 4$, and enhanced abundances for Si, Ti, Cr, and Fe, set via extrapolation of the \teff~dependent abundances determined by \cite{2019MNRAS.483.2300S}. The line lists were cleaned in same fashion as described in Paper I. The total number of lines used in the analysis for each star is reported in Table \ref{newstarstab}. 

\begin{figure}
   \centering

   \includegraphics[trim=0 0 0 0, width=.5\textwidth]{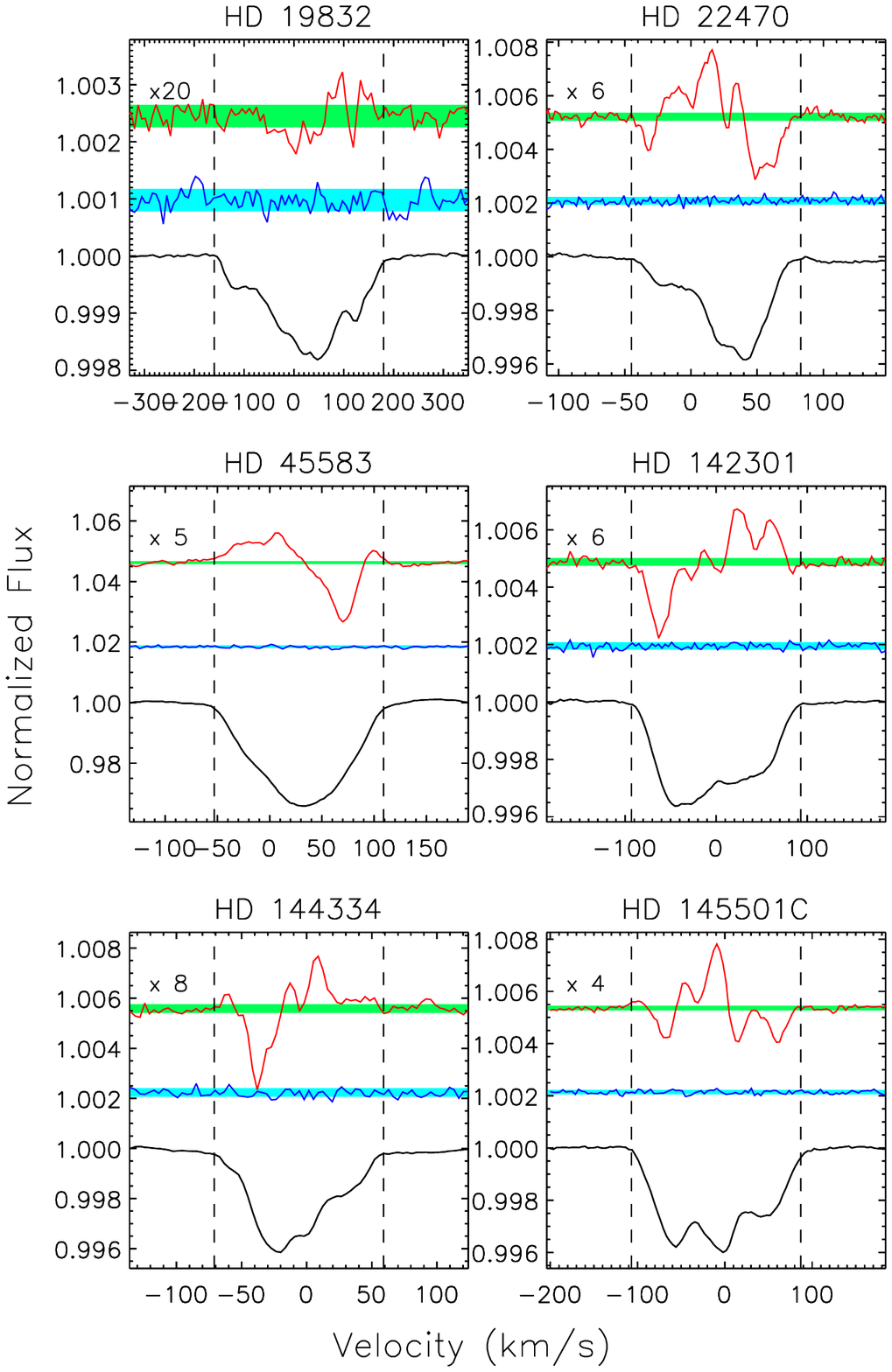}
      \caption[]{Representative LSD Stokes $I$ (black), Stokes $V$ (red), and $N$ (blue) profiles for the new sample stars. Shaded regions indicate mean uncertainties. Numbers in boxes indicate amplification factors for $N$ and Stokes $V$. Dashed lines show integration ranges.}
         \label{lsdprofs}
   \end{figure}

\begin{figure}
   \centering
\begin{tabular}{cc}

   \includegraphics[trim=50 25 25 0, width=.225\textwidth]{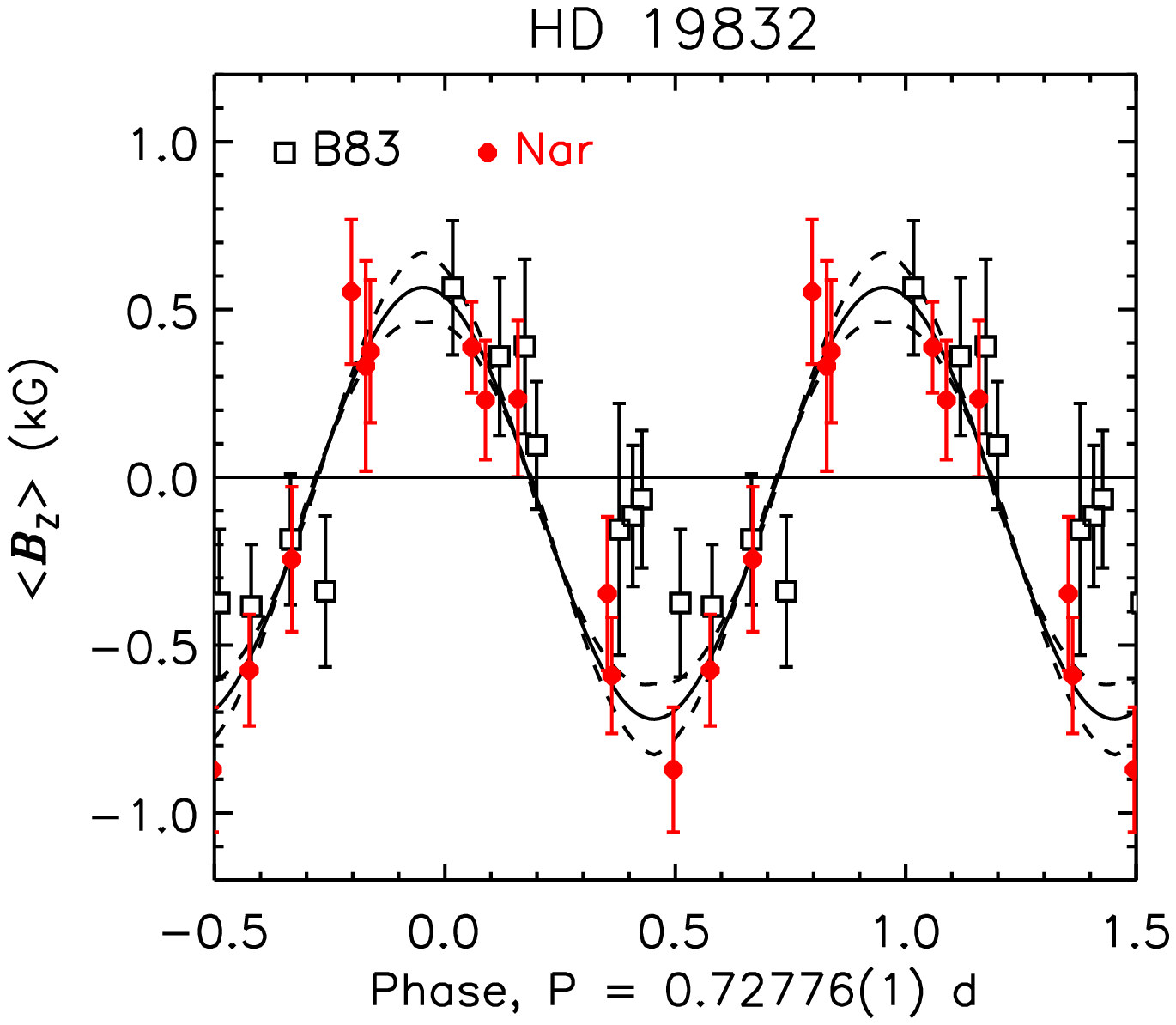} &
   \includegraphics[trim=50 25 25 0, width=.225\textwidth]{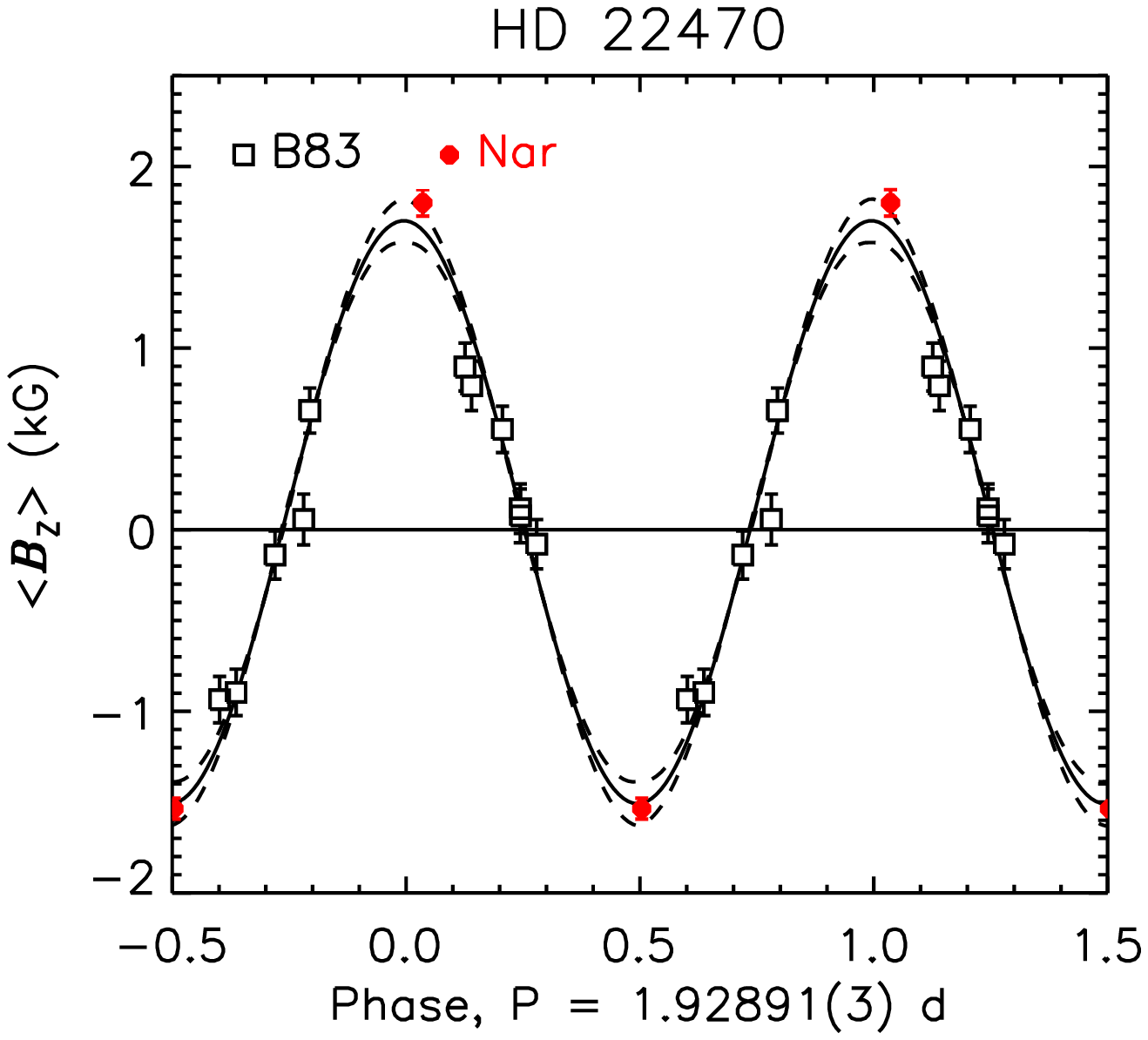} \\
   \includegraphics[trim=50 25 25 0, width=.225\textwidth]{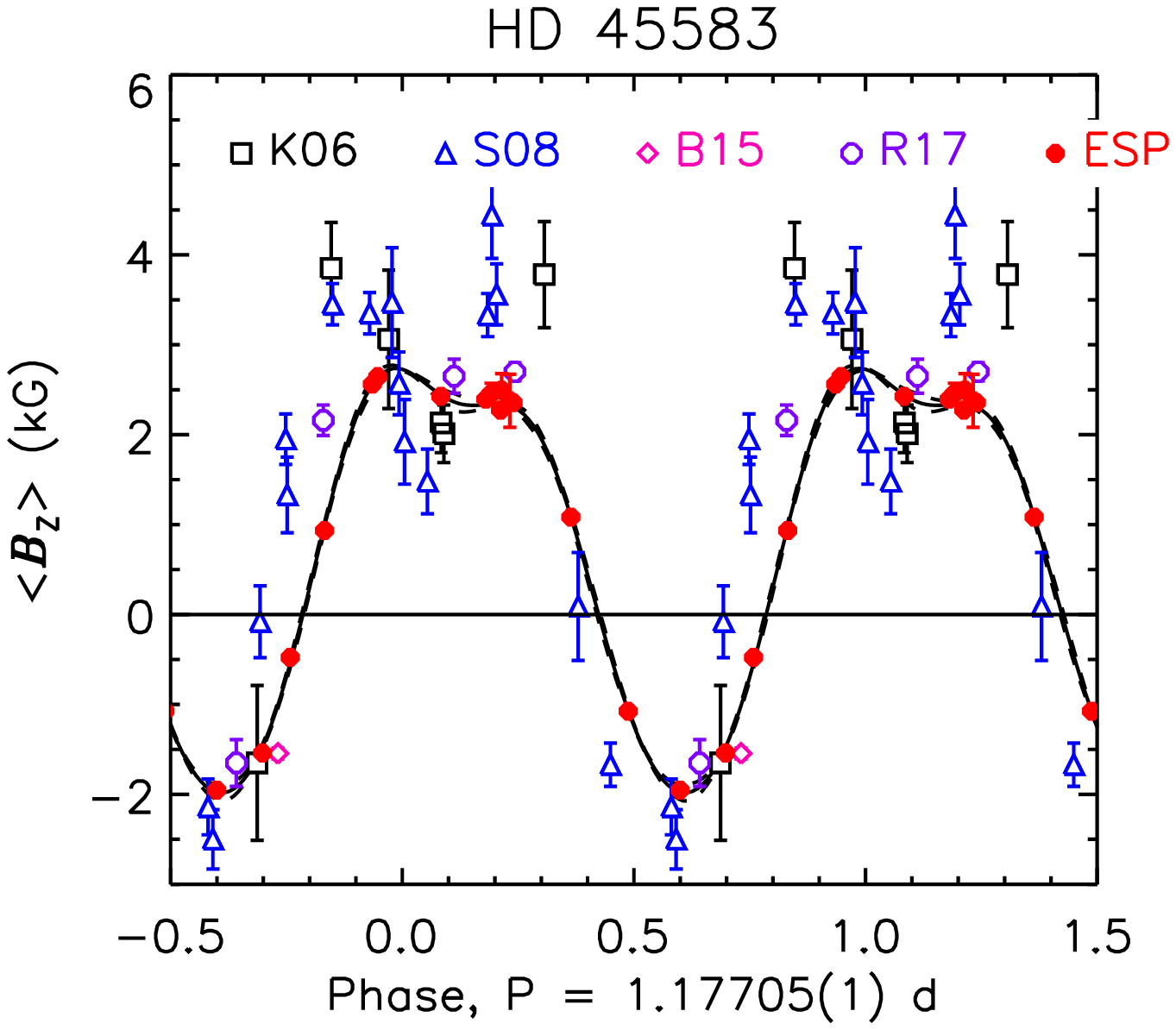} &
   \includegraphics[trim=50 25 25 0, width=.225\textwidth]{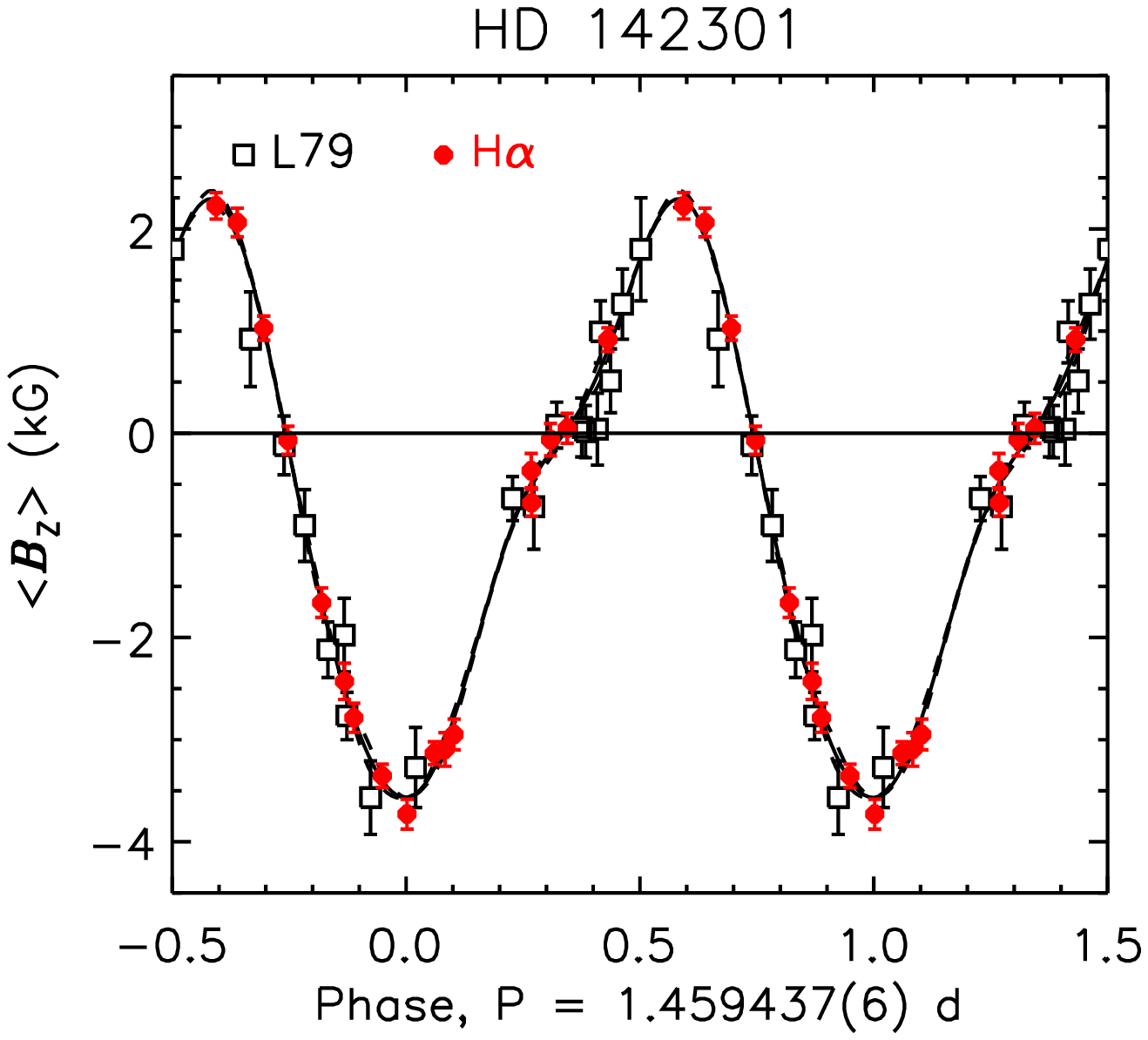} \\
   \includegraphics[trim=50 25 25 0, width=.225\textwidth]{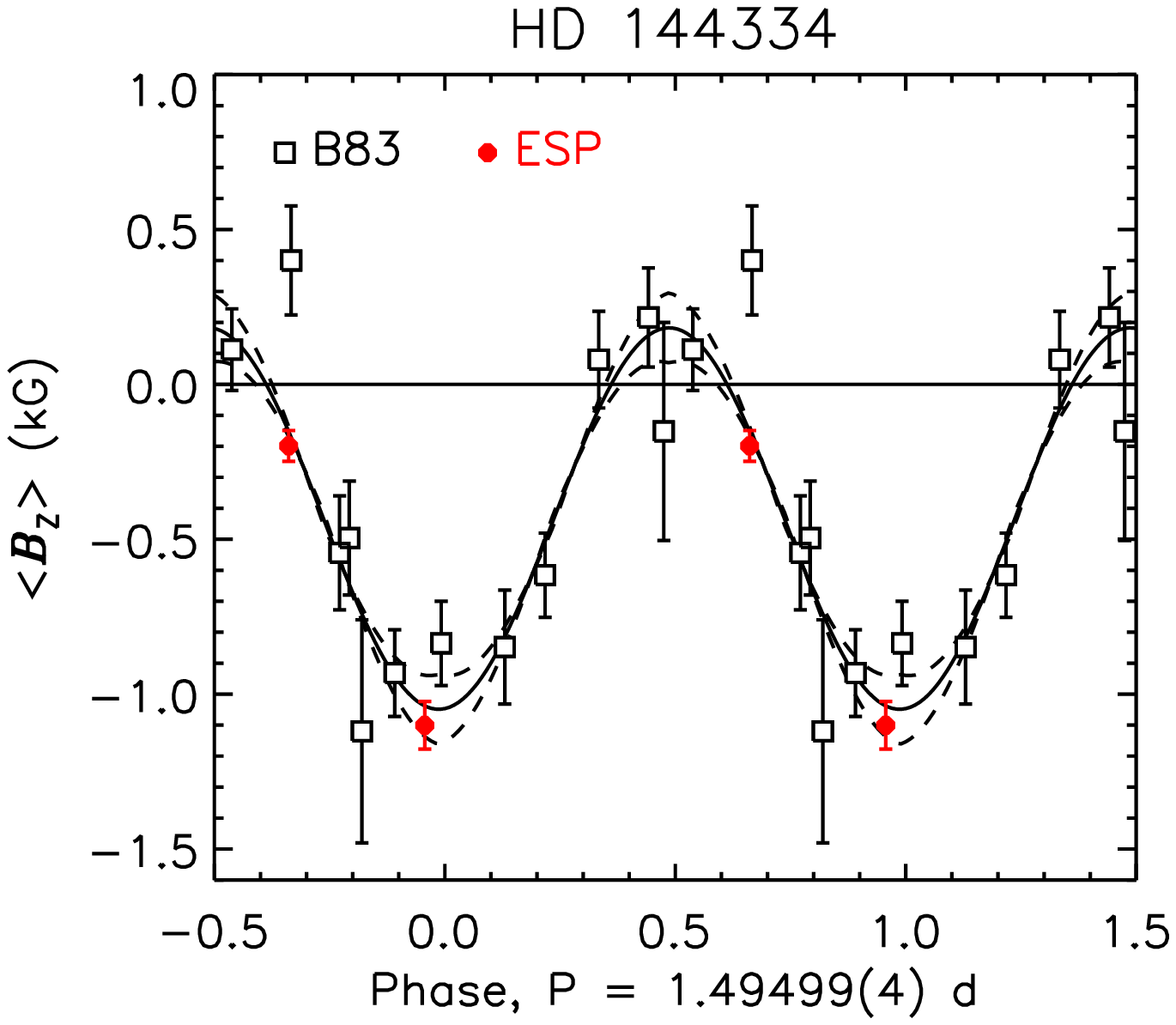} &
   \includegraphics[trim=50 25 25 0, width=.225\textwidth]{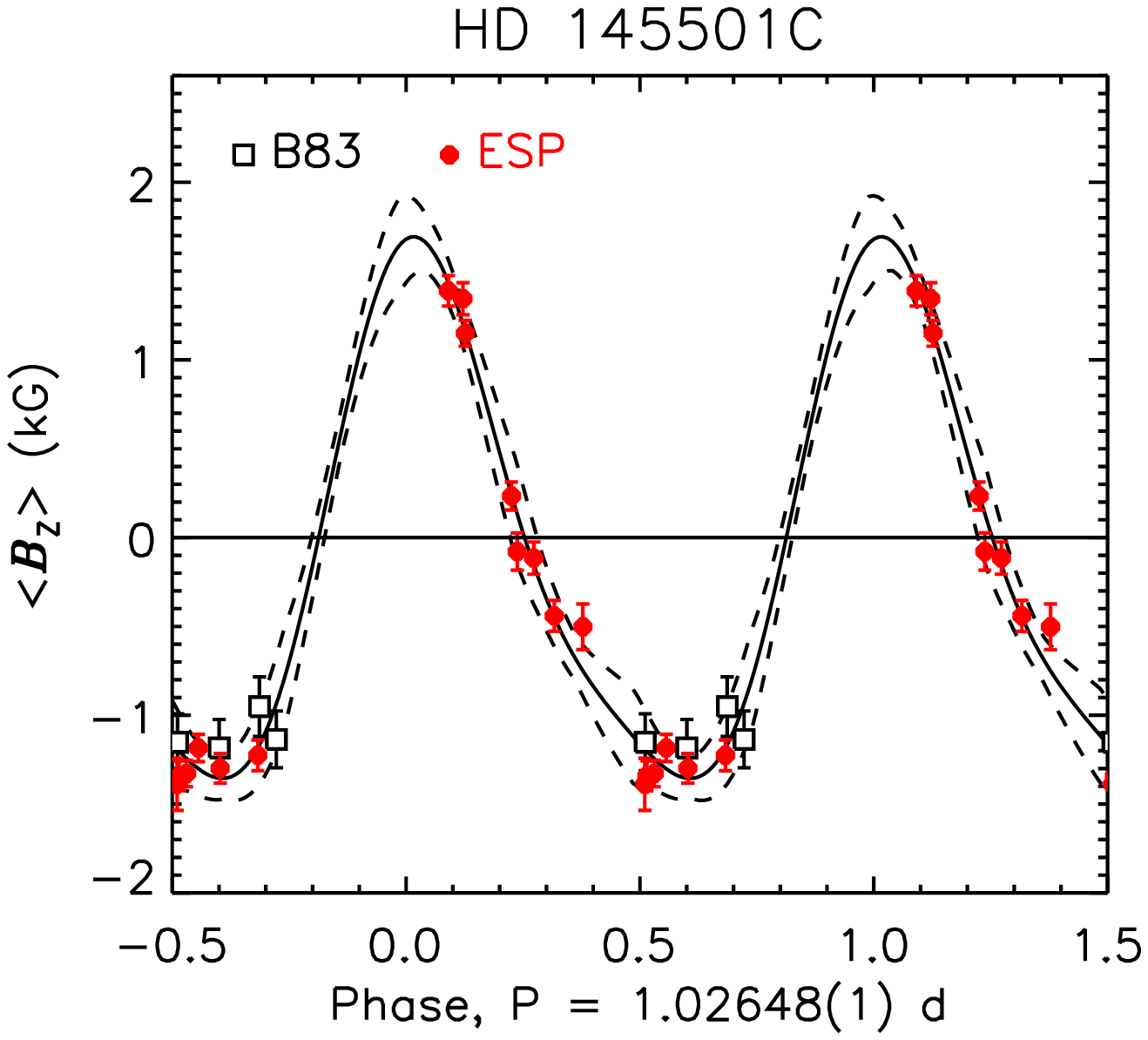} \\
\end{tabular}
      \caption[]{\bz~curves for the newly analyzed stars. Solid and dashed curves show harmonic fits and fit uncertainties, respectively. Legends give the origin of the measurements: ESP(aDOnS) or Nar(val) for the new measurements presented here; previously published measurements are from \protect\citet[][L79]{1979MNRAS.188..609L}, \protect\citet[][B83]{1983ApJS...53..151B}, \protect\citet[][K06]{2006MNRAS.372.1804K}, \protect\citet[][S08]{2008AstBu..63..128S}, \protect\citet[][B15]{2015AA...583A.115B}, and \protect\citet[][R17]{2017AstBu..72..391R}.}
         \label{bzcrvs}
   \end{figure}

Representative LSD profiles are shown in Fig.\ \ref{lsdprofs}. The numbers of `definite detections' (DD) in Stokes $V$, according to the False Alarm Probability (FAP) criteria described by \cite{1992AA...265..669D,d1997}, are reported in Table \ref{newstarstab}. \bz~was measured from the LSD profiles in the usual fashion (see Paper I and references therein). The integration ranges for calculation of FAPs and \bz~are indicated in Fig.\ \ref{lsdprofs}. These measurements were then combined with \bz~measurements from the literature to refine the rotation periods. In no case save HD\,145501C (see below) does $P_{\rm rot}$ differ significantly from the previously reported values. $P_{\rm rot}$ and $T_0$ are given in Table \ref{newstarstab}, with $T_0$ set in the same fashion as described in Paper I. Sinusoidal fitting parameters (see Paper I) $B_0$ and $B_1$ were obtained via harmonic fits to \bz. In three cases higher harmonics were required to obtain an acceptable fit to the \bz~curve; the semi-amplitudes $B_2$ and $B_3$ of these harmonics are also reported in Table \ref{newstarstab}, although they were not used for determining oblique rotator model (ORM) parameters. \bz~curves folded with these rotation periods are shown in Fig.\ \ref{bzcrvs}.

While \vsini~is typically available in the literature, we determined \vsini~from the LSD profiles (see Fig.\ \ref{lsdprofs} and Table \ref{newstarstab}). 

To determine the luminosities, we used Gaia DR2 parallaxes \citep{2018A&A...616A...1G}, bolometric corrections from the empirical relationship developed for chemically peculiar stars by \cite{2008AA...491..545N}, and extinctions $A_V$ from the literature (see Table \ref{newstarstab}). For HD\,45583, $A_V$ was determined from the observed $B-V$ colour compared to the empirical intrinsic colour table determined by \cite{2013ApJS..208....9P}. 

Fundamental stellar, rotational, ORM, wind, and magnetospheric parameters were determined using the Hertzsprung-Russell Diagram Monte Carlo sampler described in Paper III, and are given in Table \ref{newstarstab}. The definitions and details of the calculation of the parameters in Table \ref{newstarstab}, are also given in Paper III. Four stars are members of open clusters, as reported in Table \ref{newstarstab}, and the main sequence turnoff ages of these clusters were used as additional constraints in the same fashion as was done for cluster stars in Paper III, greatly increasing the precision with which the stellar radius can be constrained. Mass-loss rates and wind terminal velocities were calculated using the \cite{vink2001} recipe, using the \cite{lamers1995} ratio of \vinf~to the escape velocity, since the \cite{krticka2014} tables do not extend below 15 kK. Since the Vink mass-loss rates represent an extrapolation by 2 dex in this regime, they must be regarded as highly speculative, as must \vinf, since the ratio of wind terminal velocity to escape velocity is not constrained for these stars.

\begin{table*}
\caption[]{Table of stellar, rotational, magnetic, and magnetospheric parameters for the 6 low-luminosity stars without H$\alpha$ emission added to the sample. $N_{\rm obs}$ indicates the number of observations obtained with E(SPaDOnS) or N(arval). $N_{\rm line}$ gives the number of lines in the line mask used to extract LSD profiles. $N_{\rm DD}$ is the number of observations yielded a Definite Detection in the LSD Stokes $V$ profile. For stars that are members of open clusters, the cluster name and main-sequence turnoff age $t_{\rm cl}$ are given. Parameters are defined in Paper III. {\em Reference Key}: $a$: \protect\cite{2008AA...491..545N}; $b$: \protect\cite{2008AstBu..63..128S}; $c$: \protect\cite{2017MNRAS.468.2745N}; $d$: \protect\cite{1999AJ....117..354D}; $e$: \protect\cite{land2007}; $f$: \protect\cite{2000A&AS..146..251B}; $g$: \protect\cite{1999A&A...345..471R}; $h$: \protect\cite{2005AA...438.1163K}.}
\label{newstarstab}
\begin{tabular}{l | r r r r r r}
\hline\hline
Parameter & HD\,19832 & HD\,22470 & HD\,45583 & HD\,142301 & HD\,144334 & HD\,145501  \\
          & 56\,Ari   & 20\,Eri   & V682 Mon  & 3\,Sco     & HR\,5988   & $\nu$ Sco C \\
\hline
$N_{\rm obs}$ & 11 (N) & 2 (N) & 15 (E)    & 17 (E)     & 2 (E)      & 17 (E)      \\
$\langle {\rm S/N}_{\rm max} \rangle$ & 528 & 505 & 428 & 460 & 576 & 497 \\
$N_{\rm line}$ & 1415 & 1366 & 1663       & 1046        & 1205       & 1205         \\
$N_{\rm DD}$  & 4 & 2     & 15        & 17         & 2          & 16          \\
\hline
$B$ (mag) & 5.64 & 5.10      & 7.85      & 5.81       & 5.84       & --          \\
$V$ (mag) & 5.76 & 5.23      & 7.97      & 5.87       & 5.92       & 6.25        \\
$\pi$ (mas) & $8.0 \pm 0.2$ & $7.6 \pm 0.3$ & $3.07 \pm 0.07$ & $6.7 \pm 0.1$ & $7.5 \pm 0.1$ & $7.09 \pm 0.07$ \\
$A_V$ (mag) & $0.06^{a}$ &  $<0.09^{a}$ & $<0.03$ & $0.34 \pm 0.06^{a}$ & $0.28 \pm 0.03^{a}$ & 0.84$^{c}$ \\
$T_{\rm eff}$ (kK) & $12.8 \pm 0.4^{a}$ & $13.8 \pm 0.3^{a}$ & $13.3 \pm 0.3^{b}$ & $15.9 \pm 0.2^{a}$ & $14.8 \pm 0.4^{a}$ & $14.5 \pm 0.5^{c}$ \\
Cluster & -- & -- & NGC\,2232$^{f,g,h}$ & Upper Sco$^d$ & Upper Sco$^d$ & Upper Sco$^d$ \\
$\log{(t_{\rm cl} / {\rm yr})}$ & -- & -- & $7.55 \pm 0.1^{e}$ & $6.7 \pm 0.1^{e}$ & $6.7 \pm 0.1^{e}$ & $6.7 \pm 0.1^{e}$ \\
$\log{(L/L_\odot)}$ & $2.08 \pm 0.16$ & $2.43 \pm 0.15$ & $2.07 \pm 0.12$ & $2.56 \pm 0.07$ & $2.34 \pm 0.12$ & $2.46 \pm 0.15$ \\
$M_*$ (\msun) & $3.4 \pm 0.2$ & $3.9 \pm 0.2$ & $3.2 \pm 0.1$ & $4.4 \pm 0.1$ & $4.0 \pm 0.1$ & $4.0 \pm 0.2$ \\
$R_*$ (\rsun) & $2.3 \pm 0.3$ & $2.8 \pm 0.4$ & $2.12 \pm 0.06$ & $2.42 \pm 0.02$ & $2.27 \pm 0.04$ & $2.26 \pm 0.06$ \\
\hline
$P_{\rm rot}$ (d) & 0.72776(1) & 1.92891(3) & 1.17705(1) & 1.459437(7) & 1.49499(4) & 1.02648(1) \\
$T_0$ - 2440000 (HJD) & 2625.59(9) & 5298.35(7) & 15521.75(6) & 2594.92(8) & 3702.5(2) & 4774.97(9) \\
\vsini~(\kms) & $153 \pm 10$ & $54 \pm 3$ & $71 \pm 3$ & $83 \pm 6$ & $55 \pm 5$ & $87 \pm 4$ \\
$\log{W}$ & $-0.50^{+0.08}_{-0.02}$ & $-0.82 \pm 0.05$ & $-0.77 \pm 0.01$ & $-0.845 \pm 0.003$ & $-0.874 \pm 0.004$ & $-0.698 \pm 0.005$ \\
$R_{\rm p}/R_{\rm e}$ & $0.952^{+0.004}_{-0.02}$ & $0.991 \pm 0.006$ & $0.986 \pm 0.001$ & $0.990 \pm 0.001$ & $0.991 \pm 0.001$ & $0.981 \pm 0.001$ \\
$v_{\rm rot}$ (\kms) & $167^{+27}_{-6}$ & $73^{+11}_{-5}$ & $94 \pm 2$ & $84.4 \pm 0.5$ & $77 \pm 1$ & $114 \pm 2$ \\
$R_{\rm K}$ ($R_*$) & $2.0^{+0.1}_{-0.2}$ & $3.5 \pm 0.3$ & $3.21 \pm 0.05$ & $3.65 \pm 0.02$ & $3.82 \pm 0.02$ & $2.92 \pm 0.03$ \\
\hline
$B_0$ (kG) & $0.01 \pm 0.04$ & $0.09 \pm 0.05$ & $0.77 \pm 0.02$ & $-0.67 \pm 0.03$ & $-0.43 \pm 0.05$ & $-0.04 \pm 0.09$ \\
$B_1$ (kG) & $0.50 \pm 0.05$ & $1.61 \pm 0.07$ & $2.34 \pm 0.03$ & $-2.66 \pm 0.04$ & $-0.62 \pm 0.06$ & $1.47 \pm 0.08$ \\
$B_2$ (kG) & -- & -- & $-0.50 \pm 0.02$ & $-0.58 \pm 0.03$ & -- & $-0.30 \pm 0.06$ \\
$B_3$ (kG) & -- & -- & $-0.12 \pm 0.03$ & $-0.16 \pm 0.02$ & -- & -- \\
$|\langle B_z \rangle|_{\rm max}$ (kG) & $0.8 \pm 0.2$ & $1.80 \pm 0.07$ & $2.65 \pm 0.03$ & $3.7 \pm 0.2$ & $1.10 \pm 0.08$ & $1.39 \pm 0.09$ \\
$r$        & $-0.89 \pm 0.11$ & $-0.89 \pm 0.06$ & $-0.51 \pm 0.01$ & $-0.60 \pm 0.02$ & $-0.2 \pm 0.1$ & $-0.76 \pm 0.12$ \\
$i_{\rm rot}$ ($^\circ$) & $55^{+8}_{-6}$ & $44^{+8}_{-5}$ & $48 \pm 2$ & $68 \pm 5$ & $45 \pm 3$ & $49 \pm 3$ \\
$\beta$ ($^\circ$) & $89^{+1}_{-5}$ & $87.4^{+0.7}_{-3}$ & $70 \pm 2$ & $58^{+4}_{-12}$ & $56 \pm 5$ & $89 \pm 1$ \\
$B_{\rm d}$ (kG) & $2.7^{+0.6}_{-0.3}$ & $7.5^{+1.2}_{-0.5}$ & $9.1 \pm 0.3$ & $12.5^{+9}_{-0.3}$ & $3.6 \pm 0.3$ & $5.8 \pm 0.3$ \\
\hline
$\log{(\dot{M}/(M_\odot/{\rm yr}))}$ & $-11.7 \pm 0.2$ & $-11.1 \pm 0.2$ & $-11.77 \pm 0.06$ & $-10.86 \pm 0.03$ & $-11.22 \pm 0.07$ & $-11.2 \pm 0.1$ \\
\vinf~(\kms) & $925 \pm 25$ & $920 \pm 30$ & $997 \pm 6$ & $1089 \pm 3$ & $1060 \pm 6$ & $1060 \pm 7$ \\
$\log{\eta_*}$ & $6.7^{+0.2}_{-0.4}$ & $7.08 \pm 0.06$ & $7.65 \pm 0.05$ & $7.05^{+0.26}_{-0.03}$ & $6.34 \pm 0.08$ & $6.70 \pm 0.08$ \\
$R_{\rm A}$ ($R_*$) & $45 \pm 4$ & $58 \pm 2$ & $82 \pm 2$ & $59^{+13}_{-1}$ & $39 \pm 2$ & $48 \pm 2$ \\
$\log{(R_{\rm A}/R_{\rm K})}$ & $1.35 \pm 0.08$ & $1.22 \pm 0.05$ & $1.41 \pm 0.01$ & $1.21^{+0.07}_{-0.01}$ & $1.00 \pm 0.02$ & $1.22 \pm 0.02$ \\
$\log{(B_{\rm K} / G)}$ & $2.2^{+0.2}_{-0.1}$ & $1.9^{+0.2}_{-0.1}$ & $2.12 \pm 0.03$ & $2.11^{+0.13}_{-0.01}$ & $1.52 \pm 0.03$ & $2.06 \pm 0.03$ \\
\hline\hline
\end{tabular}

\end{table*}

\noindent {\bf HD\,19832}: The only \bz~measurements in the literature for 56\,Ari are the photopolarimetric measurements published by \cite{1983ApJS...53..151B}. The most recent photometric period of $0.727902$~d was determined by \cite{1993AJ....105.1103A}, while \cite{1993A&A...268..187S} determined a similar spectrophotometric period of 0.7278925~d. In keeping with this short period, the star has broad spectral lines (\vsini~$=153 \pm 10$~\kms; see Fig.\ \ref{lsdprofs}). To compensate for the broad lines, LSD profiles were extracted using larger than normal velocity pixels (14.4 \kms~rather than the standard 1.8~\kms). Despite this, only 4 of the 11 Narval observations yield a DD in Stokes $V$. While either period from the literature provides an acceptable phasing of either magnetic dataset individually, neither provides an acceptable phasing of the combined data. A Lomb-Scargle period search of the combined \bz~datasets, restricted to a narrow window centred on 0.728~d, yields a period of 0.72776(1) d. We used the combined datasets to fit \bz, since the uncertainities of the photopolarimetric and LSD profile measurements are comparable.

\noindent {\bf HD\,22470}: While only 2 Narval measurements were obtained for 20\,Eri, \cite{1983ApJS...53..151B} published a large number of \bz~measurements with small relative uncertainties. \citeauthor{1983ApJS...53..151B} gave $P_{\rm rot} = 0.6785$ d, while the photometric dataset analyzed by \cite{1995A&AS..114..253A} gave a period of 1.9387 d. The Hipparcos dataset favours the longer period \citep{2011MNRAS.414.2602D}, and furthermore provides an acceptable phasing of the Narval \bz~measurements with those of \citeauthor{1983ApJS...53..151B}. The fit shown in Fig.\ \ref{bzcrvs} is to the combined dataset, since the Narval dataset consists of only 2 observations. The \cite{2005yCat.3244....0G} catalogue of \vsini~measurements lists 3 values for this star between 60 and 80 \kms. From the LSD profiles we derived \vsini~$= 54 \pm 3$~\kms. 

\noindent {\bf HD\,45583}: There is an extensive magnetic dataset for V682\,Mon \citep{2006MNRAS.372.1804K,2008AstBu..63..128S,2015AA...583A.115B,2017AstBu..72..391R}. \cite{2008AstBu..63..128S} reported the clear double-wave variation of \bz, indicative of a somewhat complex surface magnetic field. The LSD profiles indicate \vsini~$=71 \pm 3$~\kms, consistent with the results reported by \cite{2008AstBu..63..128S}. \cite{2008AstBu..63..128S} found a period of 1.177000~d, which does not quite phase the combined datasets. Lomb-Scargle analysis of the combined data yields 1.17705(1)~d. The positive amplitude of the ESPaDOnS LSD profile \bz~measurements is somewhat less than that obtained from previous datasets, although the negative amplitude is similar; however, the ESPaDOnS data confirm the double-wave variation. The fit in Fig.\ \ref{bzcrvs} is to the ESPaDOnS data only. 

\noindent {\bf HD\,142301}: the most recent magnetic dataset for 3\,Sco was published by \cite{1979MNRAS.188..609L}, who found a period of 1.45955(3) d. Combining magnetic data from \cite{1979MNRAS.188..609L} with the new ESPaDOnS dataset, $P_{\rm rot} = 1.459437(6)$ d. This is close to the period found by \citeauthor{1979MNRAS.188..609L}, although it is different by close to $4\sigma$ according to the larger \citeauthor{1979MNRAS.188..609L} uncertainty. Whether this difference is real, representing e.g.\ an evolution of the period, is beyond the scope of this work. The old and new \bz~measurements are shown phased with the new period in Fig.\ \ref{bzcrvs}, where they are fit with a 3$^{rd}$-order harmonic function. The \bz~measurements shown in Fig.\ \ref{bzcrvs} were obtained from the H$\alpha$ line in the manner described in Paper I; the anharmonicity is therefore most likely a consequence of departures of the surface magnetic field from a pure dipole, rather than due to chemical spots. The LSD profiles yield \vsini~$=83 \pm 6$~\kms, close to the top of the range of values (between 50 and 90 \kms) given by \cite{2005yCat.3244....0G}.

\noindent {\bf HD\,144334}: A large \bz~dataset for HR\,5988 was published by \cite{1983ApJS...53..151B}. There are two periods in the literature: a photometric period of 1.49497 d \citep{1987A&AS...69..371N}, and a magnetic period of 3.61 d \citep{1983ApJS...53..151B}. Period analysis of the Hipparcos photometry confirms the photometric period \citep{2011MNRAS.414.2602D}. Lomb-Scargle analysis of the combined \bz~datasets yields $P_{\rm rot} = 1.49499(4)$ d, identical to the \cite{1987A&AS...69..371N} period within uncertainty. The old and new \bz~measurements are shown phased with this period in Fig.\ \ref{bzcrvs}. Since there are only 2 ESPaDOnS measurements, the harmonic fit in Fig.\ \ref{bzcrvs} was performed using the combined dataset. The \cite{2005yCat.3244....0G} \vsini~catalogue gives four \vsini~values, three around 50 \kms~and one of 140 \kms. The ESPaDOnS LSD profiles are consistent with a value of $55 \pm 5$~\kms. 

\noindent {\bf HD\,145501C}: The only previously published magnetic data for $\nu$ Sco C were obtained by \cite{1983ApJS...53..151B}; all 4 of their observations were around $-1$~kG. \cite{2012MNRAS.420..757W} determined a period of about 0.585~d via their analysis of STEREO data, which if accurate would make this one of the most rapidly rotating magnetic stars known. Period analysis of the ESPaDOnS data yields two peaks, at about 0.5 and 1.0 d. Combining the ESPaDOnS data with the \citeauthor{1983ApJS...53..151B} measurements favours the longer period, 1.02648(1)~d. Phasing \bz~with this period requires the second harmonic to achieve an acceptable fit, indicating slight departures from a pure dipole. Two \vsini~values, 60 and 70 \kms, are given in the \cite{2005yCat.3244....0G} catalogue; LSD profiles instead yield $87 \pm 4$~\kms. It is worth noting that the shorter 0.5 d period produces an almost indistinguishable phasing of \bz. If this period is adopted instead, the star's derived parameters are dramatically different: $i_{\rm rot} \sim 19^\circ$, $B_{\rm d} \sim 14$~kG, $R_{\rm K} \sim 1.4~R_*$, and $\log{B_{\rm K}} \sim 3.4$. In this case, the star would have a higher $\log{B_{\rm K}}$ value than that of any of the emission line stars, and the lack of strong H$\alpha$ emission would therefore provide very strong evidence for a luminosity cutoff in H$\alpha$ emission. Since this would be too good to be true, we have adopted the longer period as the more conservative option.

\begin{center}

\onecolumn
\thispagestyle{plain}
\vspace*{\fill}
\textbf{\Huge Online Material}
\vspace*{\fill}
\end{center}

\pagebreak
\clearpage

\twocolumn

\section{Notes on individual stars with emission}\label{appendix:indstars}

   \begin{figure*}
   \centering
\begin{tabular}{cccc}

   \includegraphics[trim=50 0 0 0, width=0.225\textwidth]{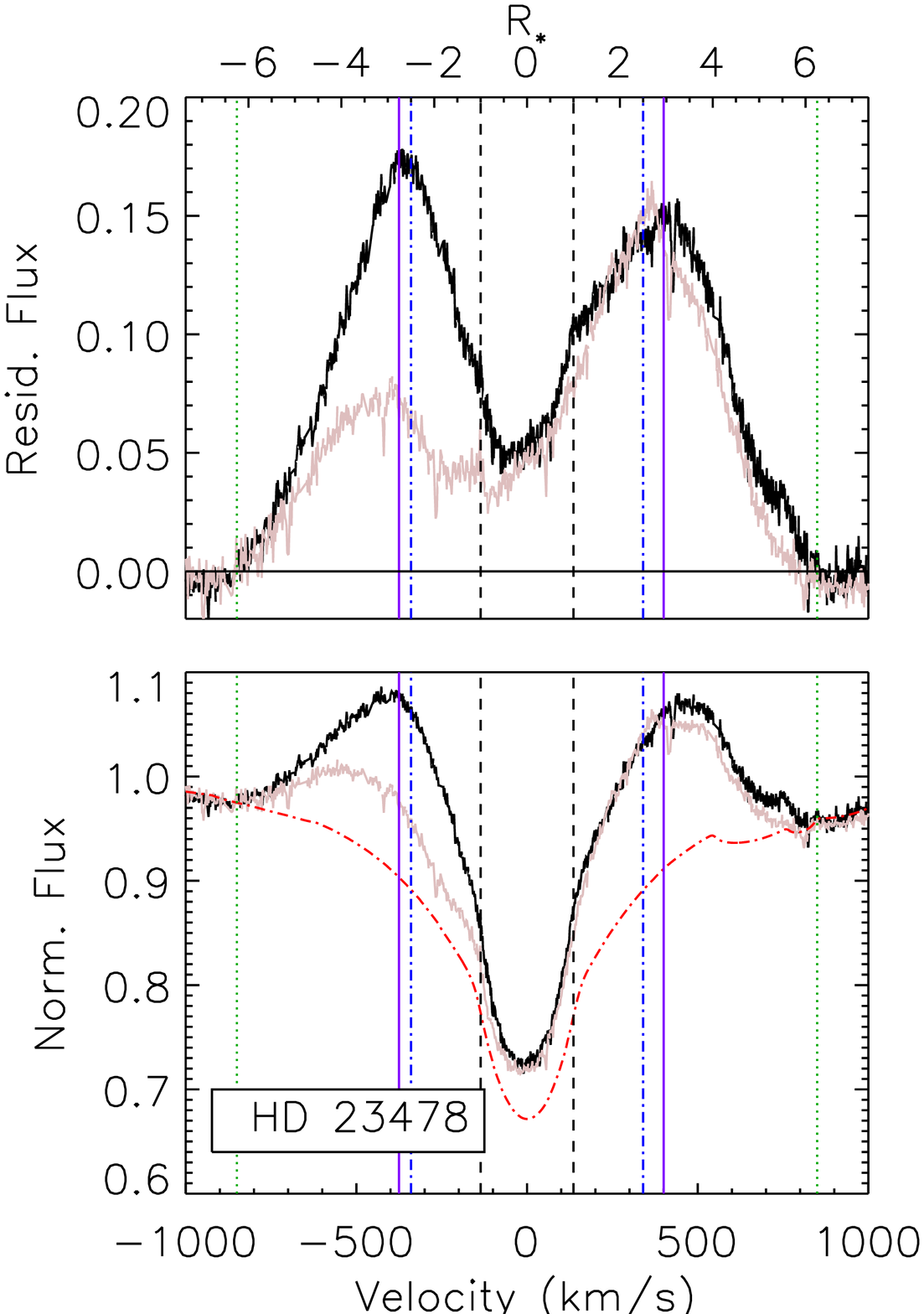} &
   \includegraphics[trim=50 0 0 0, width=0.225\textwidth]{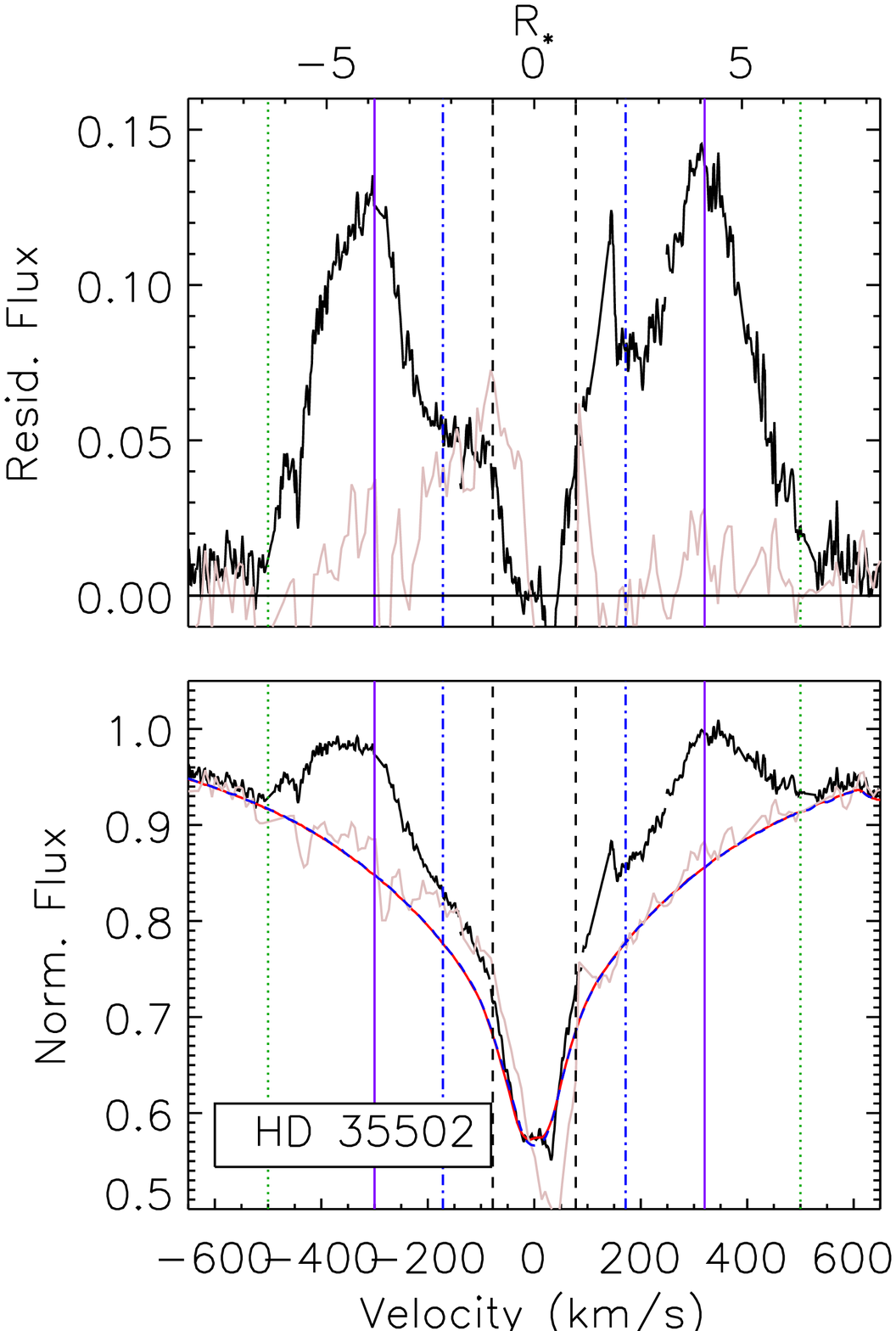} &
   \includegraphics[trim=50 0 0 0, width=0.225\textwidth]{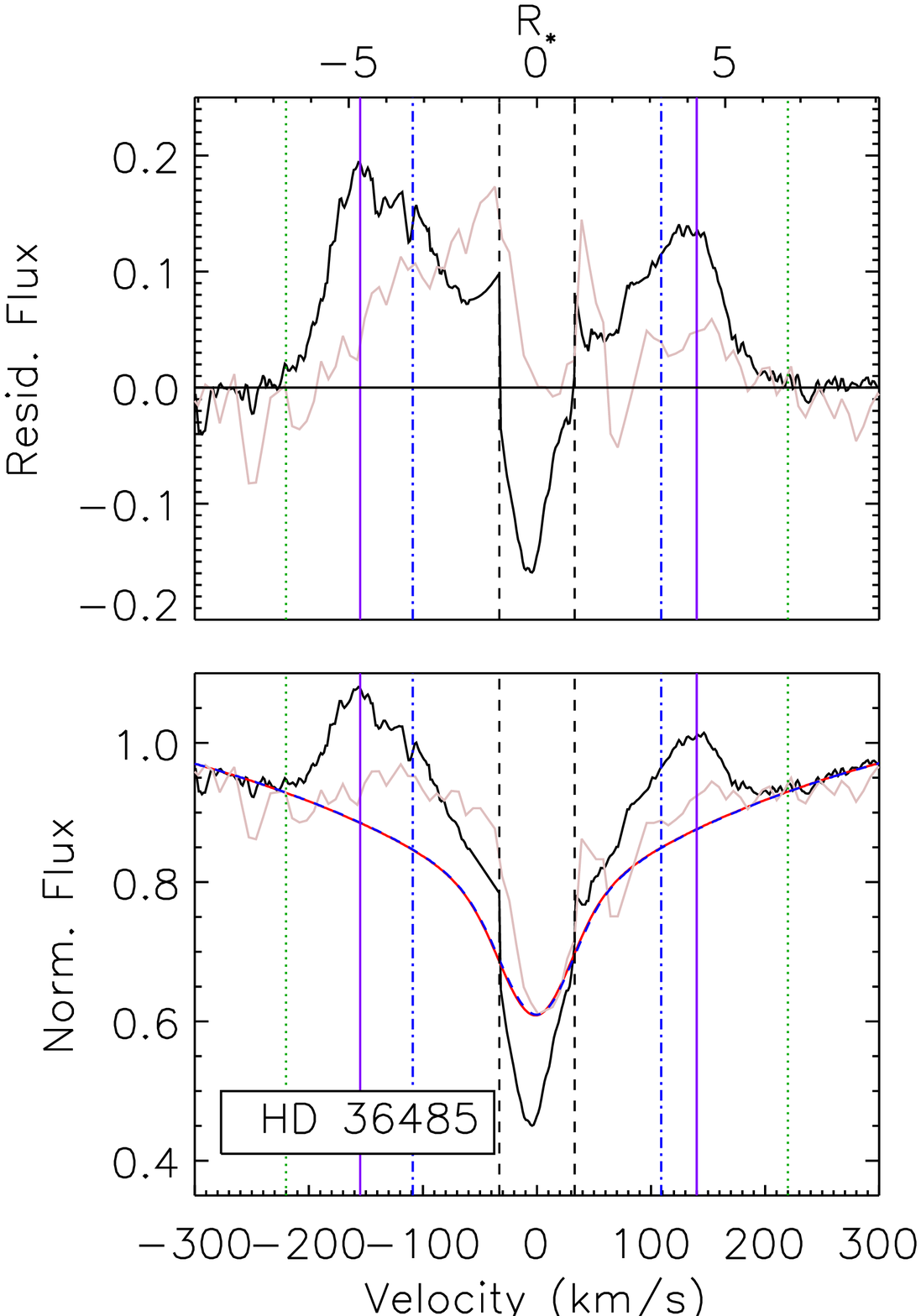} &
   \includegraphics[trim=50 0 0 0, width=0.225\textwidth]{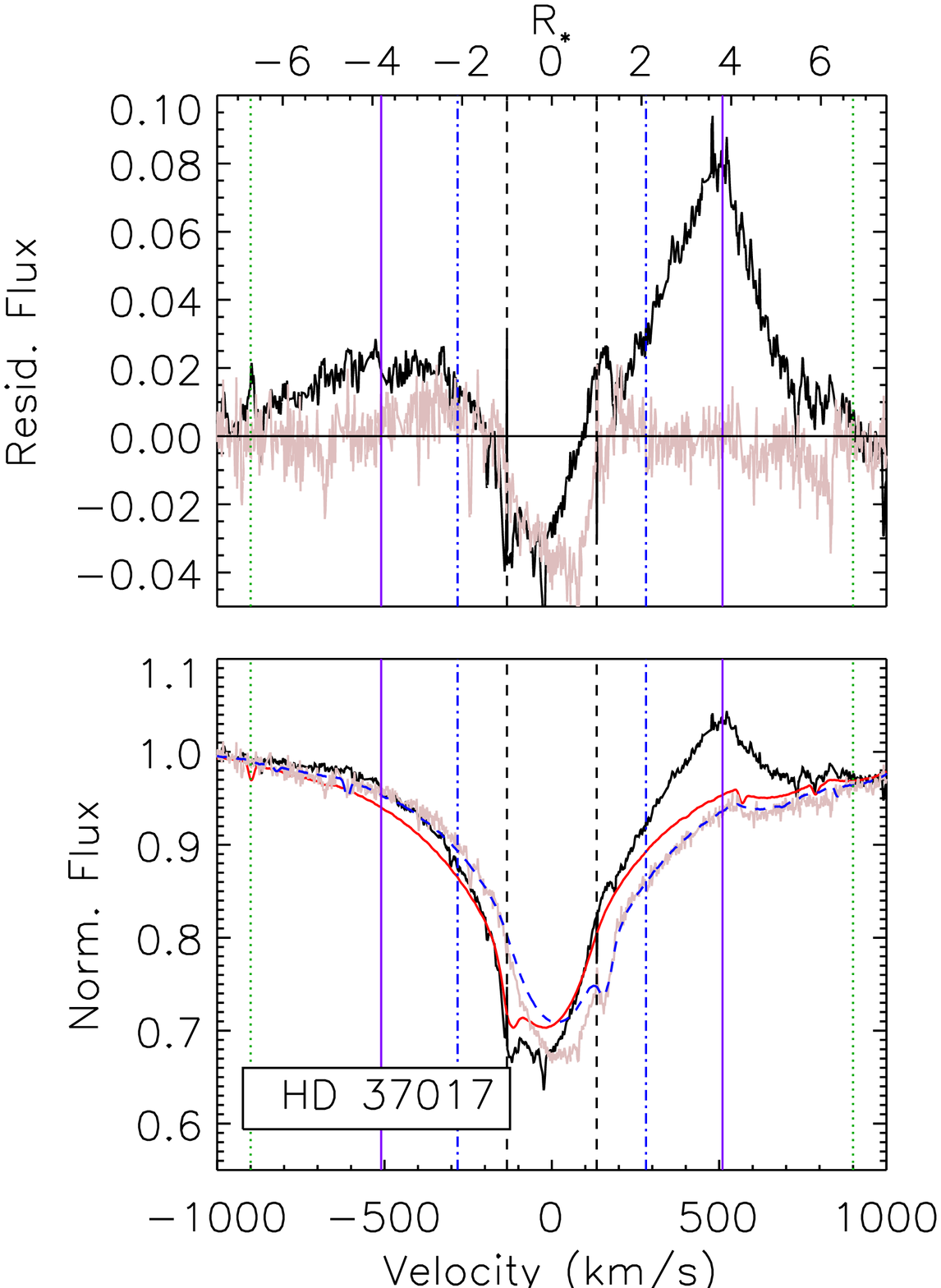} \\

   \includegraphics[trim=50 0 0 0, width=0.225\textwidth]{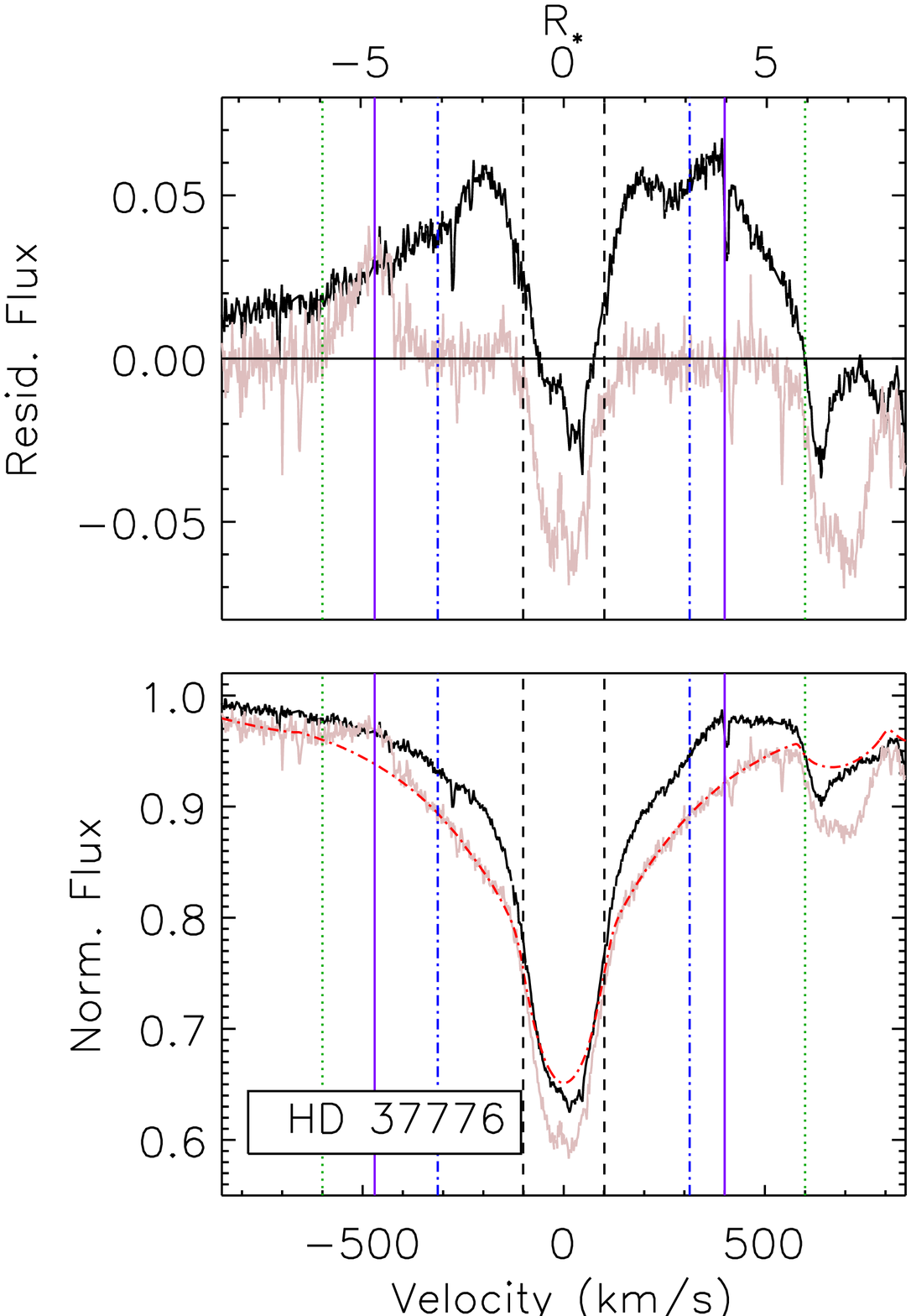} &
   \includegraphics[trim=50 0 0 0, width=0.225\textwidth]{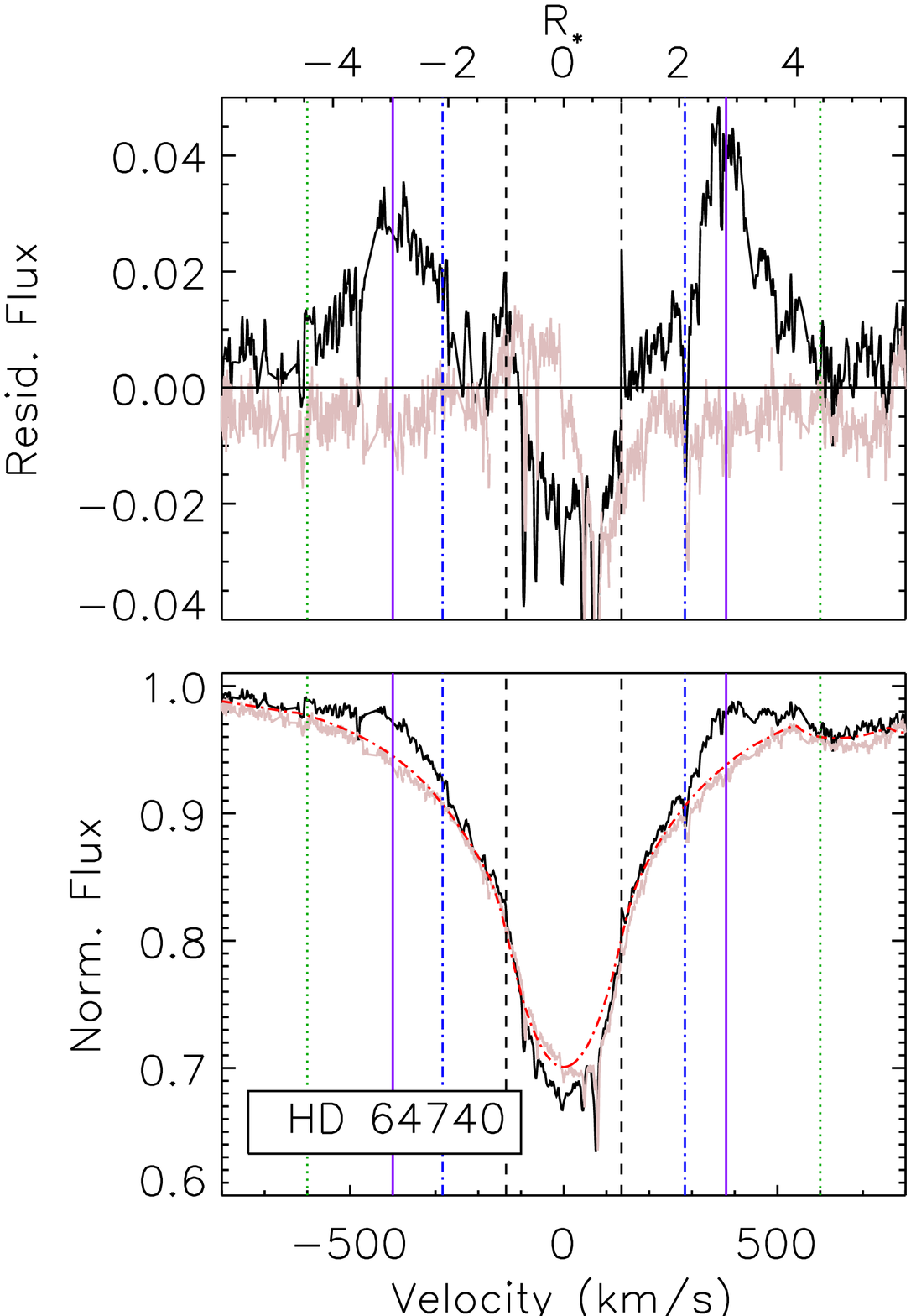} &
   \includegraphics[trim=50 0 0 0, width=0.225\textwidth]{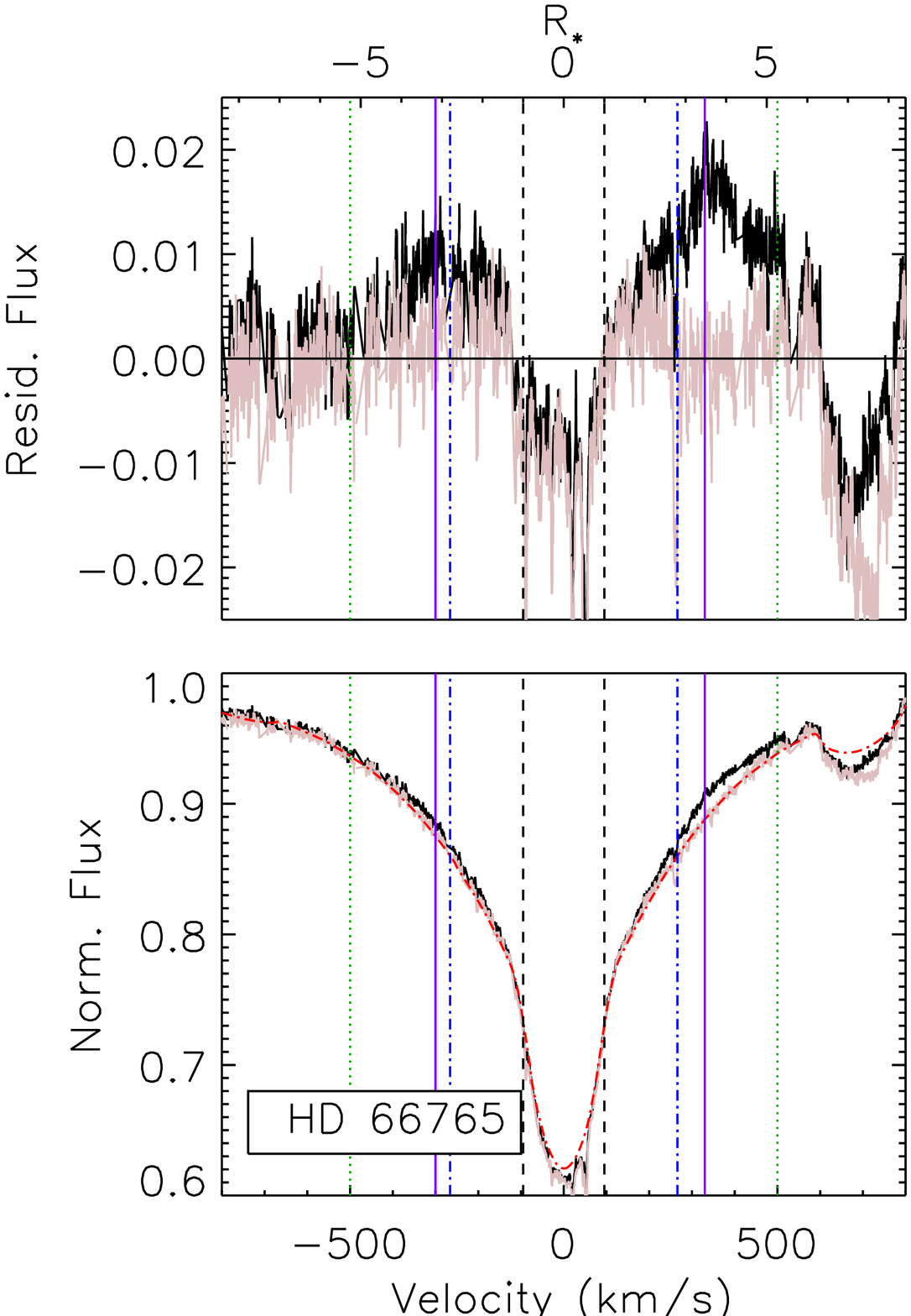} &
   \includegraphics[trim=50 0 0 0, width=0.225\textwidth]{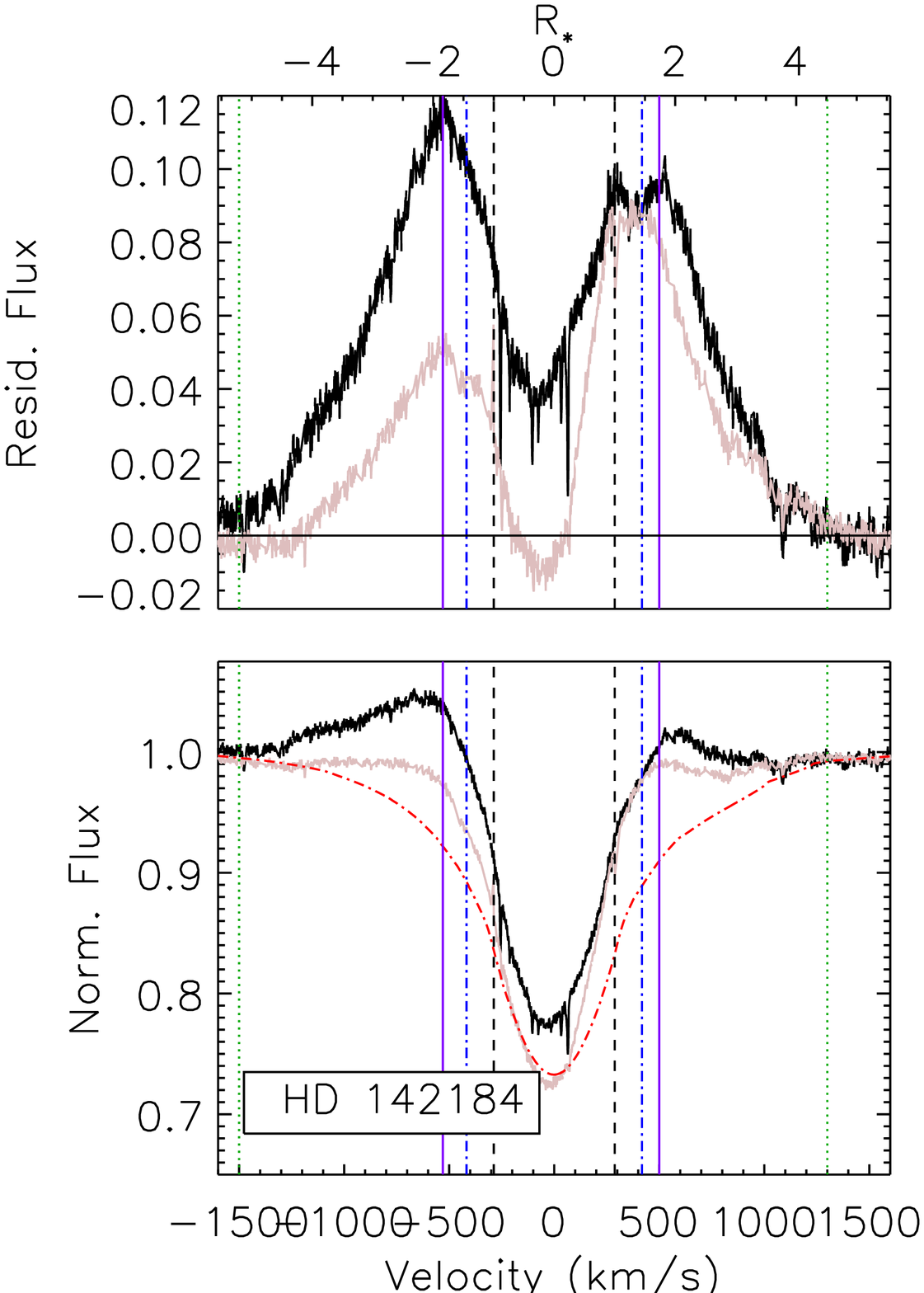} \\

   \includegraphics[trim=50 0 0 0, width=0.225\textwidth]{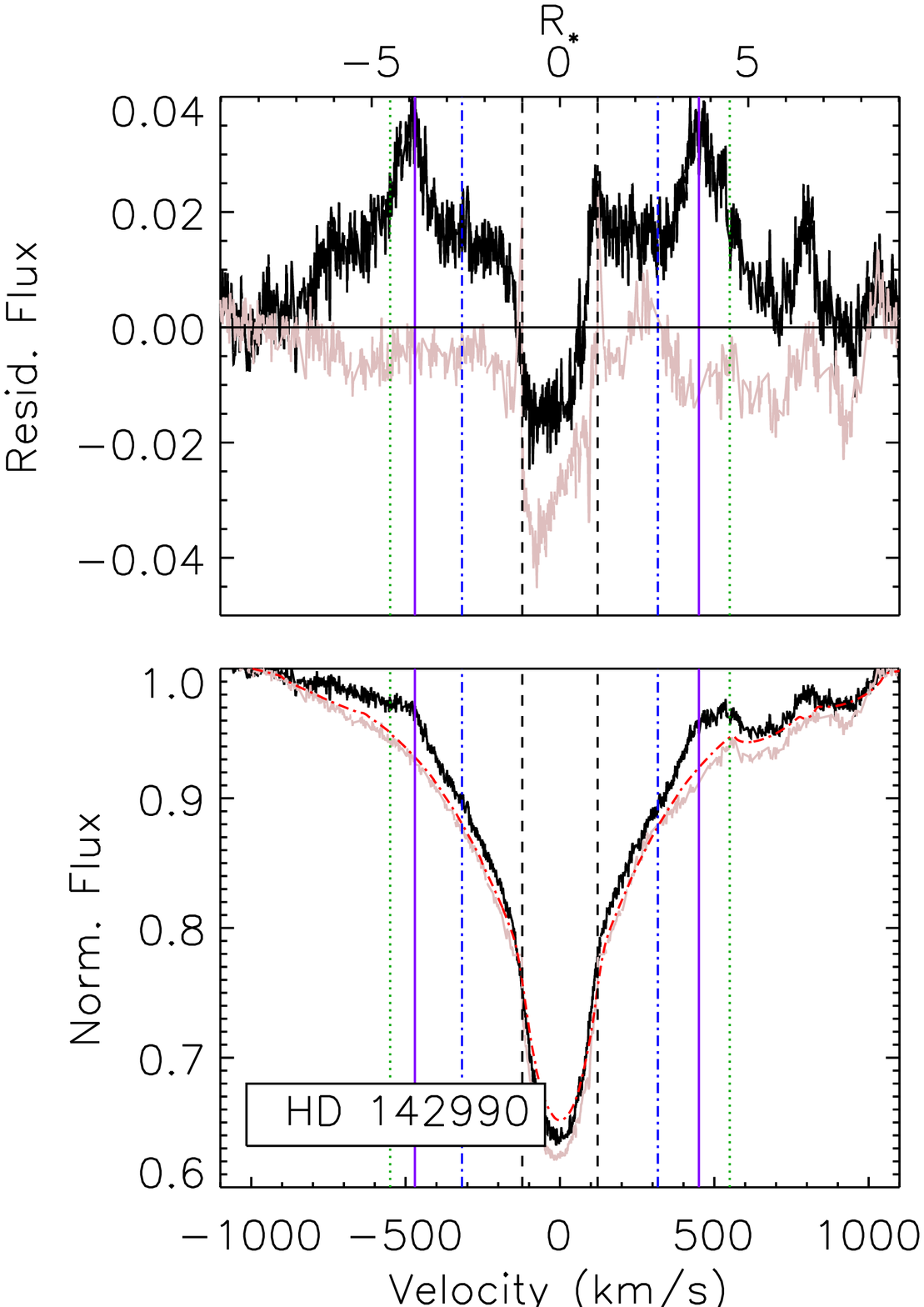} &
   \includegraphics[trim=50 0 0 0, width=0.225\textwidth]{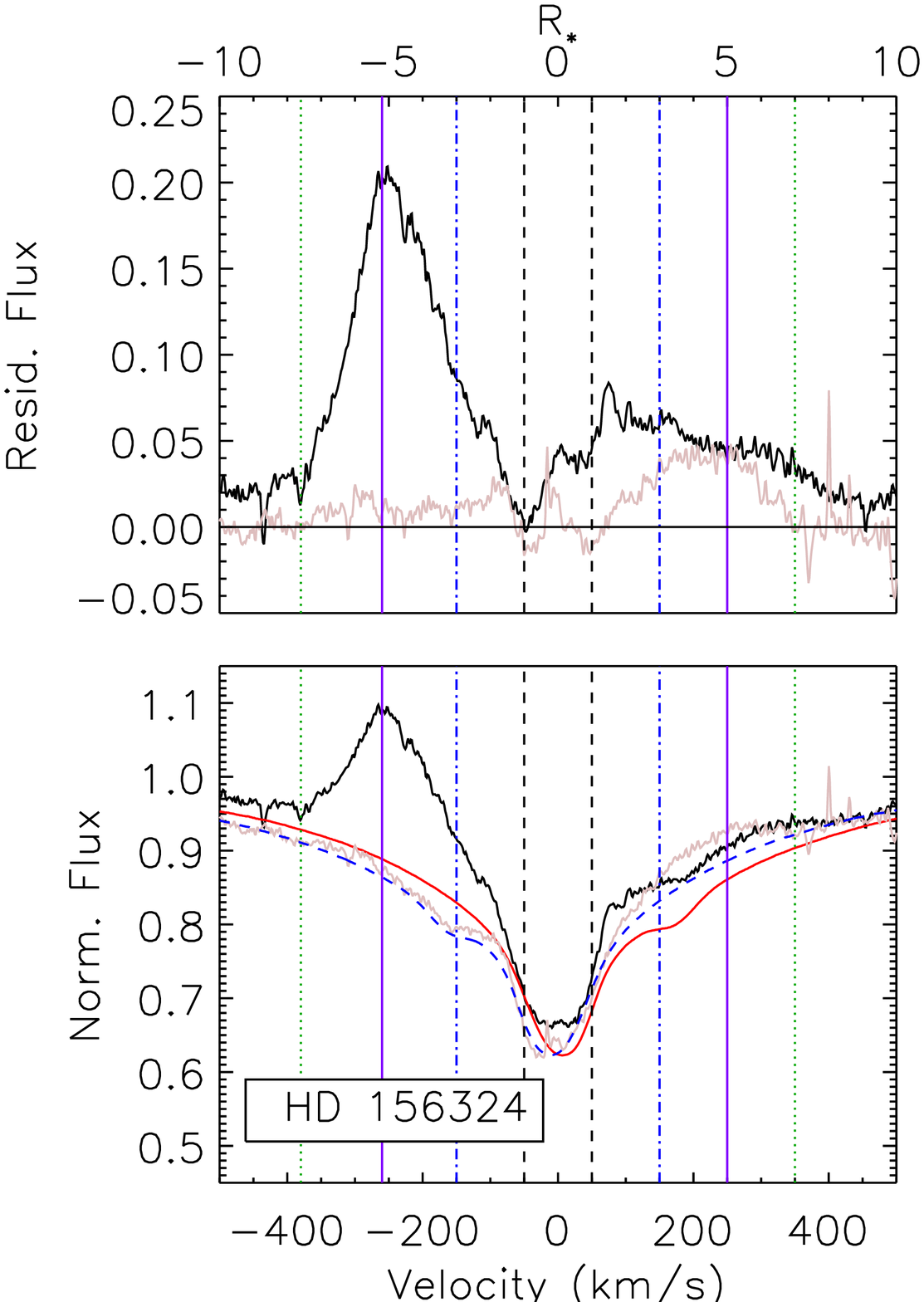} &
   \includegraphics[trim=50 0 0 0, width=0.225\textwidth]{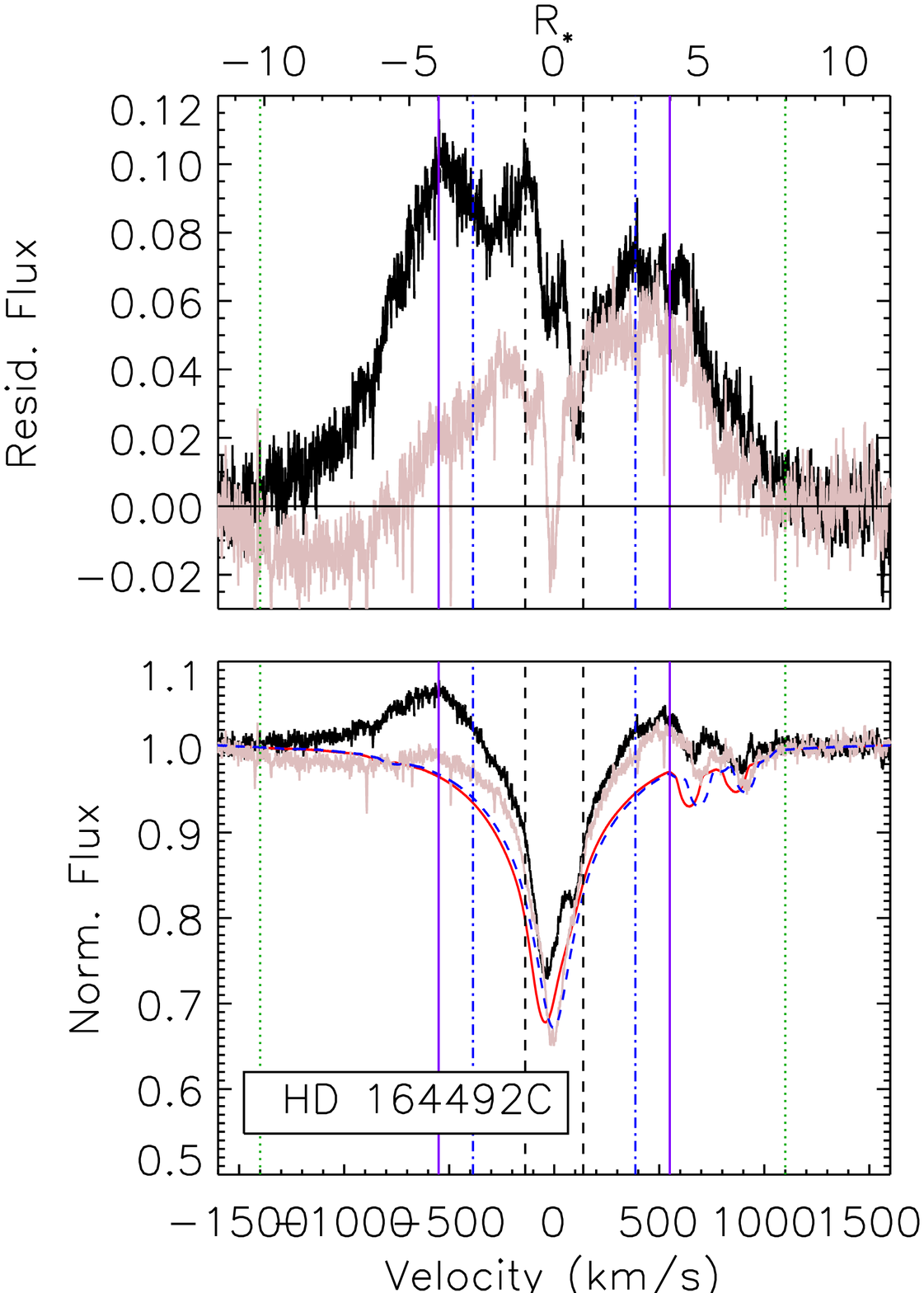} &
   \includegraphics[trim=50 0 0 0, width=0.225\textwidth]{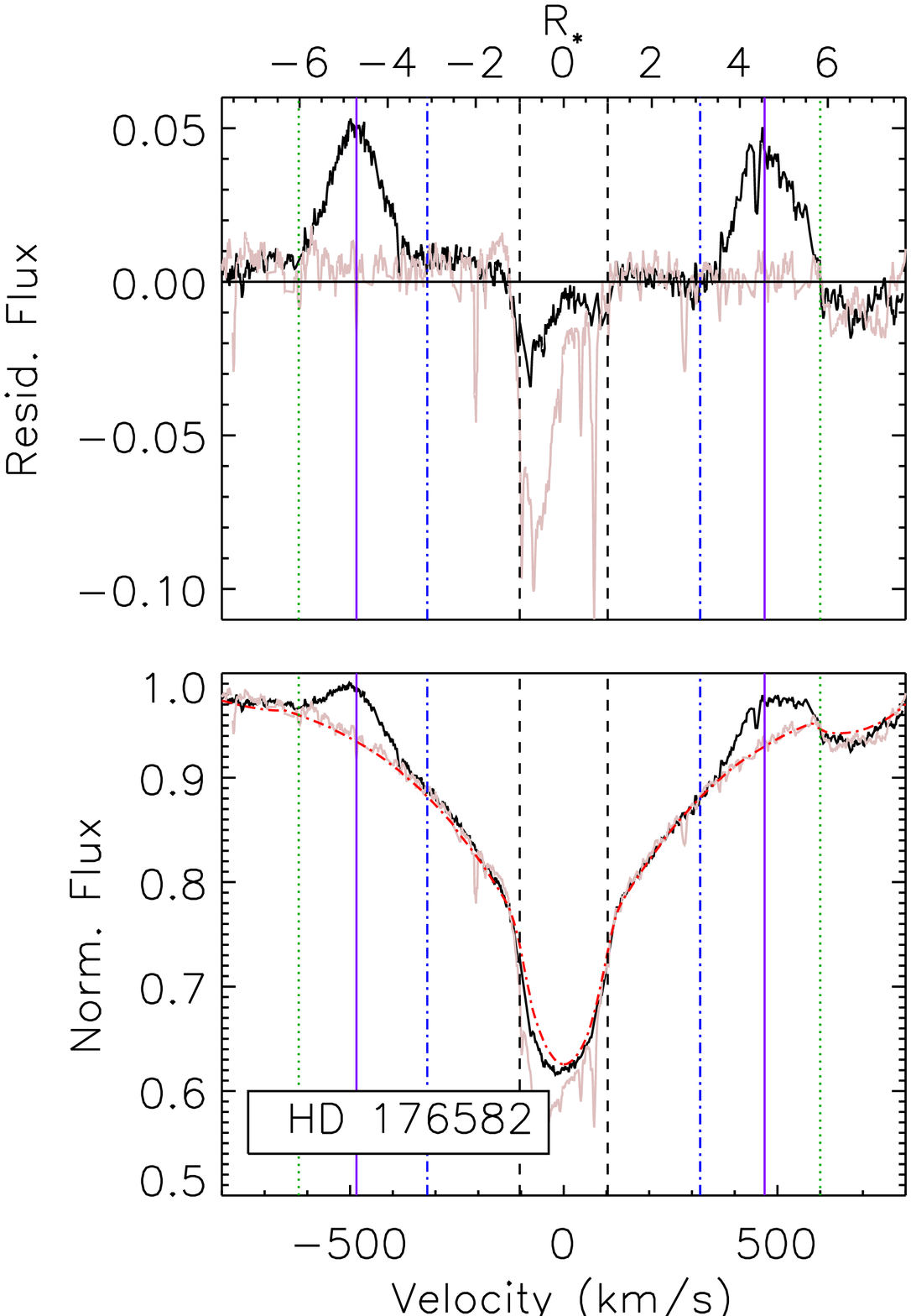} \\

\end{tabular}
      \caption[]{As Fig.\ \protect\ref{sigOriE_halpha_minmax} for HD\,23478, HD\,35502, HD\,36485, HD\,37017, HD\,37776, HD\,64740, HD\,66765, HD142184, HD\,142990, HD\,156324, HD\,164492C, and HD\,176582. For binary stars, synthetic spectra are shown for emission maximum (solid red) and emission minimum (dashed pink). }
         \label{halpha_ind1}
   \end{figure*}

   \begin{figure}
   \centering
\begin{tabular}{cc}

   \includegraphics[trim=50 0 0 0, width=0.225\textwidth]{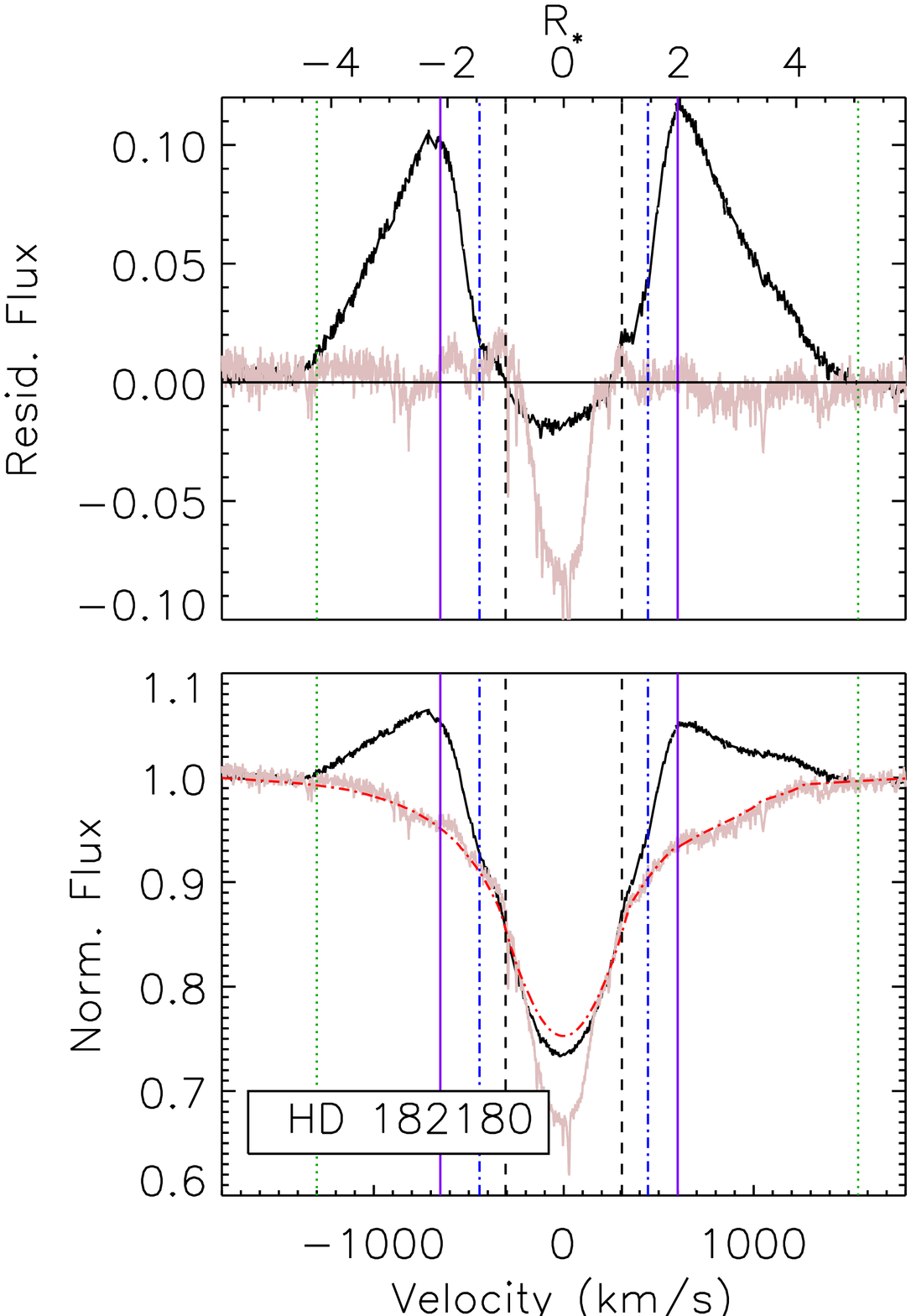} &
   \includegraphics[trim=50 0 0 0, width=0.225\textwidth]{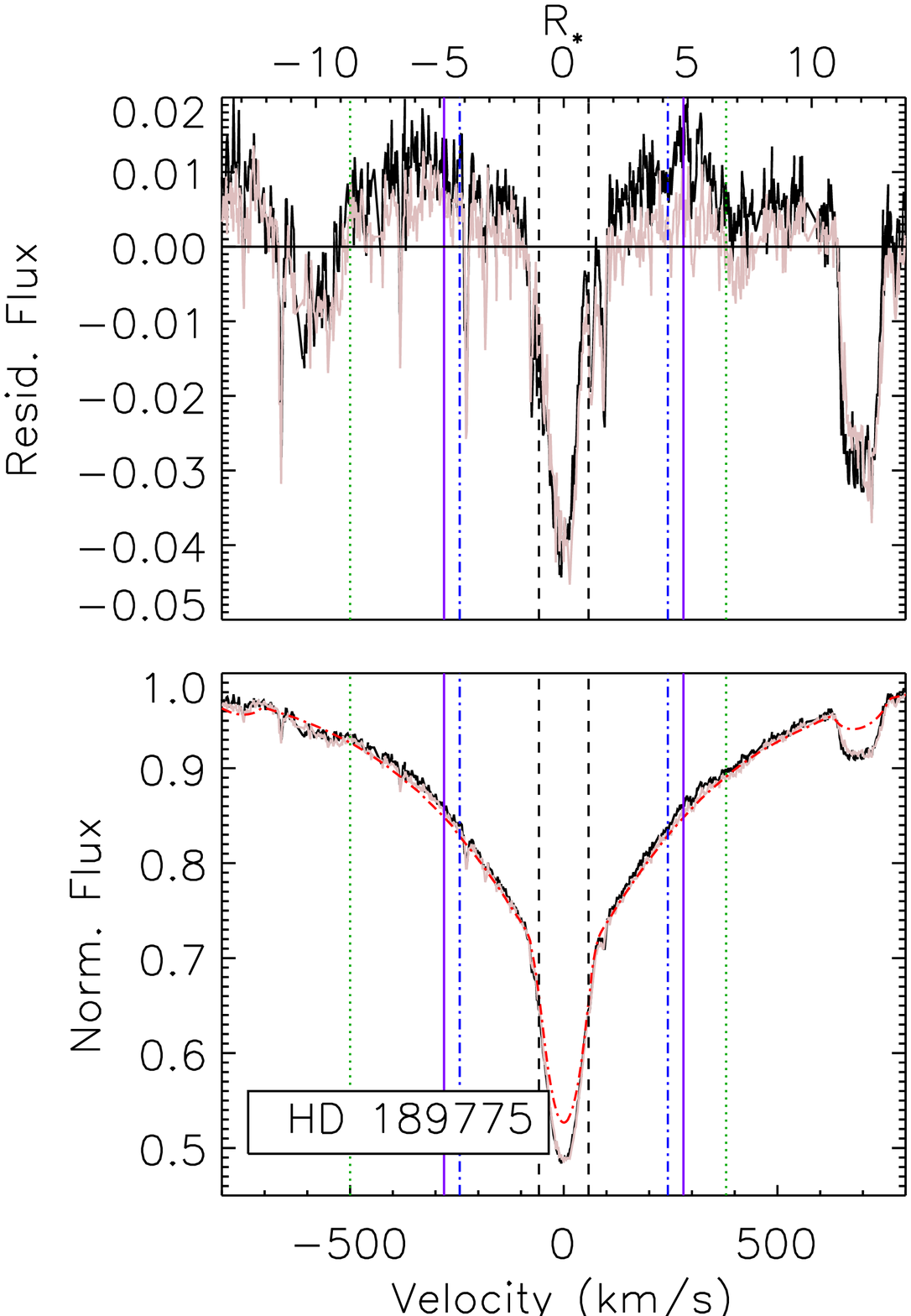} \\
   \includegraphics[trim=50 0 0 0, width=0.225\textwidth]{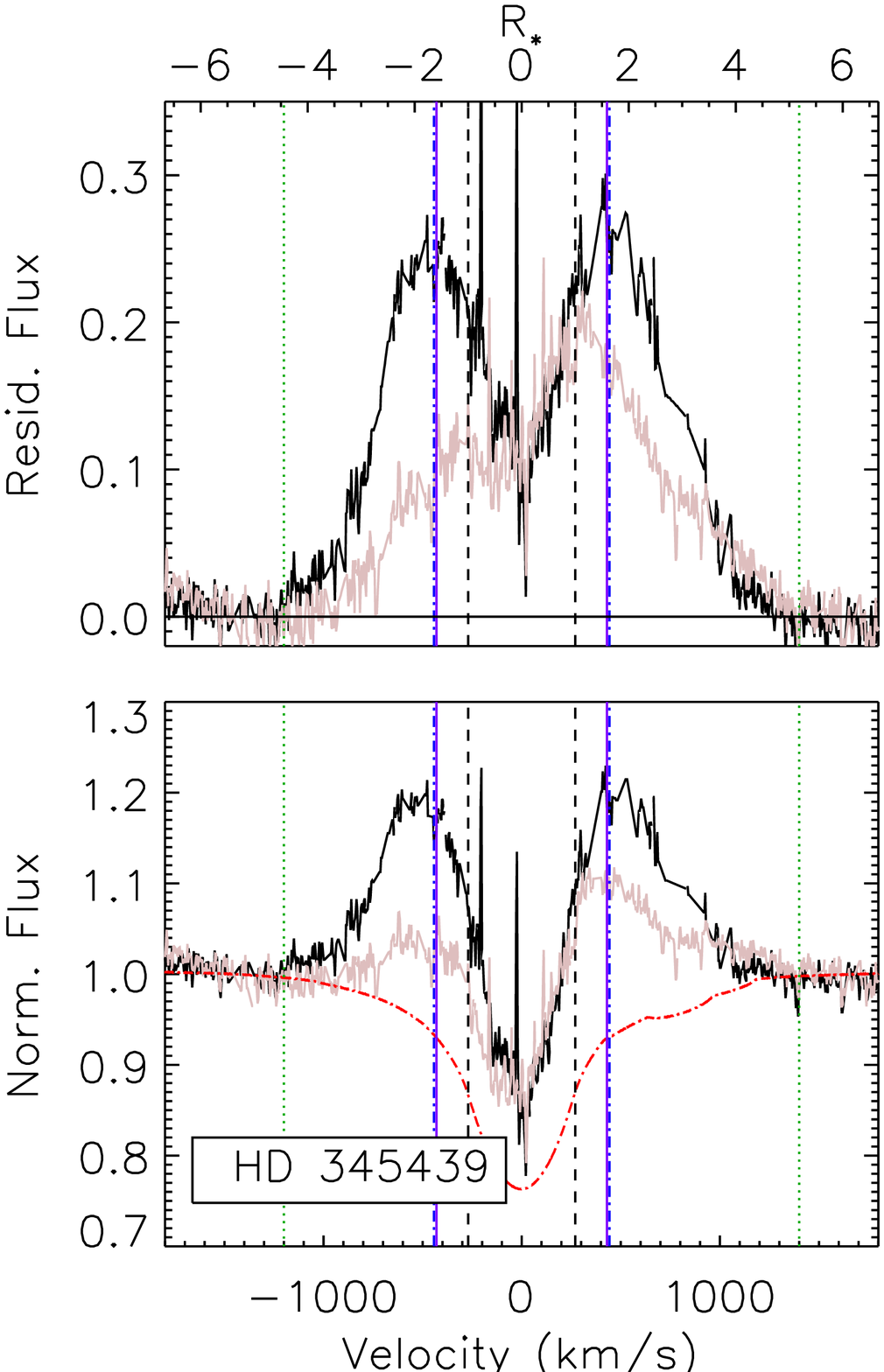} &
   \includegraphics[trim=50 0 0 0, width=0.225\textwidth]{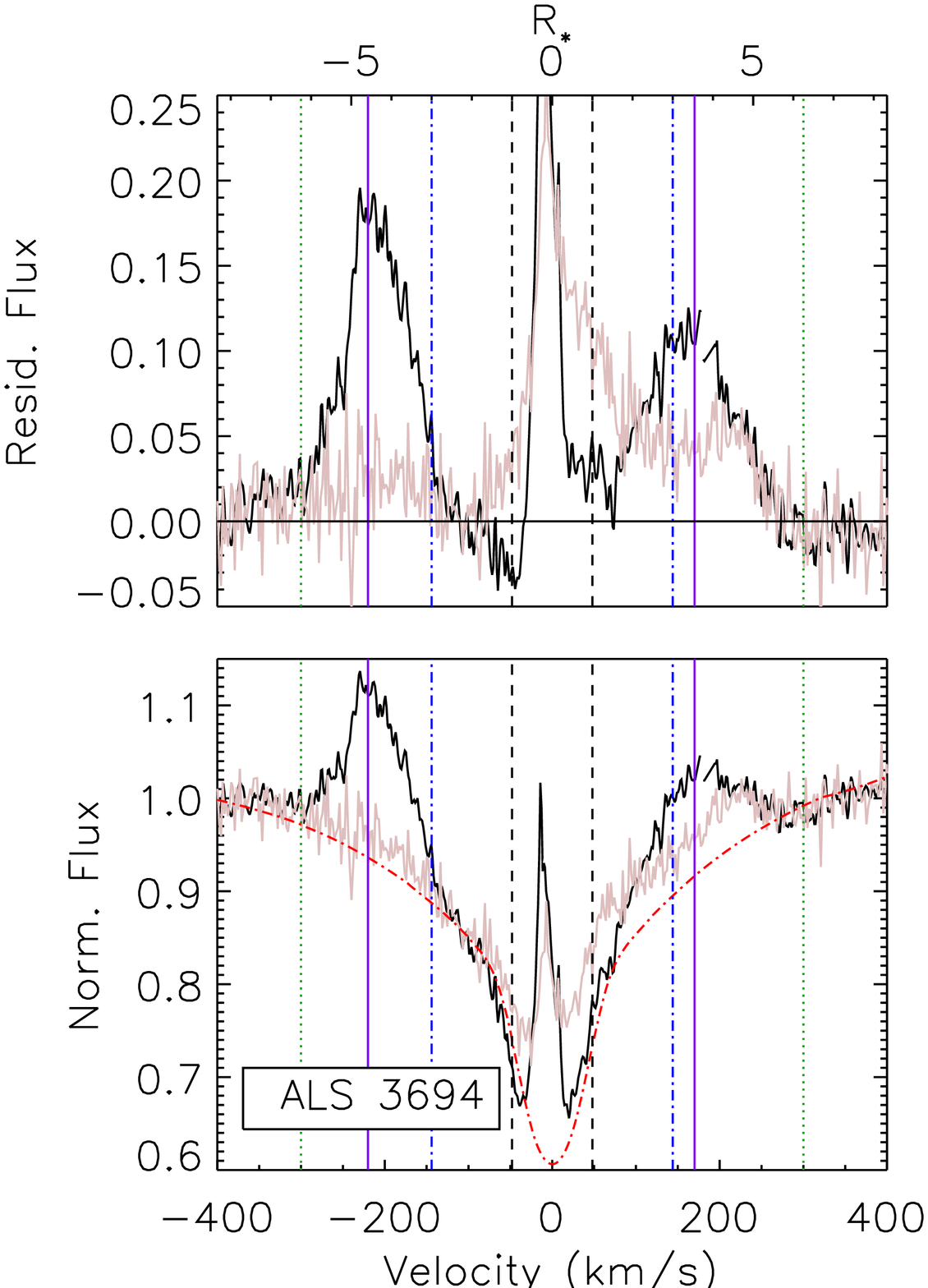} \\
   \includegraphics[trim=50 0 0 0, width=0.225\textwidth]{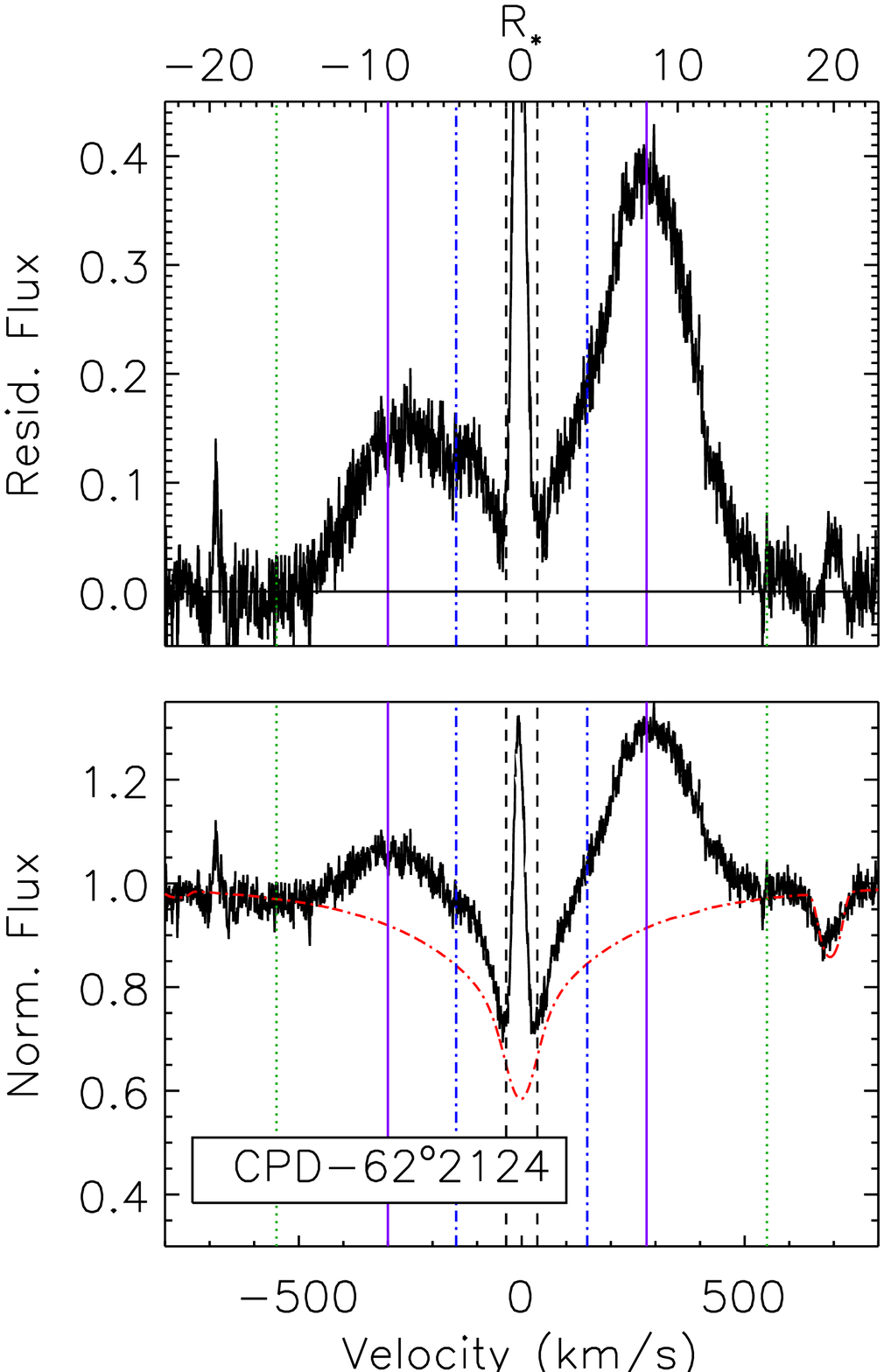} \\

\end{tabular}
      \caption[]{As Fig.\ \protect\ref{sigOriE_halpha_minmax} for HD\,182180, HD189775, HD\,345439, ALS\,3694, and CPD $-62^\circ 2124$.}
         \label{halpha_ind2}
   \end{figure}

   \begin{figure*}
   \centering
\begin{tabular}{ccc}

   \includegraphics[trim=50 50 25 0, width=0.3\textwidth]{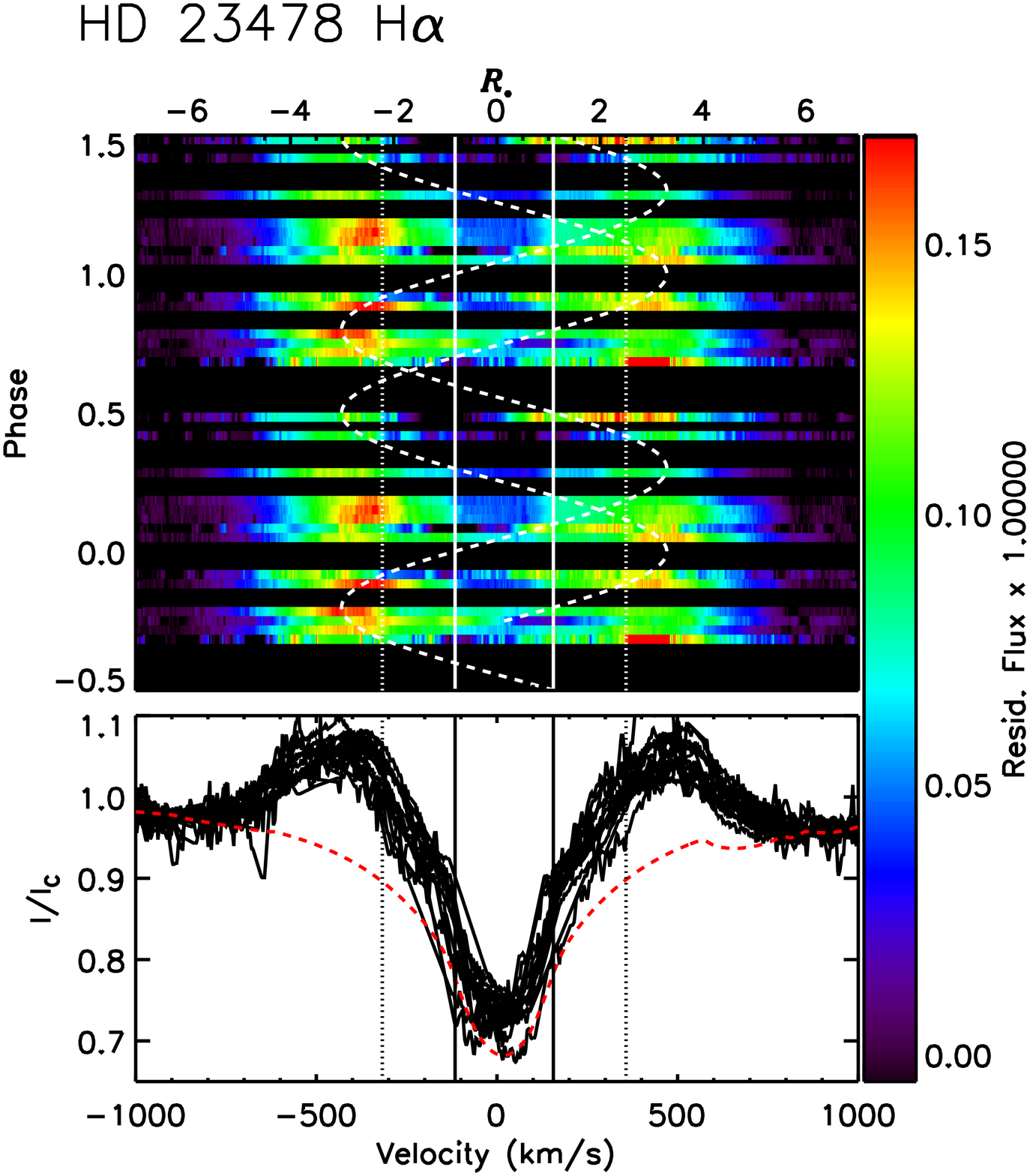} &
   \includegraphics[trim=50 50 25 0, width=0.3\textwidth]{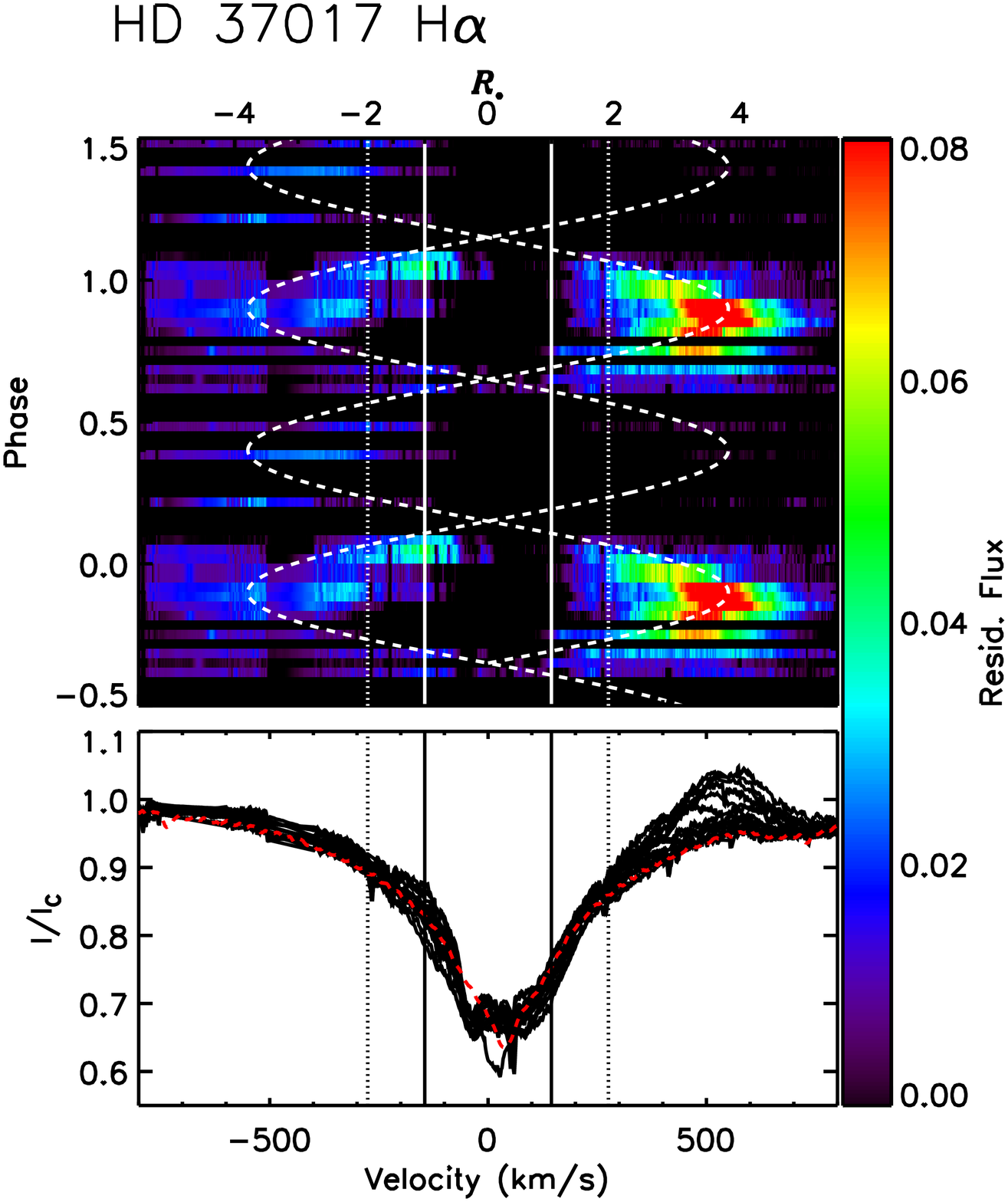} &
   \includegraphics[trim=50 50 25 0, width=0.3\textwidth]{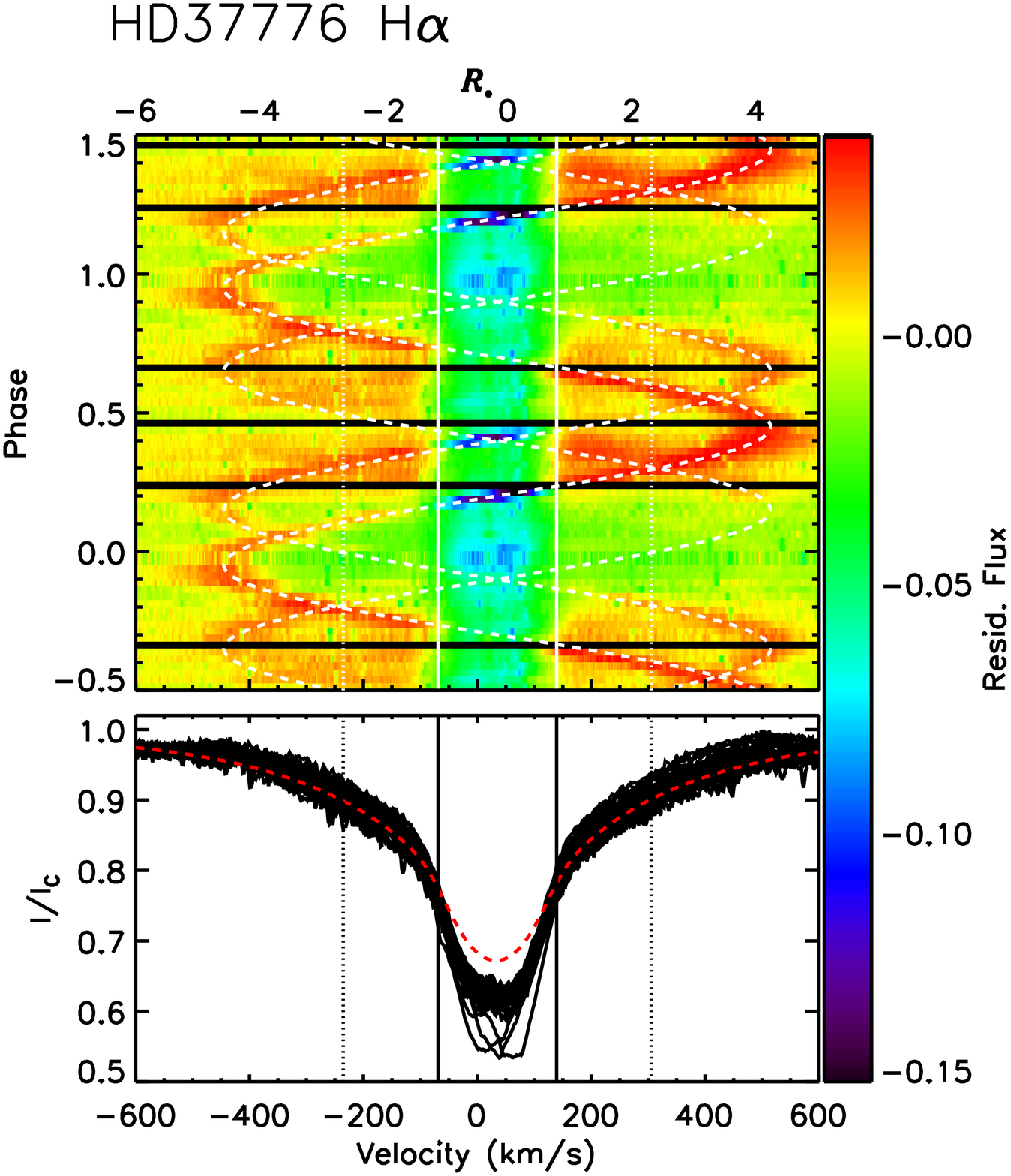} \\

   \includegraphics[trim=50 50 25 0, width=0.3\textwidth]{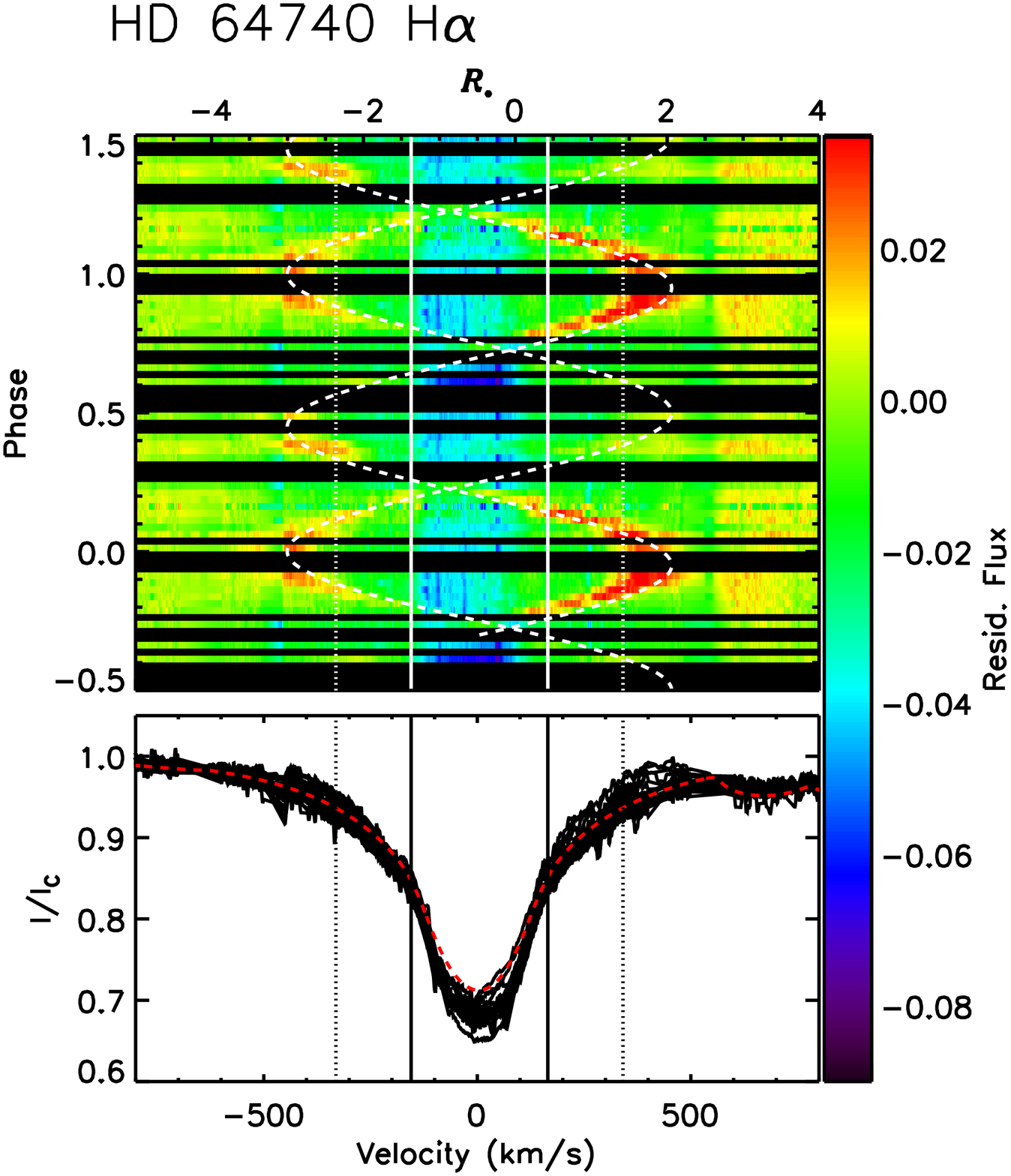} &
   \includegraphics[trim=50 50 25 0, width=0.3\textwidth]{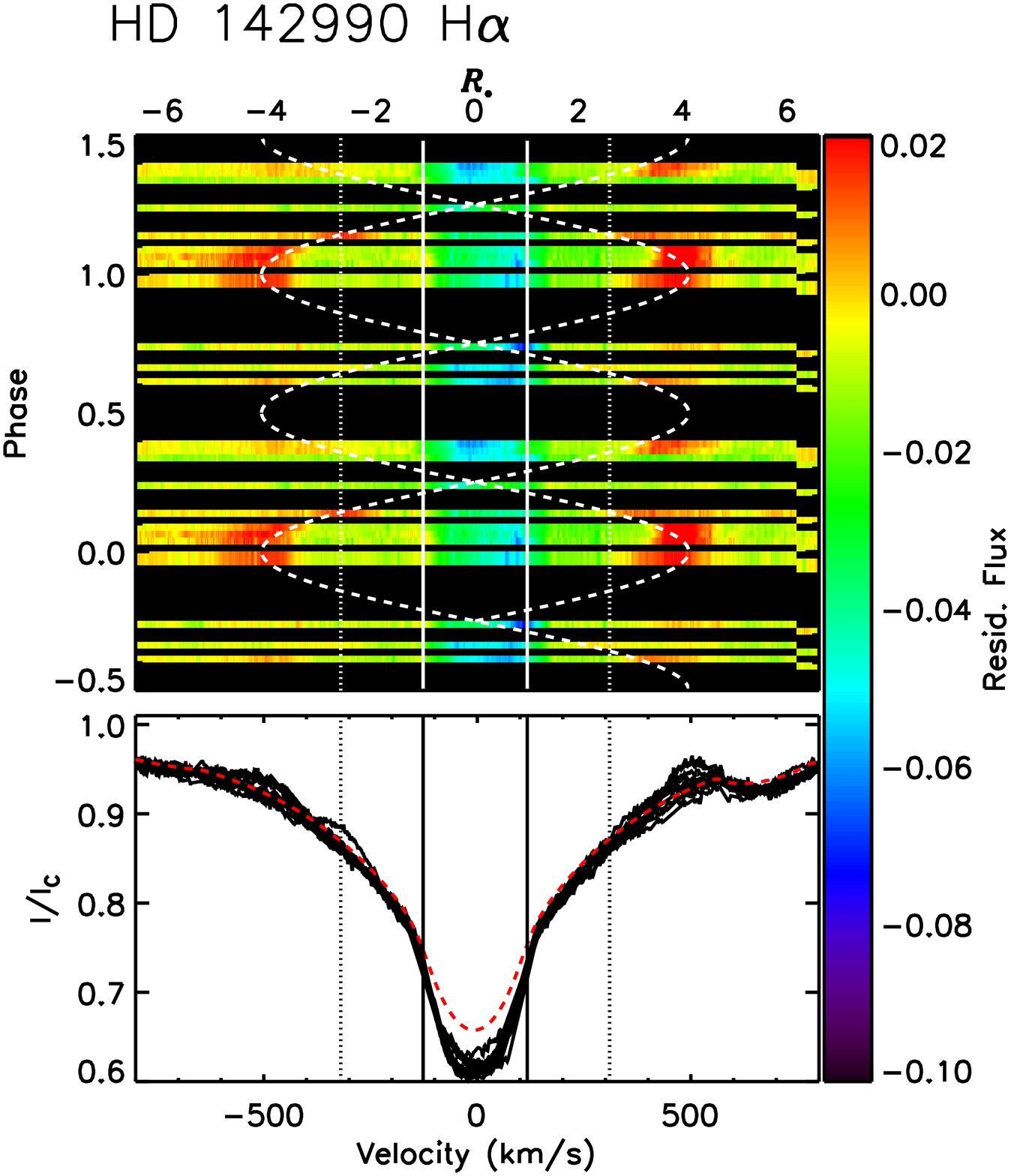} &
   \includegraphics[trim=50 50 25 0, width=0.3\textwidth]{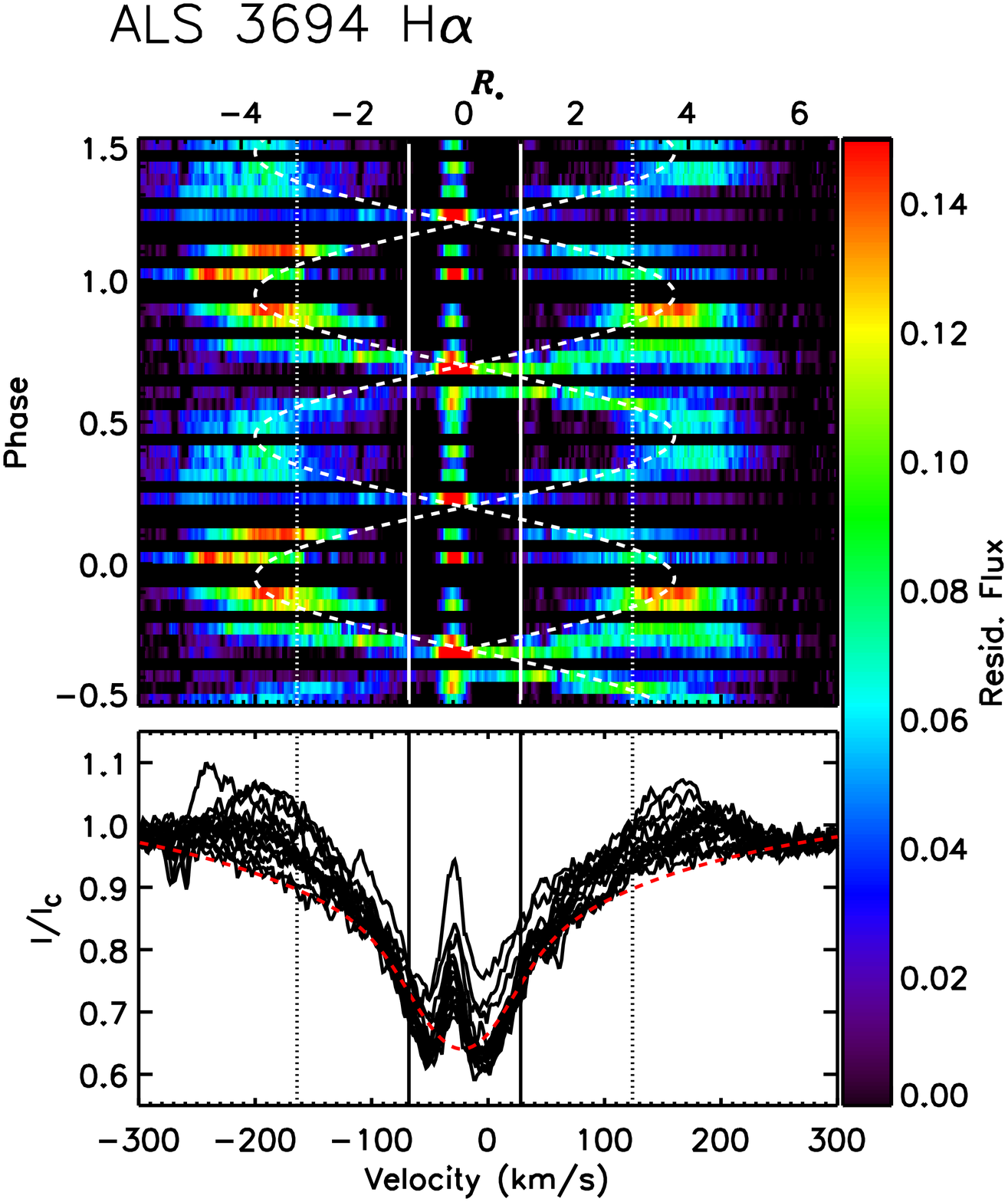} \\
\end{tabular}
      \caption[]{Dynamics spectra for stars for which these have not previously been published. Bottom panels show the individual intensity spectra (black) compared to the synthetic reference spectrum (red). Upper panels show the residual (observed minus reference) flux, mapped to the colour bars on the right, and phased with the rotation period. Vertical solid lines show $\pm$\vsini; dotted lines indicate $R_{\rm K}$. Dashed curves in the upper panels guide the eye, approximately tracing the path of the emission across the profile. In the case of the SB2 HD\,37017, the reference spectrum is for display only: residual flux was calculated using binary models.}
         \label{halpha_dyn}
   \end{figure*}

   \begin{figure*}
   \centering
\begin{tabular}{cccc}

   \includegraphics[trim=50 0 25 0, width=0.225\textwidth]{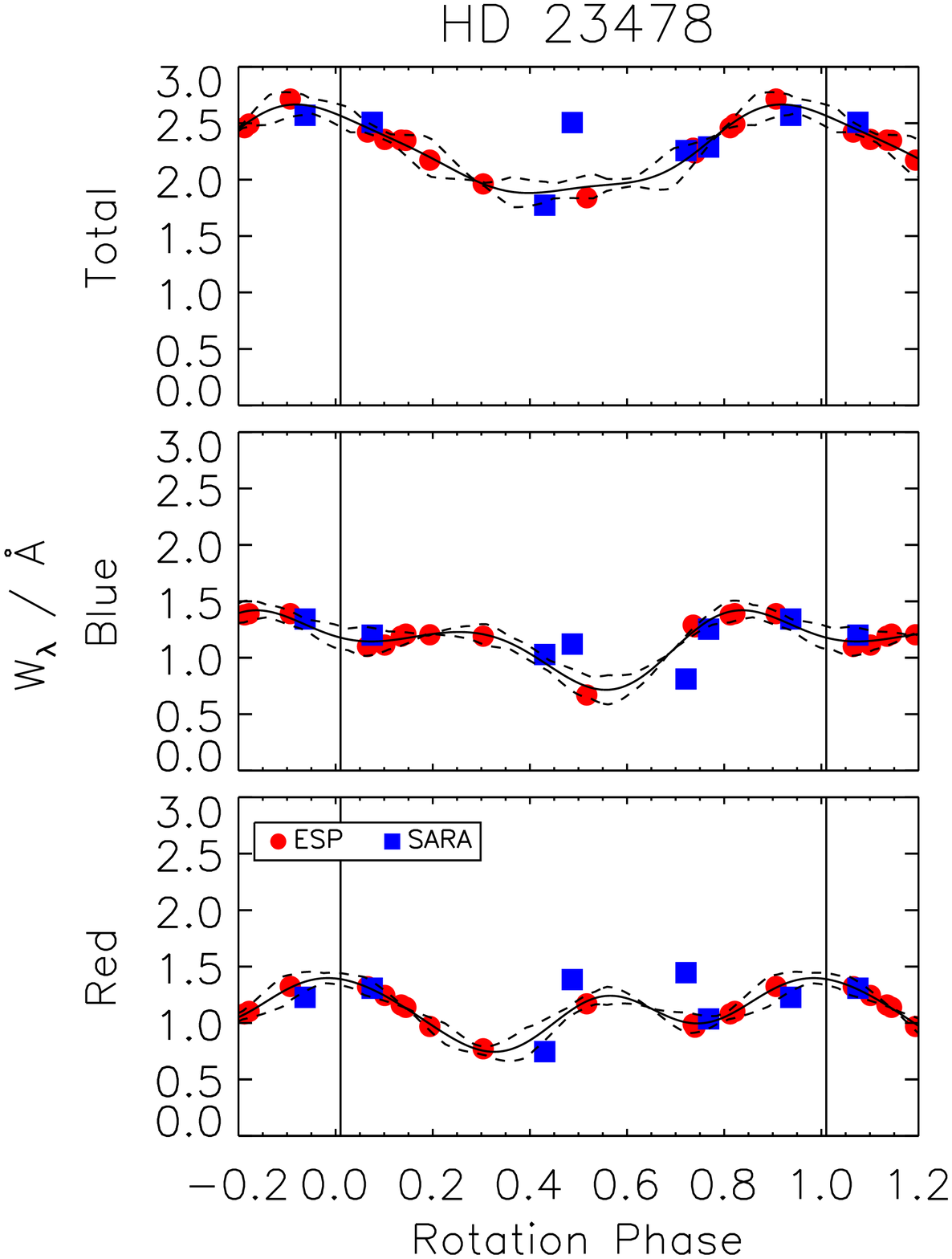} &
   \includegraphics[trim=50 0 25 0, width=0.225\textwidth]{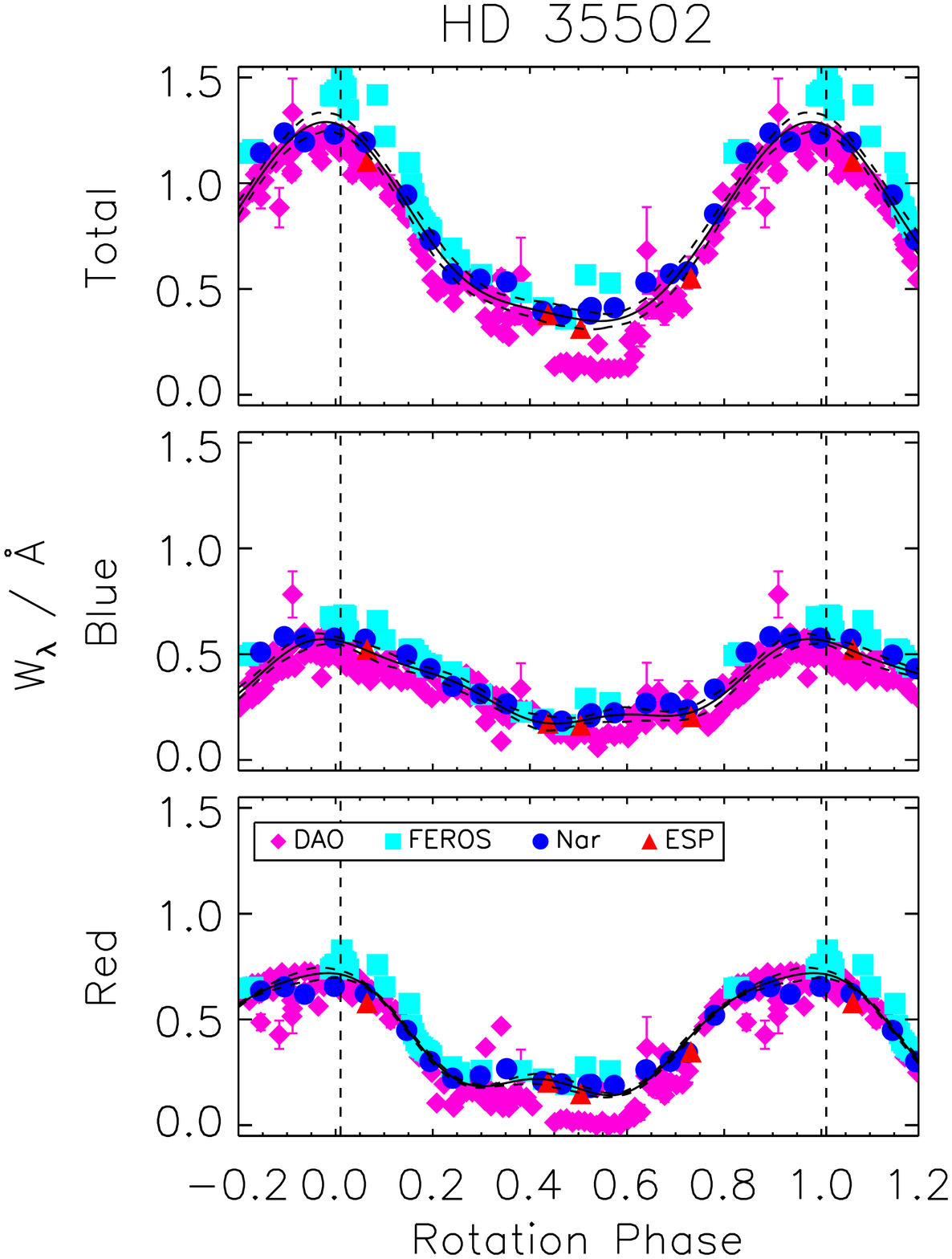} &
   \includegraphics[trim=50 0 25 0, width=0.225\textwidth]{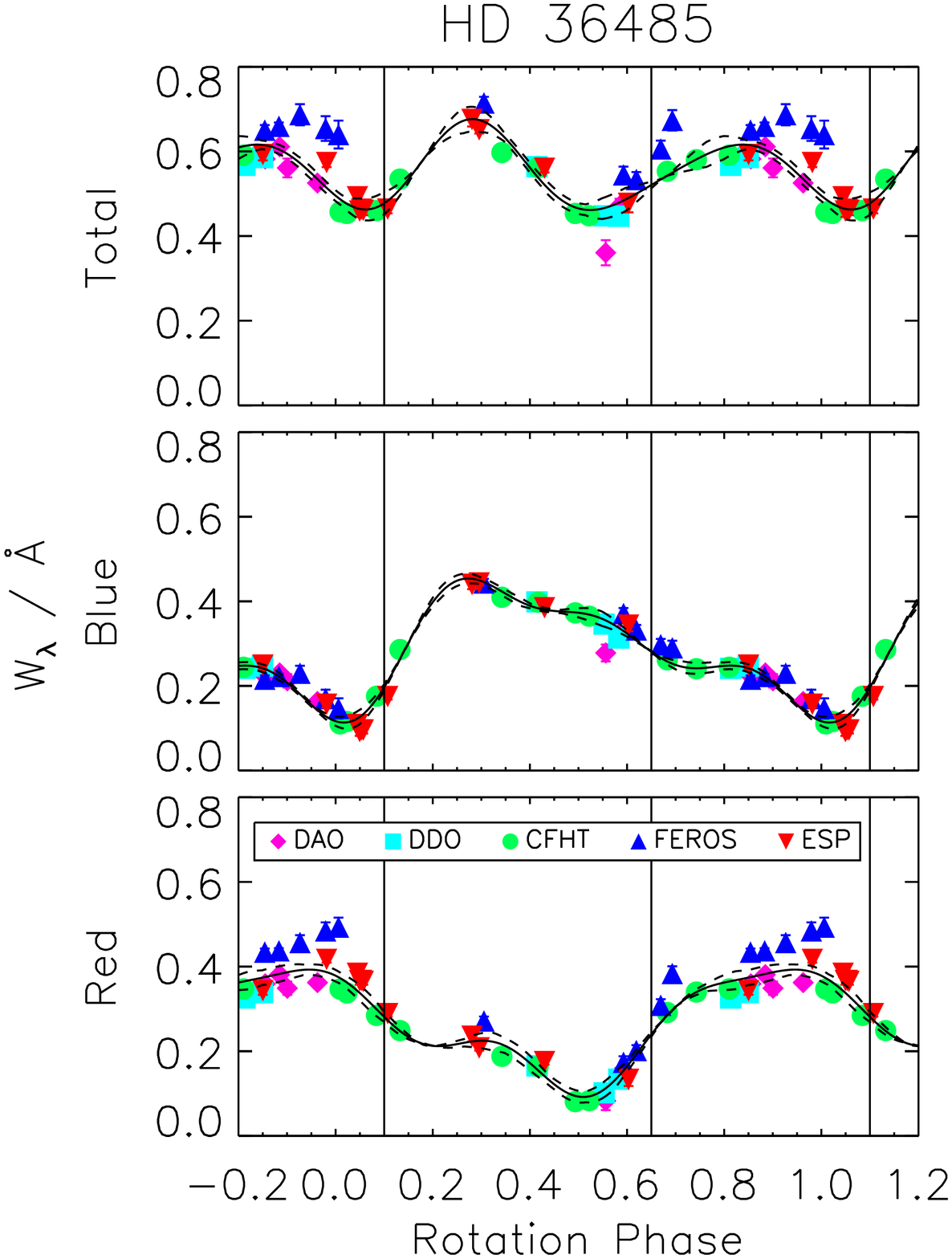} &
   \includegraphics[trim=50 0 25 0, width=0.225\textwidth]{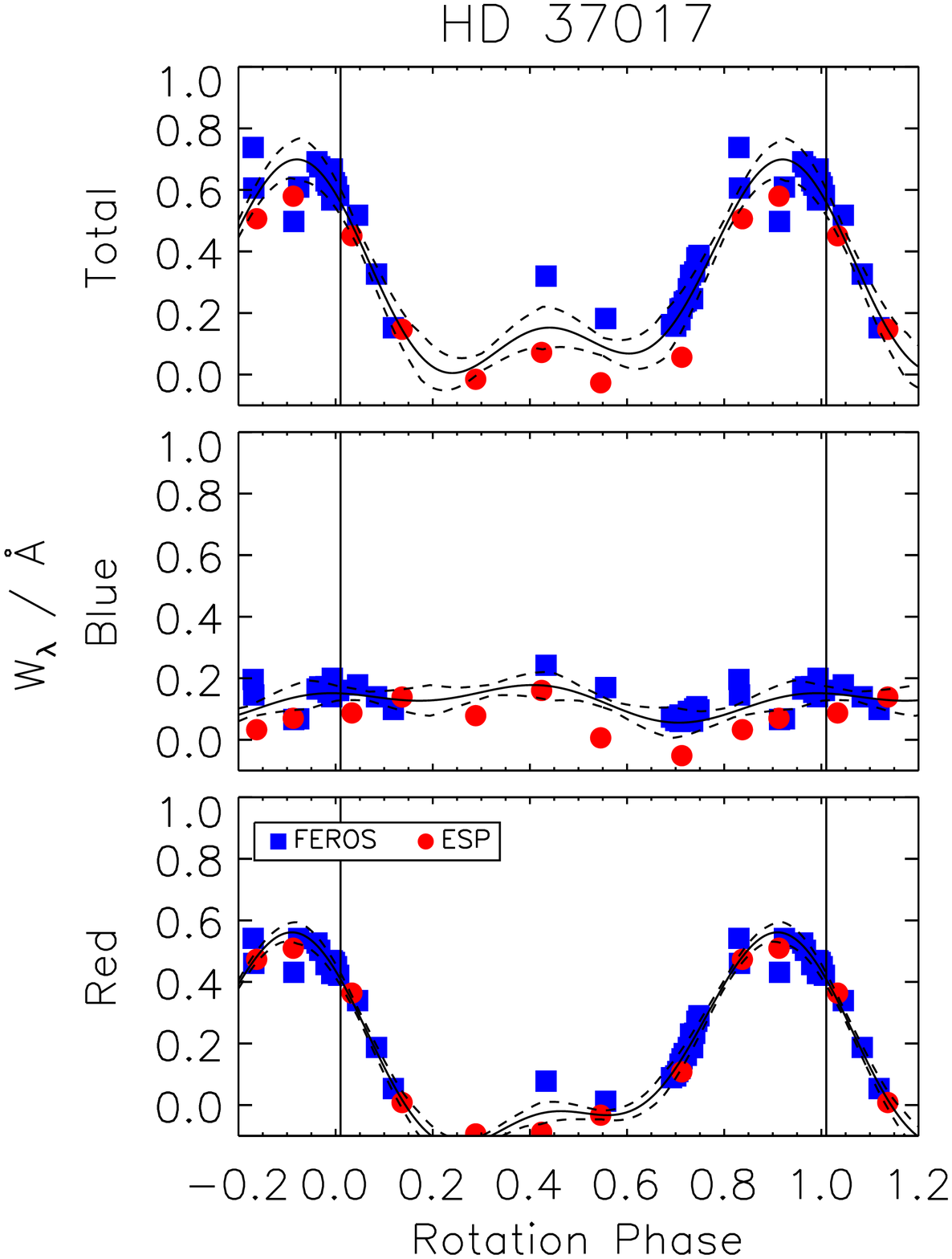} \\

   \includegraphics[trim=50 0 25 0, width=0.225\textwidth]{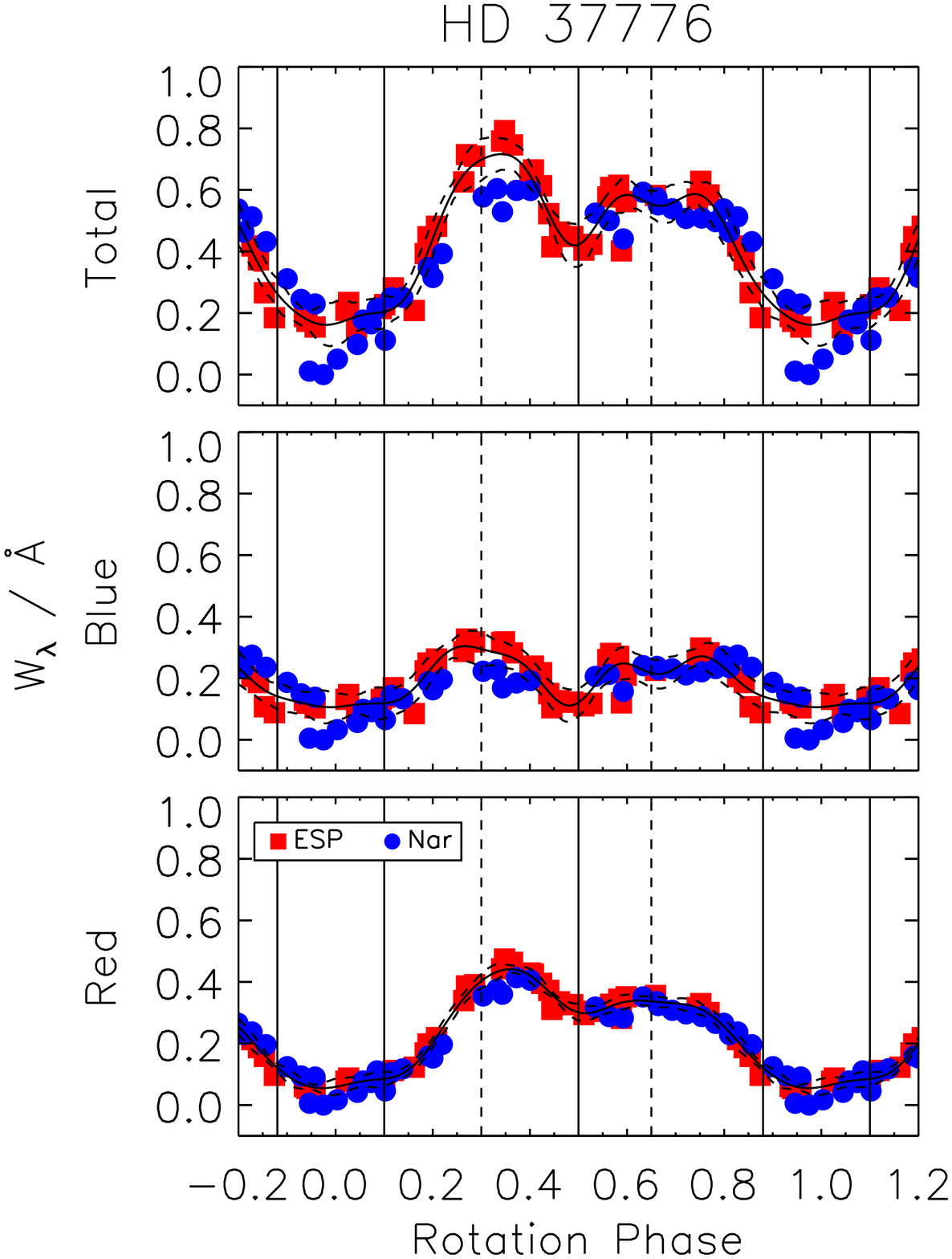} &
   \includegraphics[trim=50 0 25 0, width=0.225\textwidth]{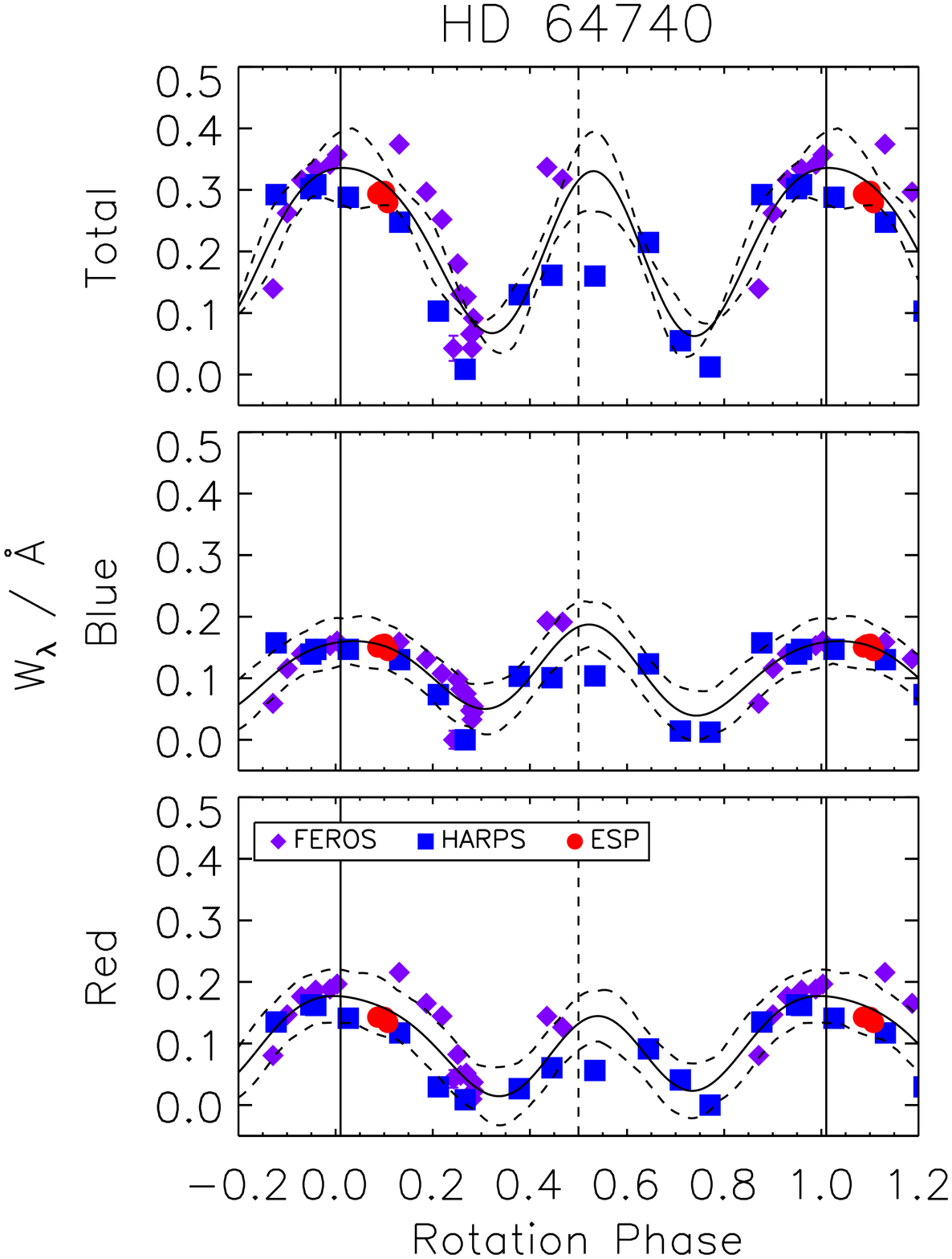} &
   \includegraphics[trim=50 0 25 0, width=0.225\textwidth]{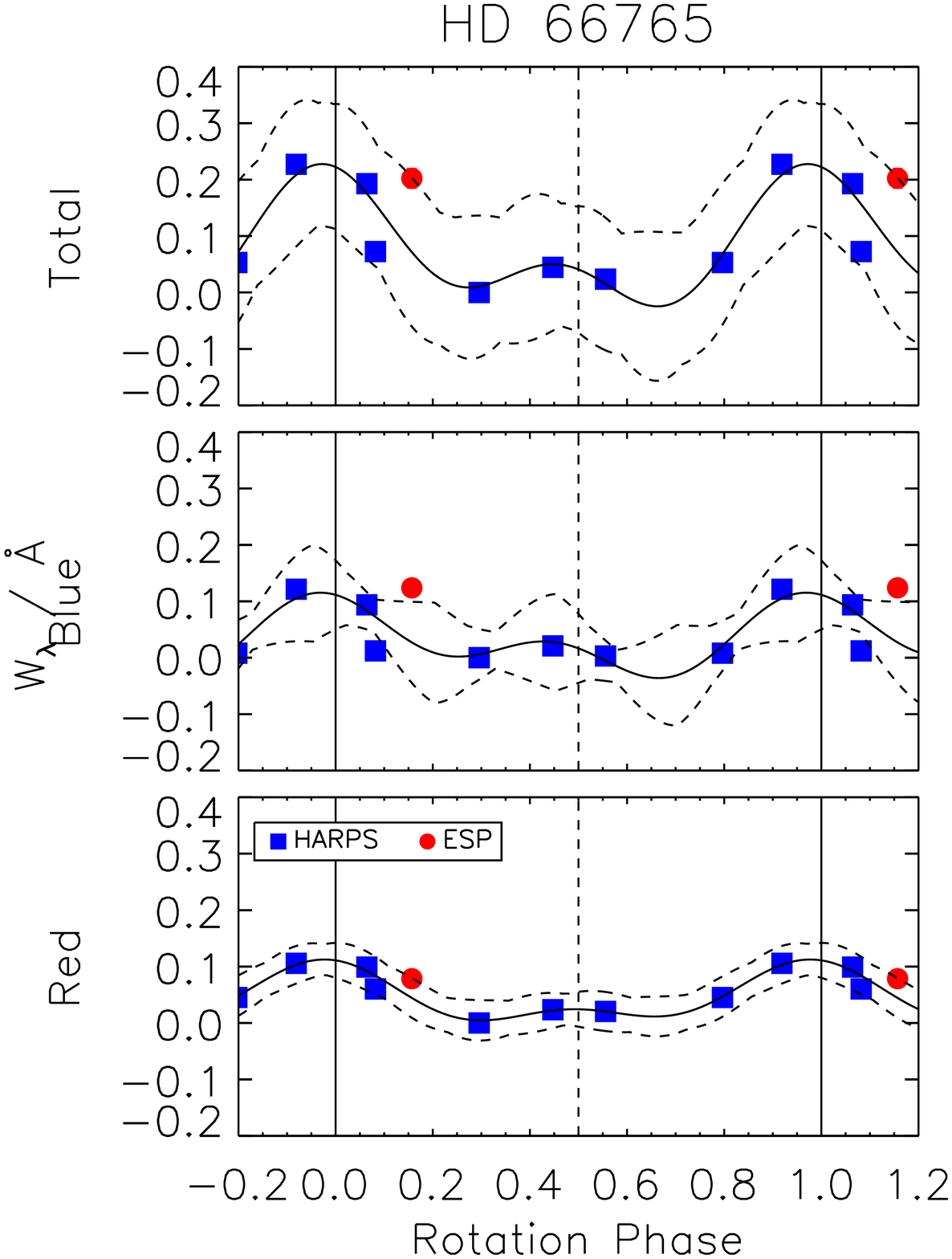} &
   \includegraphics[trim=50 0 25 0, width=0.225\textwidth]{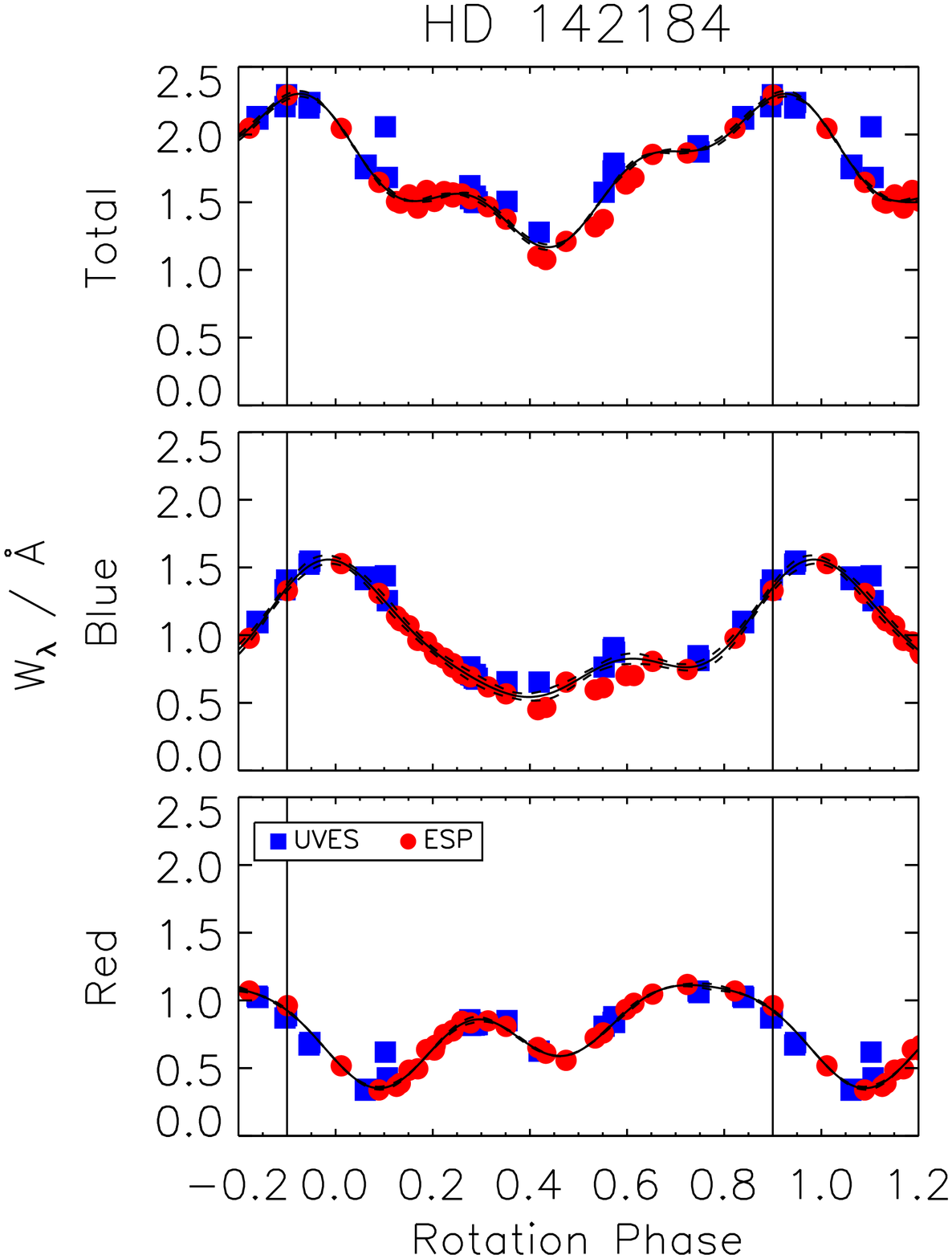} \\
\end{tabular}
      \caption[]{As Fig.\ \ref{sigOriE_halpha_minmax} for HD\,23478, HD\,35502, HD\,36485, HD\,37017, HD\,37776, HD\,64740, HD\,66765, and HD\,142184}
         \label{halpha_ew1}
   \end{figure*}

   \begin{figure*}
   \centering
\begin{tabular}{cccc}

   \includegraphics[trim=50 0 25 0, width=0.225\textwidth]{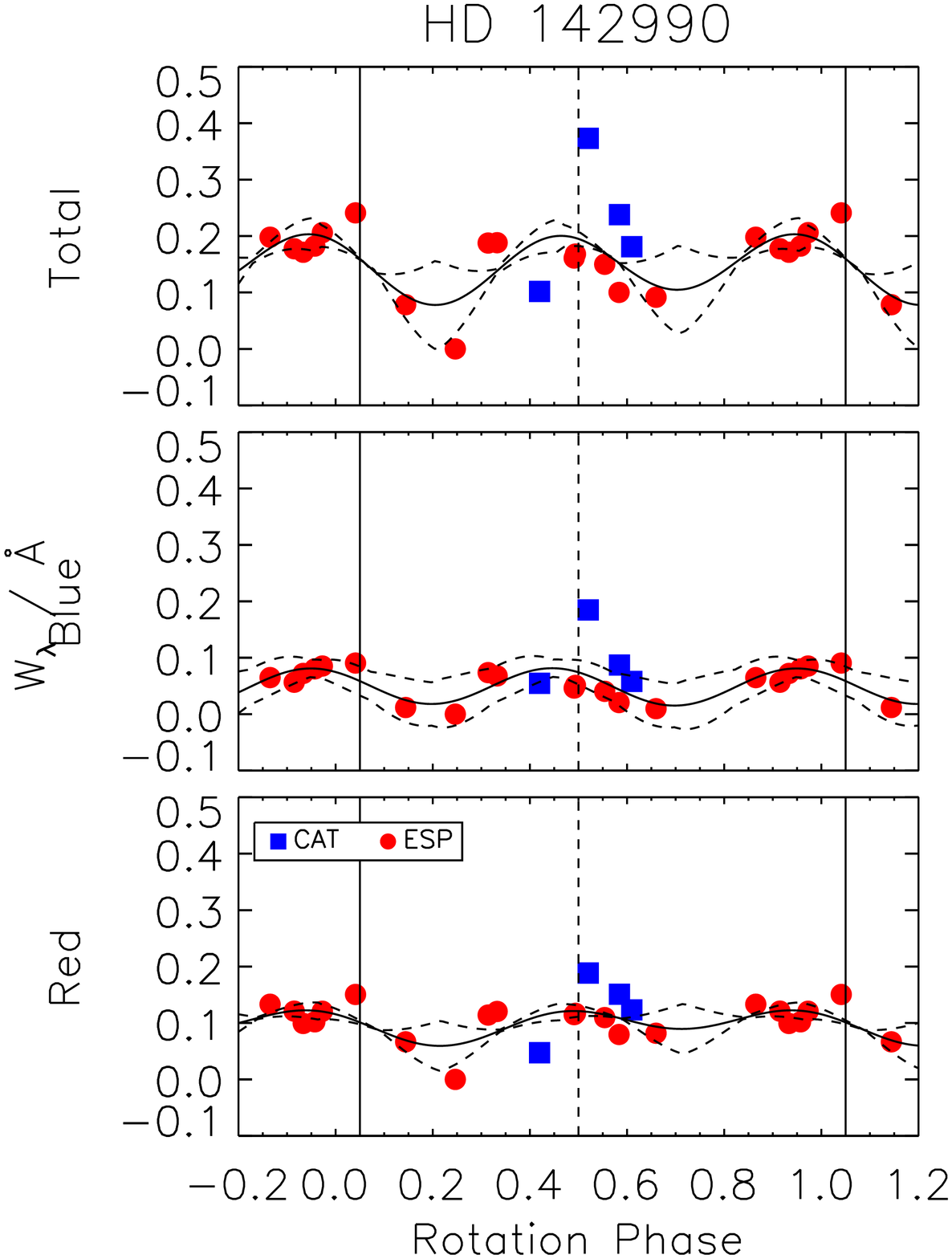} &
   \includegraphics[trim=50 0 25 0, width=0.225\textwidth]{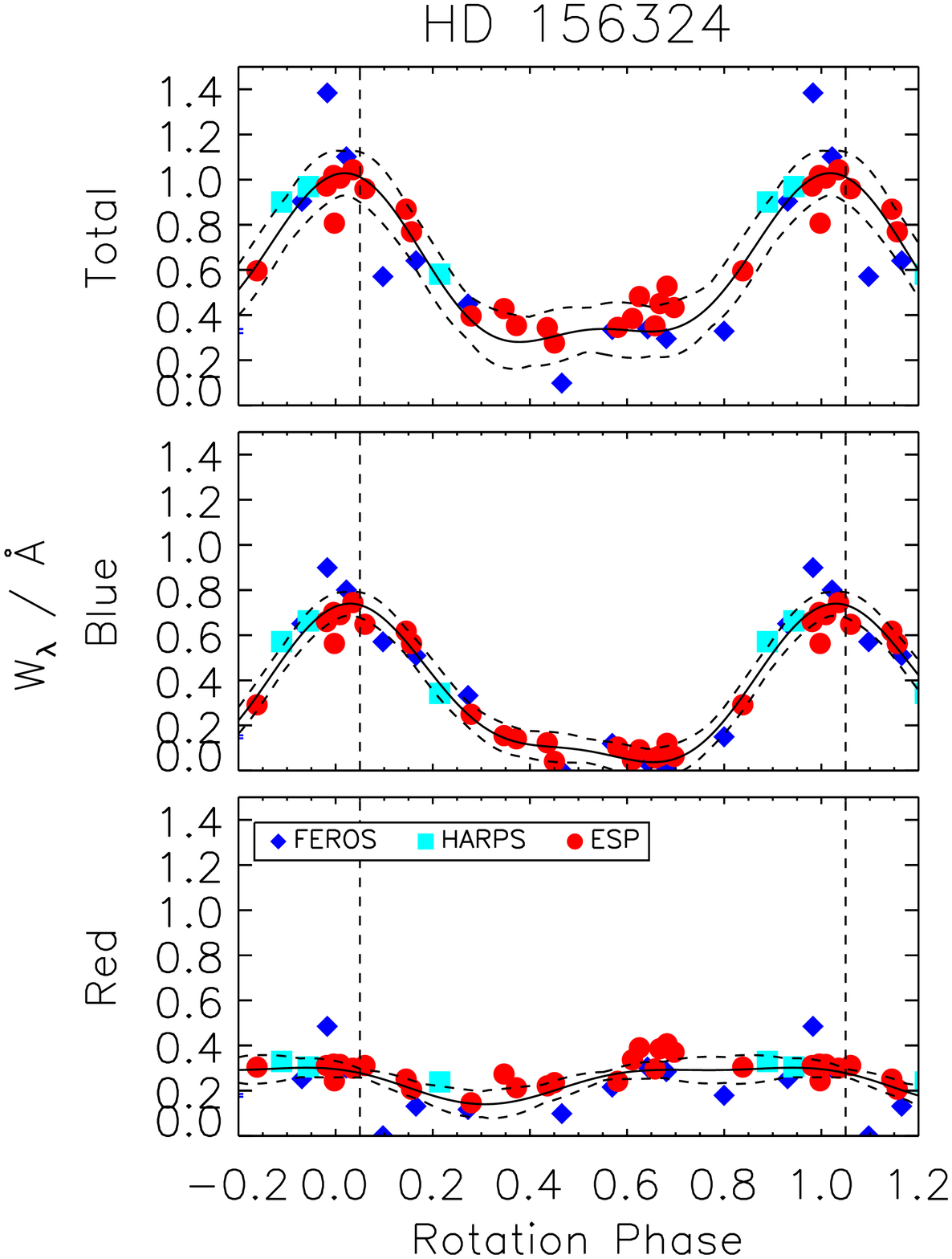} &
   \includegraphics[trim=50 0 25 0, width=0.225\textwidth]{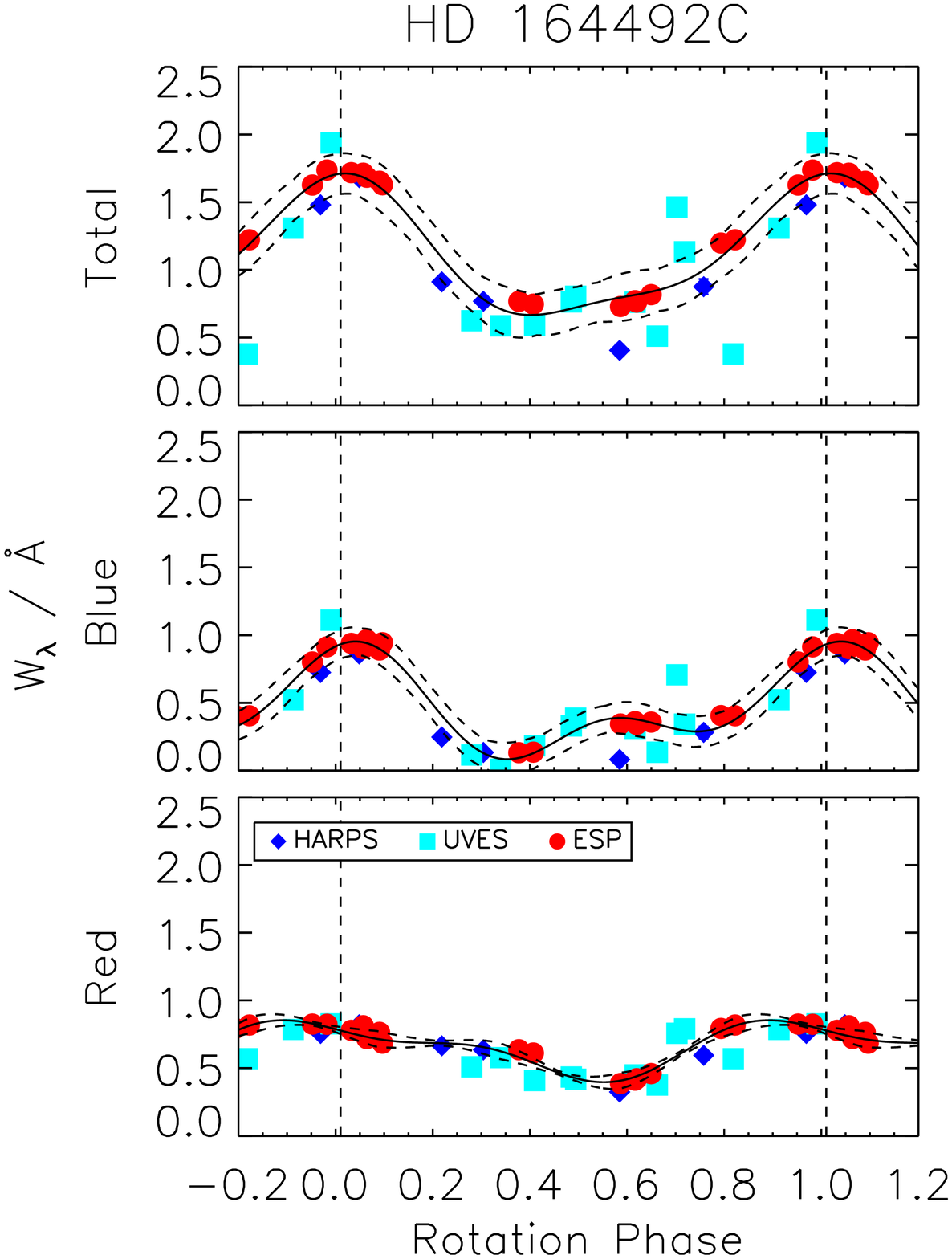} &
   \includegraphics[trim=50 0 25 0, width=0.225\textwidth]{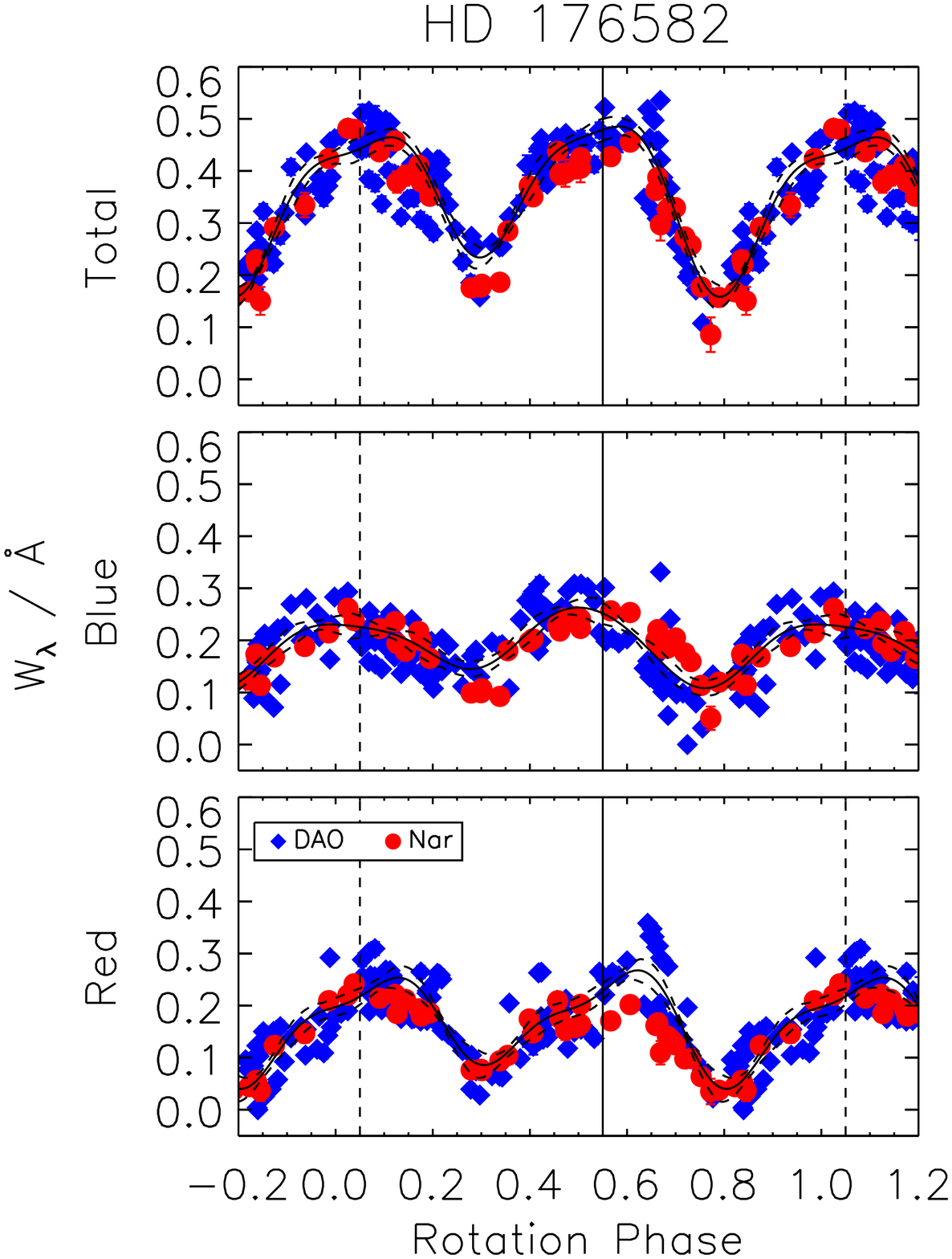} \\

   \includegraphics[trim=0 0 0 0, width=0.225\textwidth]{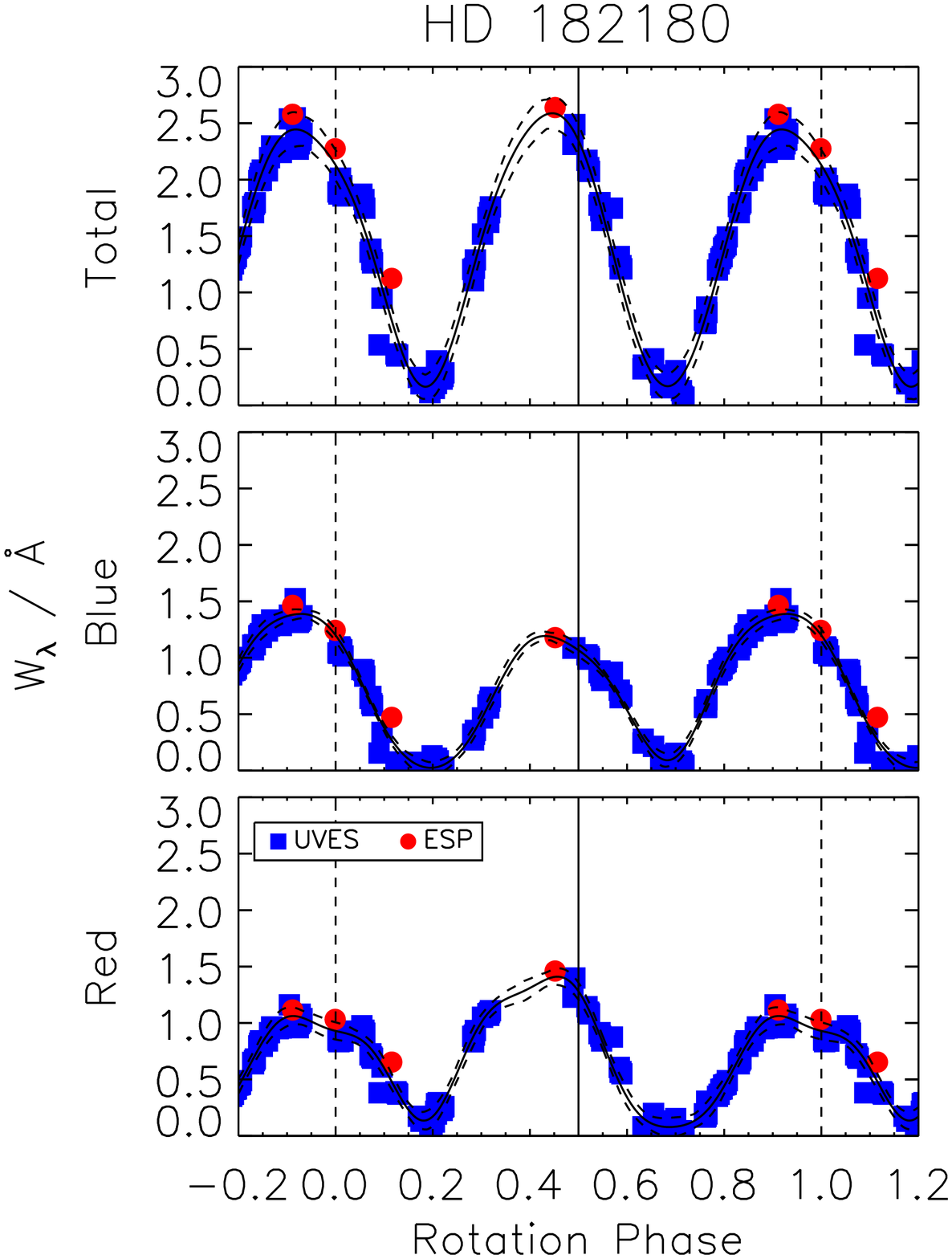} &
   \includegraphics[trim=0 0 0 0, width=0.225\textwidth]{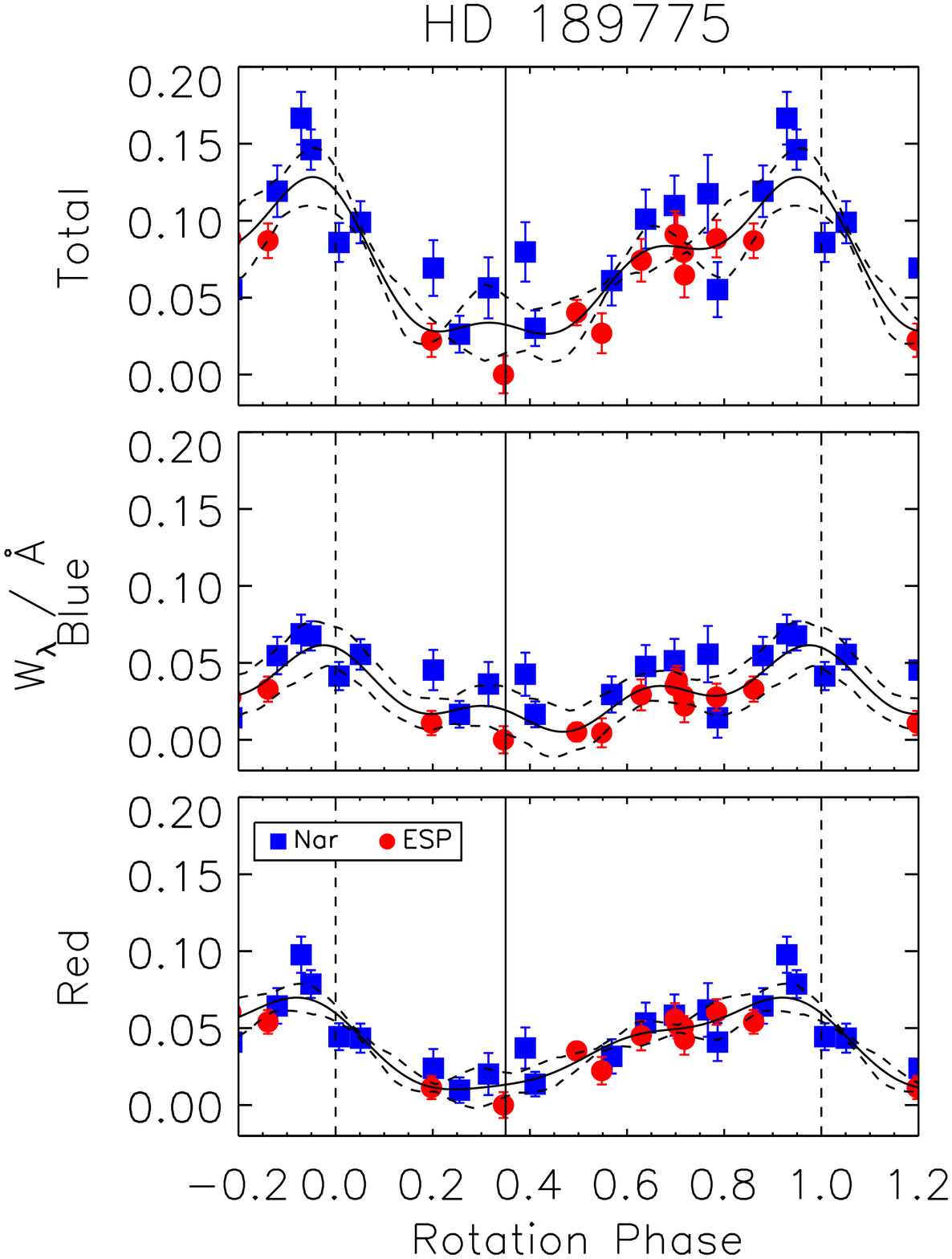} &
   \includegraphics[trim=0 0 0 0, width=0.225\textwidth]{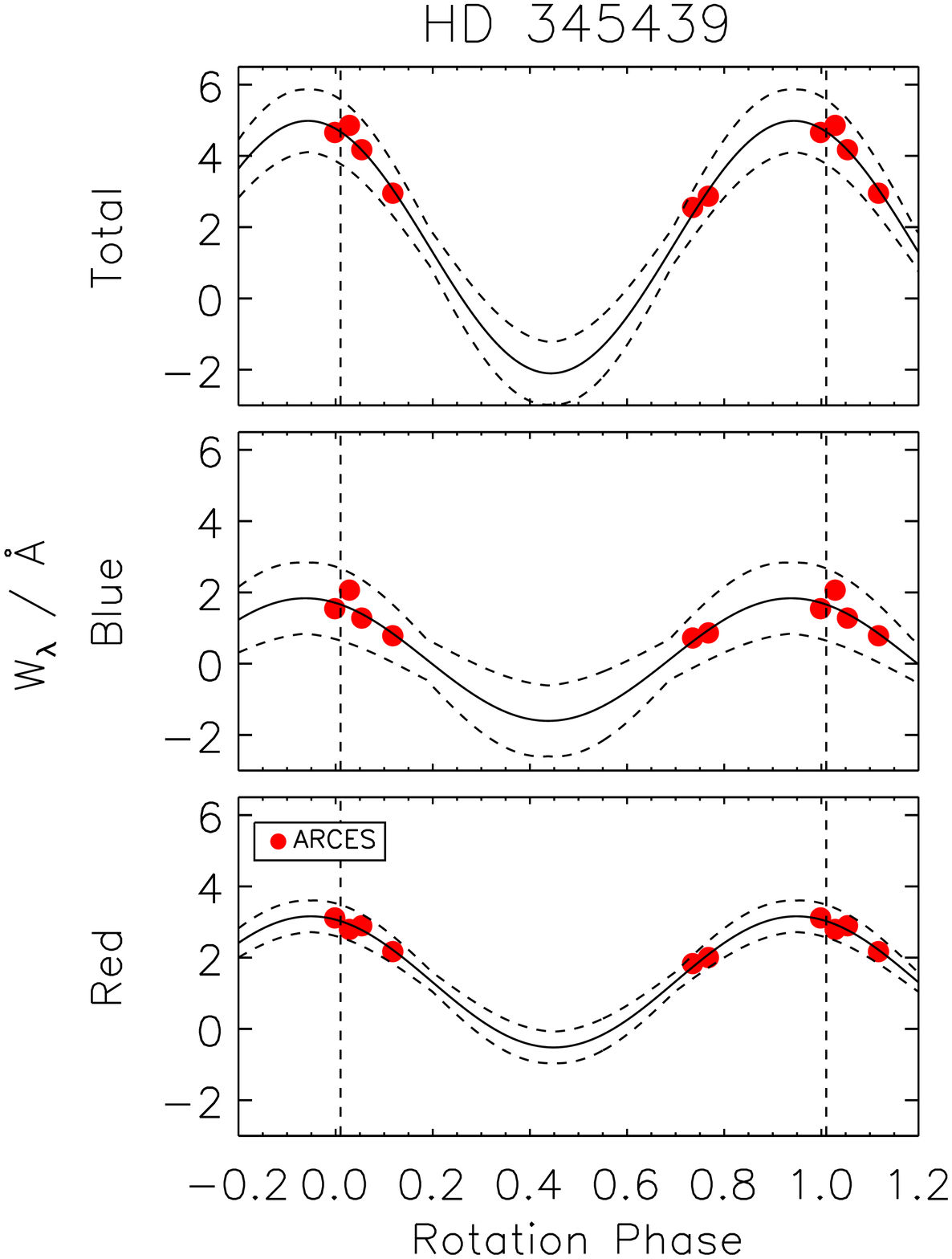} &
   \includegraphics[trim=0 0 0 0, width=0.225\textwidth]{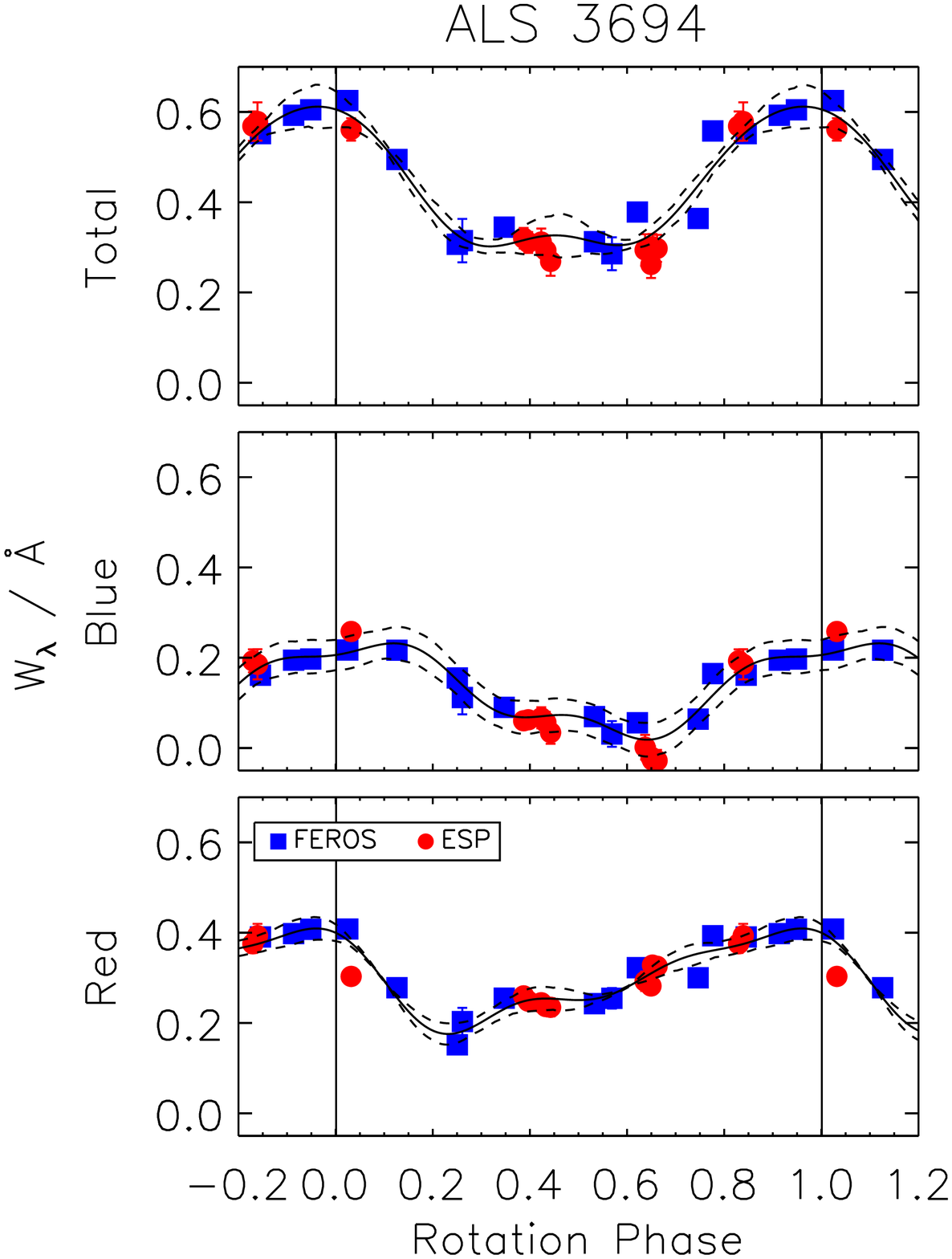} \\

\end{tabular}
      \caption[]{As Fig.\ \ref{sigOriE_halpha_minmax} for HD\,142990, HD\,156324, HD\,164492C, HD\,176582, HD\,182180, HD\,189775, HD\,345439, and ALS\,3694.}
         \label{halpha_ew2}
   \end{figure*}

\noindent {\bf HD\,23478}: This star's CM-type emission was discovered in infrared spectra by \cite{2014ApJ...784L..30E}. Magnetic and spectroscopic followup was presented by \cite{2015A&A...578L...3H}, \cite{2015ApJ...811L..26W}, and \cite{2015MNRAS.451.1928S}. H$\alpha$ profiles from the ESPaDOnS data are shown in Fig.\ \ref{halpha_ind1}. The star shows no indication of eclipses, and its H$\alpha$ profile is minimally variable (Figs.\ \ref{halpha_dyn} and \ref{halpha_ew1}). As can be seen in the dynamic spectrum in Fig.\ \ref{halpha_dyn}, a two-cloud model does a poor job of reproducing the variability. This is similar to the pattern seen in HD\,142184 \citep{grun2012}; notably, the star has a moderate $i_{\rm rot}$ and a very small $\beta$, similar to HD\,142184's configuration. Since $\beta$ is small the RRM model predicts that the CM should be a slightly warped torus without necessarily having distinct clouds. This is consistent with the emission being detectable at all rotational phases, without a large amount of variation in the location or intensity of peak emission.

\noindent {\bf HD\,35502}: This star is a hierarchical SB3, with two identical A-type companions in a close orbit \citep[$P_{\rm orb} \sim 5.67$~d][]{2016MNRAS.460.1811S}, which in turn orbit the magnetic B-type primary as distant companions. \cite{2016MNRAS.460.1811S} performed a comprehensive analysis of the star's orbital, fundamental, magnetic, and emission properties. The star is seen at a small inclination ($i_{\rm rot} \sim 26^\circ$), and therefore does not show eclipses. It's emission profile is highly variable due to the large ($\beta = 70^\circ$) obliquity. H$\alpha$ profiles are shown in Fig.\ \ref{halpha_ind1}, and EW curves are shown in Fig.\ \ref{halpha_ew1}. Its emission profile is relatively symmetrical. Its EW curve exhibits a single-wave variation, consistent with its small $i+\beta$.

\noindent {\bf HD\,36485}: \cite{leone2010} provided a detailed investigation of this star's properties. It is an SB2 system with an A-type companion on a highly elliptical 30~d orbit. The orbital and rotational inclinations are both very small ($i \sim 20^\circ$), and the magnetic axis is nearly aligned with the rotational axis ($\beta \sim 4^\circ$; Paper III). Despite this, H$\alpha$ is highly variable (Fig.\ \ref{halpha_ind1}), and exhibits a double-wave variation (Fig.\ \ref{halpha_ew1}), a phenomenon which was also noted by \cite{leone2010}, who speculated that this may be due to a highly complex magnetic field. While a dipole is formally a good fit to \bz, there is some indication that \bz~may actually be a double-wave (Paper I). The very small level of variation in \bz~due to the small $i$ and $\beta$ makes this hypothesis difficult to verify using \bz~measurements at the current level of precision. Due to the small amplitude of the \bz~variation, the two magnetic maxima indicated in Fig.\ \ref{halpha_ew1} are somewhat tentatively identified, which may explain why they do not coincide with the maxima of the EW curve. 

\noindent {\bf HD\,37017}: This is an SB2, with a late B-type companion on a highly elliptical 18~d orbit \citep{1998AA...337..183B}. Its H$\alpha$ emission was first detected by \cite{walborn1974}. H$\alpha$ is highly variable and extremely asymmetric (Figs.\ \ref{halpha_ind1}, \ref{halpha_ew1}, and \ref{halpha_dyn}), with essentially no indication of a second cloud. This asymmetry was also noted by \citeauthor{1993A&A...273..509L}, who interpreted the red-shifted emission as a jet above the magnetic pole. No such jets have been found in other magnetic early B-type stars. In principle the asymmetry might be attributed to a very complex surface magnetic field, however \bz~is entirely consistent with a dipolar configuration (Paper I). The only other star known to show such a highly asymmetric emission profile is the tidally locked binary HD\,156324 \citep[][see also Fig.\ \protect\ref{halpha_ind2}]{2018MNRAS.475..839S}. While HD\,37017 is not tidally locked, and its 0.9~d rotation period is not in an obvious harmonic relationship with the orbital period, the comparison of the semi-major axis \citep{1998AA...337..183B} and the Alfv\'en radius (Paper III) indicates that the companion penetrates the magnetosphere during its orbit. This suggests that the asymmetric emission profile may be a consequence of orbital disruption of the CM.

\noindent {\bf HD\,37479}: $\sigma$ Ori E was the first star in which an H$\alpha$-bright CM was detected \citep{walborn1974}, was the inspiration for the RRM model \citep{town2005b,town2005c} as well as its precursors \citep{1985ApSS.116..285N,2004AA...417..987P}, and is by far the most extensively studied star of this class \citep[e.g.][]{oks2012,2015MNRAS.451.2015O}. Unsurprisingly for the prototype of this class, its emission is amongst the strongest in the sample. Mild asymmetry in the emission profile (Fig.\ \ref{sigOriE_halpha_minmax}) is a consequence of a distorted dipole surface magnetic field geometry \citep{2015MNRAS.451.2015O}. Eclipsing of the star by the CM is easily detectable in H$\alpha$ as well as photometry, and timing of these eclipses has been used by \cite{town2010} to measure the spindown rate due to magnetic braking.

\noindent {\bf HD\,37776}: \cite{thom1985} discovered this star to have an obvious double-wave \bz~curve, and Zeeman Doppler Imaging by \cite{koch2011} demonstrated that the surface magnetic field is highly complex and non-axisymmetric. It also has an extremely strong magnetic field, with a maximum surface strength of about 30 kG. This leads to significant departures from the usual warped disk predicted by the RRM model, which is reflected in a complex pattern of H$\alpha$ variability (Fig.\ \ref{halpha_dyn}). Eclipses are clearly visible in the dynamic spectrum. Accurate modelling of this star's CM obviously requires an arbitary RRM model similar to that utilized for HD\,37479 by \cite{2015MNRAS.451.2015O}. EW extrema (Fig.\ \ref{halpha_ew1}) do not obviously correlate with the phases of magnetic extrema, which may also be a consequence of the extremely complex surface magnetic field of this star. There is almost certainly variability in the wings due to the star's strong He spots, and the EW variation (and the maximum emission strength, see Fig.\ \ref{halpha_ind1}) therefore likely also includes contributions unrelated to emission. 

\noindent {\bf HD\,64740}: No detailed investigation of the H$\alpha$ profile of this star has yet been performed, although it emission was first reported by \cite{walborn1974}. HD\,64740 is one of the hottest stars in the sample (\teff~$\sim 24.5$~kK), and has a relatively weak magnetic field ($B_{\rm d} \sim 3$~kG). Its emission is extremely weak (Fig.\ \ref{halpha_ind1}), but clearly follows the expected pattern for a CM formed by a tilted dipole (Fig.\ \ref{halpha_dyn}). Eclipses are visible, as expected given the moderate inclination and high obliquity ($i_{\rm rot} \sim 42^\circ$, $\beta \sim 72^\circ$; Paper III). Consistent with its angular parameters, the EW curve is a double-wave (Fig.\ \ref{halpha_ew1}).

\noindent {\bf HD\,66765}: the magnetic field of this He-strong star was discovered by \cite{alecian2014}, who also determined its 1.6~d rotation period and reported its extremely weak H$\alpha$ emission. Indeed, HD\,66765 displays the second-weakest emission in the sample (Fig.\ \ref{halpha_ind1}), consistent with its relatively long rotation period (1.6~d) and weak surface magnetic field ($B_{\rm d} \sim 2.8$~kG; Paper III). We consider the emission to be real because 1) the residual flux peaks near $R_{\rm K}$ in some but not all spectra, and shows a sharp emission feature that is difficult to explain as either a consequence of metallic lines not included in the synthetic spectrum, or normalization errors; and 2) the emission EW curve (Fig.\ \ref{halpha_ew1}) peaks close to the negative extremum of \bz, and shows a secondary peak near the positive \bz~extremum.  

\noindent {\bf HD\,142184}: HR\,5907 was studied in detail by \cite{grun2012}, who found it to be the most rapidly rotating magnetic B-type star ($P_{\rm rot} \sim 0.5$~d), a title it still retains. Due to its small tilt angle ($\beta \sim 9^\circ$) the star shows emission at all rotation phases, with only minor variability (Figs.\ \ref{halpha_ind1} and \ref{halpha_ew1}). Its \bz~curve shows clear indications of a depature from a purely dipolar geometry \citep{grun2012}, which likely explains the mild asymmetry in H$\alpha$.

\noindent {\bf HD\,142990}: this is a very rapidly rotating star ($P_{\rm rot} \sim 1$~d), with a moderate magnetic field strength ($B_{\rm d} \sim 4.5$~kG; Paper III). Its extremely weak emission was first reported by \cite{2004A&A...421..203S}, and is comparable to that of HD\,64740 (Figs.\ \ref{halpha_ind1} and \ref{halpha_dyn}). There is evidence for eclipses in the dynamic spectrum (Fig.\ \ref{halpha_dyn}). H$\alpha$ displays a double-wave variation (Fig.\ \ref{halpha_ew2}), consistent with the large inclination and obliquity ($i_{\rm rot} \sim 55^\circ$, $\beta \sim 84^\circ$; Paper III). While HD\,142990 has close to the weakest H$\alpha$ emission in the present sample, it is the hottest star from which electron cyclotron maser emission has been detected \citep{2019ApJ...877..123D}. 

\noindent {\bf HD\,156324}: A detailed analysis of this SB3 star was performed by \cite{2018MNRAS.475..839S}. The system is a hierarchical triple, Aab+B, with the magnetic field belonging to the B2\,V Aa component. The rotational period of HD156324Aa, and the Aab orbital period, are identical ($\sim 1.58$~d), indicating that HD156324Aa and Ab are tidally locked. The most striking feature of HD\,156324Aa's H$\alpha$ profile is its single-lobed shape (Fig.\ \ref{halpha_ind1}), first noted by \cite{alecian2014}, which is almost certainly a consequence of distortion of the gravitocentrifugal potential by the orbit of the secondary. Due to the small inclination ($i_{\rm rot} \sim 21^\circ$), H$\alpha$ does not show eclipses. The emission EW is shown in Fig.\ \ref{halpha_ew2}; note the single-lobed variation, with almost all of the variability accounted for by the blue line wing. 

\noindent {\bf HD\,164492C}: the magnetic field of this SB3 system was detected by \cite{2014AA...564L..10H}. Follow-up observations published by \cite{2017MNRAS.465.2517W} and \cite{2017MNRAS.467..437G} revealed the orbital, fundamental, magnetic, and emission properties of the star. It is a hierarchical triple, A+Bab, with the Bab pair in an eccentric 12.5 d orbit. The magnetic field belongs to the A component, which is the hottest B-type star with CM-type emission (\teff~$=26$~kK). The system is located in the Trifid nebula and is therefore very young, about 0.5 Myr old. Since it is located in a crowded field, several of the available H$\alpha$ spectra, acquired on nights with high seeing, had to be discarded due to contamination from a nearby Herbig star \citep{2017MNRAS.465.2517W}. Several of the spectra also show nebular emission in the line core, although this is not present in the two observations shown in Fig.\ \ref{halpha_ind1}. It is possible that the ``clean'' spectra still contain some residual contamination from the Herbig star, and that this may explain why HD\,164492C's emission appears to be anomalously strong for its \ra~and $B_{\rm K}$ values. On the other hand, it is mostly an outlier with respect to \ra; if its mass-loss rate is over-estimated by 1 dex, its Alfv\'en radius would be more consistent with its emission strength with respect to the rest of the sample. The star shows a single-wave variation, which is consistent with its small obliquity ($\beta \sim 30^\circ$) and moderate inclination ($i_{\rm rot} \sim 60^\circ$; Paper III). The presence of nebular emission in the line core in several spectra makes it difficult to determine if the star shows eclipses, although its angular parameters are on the edge of the region in which eclipses might be expected. The emission EW variation is shown in Fig.\ \ref{halpha_ew2}. 

\noindent {\bf HD\,176582}: \cite{bohl2011} discovered the emission properties of this star, and were the first to determine its rotational, fundamental, and magnetic parameters. The star is notable for having relatively weak but clearly defined emission bumps (Fig.\ \ref{halpha_ind1}). Reasoning from basic geometrical arguments, \cite{bohl2011} showed that the high degree of symmetry in its emission variability and eclipse behaviour necessitate that both $i_{\rm rot}$ and $\beta$ are very close to $90^\circ$. Consistent with these large angular parameters, the EW curve is a double-wave variation (Fig.\ \ref{halpha_ew2}). 

\noindent {\bf HD\,182180}: the magnetic field, rapid rotation, and H$\alpha$ emission of HR\,7355 were discovered by \cite{2010MNRAS.405L..46R} and \cite{2010MNRAS.405L..51O}. A detailed analysis of a large spectroscopic dataset was performed by \cite{rivi2013}. HD\,182180 is almost as rapidly rotating as HD\,142184 ($P_{\rm rot} \sim 0.5$~d), but has a much larger obliquity ($\beta \sim 82^\circ$). As a consequence it has two well-defined emission bumps (Fig.\ \ref{halpha_ind2}), a clear double-wave variation (Fig.\ \ref{halpha_ew2}), and obvious eclipses \citep{rivi2013}. The high degree of symmetry in the H$\alpha$ profile suggests that the star's magnetic field is probably very close to perfectly dipolar, however the relatively small number of ESPaDOnS and FORS2 magnetic measurements makes this difficult to verify. 

\noindent {\bf HD\,189775}: the \bz~curve of this star was first reported in Paper I and displays clear indications of departures from a purely dipolar surface magnetic field. Emission has not previously been reported. We consider that the star does in fact show extremely weak H$\alpha$ emission based on two factors. First, the residual flux (see Fig.\ \ref{halpha_ind2}) of some, but not all spectra show enhancements peaking near $R_{\rm K}$, which do not seem to be consistent with the effects of e.g.\ metallic lines not included in the line mask. Second, the EW curve (see Fig.\ \ref{halpha_ew2}) peaks at a phase corresponding to the negative magnetic extremum, which is also the strongest of the two extrema (Paper I), and also shows evidence for a local EW extremum close to the positive extremum. The presence of emission in this star should however be treated as tentative.

\noindent {\bf HD\,345439}: the discovery of CM-type infrared emission lines in this star by \cite{2014ApJ...784L..30E} motivated spectroscopic and FORS2 magnetic follow-up observations by \cite{2015ApJ...811L..26W} and \cite{2015A&A...578L...3H,2017MNRAS.467L..81H}. This is the only star in the sample for which high-resolution spectropolarimetry is not available, and the only high-resolution visual spectra are the ARCES spectra published by \cite{2015ApJ...811L..26W}. The star has an extremely short rotational period ($P_{\rm rot} \sim 0.77$~d) and a strong magnetic field ($B_{\rm d} \sim 9$~kG; Paper III), making it comparable to HD\,142184 and HD\,182180. Its emission strength is apparently even stronger than that of $\sigma$ Ori E, the previous record-holder. We note this status is not certain, as the ARCES data are very noisy compared to the data available for other stars in the sample (Fig.\ \ref{halpha_ind2}), and it is furthermore possible that there may be systematic differences between ESPaDOnS, Narval, and HARPSpol on the one hand (all of which achieve comparable results), and ARCES, on the other. Given its rapid rotation and strong magnetic field, however, the star's strong emission is not anomalous. The EW variation in Fig.\ \ref{halpha_ew2} is consistent with a single-wave variation, however the small number of measurements make this uncertain.

\noindent {\bf ALS\,3694}: the magnetic field of this star was first discoverd by \cite{bagn2006} with FORS2. Follow-up ESPaDOnS observations revealed the star's H$\alpha$ emission \citep{2016ASPC..506..305S}. This star is located in the open cluster NGC 6193, and the H$\alpha$ line core is affected by nebular emission (Fig.\ \ref{halpha_ind2}). Both $i_{\rm rot}$ and $\beta$ are small (about $20^\circ$), and emission is present at all phases with no indication of eclipsing (Fig.\ \ref{halpha_dyn}). Consistent with this, the EW shows a single-wave variation (Fig.\ \ref{halpha_ew2}). 

\noindent {\bf CPD${\mathbf -62^\circ 2124}$}: the extremely strong ($B_{\rm d} \sim 25$~kG) magnetic field and H$\alpha$ emission of this star were first reported by \cite{2017A&A...597L...6C}. \cite{2017MNRAS.472..400H} published follow-up observations with FORS2 and determined the rotational period. The star is a member of the IC 2944 open cluster, and the line core is contaminated with nebular emission. There is only one H$\alpha$ observation (Fig.\ \ref{halpha_ind2}), meaning that the empirical quantities derived from it are technically lower limits. However, the rotational phase of the HARPSpol observation is about 0.6; comparing to the H$\beta$ EW curve in Fig.\ 6 of \cite{2017A&A...597L...6C}, this is very close to magnetic maximum. CPD$-62^\circ 2124$'s H$\alpha$ profile is notable for extending to the largest distance of any star in the sample, out to about 16 $R_*$, making it the largest CM in the sample. Its emission strength is exceeded only by $\sigma$ Ori E and HD\,345439. 

\section{Notes on individual stars without emission}\label{appendix:noem}

   \begin{figure*}
   \centering
\begin{tabular}{cccc}

   \includegraphics[trim=50 0 0 0, width=0.225\textwidth]{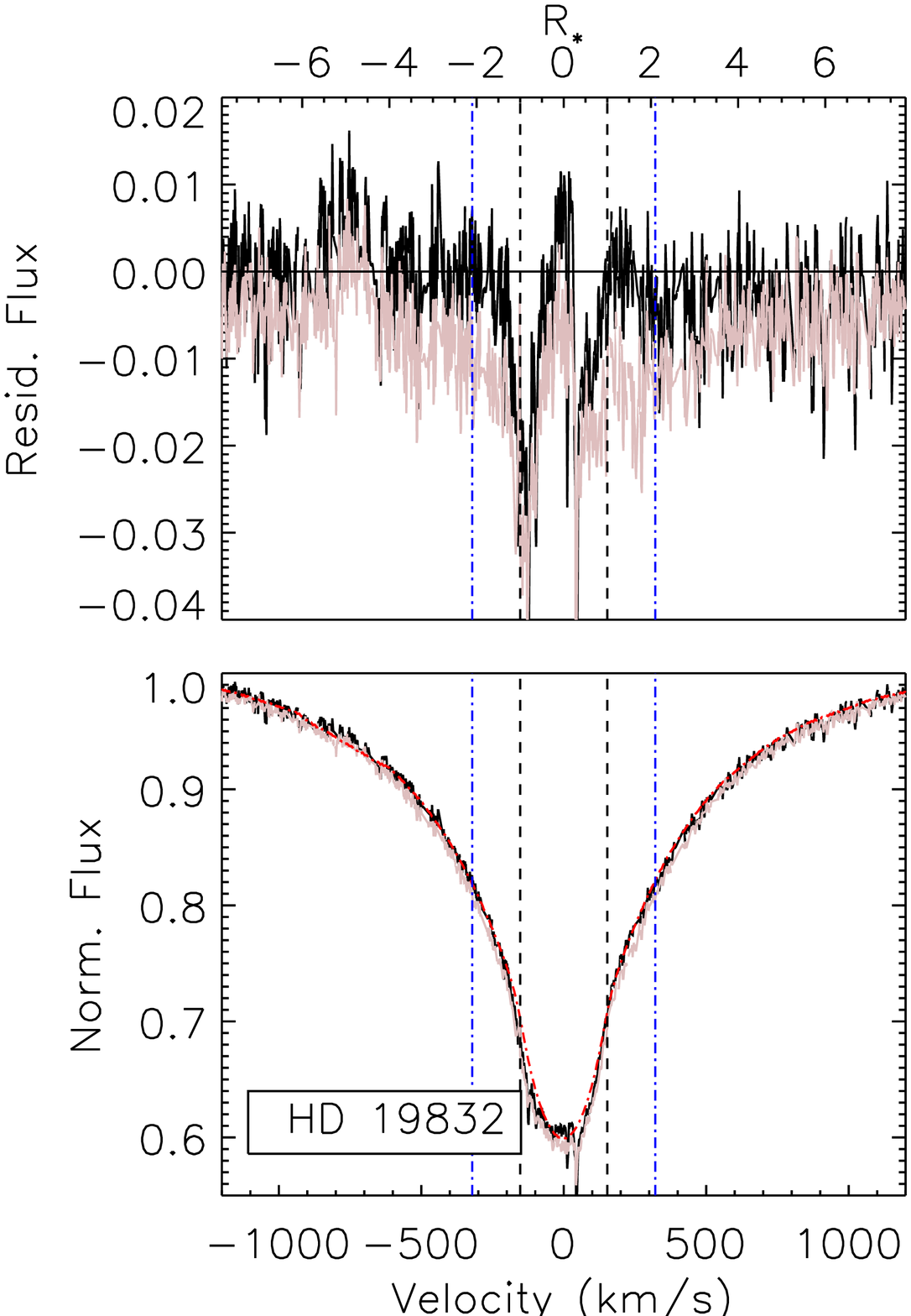} &
   \includegraphics[trim=50 0 0 0, width=0.225\textwidth]{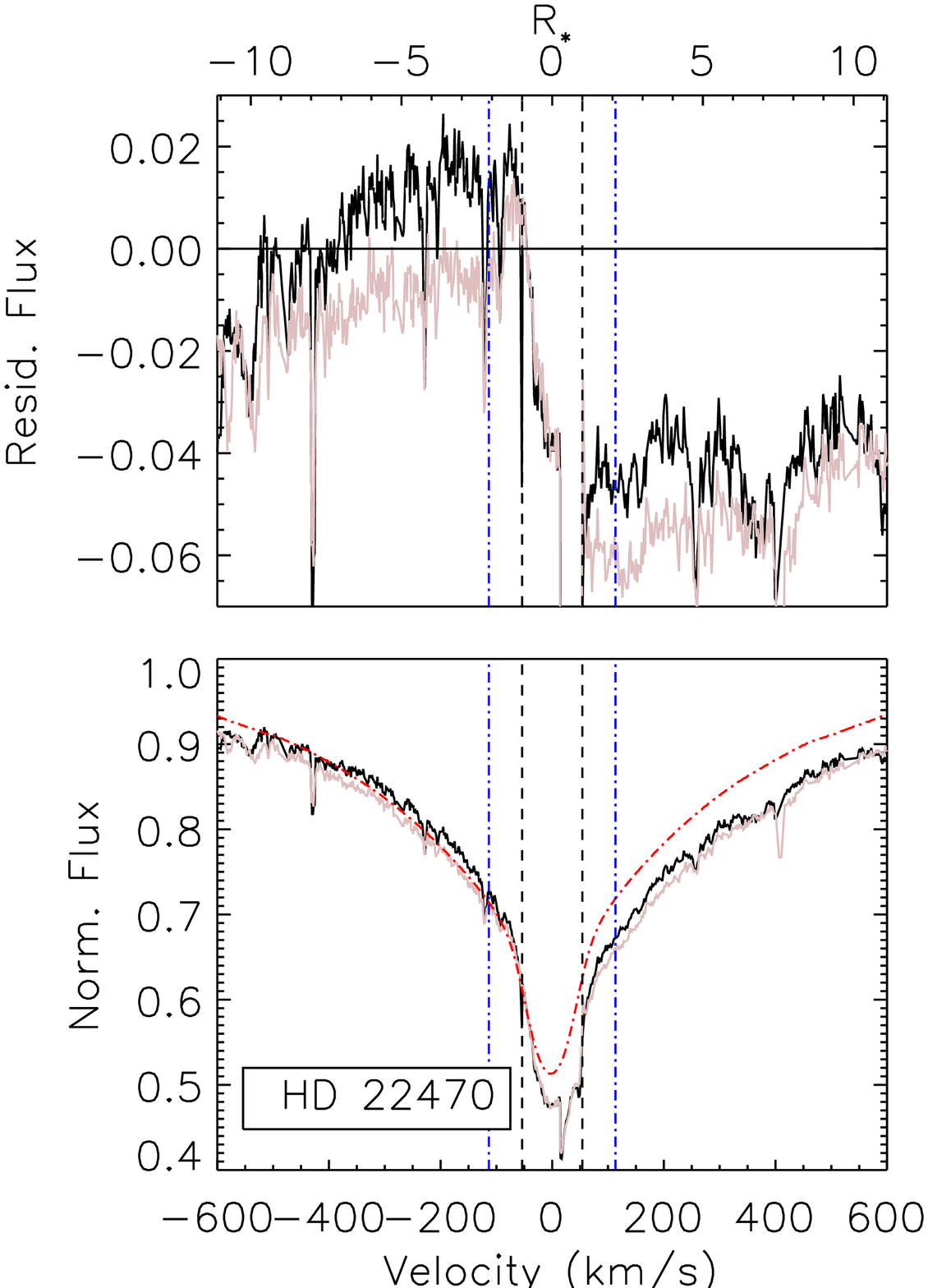} &
   \includegraphics[trim=50 0 0 0, width=0.225\textwidth]{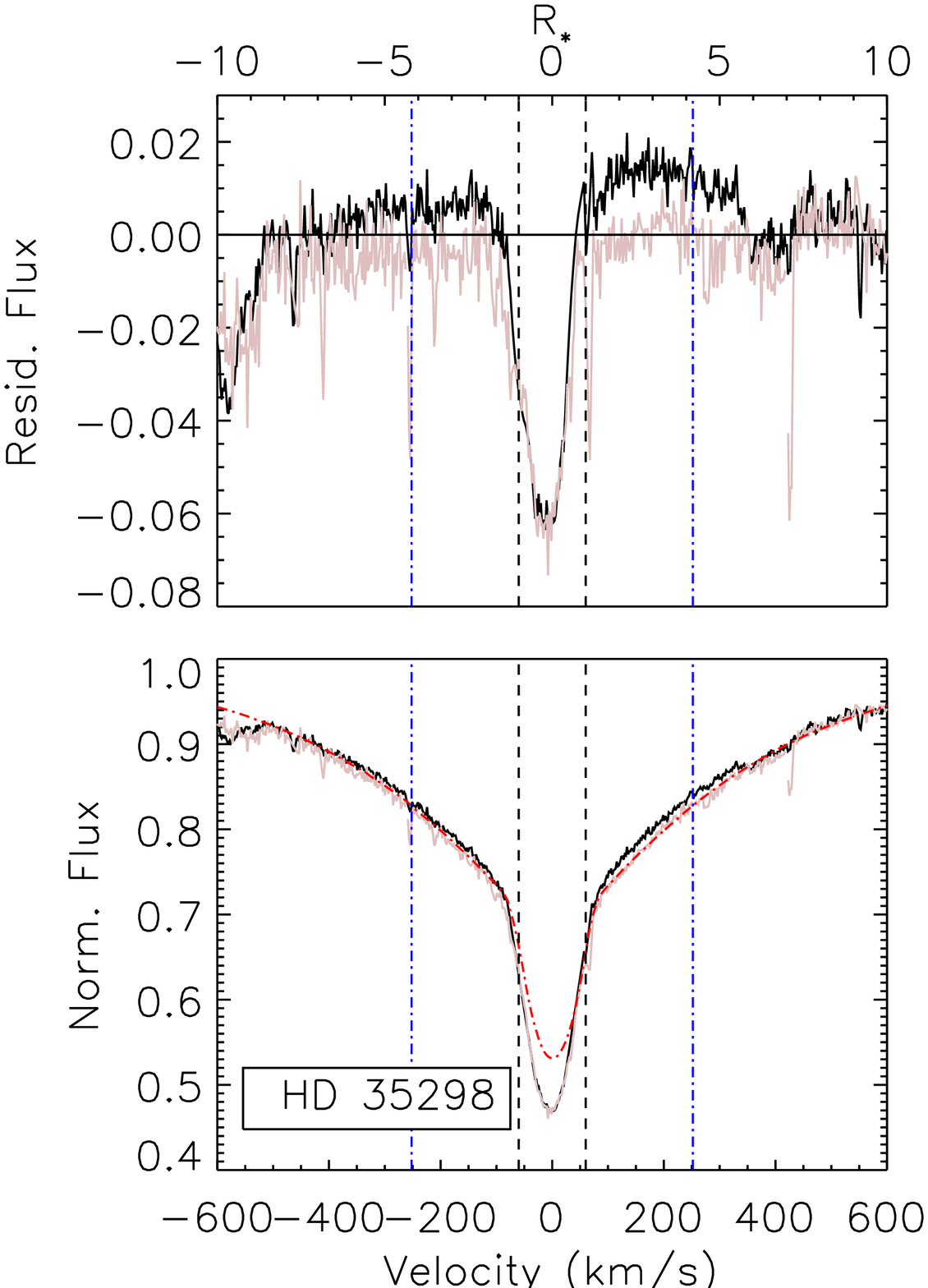} &
   \includegraphics[trim=50 0 0 0, width=0.225\textwidth]{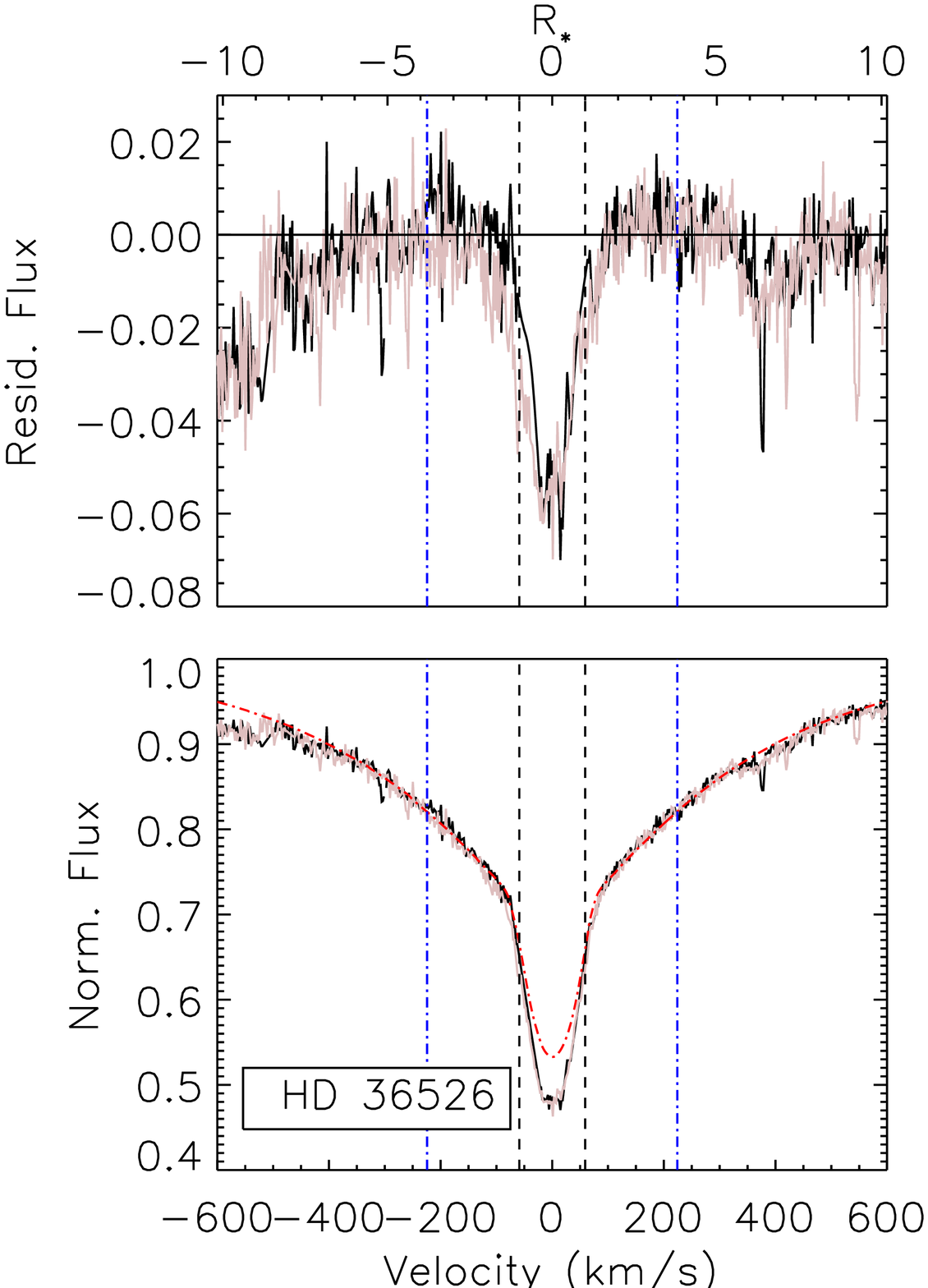} \\

   \includegraphics[trim=50 0 0 0, width=0.225\textwidth]{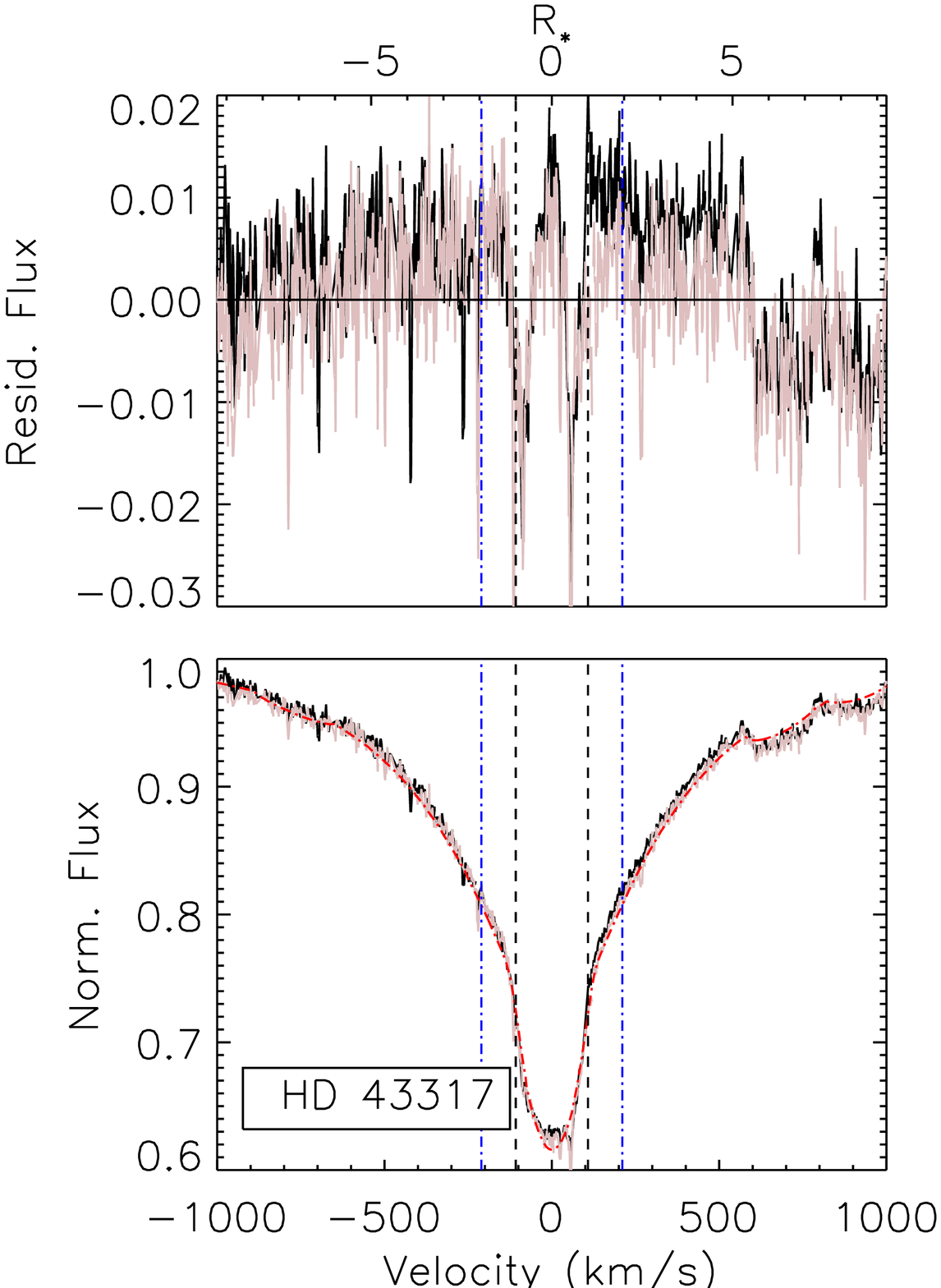} &
   \includegraphics[trim=50 0 0 0, width=0.225\textwidth]{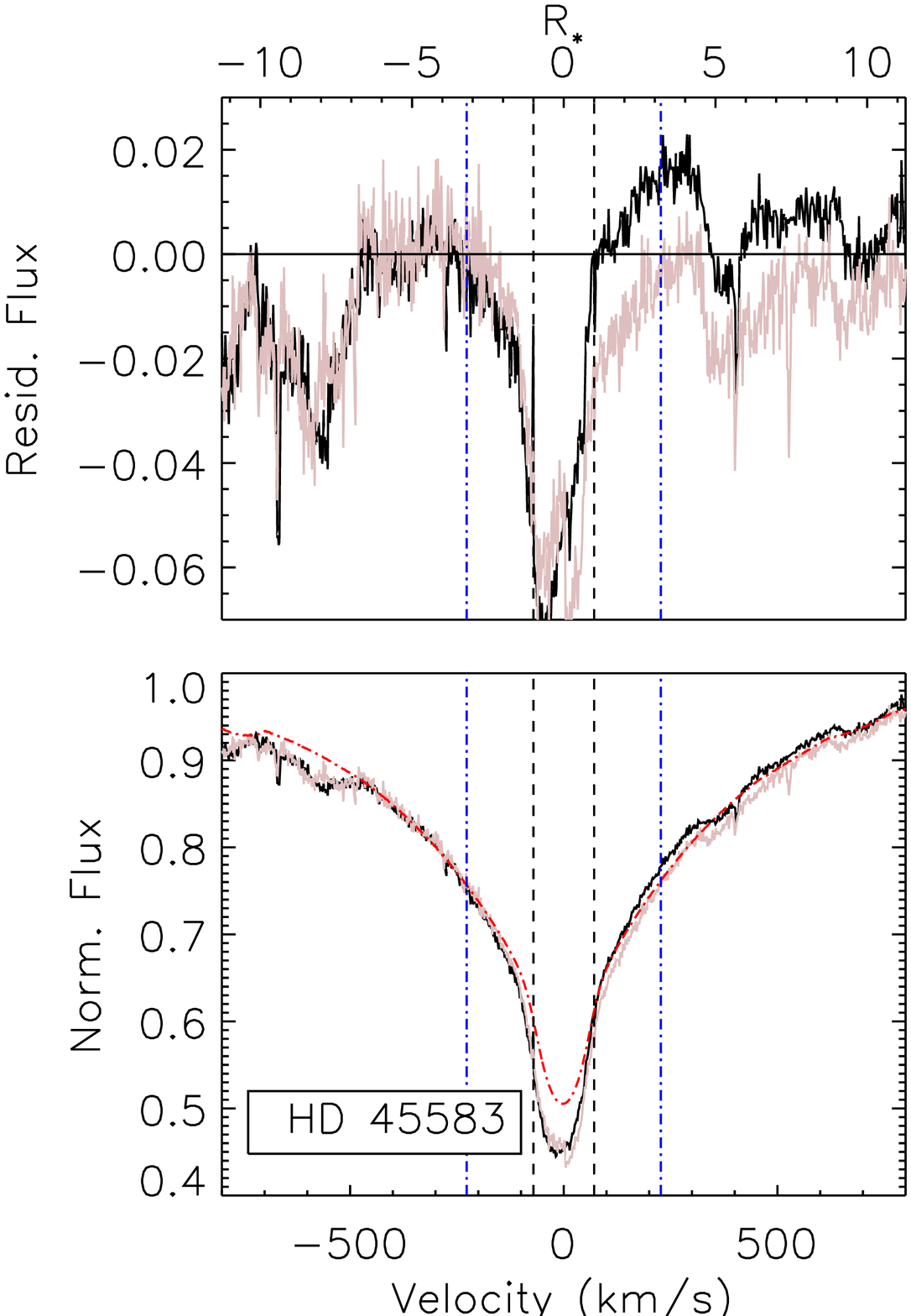} &
   \includegraphics[trim=50 0 0 0, width=0.225\textwidth]{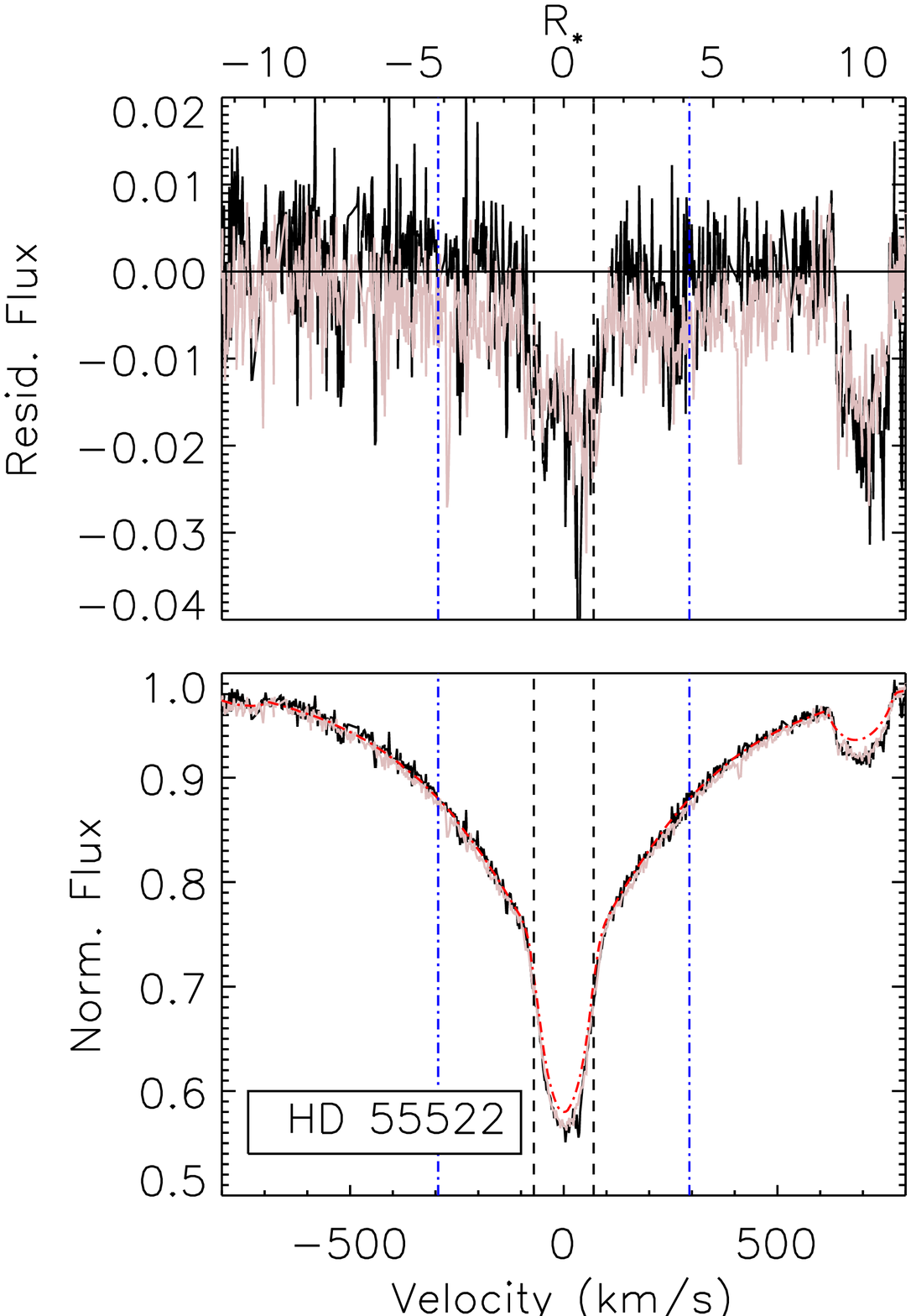} &
   \includegraphics[trim=50 0 0 0, width=0.225\textwidth]{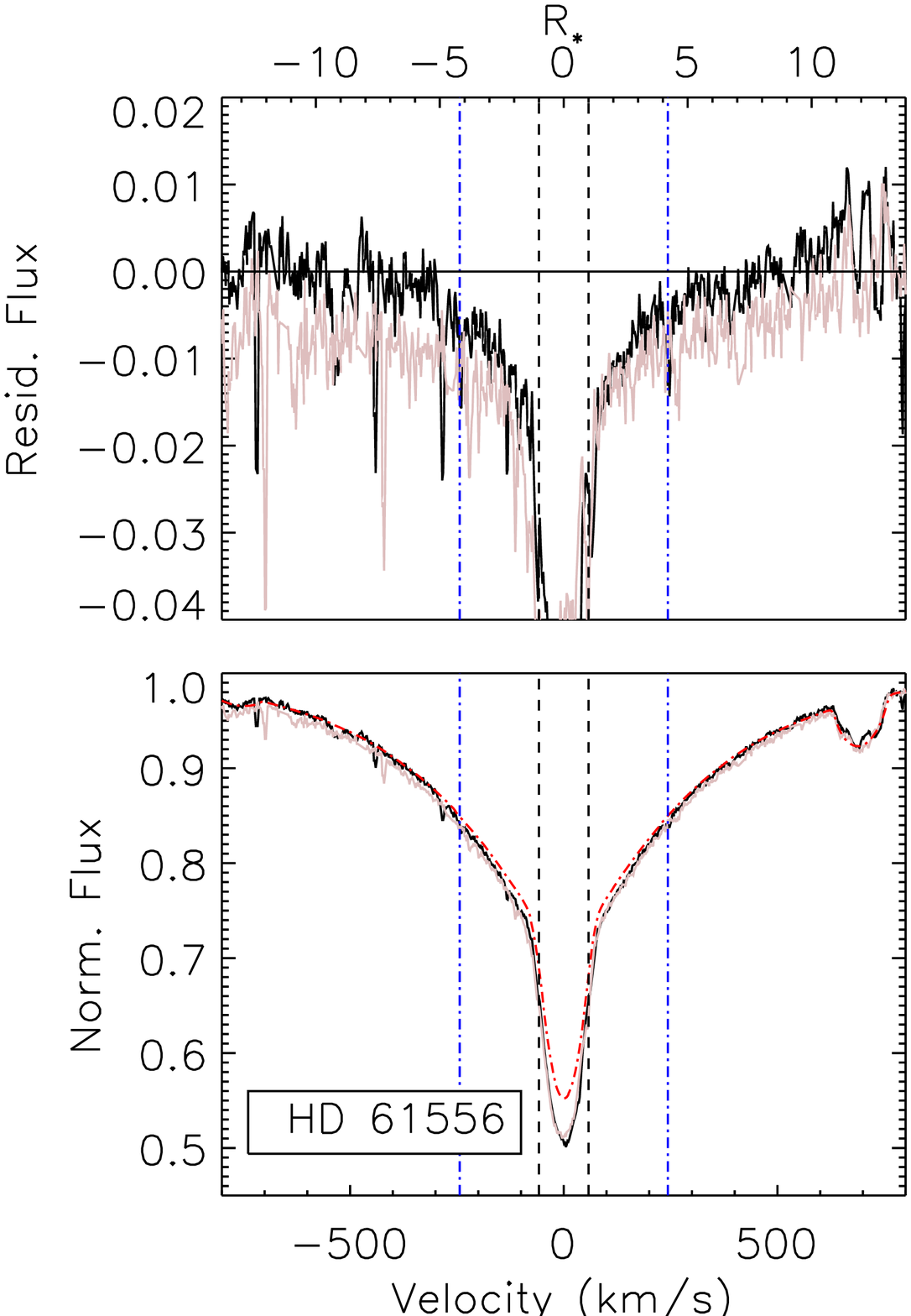} \\
\end{tabular}
      \caption[]{As Fig.\ \protect\ref{sigOriE_halpha_minmax} for stars without detected emission.}
         \label{halpha_ind3}
   \end{figure*}

   \begin{figure*}
   \centering
\begin{tabular}{cccc}

   \includegraphics[trim=50 0 0 0, width=0.225\textwidth]{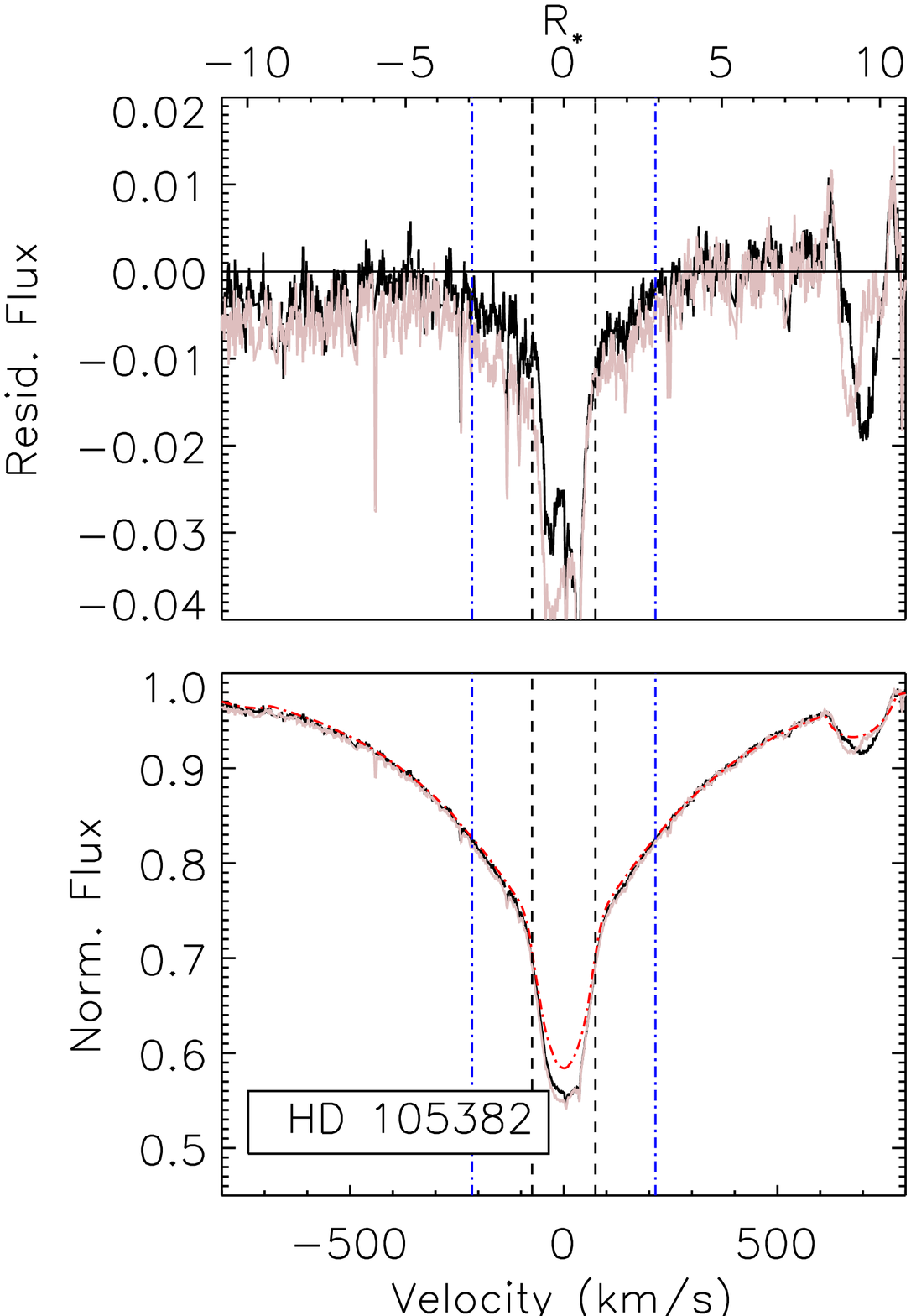} &
   \includegraphics[trim=50 0 0 0, width=0.225\textwidth]{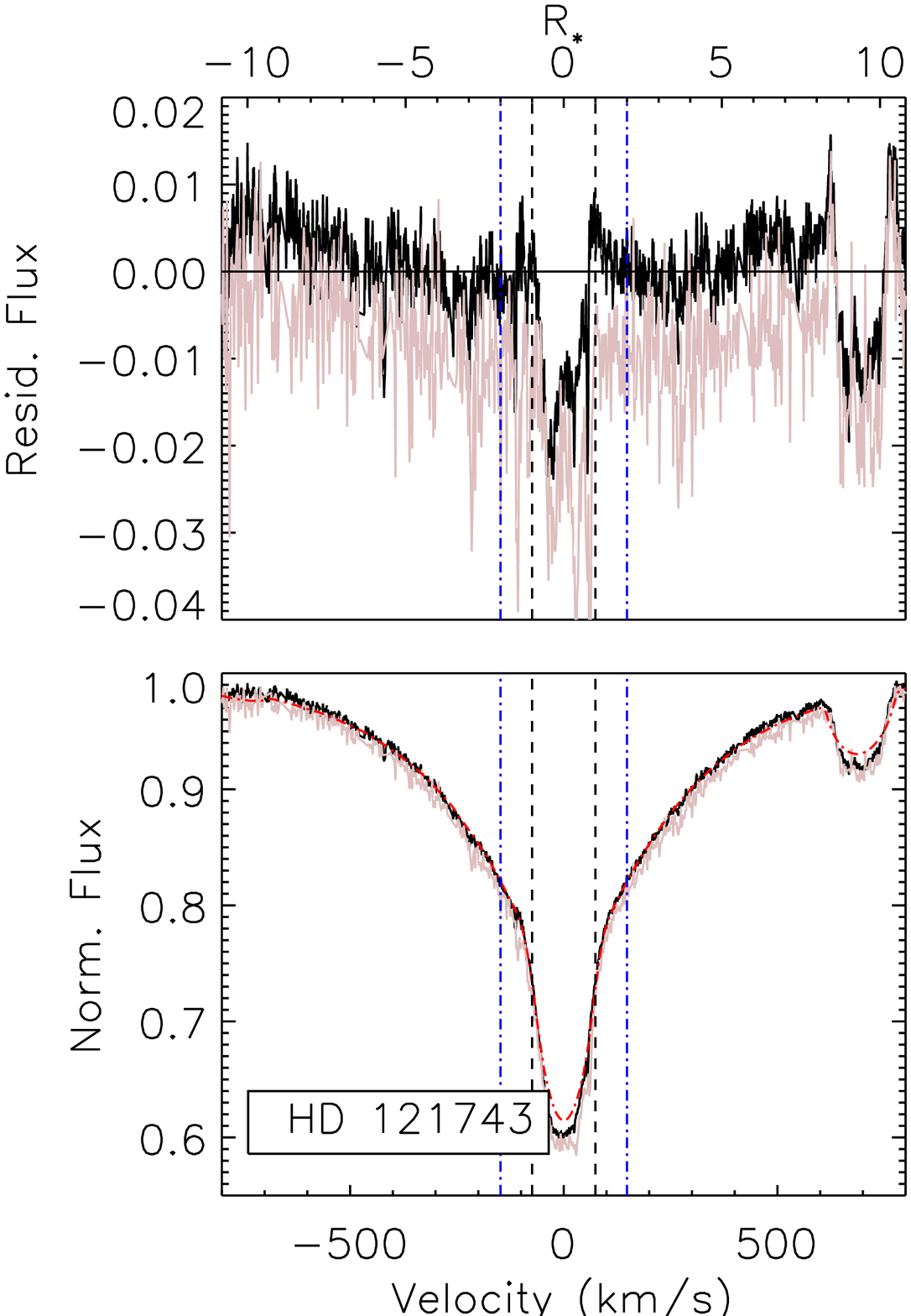} &
   \includegraphics[trim=50 0 0 0, width=0.225\textwidth]{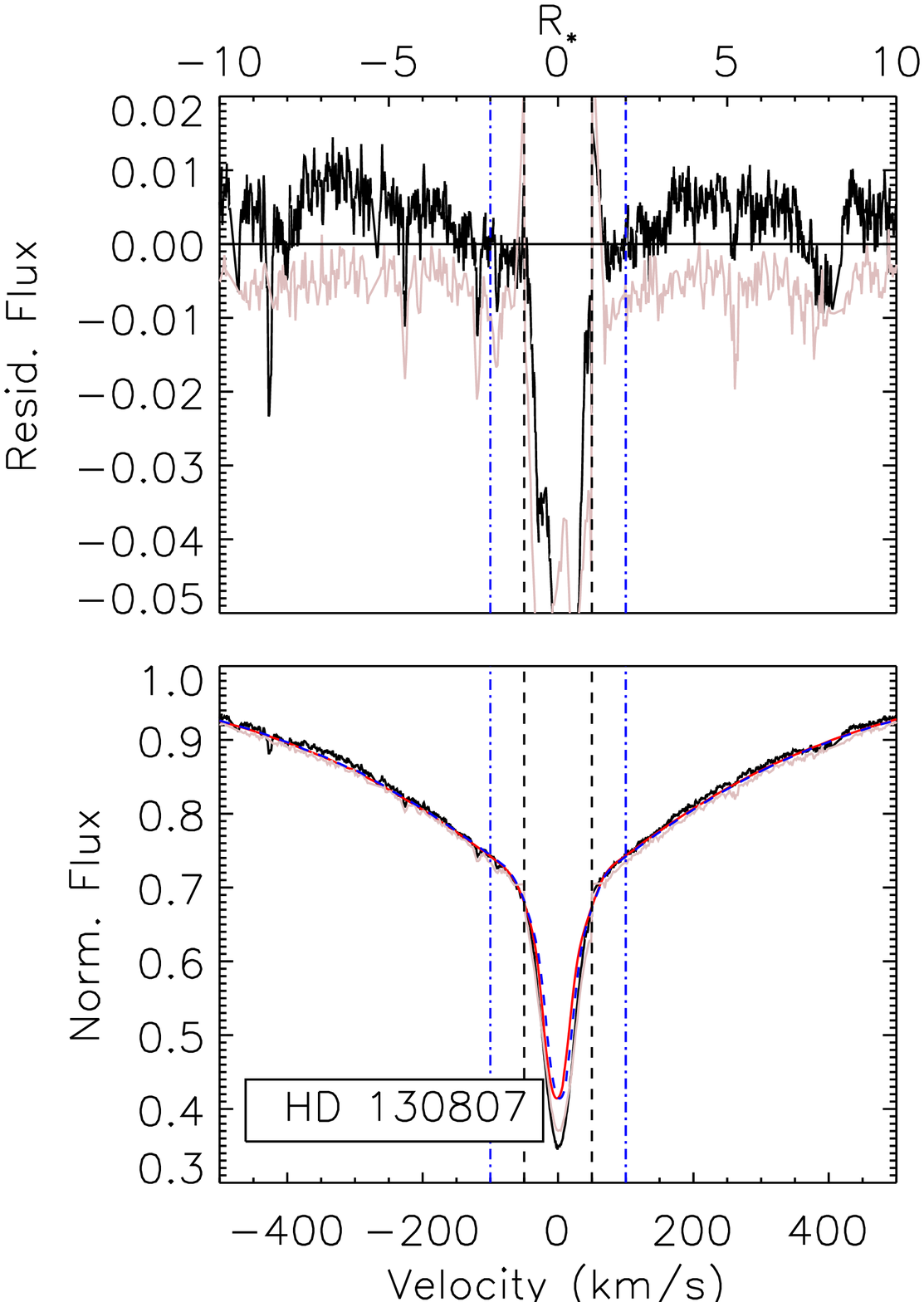} &
   \includegraphics[trim=50 0 0 0, width=0.225\textwidth]{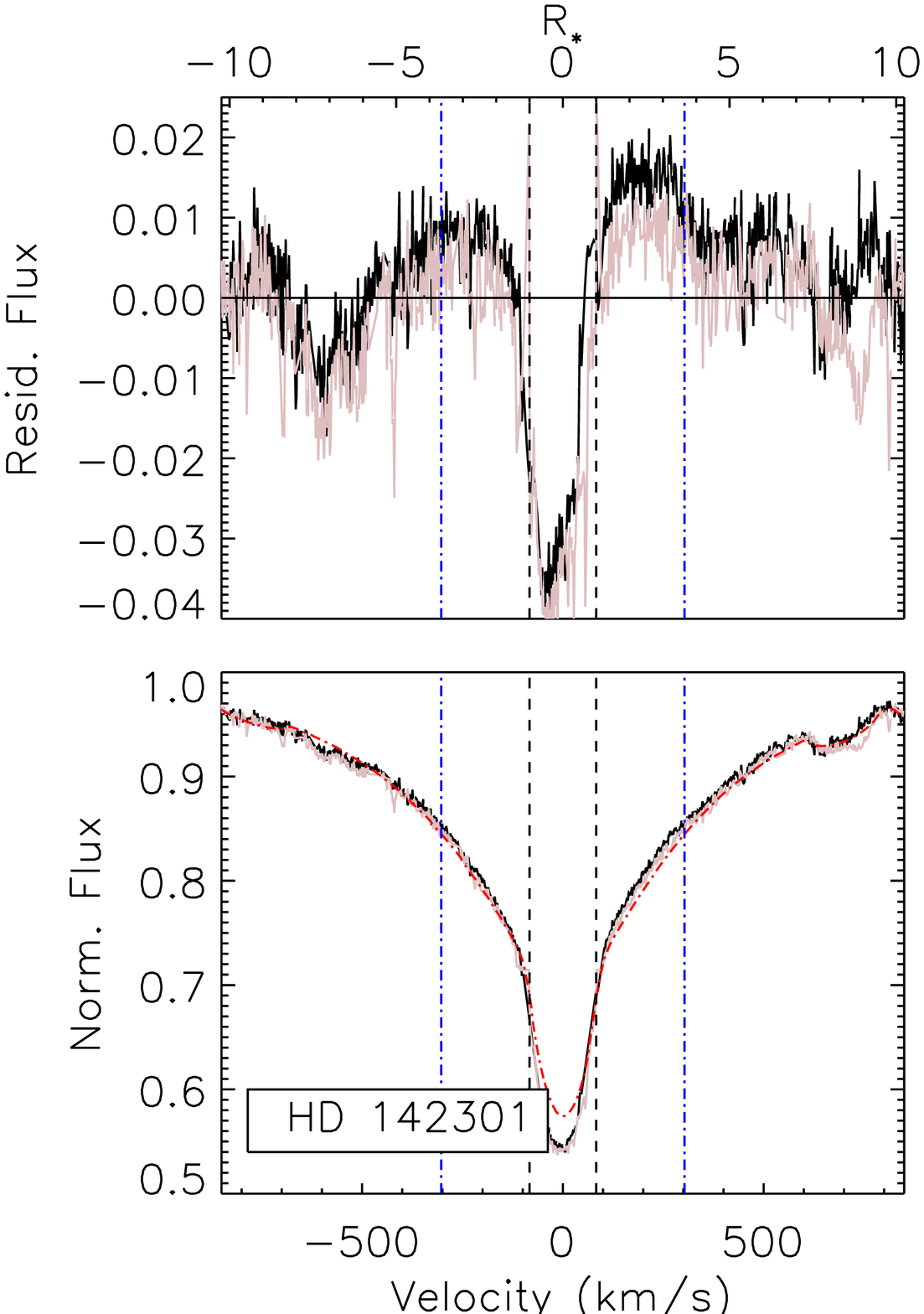} \\

   \includegraphics[trim=50 0 0 0, width=0.225\textwidth]{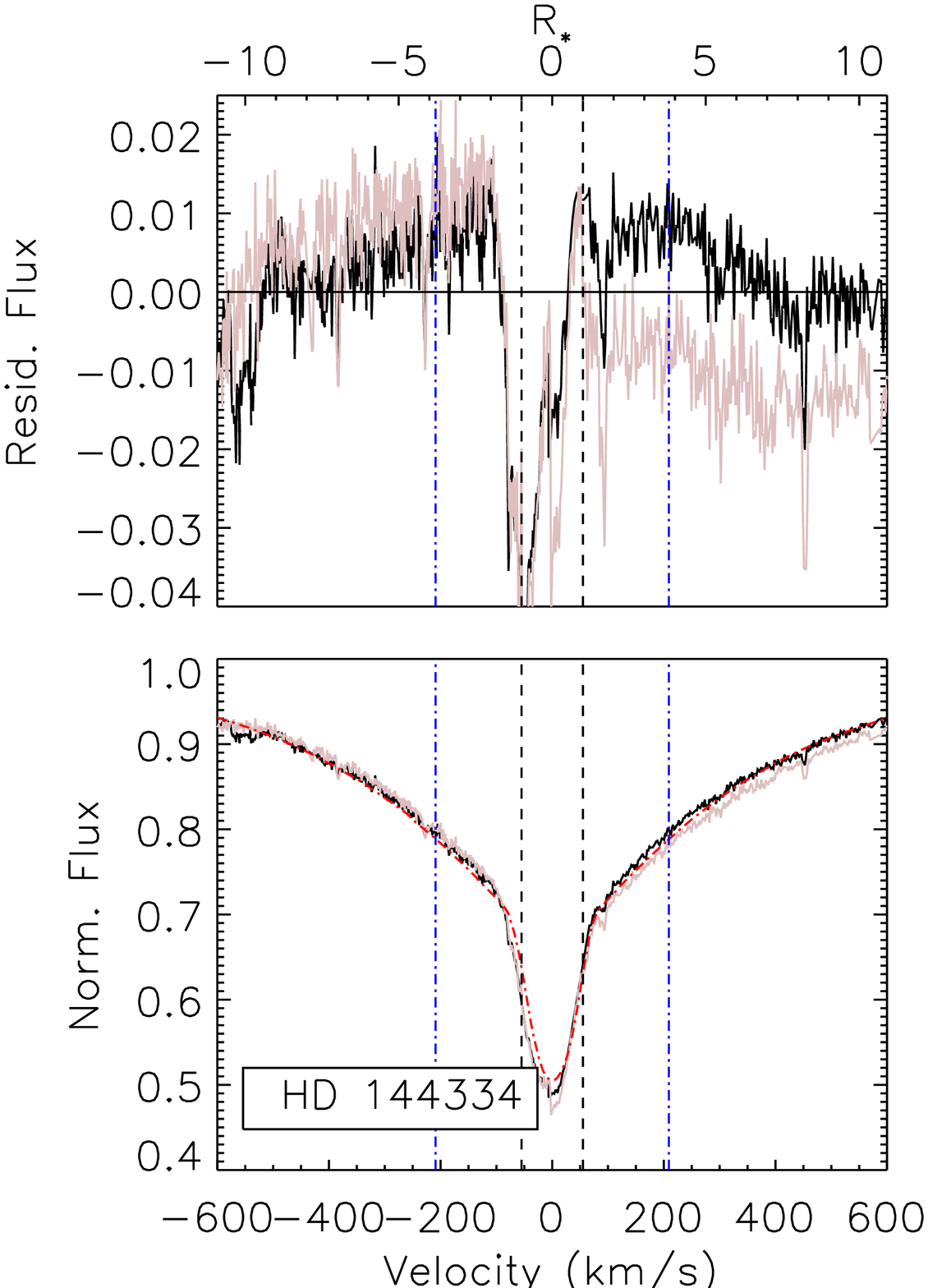} &
   \includegraphics[trim=50 0 0 0, width=0.225\textwidth]{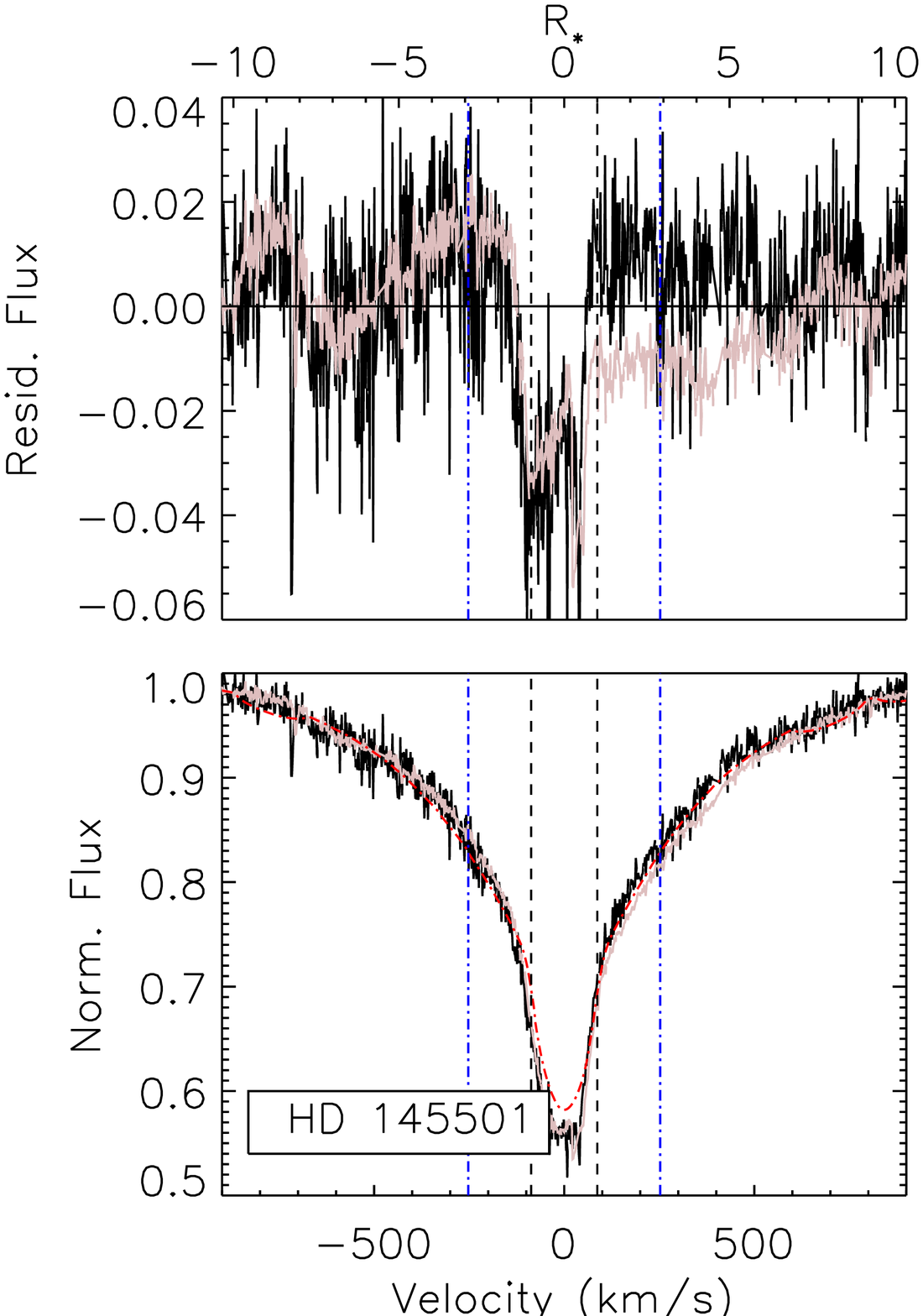} &
   \includegraphics[trim=50 0 0 0, width=0.225\textwidth]{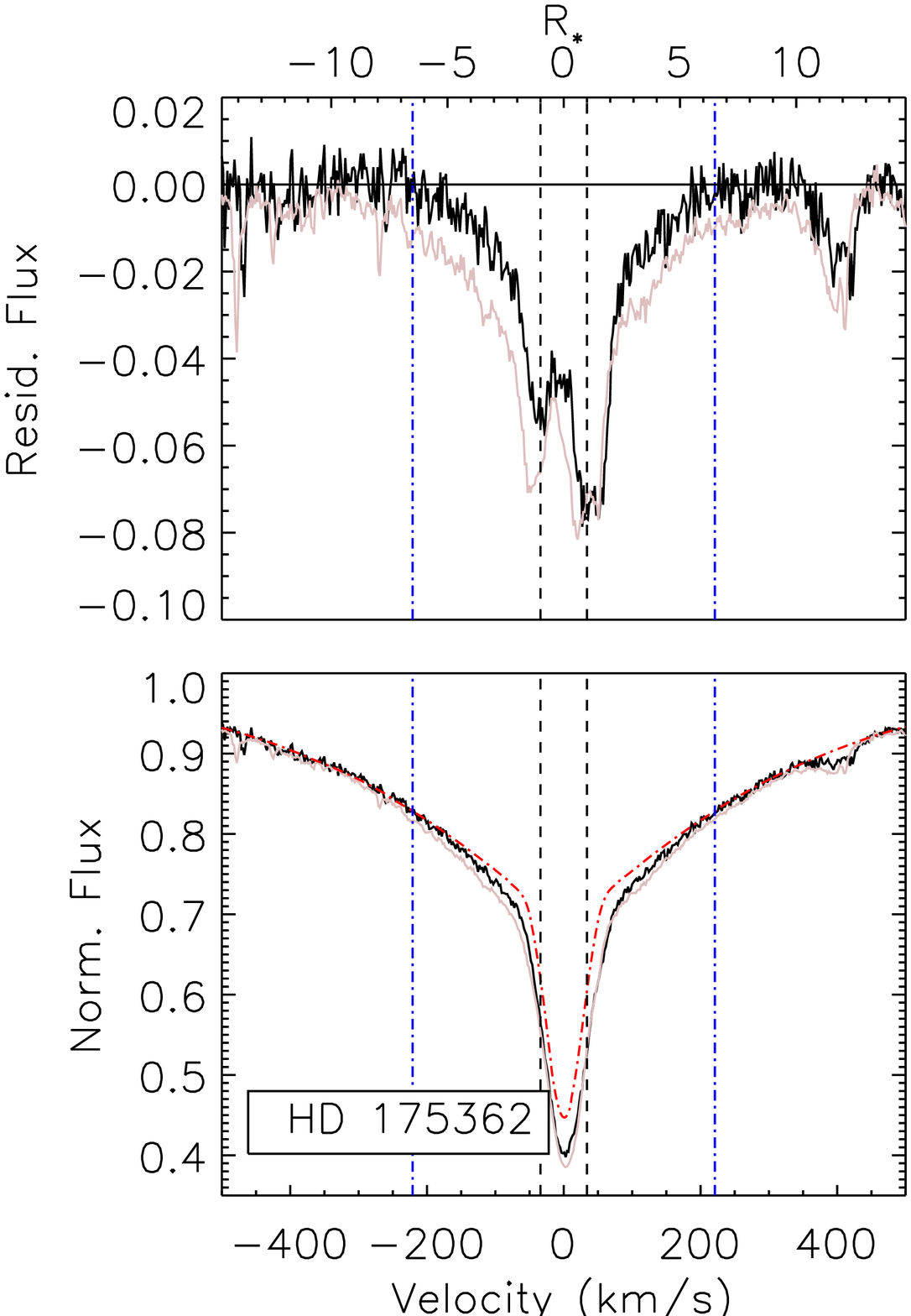} \\

\end{tabular}
      \caption[]{As Fig.\ \protect\ref{sigOriE_halpha_minmax} for stars without detected emission.}
         \label{halpha_ind4}
   \end{figure*}

   \begin{figure*}
   \centering
\begin{tabular}{cccc}

   \includegraphics[trim=50 0 25 0, width=0.225\textwidth]{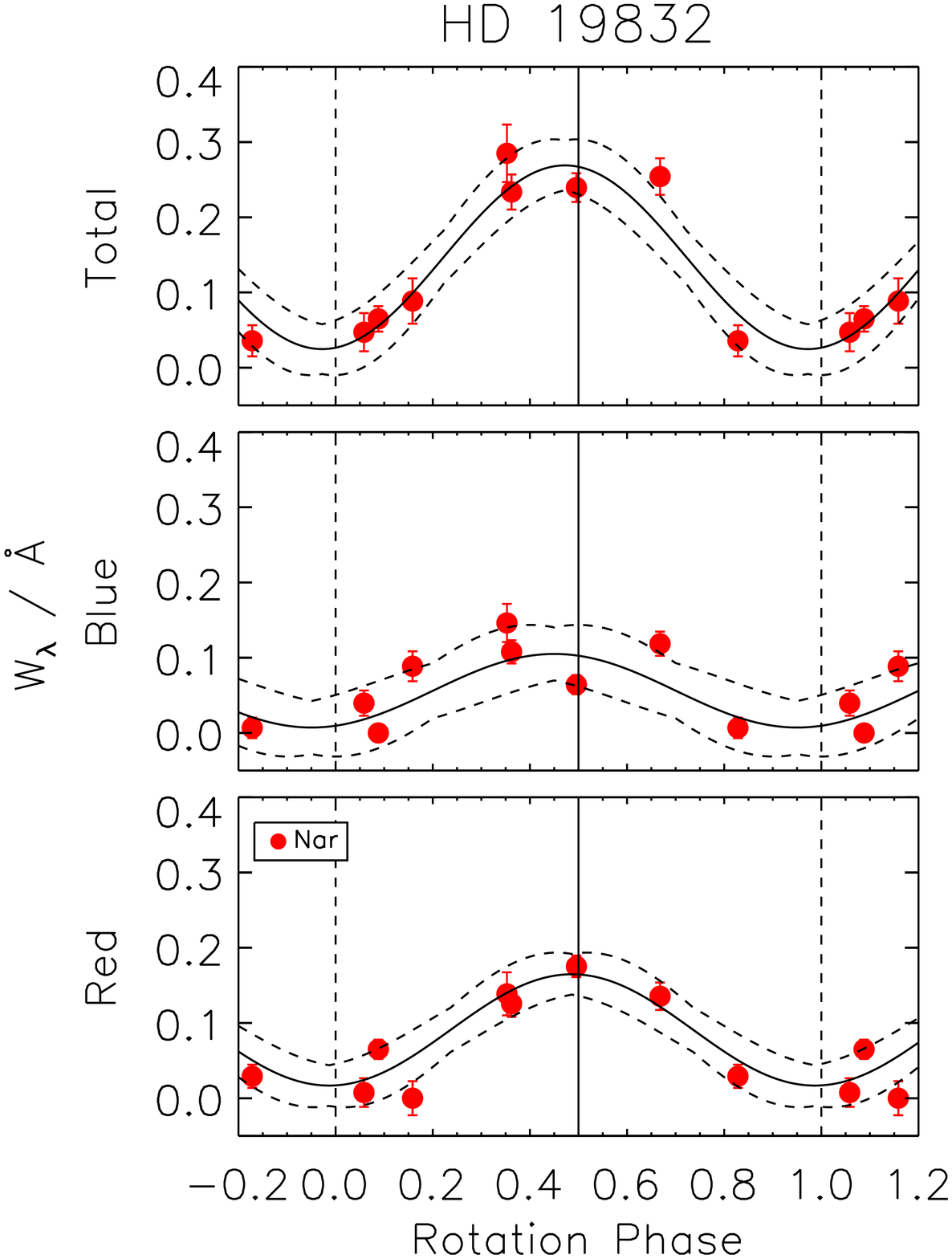} &
   \includegraphics[trim=50 0 25 0, width=0.225\textwidth]{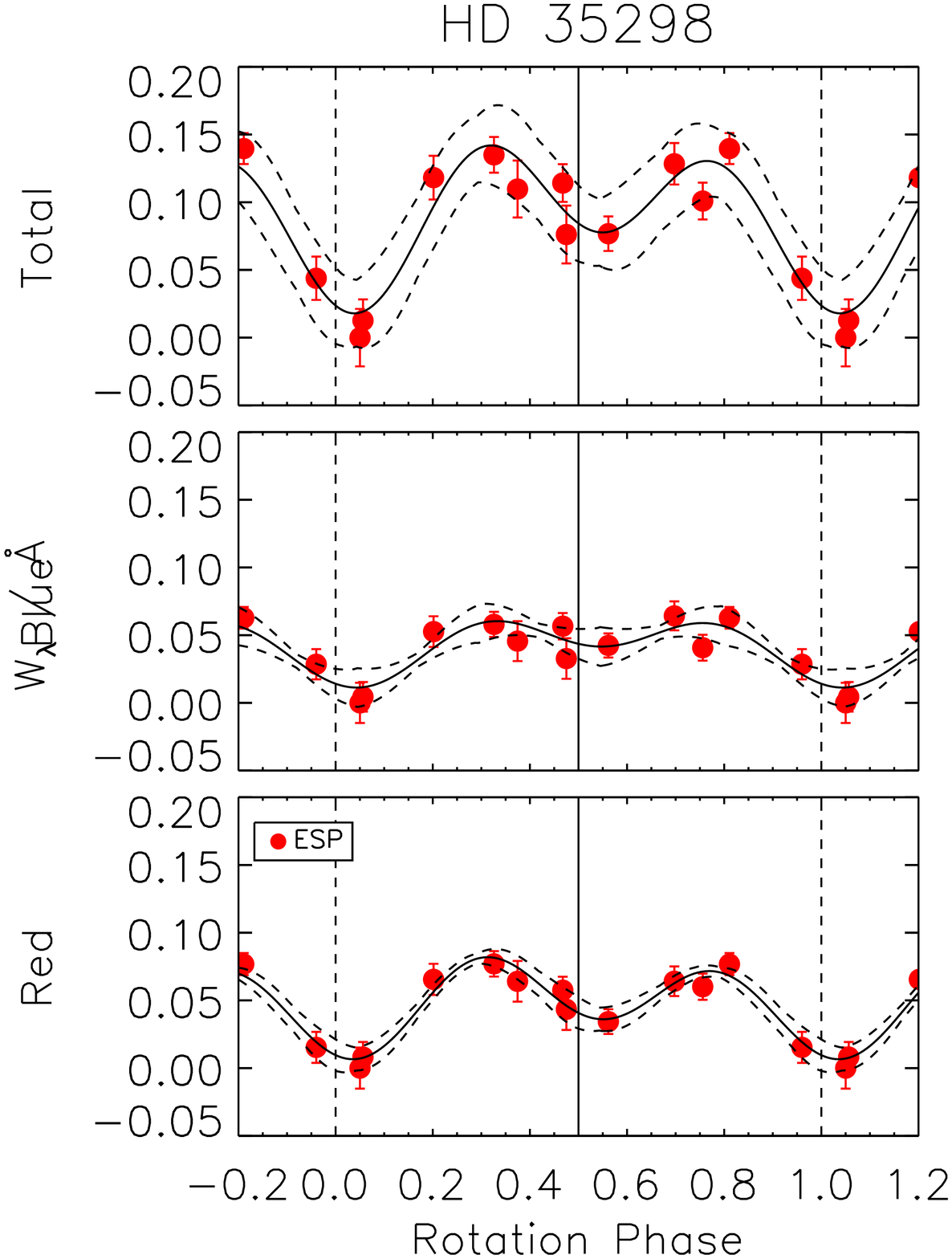} &
   \includegraphics[trim=50 0 25 0, width=0.225\textwidth]{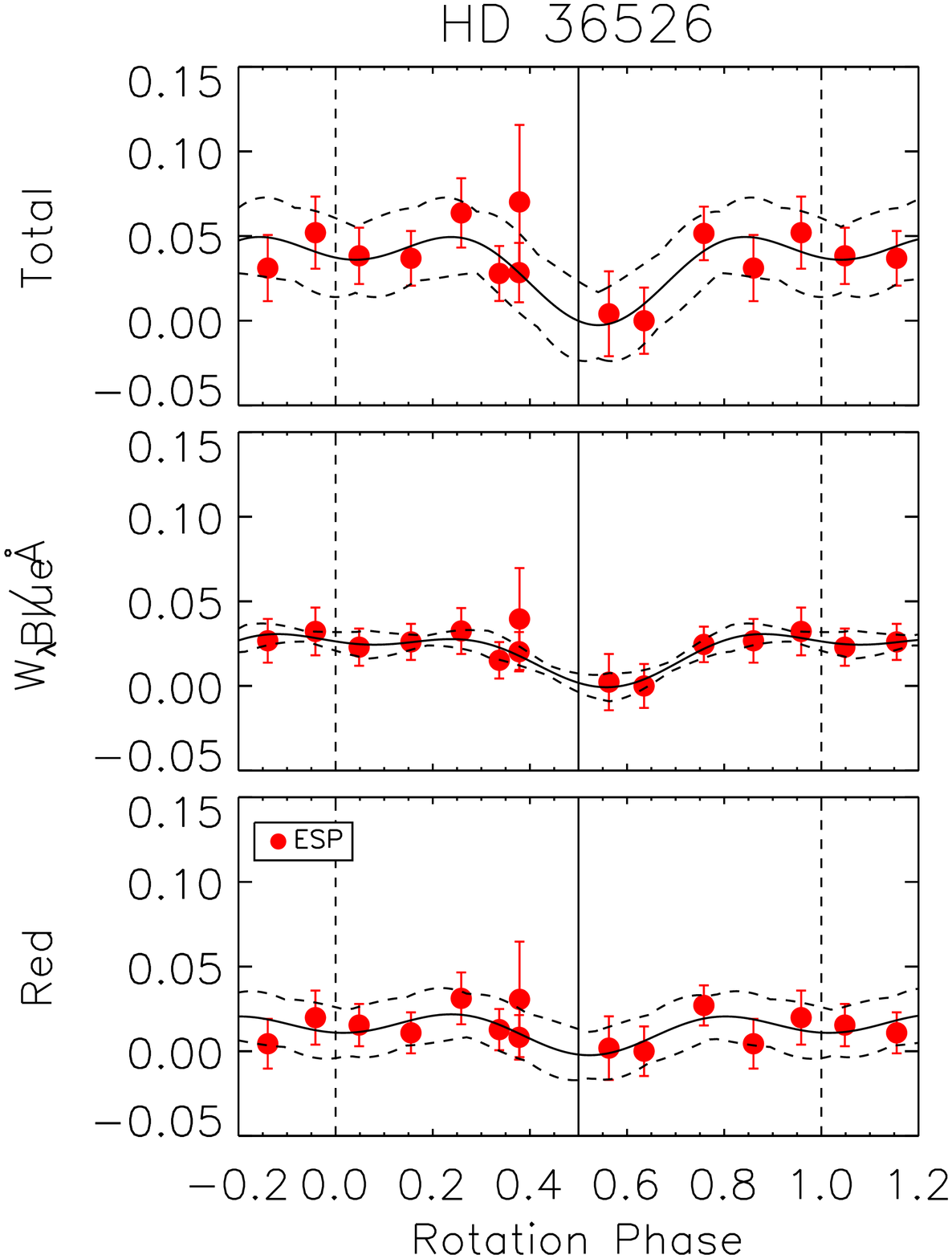} &
   \includegraphics[trim=50 0 25 0, width=0.225\textwidth]{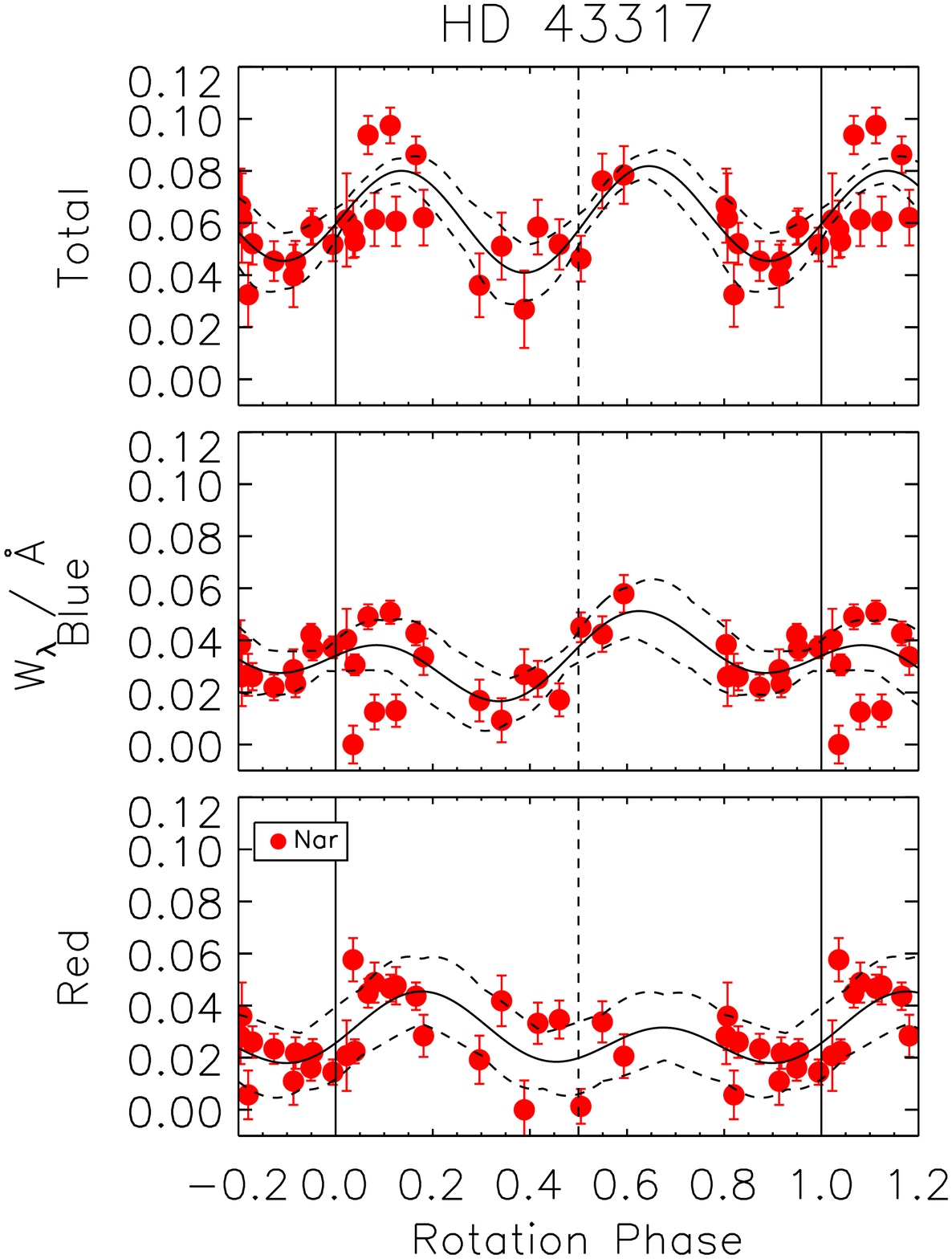} \\

   \includegraphics[trim=50 0 25 0, width=0.225\textwidth]{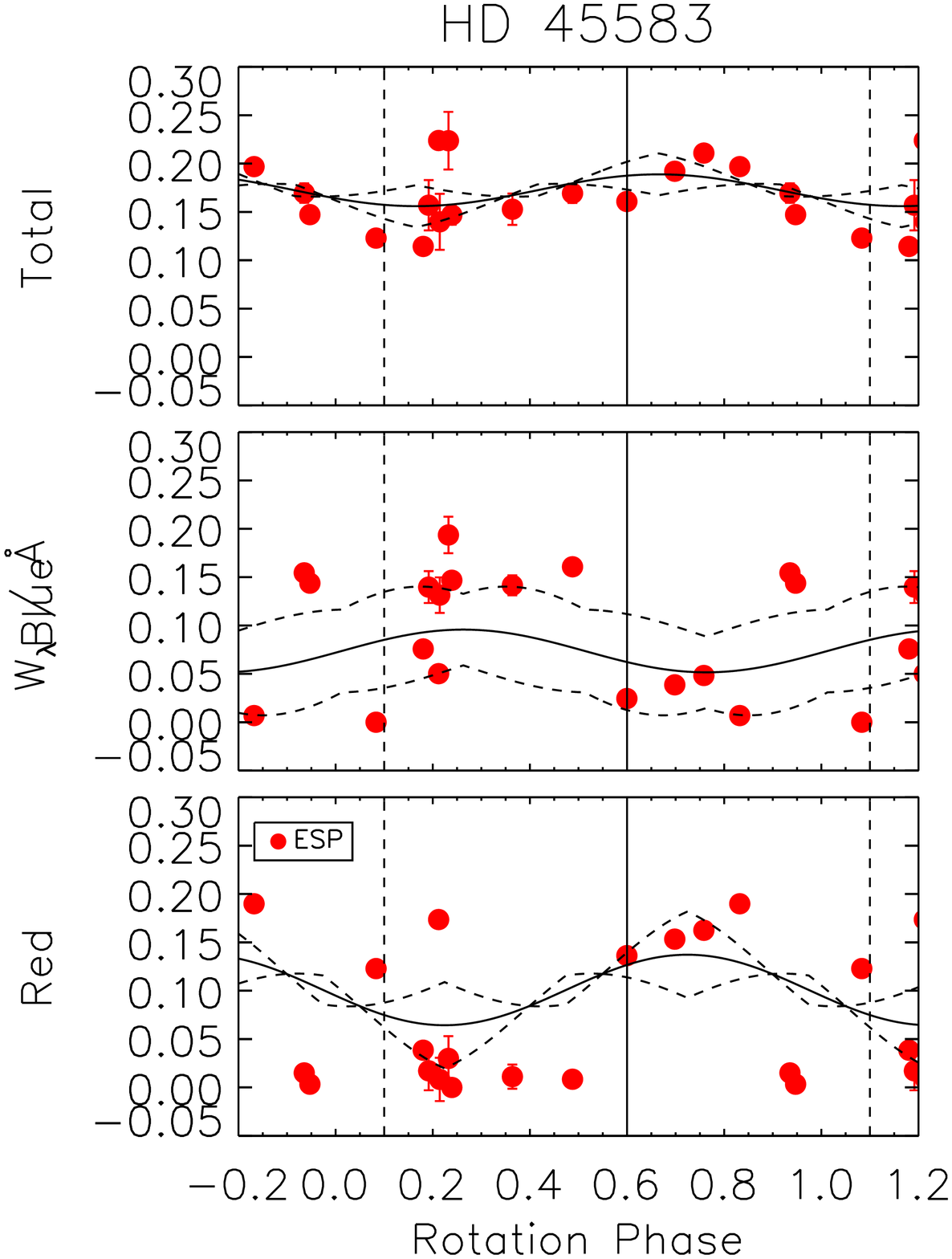} &
   \includegraphics[trim=50 0 25 0, width=0.225\textwidth]{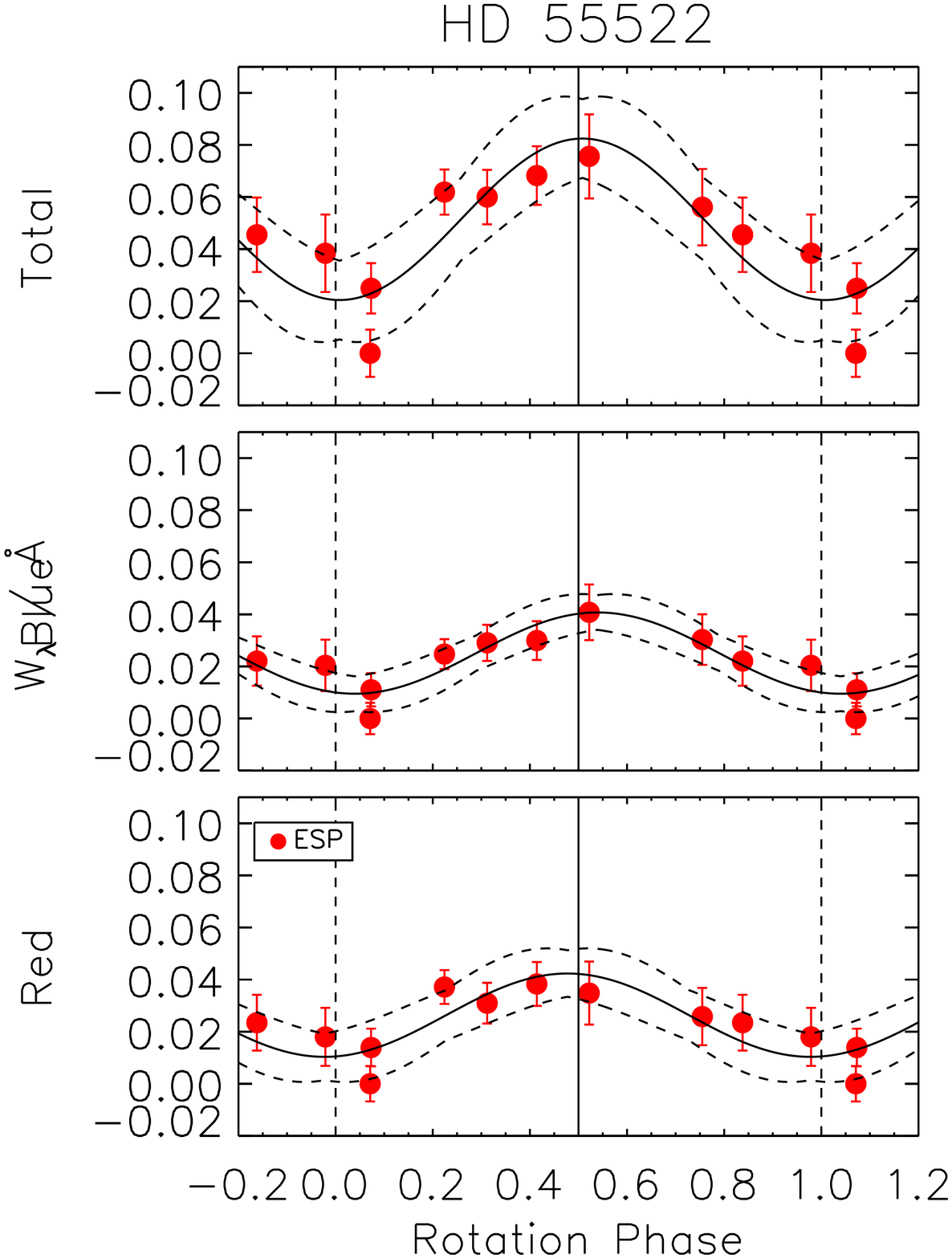} &
   \includegraphics[trim=50 0 25 0, width=0.225\textwidth]{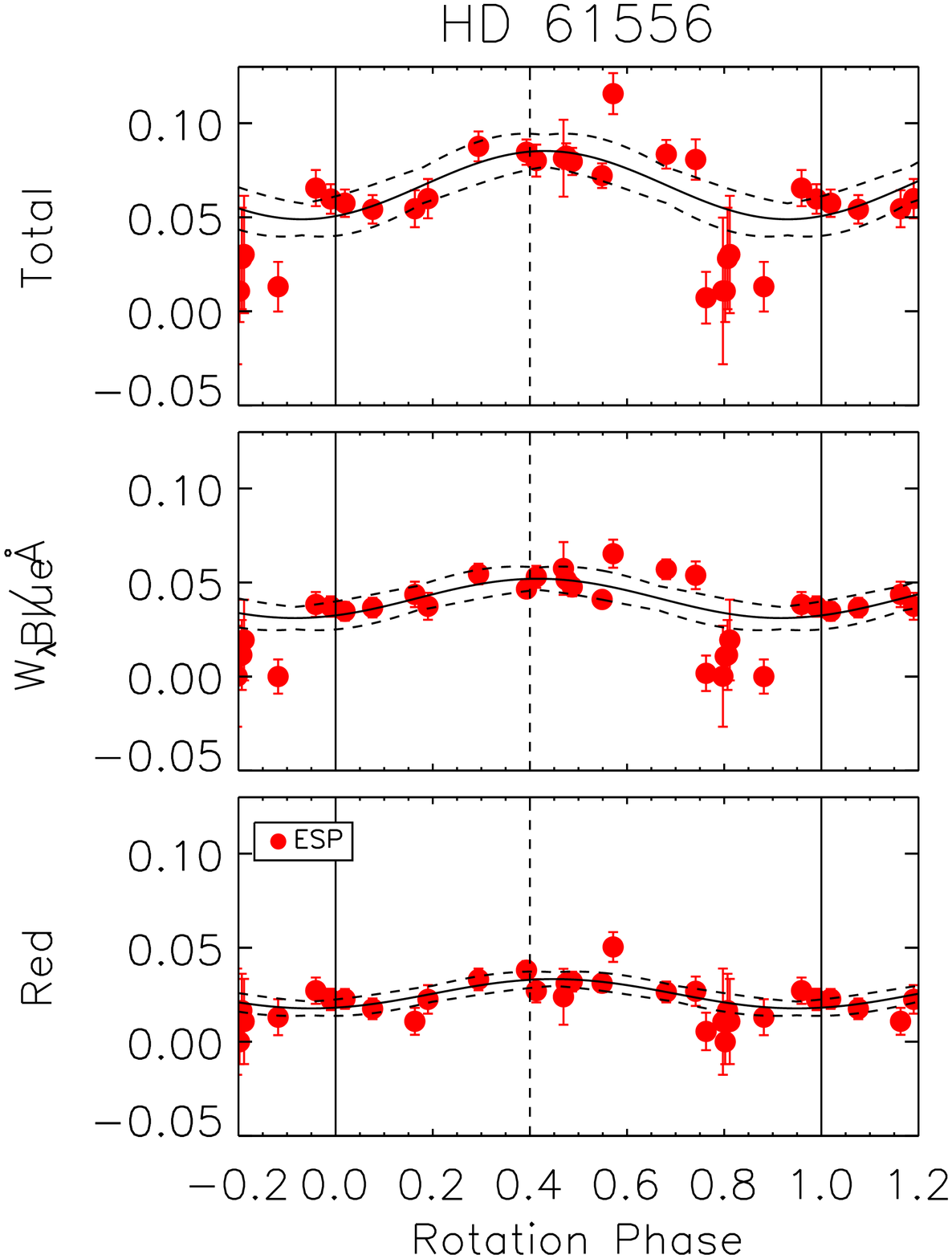} &
   \includegraphics[trim=50 0 25 0, width=0.225\textwidth]{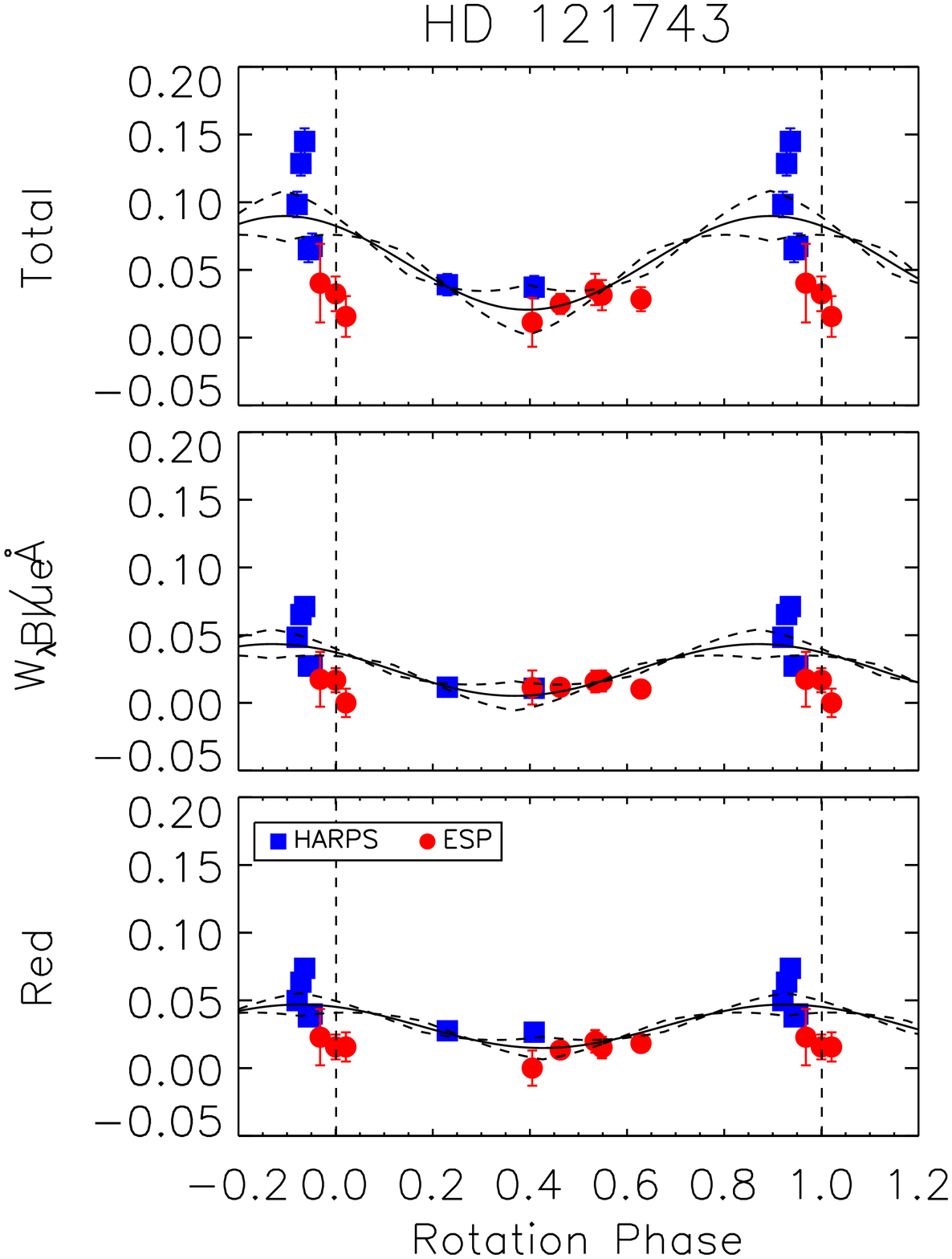} \\

   \includegraphics[trim=50 0 25 0, width=0.225\textwidth]{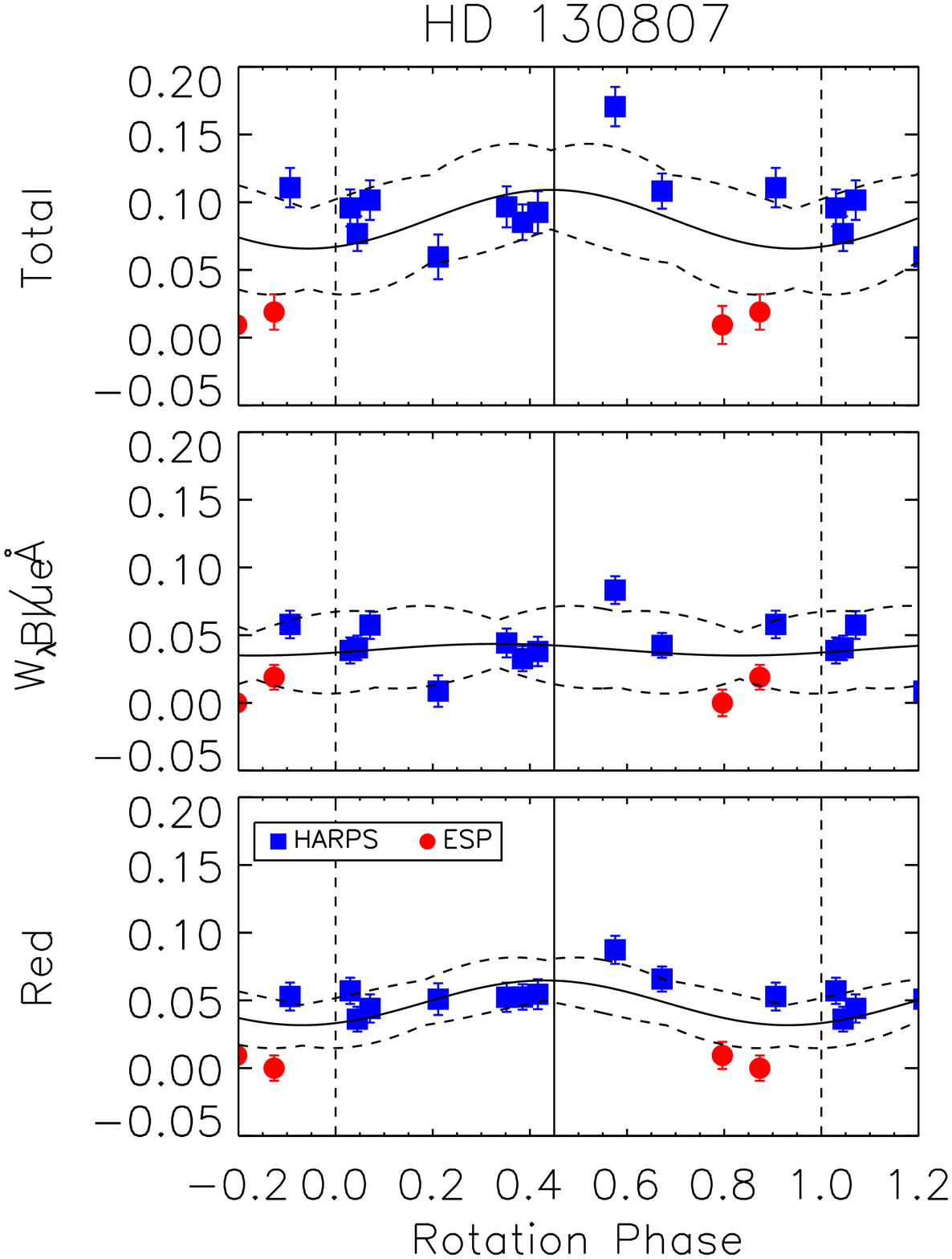} &
   \includegraphics[trim=50 0 25 0, width=0.225\textwidth]{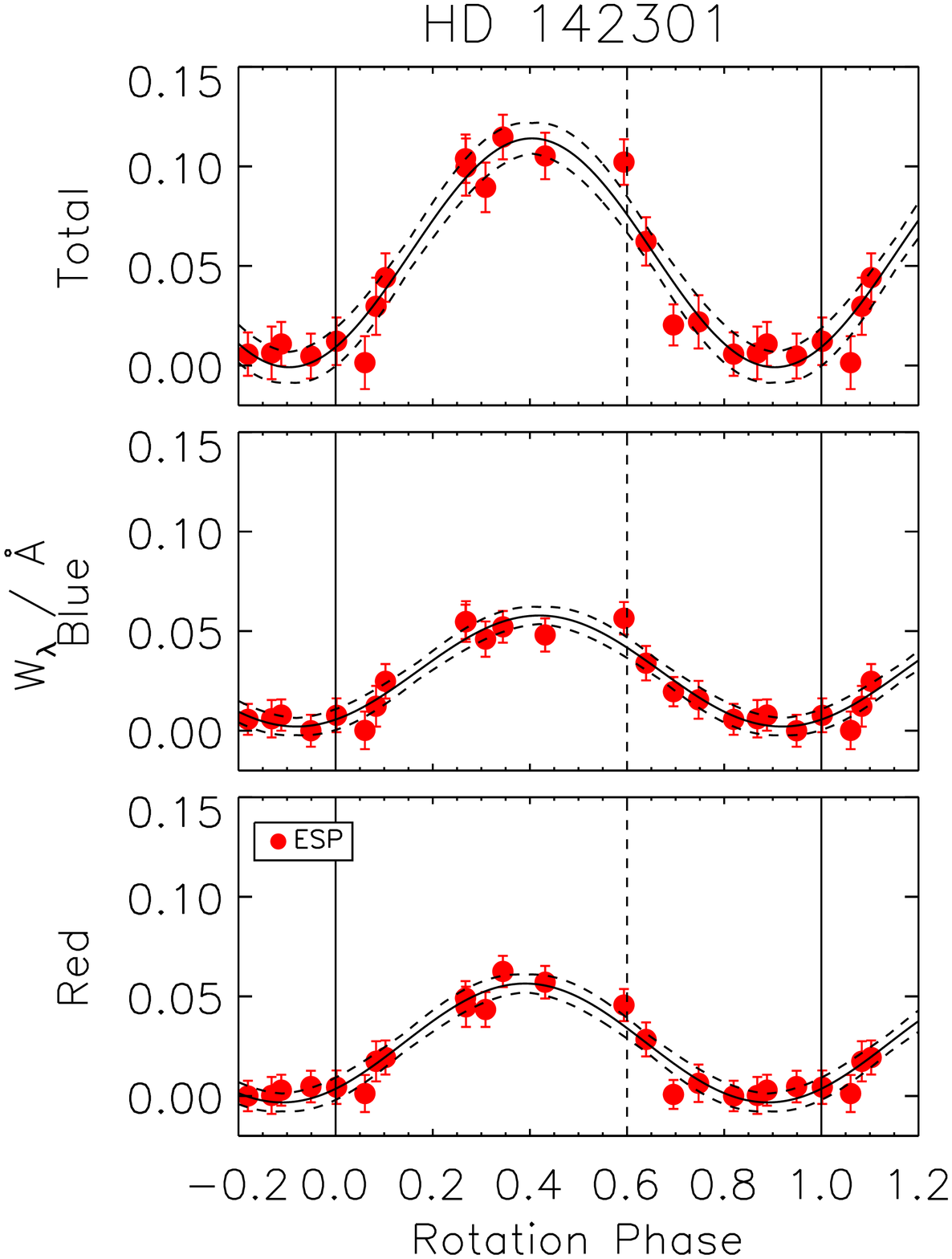} &
   \includegraphics[trim=50 0 25 0, width=0.225\textwidth]{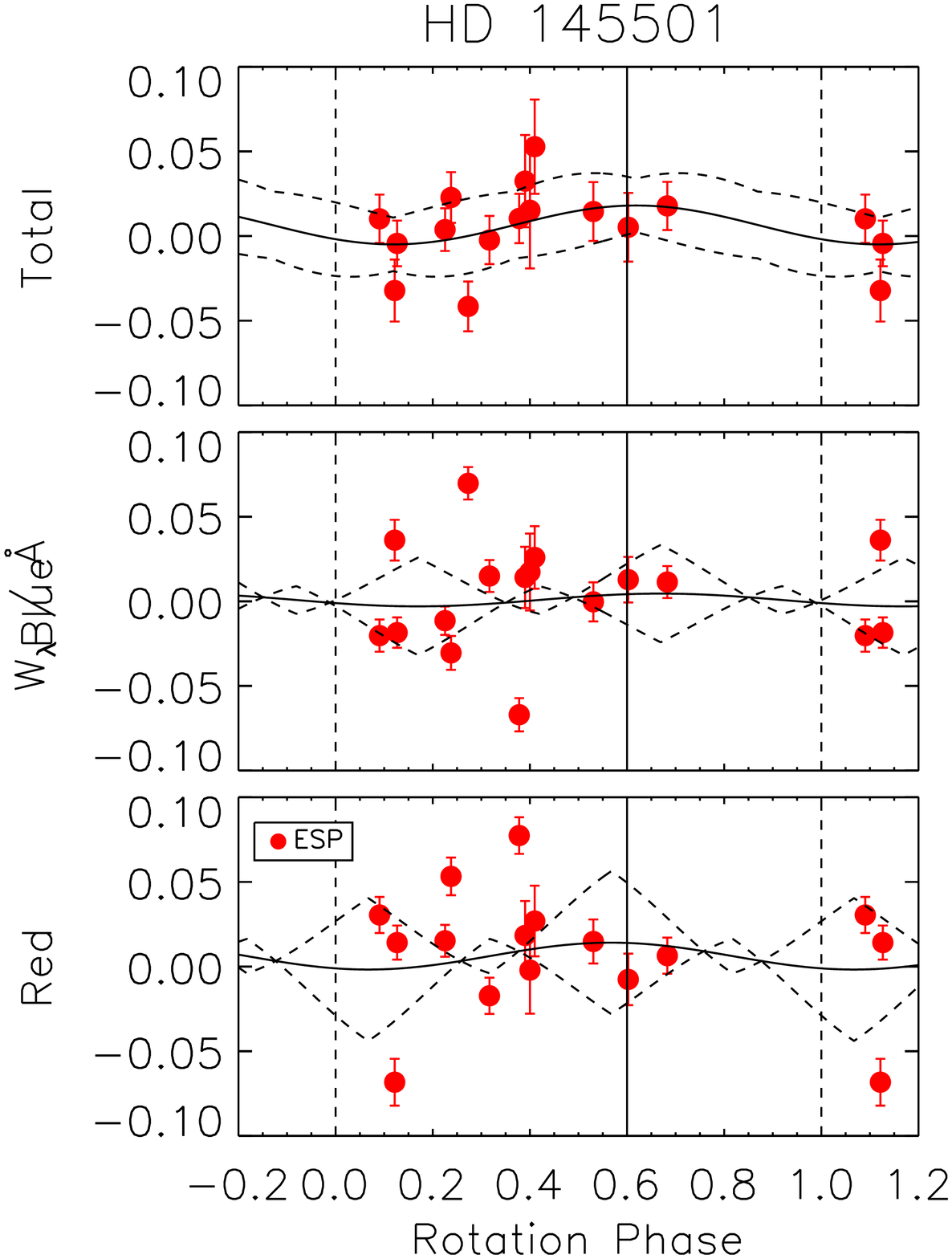} &
   \includegraphics[trim=50 0 25 0, width=0.225\textwidth]{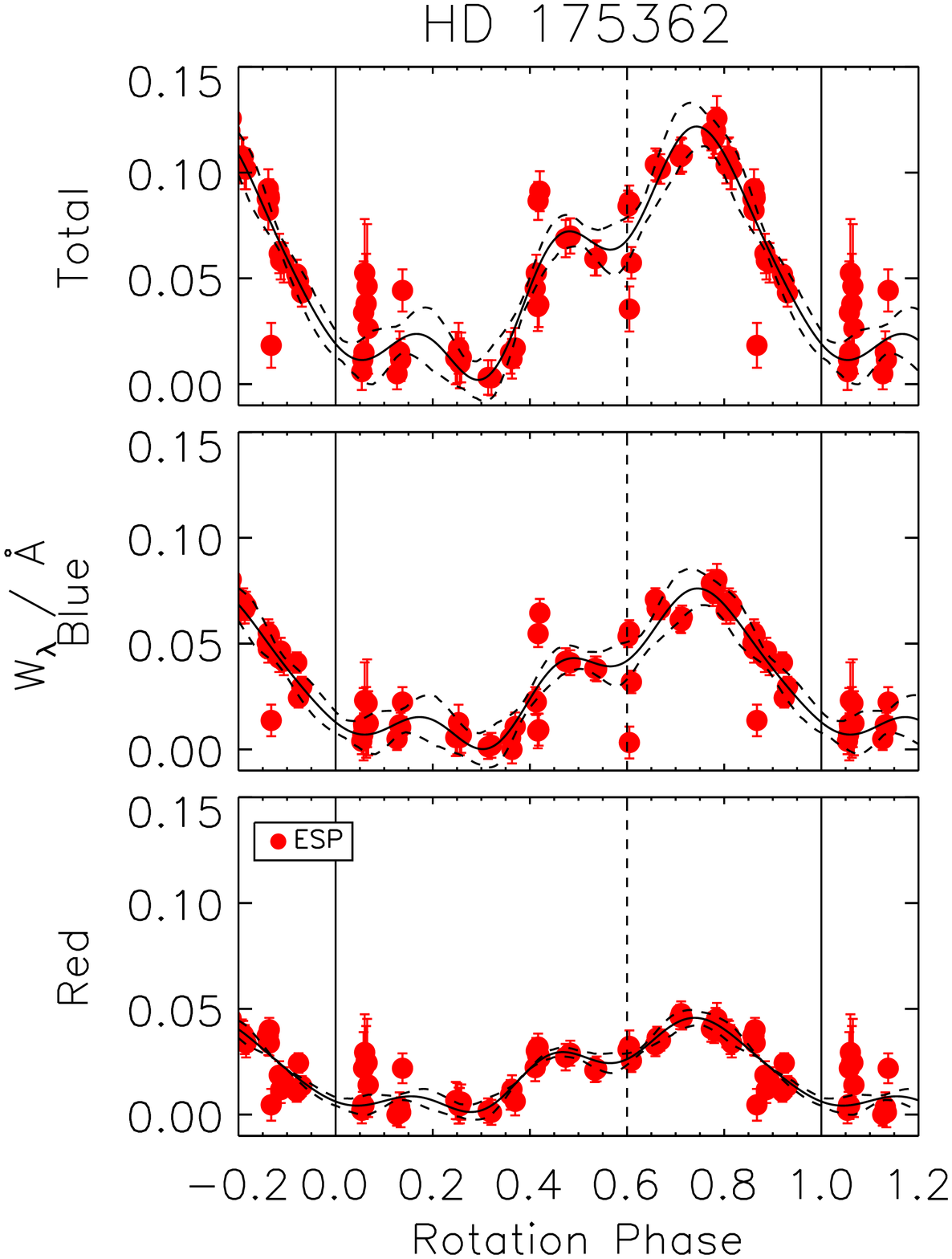} \\
\end{tabular}
      \caption[]{As Fig.\ \ref{sigOriE_halpha_minmax} for stars without emission.}
         \label{halpha_ew3}
   \end{figure*}

\noindent {\bf HD\,19832}: \cite{2010A&A...509A..28S} reported weak variability in the wings of the H Balmer lines of 56\,Ari, which they attributed to magnetic pressure due to induced electric currents in the upper atmosphere. The residual flux is almost flat (Fig.\ \ref{halpha_ind3}), however there is some evidence for coherent EW variations, with correlated variations in the red and blue halves of the line wings and an overall amplitude of about 0.02 nm (Fig.\ \ref{halpha_ew3}).

\noindent {\bf HD\,22470}: The synthetic spectrum is a poor match to both of the Narval H$\alpha$ spectra (Fig.\ \ref{halpha_ind3}). Matching the spectrum would require a RV of about 100 \kms, which is clearly ruled out by the metallic lines. Assuming there is no intrinsic problem with the spectra, it may be that this discrepancy is due to the light of a binary companion. Since a clean measurement of the H$\alpha$ EW cannot be obtained, the upper limit was taken as $3\times$ the uncertainty.

\noindent {\bf HD\,35298}: There is some weak variability in the wings of H$\alpha$ (Fig.\ \ref{halpha_ind3}), as well as coherent variations in the EWs (Fig.\ \ref{halpha_ew3}). The variations in the red and blue halves of the line wings show a similar double-wave variation. Since the minima of the pseudo-emission correspond to the \bz~extrema, this variation is probably not circumstellar but rather related to chemical spots \citep[e.g.]{2015MNRAS.447.1418Y} or the pressure changes due to the Lorentz force \citep[e.g.]{shulyak2007,2010A&A...509A..28S}.

\noindent {\bf HD\,36526}: The residual flux is almost flat (Fig.\ \ref{halpha_ind3}), but there is some evidence for a coherent variation in the EWs (Fig.\ \ref{halpha_ew3}). This variation is correlated in the red and blue wings, and the minima correspond to \bz~maxima; it is therefore unlikely to be circumstellar in origin.  

\noindent {\bf HD\,43317}: There is no convincing evidence of emission in the residual flux (Fig.\ \ref{halpha_ind2}). The EWs can be fit with a second-harmonic curve (Fig.\ \ref{halpha_ew3}), albeit not convincingly. Since this star is a Slowly Pulsating B-type star with multiple excited modes \citep{2012A&A...542A..55P,2019A&A...627A..64P}, it is probable that the EW variations are scatter due to pulsation and are unrelated to rotation.

\noindent {\bf HD\,45583}: There are features in the residual flux line wings that might possibly be weak emission (Fig.\ \ref{halpha_ind3}); however, it is more likely that these are due to spectral lines not included in the synthetic spectrum. The EWs (Fig.\ \ref{halpha_ew3}) show no convincing evidence for coherent variations.

\noindent {\bf HD\,55522}: The residual flux line wings are apparently flat (Fig.\ \ref{halpha_ind3}), however there is an apparently coherent EW variation (Fig.\ \ref{halpha_ew3}), correlated between the red and blue wings, with extrema corresponding to the \bz~extrema. 

\noindent {\bf HD\,61556}: A close examination of H Balmer line variations was conducted by \cite{2015MNRAS.449.3945S}, who found that the Balmer line wings exhibited weak variability, correlated between the red and blue wings, with similar amplitudes between the different Balmer lines. They concluded that these variations are probably related to pressure changes due to e.g.\ chemical spots or the Lorentz force. The star's H$\alpha$ line is shown in Fig.\ \ref{halpha_ind3}. The EW is variable, but the scatter in the measurements is apparently larger than any coherent variation (Fig.\ \ref{halpha_ew3}).

\noindent {\bf HD\,105382}: There is some evidence of variability in the line wings (Fig.\ \ref{halpha_ind4}), but the shape of the variations seems inconsistent with the expectations for a CM. The small size of the dataset precludes a search for coherent EW variability.

\noindent {\bf HD\,121743}: The residual flux is basically flat (Fig.\ \ref{halpha_ind4}). While the EW seems to be variable, it does not seem to be coherent with the rotation period. \cite{2006AA...452..945T} list this star as a $\beta$ Cep variable, so the small EW variations (Fig. \ref{halpha_ew3}) are probably related to pulsation.

\noindent {\bf HD\,130807}: There is detectable variability in the line wings of H$\alpha$ (Fig.\ \ref{halpha_ind4}), but no coherent variation in the EWs (Fig.\ \ref{halpha_ew3}). Since this star is both an SB2 and a Slowly Pulsating B-type star \citep{2019A&A...622A..67B}, it is probable that both EW and line profile variations are related to these factors. 

\noindent {\bf HD\,142301}: \cite{2004A&A...421..203S} found no evidence for H$\alpha$ emission in 3\,Sco. The residual flux does show some variability in the wings (Fig.\ \ref{halpha_ind4}), which might possibly be ascribed to emission. However, there is also a very clear, coherent EW variation (Fig. \ref{halpha_ew3}), which is correlated between the red and blue wings, and does not peak at either of the \bz~maxima (in fact, the EW is at a minimum at one of the \bz~maxima). The H$\alpha$ variations are therefore probably not related to circumstellar emission, but instead to chemical spots or to the Lorentz force. 

\noindent {\bf HD\,144334}: \cite{2004A&A...421..203S} found no evidence for H$\alpha$ emission. The line wing residual flux shows some evidence for variability (Fig.\ \ref{halpha_ind4}), but it is too broad to be convincingly ascribed to circumstellar emission. The small size of the dataset precludes looking for coherent EW variability. 

\noindent {\bf HD\,145501}: The residual flux is essentially flat (Fig.\ \ref{halpha_ind4}), and there is no evidence for coherent variation in the EWs (Fig.\ \ref{halpha_ew3}). 

\noindent {\bf HD\,175362}: \cite{2004A&A...421..203S} found no evidence for H$\alpha$ emission. The line wings show broad variations, correlated between the red and blue wings (Fig.\ \ref{halpha_ind4}). This is reflected in the EW curves (Fig.\ \ref{halpha_ew3}), which are similar between the red and blue line wing measurements, and show a coherent variation which is fit by a $4^{th}$ degree harmonic function. The EW curve shows minima at the \bz~maxima, and is therefore very unlikely to be due to circumstellar emission, but instead due to local pressure variations due to chemical spots or the Lorentz force. 

\end{document}